\documentclass[12pt,nohyper,letterpaper]{package}         % For preprint
\usepackage{amssymb,amsfonts}
\usepackage{epsfig}
\def\be{\begin{eqnarray}}
\def\ee{\end{eqnarray}}
\newcommand{\nn}{\nonumber}
\newcommand\para{\paragraph{}}
\newcommand{\ft}[2]{{\textstyle\frac{#1}{#2}}}
\newcommand{\eqn}[1]{(\ref{#1})}

\newcommand{\ppp}[2]{\frac{\partial {#1}}{\partial {#2}}}

\def\Dslash{\,\,{\raise.15ex\hbox{/}\mkern-13mu D}}
\def\Dbarslash{\,\,{\raise.15ex\hbox{/}\mkern-12mu {\bar D}}}
\def\delslash{\,\,{\raise.15ex\hbox{/}\mkern-10mu \partial}}
\def\delbarslash{\,\,{\raise.15ex\hbox{/}\mkern-9mu {\bar\partial}}}
\def\pslash{\,\,{\raise.15ex\hbox{/}\mkern-11mu p}}
\def\qslash{\,\,{\raise.15ex\hbox{/}\mkern-9mu q}}
\def\kslash{\,\,{\raise.15ex\hbox{/}\mkern-11mu k}}
\def\eslash{\,\,{\raise.15ex\hbox{/}\mkern-9mu \epsilon}}

\newcommand{\slsh}[1]{\,\,{\raise.15ex\hbox{/}\mkern-12mu {#1}}}

\def\ket#1{\left| #1 \right\rangle}
\def\bra#1{\left\langle #1 \right|}
\def\vev#1{\left\langle #1 \right\rangle}

\newcommand\DFP{\Delta_{FP}}
\newcommand\DFPI{\Delta^{-1}_{FP}}

\renewcommand\ap{\alpha^\prime}

\newcommand\gab{g_{\alpha\beta}}
\newcommand\gabi{g^{\alpha\beta}}
\newcommand\eab{\eta_{\alpha\beta}}
\newcommand\sg{\sqrt{-g}}

\newcommand\Tr{{\rm Tr}}
\newcommand\tab{T_{\alpha\beta}}
\newcommand\pa{\partial_\alpha}
\newcommand\pb{\partial_\beta}
\newcommand\sfap{\sqrt{\frac{\ap}{2}}}
\newcommand\an{\alpha^\mu_n}
\newcommand\tilan{\tilde{\alpha}^\mu_n}
\newcommand\vac{|0\rangle}
\newcommand\bz{{\bar{z}}}
\newcommand\p{\partial}
\def\pb{\bar{\partial}}
\newcommand\bp{\bar{\partial}}
\newcommand\bt{\bar{T}}
\newcommand\is{\int d^2\sigma\,}
\newcommand\bw{{\bar{w}}}
\newcommand\Res{{\rm Res}\,}
\newcommand\e{\epsilon}
\newcommand\calo{{\cal O}}
\newcommand\amp{{\cal A}^{(4)}}
\newcommand\bmn{\beta_{\mu\nu}}

\title{{\Huge String Theory} \\
{\large University of Cambridge Part III Mathematical Tripos}}

\author{David Tong\\
Department of Applied Mathematics and Theoretical Physics, \\
Centre for Mathematical Sciences, \\
Wilberforce Road, \\
Cambridge, CB3 OWA, UK \\ {}\\
{\rm http://www.damtp.cam.ac.uk/user/tong/string.html}\\
\email{d.tong@damtp.cam.ac.uk}
\\{}\\{}\\{}\\{}\\{}\\{}\\{}\\{}\\{}\\{}\\{}\\{}\\{}\\{}
\\{}\\{}\\{}\\{}\\{}\\{}
}

\preprint{January 2009}

\abstract{

\begin{itemize}
\item J. Polchinski, {\it String Theory}
\end{itemize}
This two volume work is the standard introduction to the subject.
Our lectures will more or less follow the path laid down in volume one covering the
bosonic string. The book contains explanations and descriptions of many details
that have been deliberately (and, I suspect, at times inadvertently) swept under a very
large rug in these lectures. Volume two covers the superstring.

\begin{itemize}
\item M. Green, J. Schwarz and E. Witten, {\it Superstring Theory}
\end{itemize}
Another two volume set. It is now over 20 years old and takes a slightly
old-fashioned route through the subject with no explicit mention of
conformal field theory. However, it does contain
much good material and the explanations are uniformly excellent. Volume
one is most relevant for these lectures.
\begin{itemize}
\item B. Zwiebach, {\it A First Course in String Theory}
\end{itemize}
This book grew out of a course given to undergraduates who had no previous exposure to general
relativity or quantum field theory. It has wonderful pedagogical discussions of the basics of lightcone
quantization. More surprisingly, it also has  clear descriptions of several
advanced topics, even though it misses out all the bits in between.

\begin{itemize}
\item P. Di Francesco, P. Mathieu and D. S\'en\'echal, {\it Conformal Field Theory}
\end{itemize}
This big yellow book is affectionately known as the yellow pages. It's a great way to
learn conformal field theory. At first glance, it comes across as
slightly daunting because it's big. (And yellow). But you soon realise that it's big because
it starts at the beginning and provides detailed explanations at every step. The material necessary for
this course can be found in chapters 5 and 6.

\para {\ }\para
Further References: ``{\it String Theory and M-Theory}" by Becker, Becker and
Schwarz and ``{\it String Theory in a Nutshell}" by Kiritsis both deal
with the bosonic string fairly quickly, but include many more advanced topics.
The book ``{\it D-Branes}" by Johnson  has lively and clear discussions
about the many joys of D-branes.
Links to several excellent online resources are listed on the course webpage:
http://www.damtp.cam.ac.uk/user/tong/string.html}

\begin{document}
%%%%%%%%%%%%%%%%%%%%%%%%%%%%%%%%%%%%%%%%%%%%%%%%%%%%%%%%%%%%%%%

\newpage
{\ }\\{}\\{}\\
%\\{}\\{}\\{}\\{}\\{}\\{}\\{}\\{}
\subsection*{\center{Acknowledgements}}
{\ }\\
These lectures are aimed at beginning graduate students. They assume a working
knowledge of quantum field theory and general relativity. The lectures were
given over one semester and are based broadly on Volume one of the book by Joe Polchinski.
I inherited the course from Michael Green whose notes were extremely useful. I
also benefited enormously from the insightful and entertaining video lectures by Shiraz Minwalla.

\para
To first approximation, these lecture notes contain no references to original work. Detailed
bibliographies can be found in the text books listed at the beginning.

\para
I'm grateful to Anirban Basu, Joe Bhaseen, Diego Correa, Nick Dorey, Michael Green,
Anshuman Maharana, Malcolm Perry and Martin Schnabl for discussions and help with various aspects
of these notes. I'm also grateful to the students, especially Carlos Guedes,
for their excellent questions
and superhuman typo-spotting abilities. Finally, my thanks to Alex Considine for infinite patience and
understanding over the weeks these notes were written. I am supported by the Royal Society.

%\newpage
%
%\begin{figure}[htb]
%\begin{center}
%\epsfxsize=6in\leavevmode\epsfbox{bullshit.eps}
%\end{center}
%\caption{New Yorker cartoon by Lee Lorenz (2007).}
%%%\label{mir}
%\end{figure}

\newpage
\setcounter{page}{1} \setcounter{section}{-1}
\section{Introduction}

%\begin{quote}
%``There are no real one-particle systems in nature, not even
%few-particle systems. The existence of
%virtual pairs and of pair fluctuations shows that the days of
%fixed particle numbers are over."
%\end{quote}
%\vspace{-0.3in}\hfill{\it Viki Weisskopf}

String theory is an ambitious project. It purports to be an all-encompassing theory of the universe,
unifying the forces of Nature, including gravity, in a single quantum mechanical framework.

\para
The premise of string theory is that, at the fundamental level, matter does not
consist of point-particles but rather of tiny loops of string. From this slightly absurd beginning,
the laws of physics emerge. General relativity, electromagnetism and Yang-Mills gauge theories
all appear in a surprising fashion. However, they come with baggage. String theory
gives rise to a host of other ingredients, most strikingly extra spatial dimensions of the
universe beyond the three that we have observed.
The purpose of this course is to understand these statements in detail.

\para
These lectures differ from most other courses that you will take in a physics degree.
String theory is speculative science. There is no experimental evidence that string theory is the
correct description of our world and scant hope that hard evidence will arise in the near future.
Moreover, string theory is very much a work in progress and certain aspects of the theory are far from
understood. Unresolved issues abound and it seems likely that the final formulation has yet
to be written. For these reasons, I'll begin this introduction by suggesting some answers
to the question: Why study string theory?

\subsubsection*{Reason 1. String theory is a theory of quantum gravity}

String theory unifies Einstein's theory of general relativity with quantum mechanics. Moreover,
it does so in a manner that retains the explicit connection with both quantum theory and the low-energy
description of spacetime.

\para
But quantum gravity contains many puzzles, both technical and conceptual. What does spacetime
look like at the shortest distance scales? How can we understand physics if the causal
structure fluctuates quantum mechanically? Is the big bang truely the beginning of time? Do
singularities that arise in black holes really signify the end of time? What is the
microscopic origin of black hole entropy and what is it telling us? What is the resolution
to the information paradox? Some of these issues will be reviewed later in this introduction.

\para
Whether or not string theory is the true description of reality, it offers a framework in which
one can begin to explore these issues. For some questions, string theory has
given very impressive and compelling answers. For others, string theory has been almost silent.

\subsubsection*{Reason 2. String theory may be {\it the} theory of quantum gravity}

With broad brush, string theory looks like an extremely good candidate to describe the real world.
At low-energies it naturally gives rise to general relativity, gauge theories, scalar fields and chiral fermions. In other words, it contains all the ingredients that make up our universe. It also gives the only presently credible explanation for the value of the cosmological constant although, in
fairness, I should add that the explanation is so distasteful to some that the community is rather amusingly split between whether this is a good thing or a bad thing. Moreover, string theory
incorporates several ideas which do not yet have experimental evidence but which are considered to
be likely candidates for physics beyond the standard model. Prime examples are supersymmetry and axions.

\para
However, while the broad brush picture looks good, the finer details have yet to be painted.
String theory does not provide unique predictions for low-energy physics but instead offers a bewildering array of possibilities, mostly dependent on what is hidden in those extra dimensions. Partly, this problem is inherent to any theory of quantum gravity: as we'll review shortly, it's a long way down from the Planck scale to the domestic energy scales explored at the LHC. Using quantum gravity to extract predictions for particle physics is akin to using QCD to extract predictions for how coffee makers
work. But the mere fact that it's hard is little comfort if we're looking for convincing evidence that string theory describes the world in which we live.

\para
While string theory cannot at present offer falsifiable predictions, it has nonetheless
inspired new and imaginative proposals for solving
outstanding problems in particle physics and cosmology. There are
scenarios in which string theory might reveal itself in forthcoming experiments. Perhaps we'll
find extra dimensions at the LHC, perhaps we'll see a network of fundamental strings stretched
across the sky, or perhaps we'll detect some feature of non-Gaussianity in the CMB that is
characteristic of D-branes at work during inflation. My personal feeling however is that
each of these is a long shot and we may not know whether string theory is right or wrong
within our lifetimes. Of course, the history of physics is littered with
naysayers, wrongly suggesting that various theories will never be testable. With luck,
I'll be one of them.

%(Personally, I'm quite fond of the latter).

%\para
%Before moving on, let me finally mention the close relationship between string theory and supersymmetry.
%To the best of our knowledge, string theory requires supersymmetry at the Planck scale. It does not
%necessarily require supersymmetry at the TeV scale, where it would play a role in the solution of the
%hierarchy problem. Indeed, the existence of supersymmetry at the LHC and string theory are logically
%independent. Nonetheless, if supersymmetry was discovered at the LHC, it would presumably mean that
%supergravity is the correct description of gravity and I would personally find it a
%very encouraging sign that string theory is on the right track.

\subsubsection*{Reason 3. String theory provides new perspectives on gauge theories}

String theory was born from attempts to understand the strong force. Almost forty
years later, this remains one of the prime motivations for the subject. String theory provides
tools with which to analyze down-to-earth aspects of quantum field theory
that are far removed from high-falutin' ideas about gravity and black holes.

\para
Of immediate relevance to this course are the pedagogical reasons to invest time in
string theory. At heart, it is the study of conformal field theory and gauge symmetry.
The techniques that we'll learn are not isolated to string theory, but apply to countless
systems which have direct application to real world physics.

%the techniques of string theory can be applied to understand
%quantum field theories in new and interesting ways. For example, as we'll see in this
%course, the computation of string scattering amplitudes is equivalent to a computation of
%an infinite number of Feynman diagrams. In certain contexts, the simplest way to compute
%Feynamn diagrams of interest in field theory is actually to perform the computation in
%string theory and subsequently take a limit. There is now an impressive industry applying
%string techniques to compute Feynman diagrams in field theory.

\para
On a deeper level, string theory provides new and very surprising methods to understand
aspects of quantum gauge theories. Of these, the most startling is the
{\it AdS/CFT correspondence}, first conjectured by Juan Maldacena, which gives a relationship
between strongly coupled quantum field theories and  gravity in higher dimensions.
These ideas have been applied in areas ranging from nuclear physics to condensed matter physics,
and have provided qualitative (and arguably quantitative) insights into strongly coupled phenomena.

\subsubsection*{Reason 4. String theory provides new results in mathematics}

For the past 250 years, the close relationship between mathematics and physics has been almost
a one-way street: physicists borrowed many things from mathematicians but, with a few noticeable exceptions, gave little back. In recent times, that has changed. Ideas and techniques from
string theory and quantum field theory have been employed to give new ``proofs" and, perhaps
more importantly, suggest new directions and insights in mathematics. The most
well known of these is {\it mirror symmetry}, a relationship between topologically
different Calabi-Yau manifolds.

\para
The four reasons described above also crudely characterize the string theory community: there
are ``relativists" and ``phenomenologists" and ``field theorists" and ``mathematicians". Of
course, the lines between these different sub-disciplines are not fixed and one of the great
attractions of string theory is its ability to bring together people working in different
areas --- from cosmology to condensed matter to pure mathematics --- and provide a
framework in which they can profitably communicate. In my opinion, it is this cross-fertilization
between fields which is the greatest strength of string theory.

\subsection{Quantum Gravity}

This is a starter course in string theory. Our focus will be on the perturbative approach
to the bosonic string and, in particular, why this gives a consistent theory of quantum
gravity. Before we leap into this, it is probably best to say a few words about
quantum gravity itself. Like why it's hard. And why it's important. (And why it's not).

\para
The Einstein Hilbert action is given by
\be
S_{EH} = \frac{1}{16\pi G_N} \int d^4x \sqrt{-g} {\cal R}\nn\ee
Newton's constant $G_N$ can be written as
\be 8\pi G_N = \frac{\hbar c}{M_{pl}^2} \nn\ee
Throughout these lectures we work in units with $\hbar=c=1$. The Planck mass $M_{pl}$  defines
an energy scale
\be M_{pl} \approx 2 \times 10^{18} \  \mbox{GeV}\ . \nn\ee
(This is sometimes referred to as the reduced Planck mass, to distinguish it from the scale without
the factor of $8\pi$, namely $\sqrt{1/G_N} \approx 1 \times 10^{19}$ GeV).

\para
There are a couple of simple lessons that we can already take from this. The first is that the
relevant coupling in the quantum theory is $1/M_{pl}$.
To see that this is indeed the case from the perspective of the
action, we consider small perturbations around flat Minkowski space,
\be g_{\mu\nu}=\eta_{\mu\nu} + \frac{1}{M_{pl}} h_{\mu\nu} \nn\ee
The factor of $1/M_{pl}$ is there to ensure that when we expand out the Einstein-Hilbert
action, the kinetic term for $h$ is canonically normalized, meaning
that it comes with no powers of $M_{pl}$. This then gives the kind of theory that
you met in your first course on quantum field theory, albeit with an
infinite series of interaction terms,
\be S_{EH} = \int d^4x \ (\partial h)^2  + \frac{1}{M_{pl}}\,h\, (\partial h)^2 + \frac{1}{M_{pl}^2}\,h^2\, (\partial h)^2 + \ldots \nn\ee
Each of these terms is schematic: if you were to do this explicitly, you would find
a mess of indices contracted in different ways. We see that the interactions are
suppressed by powers of $M_{pl}$. This means that quantum perturbation theory is an expansion in the
dimensionless ratio  $E^2/M_{pl}^2$, where $E$ is the energy associated to the process of interest.
We learn that gravity is weak, and therefore under control, at low-energies.
But gravitational interactions become strong as the energy involved approaches the Planck scale.
In the language of the renormalization group, couplings of this type are known as {\it irrelevant}.

\para
The second lesson to take away is that the Planck scale $M_{pl}$ is very very large. The LHC will probe the electroweak scale, $M_{EW} \sim 10^3$ GeV. The ratio is $M_{EW}/M_{pl} \sim 10^{-15}$. For this reason, quantum gravity will not affect your daily life, even
if your daily life involves the study of the most extreme observable conditions in the
universe.

\subsubsection*{Gravity is Non-Renormalizable}

Quantum field theories with irrelevant couplings are typically ill-behaved at high-energies,
rendering the theory ill-defined. Gravity is no exception. Theories of this
type are called {\it non-renormalizable}, which means that the divergences that appear in
the Feynman diagram expansion cannot be absorbed by a finite number of counterterms.
In pure Einstein gravity, the symmetries of the theory are
enough to ensure that the one-loop S-matrix is finite. The first divergence occurs at two-loops
and requires the introduction of a counterterm of the form,
\be \Gamma  \sim \frac{1}{\epsilon} \frac{1}{M_{pl}^4}\int d^4x \sqrt{-g}
\,{\cal R}^{\mu\nu}_{\ \ \rho\sigma}
{\cal R}^{\rho\sigma}_{\ \ \lambda\kappa}{\cal R}^{\lambda\kappa}_{\ \ \mu\nu} \nn\ee
with $\epsilon=4-D$. All indications point towards the fact that this is the first in
an infinite number of necessary counterterms.

\para
Coupling gravity to matter requires an interaction term of the form,
\be S_{int} = \int d^4x\ \frac{1}{M_{pl}}\,h_{\mu\nu} T^{\mu\nu} + {\cal O}(h^2)\nn\ee
\EPSFIGURE{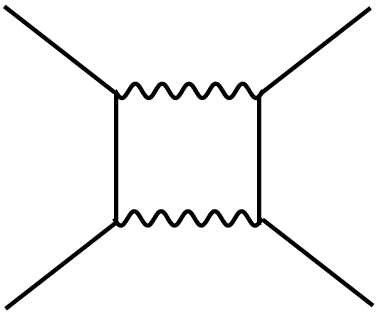,height=73pt}{}
\noindent This makes the situation marginally worse, with the first divergence
now appearing at one-loop. The Feynman diagram in the figure shows particle scattering
through the exchange of two gravitons. When the momentum $k$ running in the loop
is large, the diagram is badly divergent: it scales as
\be \frac{1}{M_{pl}^4} \, \int^\infty d^4k \nn\ee
Non-renormalizable theories are commonplace in the history of physics, the most commonly
cited example being Fermi's theory of the weak interaction. The first thing to say about them
is that they are far from useless! Non-renormalizable theories are typically viewed as
{\it effective} field theories, valid only up to some energy scale $\Lambda$. One deals with the divergences by simply
admitting ignorance beyond this scale and treating $\Lambda$ as a UV cut-off on any momentum
integral. In this way, we get results which are valid to an accuracy of $E/\Lambda$ (perhaps
raised to some power).
In the case of the weak interaction, Fermi's theory accurately predicts physics up to
an energy scale of $\sqrt{1/G_F} \sim 100$ GeV. In the case of quantum gravity, Einstein's theory
works to an accuracy of $(E/M_{pl})^2$.

\para
However, non-renormalizable theories are typically unable to describe physics at their
cut-off scale $\Lambda$ or beyond. This is because they are missing the true ultra-violet degrees of freedom which tame the high-energy behaviour. In the case of the weak force, these new
degrees of freedom are the W and Z bosons. We would like to know what missing
degrees of freedom are needed to complete gravity.

\subsubsection*{Singularities}

Only a particle physicist would  phrase all questions about the universe in
terms of scattering amplitudes. In general relativity we typically think about
the geometry as a whole, rather than bastardizing the Einstein-Hilbert action and discussing
perturbations around flat space. In this language, the question
of high-energy physics turns into one of short distance physics. Classical general relativity is
not to be trusted in regions where the curvature of spacetime approaches the Planck scale and
ultimately becomes singular. A quantum theory of gravity should resolve these singularities.

\para
The question of spacetime singularities is morally equivalent to that of high-energy
scattering. Both probe the ultra-violet nature of gravity. A spacetime
geometry is made of a coherent collection of gravitons, just as the electric and magnetic
fields in a laser are made from a collection of photons. The short distance structure
of spacetime is governed -- after Fourier transform -- by high momentum gravitons. Understanding
spacetime singularities and high-energy scattering are different sides of the same coin.

\para
There are two situations in general relativity where singularity theorems tell us
that the curvature of spacetime gets large: at the big bang and in the center of a
black hole. These provide two of the biggest challenges to any putative theory of quantum
gravity.

\subsubsection*{Gravity is Subtle}

It is often said that general relativity contains the seeds of its own
destruction. The theory is unable to predict physics at the Planck scale and
freely admits to it. Problems such as non-renormalizability and singularities
are, in a Rumsfeldian sense, known unknowns. However, the full story is more
complicated and subtle. On the one hand, the issue of non-renormalizability may
not quite be the crisis that it first appears. On the other hand, some aspects of
quantum gravity suggest that general relativity isn't as honest about its own failings
as is usually advertised. The theory hosts a number of unknown unknowns, things that we didn't
even know that we didn't know. We won't have a whole lot to say about these issues in this
course, but you should be aware of them. Here I mention only a few salient points.

\para
Firstly, there is a key difference between Fermi's theory of the weak interaction and
gravity. Fermi's theory was unable to provide predictions for any scattering process
at energies above $\sqrt{1/G_F}$. In contrast, if we scatter two objects
at extremely high-energies in gravity --- say, at energies $E\gg M_{pl}$ ---
then we know exactly what will happen: we form a big black hole. We don't need quantum
gravity to tell us this. Classical general relativity is sufficient. If we restrict attention
to scattering, the crisis of non-renormalizability is not problematic at ultra-high
energies. It's troublesome only within a window of energies around the Planck scale.

\para
Similar caveats hold for singularities. If you are foolish enough to jump into a black
hole, then you're on your own: without a theory of quantum gravity, no one can
tell you what
fate lies in store at the singularity.
Yet, if you are smart and stay outside of the black hole, you'll be hard pushed
to see any effects of quantum gravity. This is because Nature has conspired to
hide
Planck scale curvatures from our inquisitive eyes. In the case of black holes this is
achieved through cosmic censorship which is a conjecture in classical general relativity
that says singularities are hidden behind horizons. In the case of the big bang,
it is achieved through inflation, washing away any traces from the very early universe.
Nature appears to shield us from the effects of quantum gravity, whether in high-energy
scattering or in singularities. I think it's fair to say that no one knows if this conspiracy is
pointing at something deep, or is merely inconvenient for scientists trying to probe the
Planck scale.

\para
While horizons may protect us from the worst excesses of singularities, they come with
problems of their own. These are the unknown unknowns: difficulties that arise when
curvatures are small and general relativity says ``trust me".
The entropy of black holes and the associated paradox of information loss strongly
suggest that local quantum field theory breaks down at macroscopic distance scales.
Attempts to formulate quantum gravity in de Sitter space,
or in the presence of eternal inflation, hint at similar difficulties. Ideas of holography,
black hole complimentarity and the AdS/CFT correspondence all point towards non-local effects
and the emergence of spacetime. These are the deep puzzles of quantum gravity and their
relationship to the ultra-violet properties of gravity is unclear.

\para
As a final thought, let me mention the one observation that has an outside chance
of being related to quantum gravity: the cosmological constant. With an energy scale of
$\Lambda \sim 10^{-3}$ eV it appears to have little to
do with ultra-violet physics. If it does have its origins in a theory of quantum gravity,
it must either be due to some subtle ``unknown unknown", or because it is
explained away as an environmental quantity as in string theory.

\subsubsection*{Is the Time Ripe?}

Our current understanding of physics, embodied in the standard model, is valid up to energy
scales of $10^3$ GeV. This is 15 orders of magnitude away from the Planck scale. Why do we
think the time is now ripe to tackle quantum gravity? Surely we are like the ancient Greeks arguing
about atomism. Why on earth do we believe
that we've developed the right tools to even address the question?
\EPSFIGURE{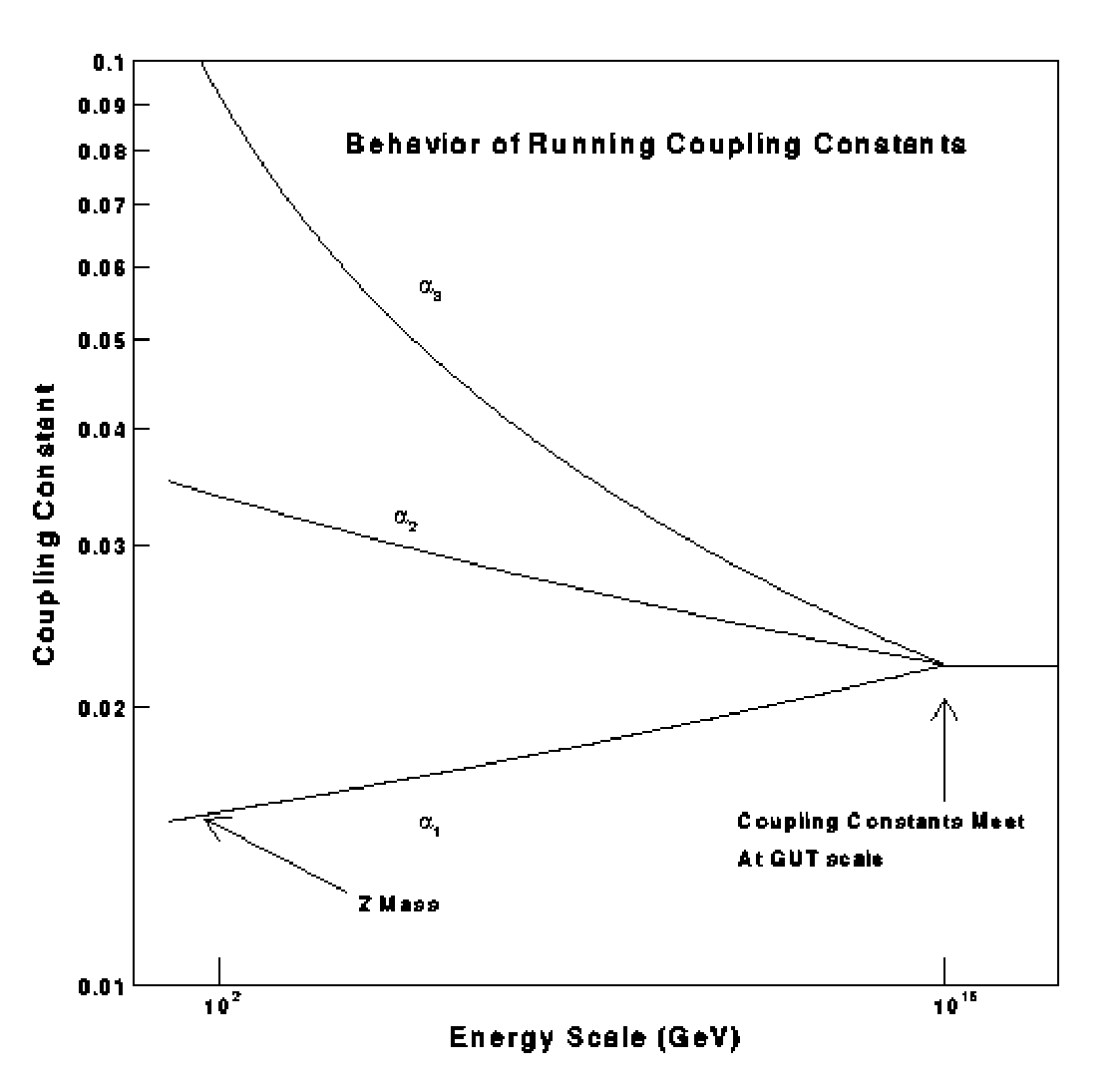,height=160pt}{}

\para
\noindent The honest answer, I think, is hubris.

\para
However, there is mild circumstantial evidence that the framework of quantum field theory
might hold all the way to the Planck scale without anything very dramatic happening in between.
The main argument is unification. The three coupling constants of Nature
run logarithmically, meeting miraculously at the GUT energy scale of $10^{15}$ GeV.
Just slightly
later, the fourth force of Nature, gravity, joins them. While not overwhelming, this
does provide a hint that perhaps quantum field theory can be taken seriously at these ridiculous
scales. Historically I suspect this was what convinced large parts
of the community that it was ok to speak about processes at $10^{18}$ GeV.

\para
Finally, perhaps the most compelling argument for studying physics at the Planck scale
is that string theory {\it does} provide a consistent unified quantum theory of gravity
and the other forces. Given that we have this theory sitting in our laps, it would be
foolish not to explore its consequences. The purpose of these lecture notes is to begin this
journey.

\newpage
\section{The Relativistic String}
\label{classical}

All lecture courses on string theory start with a discussion of the
point particle. Ours is no exception. We'll take a flying tour through the
physics of the relativistic point particle and extract a couple of
important lessons that we'll take with us as we move onto string theory.

\subsection{The Relativistic Point Particle}

We want to write down the Lagrangian describing a relativistic
particle of mass $m$. In anticipation of
string theory, we'll consider $D$-dimensional Minkowski space
${\bf R}^{1,D-1}$. Throughout
these notes, we work with signature
\be \eta_{\mu\nu} = {\rm diag}(-1,+1,+1,\ldots,+1)\nn\ee
Note that this is the opposite signature to my quantum field theory notes.

\para
If we fix a frame with coordinates $X^\mu = (t,\vec{x})$ the
action is simple:
\be S = -m \int dt\,\sqrt{1-\dot{\vec{x}}\cdot\dot{\vec{x}}} \ . \label{notsogood}\ee
To see that this is correct we can compute the momentum $\vec{p}$,
conjugate to $\vec{x}$, and the energy $E$ which is equal to the Hamiltonian,
\be \vec{p} = \frac{m\dot{\vec{x}}}{\sqrt{1-\dot{\vec{x}}\cdot\dot{\vec{x}}}}\ \ \ ,\ \ \ E=\sqrt{m^2+\vec{p}{}^2}\ ,\nn\ee
both of which should be familiar from courses on special relativity.

\para
Although the Lagrangian \eqn{notsogood} is correct, it's not fully satisfactory.
The reason is that time $t$ and space $\vec{x}$ play very different
roles in this Lagrangian. The  position $\vec{x}$ is a
dynamical degree of freedom. In contrast, time $t$ is merely a parameter
providing a label for the position. Yet Lorentz transformations are
supposed to mix up $t$ and $\vec{x}$ and such symmetries are not completely obvious
in \eqn{notsogood}. Can we find a new Lagrangian
in which time and space are on equal footing?

\para
One possibility is to treat both time and space as labels. This leads us
to the concept of field theory. However, in this course we will be more
interested in the other possibility: we will promote time to a dynamical
degree of freedom. At first glance,
this may appear odd: the number of degrees of freedom  is one of the
crudest ways we have to characterize a system. We shouldn't be able to
add more degrees of freedom at will without fundamentally changing the
system that we're talking about. Another way of saying this is that the
particle has the option to move in space, but it doesn't have the option
to move in time. It {\it has} to move in time. So we somehow need a way
to promote time to a degree of freedom without it really being a true
dynamical degree of freedom! How do we do this? The answer, as we will now show,
is gauge symmetry.
\EPSFIGURE{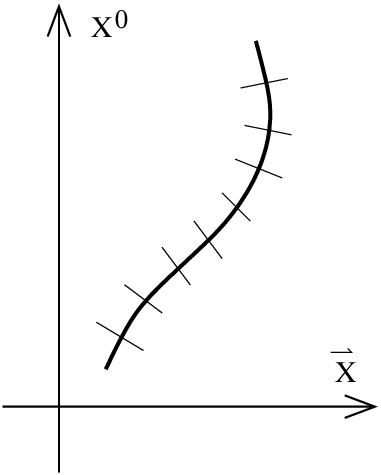,height=102pt}{}

\para
\noindent Consider the action,
\be S = -m\int d\tau\,\sqrt{-\dot{X}^\mu\dot{X}^\nu\eta_{\mu\nu}}\ ,
\label{1more}\ee
where $\mu=0,\ldots,D-1$ and $\dot{X}^\mu=dX^\mu/d\tau$.  We've introduced
a new parameter $\tau$ which labels the position along the worldline
of the particle as shown by the dashed lines in the figure.
This action has a simple
interpretation: it is just the proper time $\int ds$ along the worldline.

\para
Naively it looks as if we now have $D$ physical degrees of freedom
rather than $D-1$ because, as promised, the time direction $X^0\equiv t$
is among our dynamical variables: $X^0=X^0(\tau)$.
However, this is an illusion. To see why, we need to note that the
action \eqn{1more} has a very important property: reparameterization invariance.
This means that we can pick a different parameter $\tilde{\tau}$ on
the worldline, related to $\tau$ by any monotonic function
\be \tilde{\tau}=\tilde{\tau}(\tau)\ .\nn\ee
Let's check that the action is invariant under transformations of this type.
The integration measure in the
action changes as $d\tau = d\tilde{\tau}\,|d\tau/d\tilde{\tau}|$.
Meanwhile, the velocities change as
${dX^\mu}/{d\tau} = ({dX^\mu}/{d\tilde{\tau}})\,({d\tilde{\tau}}/{d\tau})$.
Putting this together, we see that the action can just as well be
written in the $\tilde{\tau}$ reparameterization,
\be S= -m\int d\tilde{\tau}\,\sqrt{-\frac{{dX}^\mu}{d\tilde{\tau}}\,\frac{d{X}^\nu}{d\tilde{\tau}}\eta_{\mu\nu}}\ .\nn\ee
The upshot of this is that not all $D$ degrees of freedom $X^\mu$ are
physical. For example, suppose you find a solution to this system, so
that you know how $X^0$ changes with $\tau$ and how $X^1$ changes with
$\tau$, and so on. Not all of that information is meaningful
because $\tau$ itself is not meaningful. In particular, we could use
our reparameterization invariance to simply set
\be \tau=X^0(\tau)\equiv t\ee
If we plug this choice into the action \eqn{1more} then we recover our initial action \eqn{notsogood}.
The reparameterization invariance is a {\it gauge symmetry} of the system. Like all gauge symmetries,
it's not really a symmetry at all. Rather, it is a redundancy in our description.
In the present case, it means that although we
seem to have $D$ degrees of freedom $X^\mu$, one of them is fake.

\para
The fact that one of the degrees of freedom is a fake also shows up if we look at the momenta,
\be p_\mu =\frac{\partial L}{\partial \dot{X}^\mu} = \frac{m\dot{X}^\nu\eta_{\mu\nu}}{\sqrt{-\dot{X}^\lambda\dot{X}^\rho\,\eta_{\lambda\rho}}}\ee
These momenta aren't all independent. They satisfy
\be p_\mu p^\mu +m^2 = 0 \label{pcons}\ee
This a constraint on the system. It is, of course, the mass-shell constraint for a relativistic particle of mass $m$. From the worldline perspective, it tells us that the particle isn't allowed to sit still in Minkowski space: at the very least, it had better keep moving in a timelike direction with $(p^0)^2 \geq m^2$.

\para
One advantage of the action \eqn{1more} is that the Poincar\'e symmetry of the particle is now
manifest, appearing as a global symmetry on the worldline
\be X^\mu \rightarrow \Lambda^\mu_{\ \nu}X^\nu + c^\mu \label{poincare}\ee
where $\Lambda$ is a Lorentz transformation satisfying $\Lambda^\mu_{\ \nu}\eta^{\nu\rho}\Lambda^\sigma_{\ \rho}=\eta^{\mu\sigma}$, while $c^\mu$ corresponds to a constant translation. We have made all the symmetries manifest at the price of introducing a gauge symmetry into our system. A similar gauge symmetry will arise in the relativistic string, and much of this course will be devoted to understanding its consequences.

\subsubsection{Quantization}

It's a trivial matter to quantize this action. We introduce a wavefunction $\Psi(X)$. This satisfies the usual Schr\"odinger equation,
\be i\,\frac{\partial\Psi}{\p\tau}=H\Psi\ .\nn\ee
But, computing the Hamiltonian $H=\dot{X}^\mu p_\mu -L$, we find that it vanishes: $H=0$. This shouldn't be surprising. It is simply telling us that the wavefunction doesn't depend on $\tau$. Since the wavefunction is something physical while, as we have seen, $\tau$ is not, this is to be expected. Note that this doesn't mean that time has dropped out of the problem. On the contrary, in this relativistic context, time $X^0$ is an operator, just like the spatial coordinates $\vec{x}$. This means that the wavefunction $\Psi$ is immediately a function of space and time. It is not like a static state in quantum mechanics, but more akin to the fully integrated solution to the non-relativistic Schr\"odinger equation.

\para
The classical system has a constraint given by \eqn{pcons}. In the quantum theory, we impose this constraint as an operator equation on the wavefunction, namely $(p^\mu p_\mu+ m^2)\Psi=0$. Using the usual representation of the momentum operator $p_\mu = -i\partial/\partial X^\mu$, we recognize this constraint as the Klein-Gordon equation
\be \left(-\frac{\partial}{\partial X^\mu}\,\frac{\partial}{\partial X^\nu}\,\eta^{\mu\nu}+m^2\right)\,\Psi(X)=0
\ee
Although this equation is familiar from field theory, it's important to realize that the interpretation is somewhat different. In relativistic field theory, the Klein-Gordon equation is the equation of motion obeyed by a scalar field. In relativistic quantum mechanics, it is the equation obeyed by the  wavefunction. In the early days of field theory, the fact that these two equations are the same led people to think one should view the wavefunction as a classical field and quantize it a second time. This isn't correct, but nonetheless the language has stuck and it is common to talk about the point particle perspective as ``first quantization" and the field theory perspective as ``second quantization".
\EPSFIGURE{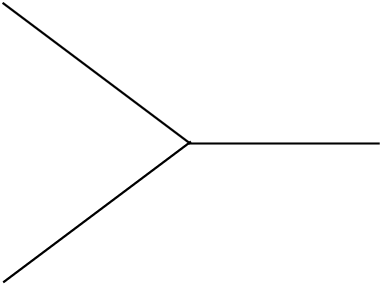,height=60pt}{}

\para
\noindent So far we've considered only a free point particle. How can we introduce
interactions into this framework? We would have to first decide which interactions are
allowed: perhaps the particle can split into two; perhaps it can fuse with other particles? Obviously, there is a huge range of options for us to choose from. We would then assign amplitudes for these processes to happen. There would be certain restrictions  coming from the requirement of unitarity which, among other things, would lead to the necessity of anti-particles. We could
draw diagrams associated to the different interactions --- an example is given in the figure ---  and in this manner we would slowly build up the Feynman diagram expansion that is familiar from field theory.
In fact, this was pretty much the way Feynman himself approached the topic of QED.
However, in practice we rarely construct particle interactions in this way because the field theory framework provides a much better way of looking at things. In contrast, this way of building up interactions is exactly what we will later do for strings.

\subsubsection{Ein Einbein}

There is another action that describes the relativistic point particle. We introduce yet another
field on the worldline, $e(\tau)$, and write
\be S=\frac{1}{2} \int d\tau\ \left(e^{-1}\dot{X}^2 - em^2\right)\ ,\label{1e}\ee
where we've used the notation $\dot{X}^2 = \dot{X}^\mu\dot{X}^\nu\eta_{\mu\nu}$. For the rest of these lectures, terms like $X^2$ will always mean an implicit contraction with the spacetime Minkowski metric.

\para
This form of the action makes it look as if we have coupled the worldline theory to 1d gravity, with the field $e(\tau)$ acting as an einbein (in the sense of vierbeins that are introduced in general relativity). To see this, note that we could change notation and write this action in the more suggestive form
\be S=-\frac{1}{2}\int d\tau \,\sqrt{-g_{\tau\tau}}\left(g^{\tau\tau}\dot{X}^2+m^2\right)\ .\ee
where $g_{\tau\tau}=(g^{\tau\tau})^{-1}$ is the metric on the worldline and $e=\sqrt{-g_{\tau\tau}}$

\para
Although our action appears to have one more degree of freedom, $e$, it can be easily checked that it has the same equations of motion as \eqn{1more}. The reason for this is that $e$ is completely fixed by its equation of motion, $\dot{X}^2+e^2m^2=0$. Substituting this into the action \eqn{1e}  recovers   \eqn{1more}

\para
The action \eqn{1e} has a couple of advantages over \eqn{1more}. Firstly, it works for massless particles with $m=0$. Secondly, the absence of the annoying square root means that it's easier to quantize in a path integral framework.

\para
The action \eqn{1e} retains invariance under reparameterizations which are now written in a form
that looks more like general relativity. For transformations parameterized by an infinitesimal $\eta$,
we have
\be \tau \rightarrow \tilde{\tau} = \tau -\eta(\tau)\ \ \ , \ \ \ \delta e &=& \frac{d}{d\tau}(\eta(\tau)e)\ \ \ ,\ \ \
\delta X^\mu = \frac{dX^\mu}{d\tau}\eta(\tau)
\ee
The einbein $e$ transforms as a density on the worldline, while each of the coordinates $X^\mu$ transforms as a worldline scalar.

\newpage
\subsection{The Nambu-Goto Action}
\EPSFIGURE{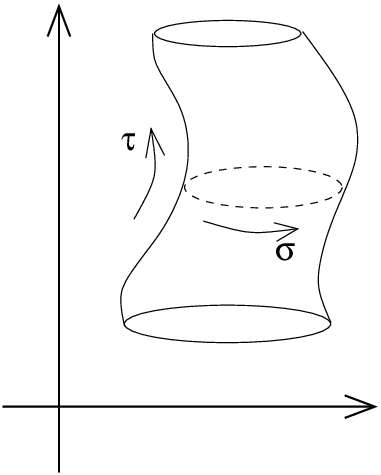,height=120pt}{}

\noindent
A particle sweeps out a worldline in Minkowski space. A string sweeps out a {\it worldsheet}. We'll parameterize this worldsheet by one timelike coordinate $\tau$, and one spacelike coordinate $\sigma$. In this section we'll focus on closed strings and take $\sigma$ to be periodic, with range
\be \sigma \in [0,2\pi)\ .\label{speriod}\ee
We will sometimes package the two worldsheet coordinates together as $\sigma^\alpha=(\tau,\sigma)$, $\alpha=0,1$. Then the string sweeps out a surface in spacetime which defines a map from the worldsheet to Minkowski space, $X^\mu(\sigma,\tau)$ with $\mu=0,\ldots,D-1$. For closed strings, we require
\be X^\mu(\sigma,\tau)=X^\mu(\sigma+2\pi,\tau)\ .\nn\ee
In this context, spacetime is sometimes referred to as the {\it target space} to distinguish it from the worldsheet.

\para
We need an action that describes the dynamics of this string. The key property that we will ask for is that nothing depends on the coordinates $\sigma^\alpha$ that we choose on the worldsheet. In other words, the string action should be reparameterization invariant. What kind of action does the trick? Well, for the point particle the action was proportional to the length of the worldline. The obvious generalization is that the action for the string should be proportional to the area, $A$, of the worldsheet. This is certainly a property that is characteristic of the worldsheet itself, rather than any choice of parameterization.

\para
How do we find the area $A$ in terms of the coordinates  $X^\mu(\sigma,\tau)$? The worldsheet is a curved surface embedded in spacetime. The induced metric, $\gamma_{\alpha\beta}$, on this surface is the pull-back of the flat metric on Minkowski space,
\be \gamma_{\alpha\beta} = \frac{\partial X^\mu}{\partial \sigma^\alpha}\,\frac{\partial X^\nu}{\partial \sigma^\beta} \eta_{\mu\nu}\ .\label{gametric}\ee
Then the action which is proportional to the area of the worldsheet is given by,
\be S=-T\int d^2\sigma\ \sqrt{-\det\,\gamma}\ .\label{ng}\ee
Here $T$ is a constant of proportionality. We will see shortly that it is the {\it tension} of the string, meaning the mass per unit length.

\para
We can write this action a little more explicitly. The pull-back of the metric is given by,
\be \gamma_{\alpha\beta} = \left(\begin{array}{cc} \dot{X}^2 & \dot{X}\cdot X^\prime \\ \dot{X}\cdot X^\prime & X^\prime{}^2\end{array}\right)\ .\nn\ee
where $\dot{X}^\mu = \partial X^\mu/\partial \tau$ and $X^{\mu\,\prime}=\partial X^\mu/\partial \sigma$. The action then takes the form,
\be S=-T\int d^2\sigma \sqrt{-(\dot{X})^2\, (X^{\prime})^2 + (\dot{X}\cdot X^\prime)^2} \ . \label{ng2}\ee
This is the {\it Nambu-Goto} action for a relativistic string.

\subsubsection*{Action = Area: A Check}
\EPSFIGURE{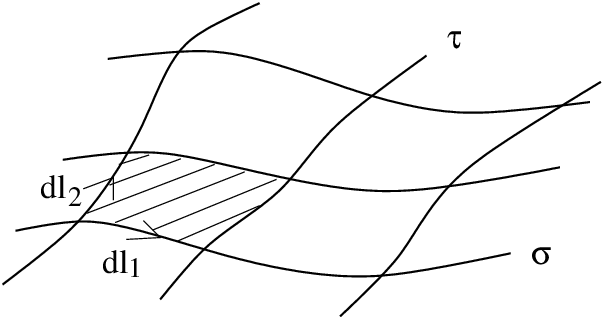,height=70pt}{}

\noindent If you're unfamiliar with differential geometry, the argument about the pull-back of the metric may be a bit slick. Thankfully, there's a more pedestrian way to see that the action \eqn{ng2} is equal to the area swept out by the worldsheet. It's slightly simpler to make this argument for a surface embedded in Euclidean space rather than Minkowski space. We choose some parameterization of the sheet in terms of $\tau$ and $\sigma$, as drawn in the figure, and we write the coordinates of Euclidean space as $\vec{X}(\sigma,\tau)$.  We'll compute the area of the infinitesimal shaded region. The vectors tangent to the boundary are,
\be \vec{dl}_1=\frac{\partial\vec{X}}{\partial\sigma}\ \ \ ,\ \ \ \vec{dl}_2=\frac{\partial\vec{X}}{\partial\tau}\ .\nn\ee
If the angle between these two vectors is $\theta$, then the area is then given by
\be ds^2 = |\vec{dl}_1||\vec{dl}_2|\sin\theta = \sqrt{{dl}_1^2\,{dl}_2^2 (1-\cos^2\theta)}
= \sqrt{{dl}_1^2\,{dl}_2^2 - (\vec{dl}_1\cdot\vec{dl}_2)^2}\ee
which indeed takes the form of the integrand of \eqn{ng2}.

\subsubsection*{Tension and Dimension}

Let's now see that $T$ has the physical interpretation of tension. We write Minkowski coordinates as $X^\mu=(t,\vec{x})$. We work in a gauge with $X^0\equiv t = R\tau$, where $R$ is a constant that is needed to balance up dimensions (see below) and will drop out at the end of the argument. Consider a snapshot of a string configuration at a time when $d\vec{x}/d\tau=0$ so that the instantaneous kinetic energy
vanishes. Evaluating the action for a time $dt$ gives
\be S = - T \int d\tau d\sigma R\sqrt{(d\vec{x}/d\sigma)^2} = -T \int dt\,
(\mbox{spatial length of string}) \ .\label{stretch}\ee
But, when the kinetic energy vanishes, the  action is proportional to the time integral of
the potential energy,
\be
\mbox{potential energy} = T\times(\mbox{spatial length of string})
\ .\nn\ee
So $T$ is indeed the energy per unit length as claimed. We learn that the string acts rather like an elastic band and its energy increases linearly with length. (This is different from the elastic bands you're used to which obey Hooke's law where energy increased quadratically with length). To minimize its potential energy, the string will want to shrink to zero size. We'll see that when we include  quantum
effects this can't happen because of the usual zero point energies.

\para
There is a slightly annoying way of writing the tension that has its origin in ancient history, but is commonly used today
\be T=\frac{1}{2\pi\alpha^\prime}\ee
where $\ap$ is pronounced ``alpha-prime". In the language of our ancestors, $\ap$ is referred to as the ``universal Regge slope". We'll explain why later in this course.

\para
At this point, it's worth pointing out some conventions that we have, until now, left implicit.  The spacetime coordinates have dimension $[X]=-1$. In contrast, the worldsheet coordinates are taken to be dimensionless, $[\sigma]=0$. (This can be seen in our identification $\sigma\equiv\sigma +2\pi$).
The tension is equal to the mass per unit length and has dimension $[T]=2$. Obviously this means that $[\ap]=-2$. We can therefore associate a length scale, $l_s$, by
\be \ap = l_s^2
\ee
The {\it string scale} $l_s$ is the natural length that appears in string theory. In fact, in a certain sense (that we will make more precise below) this length scale is the only parameter of the theory.

\subsubsection*{Actual Strings vs. Fundamental Strings}

There are several situations in Nature where string-like objects arise. Prime examples include
magnetic flux tubes in superconductors and chromo-electric flux tubes in QCD. Cosmic strings,
a popular speculation in cosmology, are similar objects, stretched across the sky.
In each of these situations, there are typically two length scales associated to the string: the tension, $T$ and the width of the string, $L$. For all these objects, the dynamics is governed by the Nambu-Goto
action as long as the curvature of the string is much greater than $L$. (In the case of superconductors, one should work with a suitable non-relativistic version of the Nambu-Goto action).

\para
However, in each of these other cases, the Nambu-Goto action is not the end of the story. There will typically be additional terms in the action that depend on the width of the string. The form
of these terms is not universal, but often includes a {\it rigidity} piece of form
$L \int \,K^2$, where $K$ is the extrinsic curvature of the worldsheet. Other terms could be added to describe fluctuations in the width of the string.

\para
The string scale, $l_s$, or equivalently the tension, $T$, depends on the kind of string that we're considering. For example, if we're interested in QCD flux tubes then we would take
\be T \sim (1\ {\rm Gev})^2\ee
In this course we will consider {\it fundamental strings} which have zero width. What this means in practice is that we take the Nambu-Goto action as the complete description
for all configurations of the string. These strings will have relevance to quantum gravity and
the tension of the string is taken to be much larger, typically an order of magnitude
or so below the Planck scale.
\be T \lesssim M_{pl}^2 = (10^{18}\ {\rm GeV})^2 \ee
However, I should point out that when we try to view string theory as a fundamental theory of quantum gravity, we don't really know what value $T$ should take. As we will see later in this course, it
depends on many other aspects, most notably the string coupling and the volume of the extra dimensions.

\subsubsection{Symmetries of the Nambu-Goto Action}

The Nambu-Goto action has two types of symmetry, each of a different nature.
\begin{itemize}
\item Poincar\'e invariance of the spacetime \eqn{poincare}. This is a global symmetry from the perspective of the worldsheet, meaning that the parameters $\Lambda^\mu_{\ \nu}$ and $c^\mu$ which label the symmetry transformation are constants and do not depend on worldsheet coordinates $\sigma^\alpha$.
\item Reparameterization invariance, $\sigma^\alpha\rightarrow \tilde{\sigma}^\alpha(\sigma)$. As for the point particle, this is a gauge symmetry. It reflects the fact that we have a redundancy in our description because the worldsheet coordinates $\sigma^\alpha$ have no physical meaning.
\end{itemize}

\subsubsection{Equations of Motion}

To derive the equations of motion for the Nambu-Goto string, we first introduce the momenta which we call $\Pi$ because there will be countless other quantities that we want to call $p$ later,
\be \Pi^\tau_\mu &=& \frac{\partial {\cal L}}{\partial \dot{X}^\mu} = -T\,\frac{(\dot{X}\cdot X^\prime)X^\prime_\mu - (X^{\prime\,2})\dot{X}_\mu}{\sqrt{(\dot{X}\cdot X^\prime)^2 - \dot{X}^2\,X^{\prime\,2}}} \ .\nn\\
\Pi^\sigma_\mu &=& \frac{\partial {\cal L}}{\partial {X}^{\prime\,\mu}} = -T\,\frac{(\dot{X}\cdot X^\prime)\dot{X}_\mu - (\dot{X}^2){X}^\prime_\mu}{\sqrt{(\dot{X}\cdot X^\prime)^2 - \dot{X}^2\,X^{\prime\,2}}} \ .\nn\ee
The equations of motion are then given by,
\be \frac{\partial \Pi_\mu^\tau}{\partial \tau}+\frac{\partial \Pi_\mu^\sigma}{\partial \sigma} = 0 \nn\ee
These look like nasty, non-linear equations. In fact, there's a slightly nicer way to write these equations, starting from the earlier action \eqn{ng}. Recall that the variation of a determinant is $\delta\sqrt{-\gamma} = \ft12\sqrt{-\gamma}\,\gamma^{\alpha\beta}\delta\gamma_{\alpha\beta}$. Using the definition of the pull-back metric $\gamma_{\alpha\beta}$, this gives rise to the equations of motion
\be \partial_\alpha(\sqrt{-\det\gamma}\,\gamma^{\alpha\beta}\partial_\beta X^\mu)=0\ ,\label{eomng}\ee
Although this notation makes the equations look a little nicer, we're kidding ourselves.
Written in terms of $X^\mu$, they are still the same equations. Still nasty.

\subsection{The Polyakov Action}

The square-root in the Nambu-Goto action means that it's rather difficult to quantize using path integral techniques. However, there is another form of the string action which is classically equivalent to the Nambu-Goto action. It eliminates the square root at the expense of introducing another field,
\be S= -\frac{1}{4\pi\alpha^\prime} \int d^2\sigma \sqrt{-g} g^{\alpha\beta}\,\partial_\alpha X^\mu\partial_\beta X^\nu\,\eta_{\mu\nu}\label{poly}\ee
where $g\equiv \det g$.
This is the {\it Polyakov} action. (Polyakov didn't discover the action, but he understood how to
work with it in the path integral, and for this reason it carries his name. The path
integral treatment of this action will be the subject of  Chapter \ref{polyakov}).

\para
The new field is $g_{\alpha\beta}$. It is a dynamical metric on the worldsheet. From the perspective
of the worldsheet, the Polyakov action is a bunch of scalar fields $X$ coupled to 2d gravity.

\para
The equation of motion for $X^\mu$ is
\be \partial_\alpha (\sqrt{-g}g^{\alpha\beta} \partial_\beta X^\mu ) = 0\ ,\label{polyeom}\ee
which coincides with the equation of motion \eqn{eomng} from the Nambu-Goto action, except that $g_{\alpha\beta}$ is now an independent variable which is fixed by its own equation of motion. To determine this, we vary the action (remembering again that $\delta\sqrt{-g} = -\ft12 \sqrt{-g}g_{\alpha\beta} \delta g^{\alpha\beta} = +\ft12 \sqrt{-g} g^{\alpha\beta} \delta g_{\alpha\beta}$),
\be \delta S = -\frac{T}{2}\int d^2\sigma\  \delta g^{\alpha\beta} \left(\sqrt{-g}\,\partial_\alpha X^\mu \partial_\beta X^\nu - \ft12\sqrt{-g}\,g_{\alpha\beta} g^{\rho\sigma}\partial_\rho X^\mu \partial_\sigma X^\nu\right)\eta_{\mu\nu} =0 \ .\label{varyg}\ee
The worldsheet metric is therefore given by,
\be g_{\alpha\beta} = 2f(\sigma)\,\partial_\alpha X \cdot\partial_\beta X\ ,\label{ggam}\ee
where the function $f(\sigma)$ is given by,
\be f^{-1} = g^{\rho\sigma}\, \partial_\rho X\cdot\partial_\sigma X \nn\ee
A comment on the potentially ambiguous notation: here, and below, any function
$f(\sigma)$ is always short-hand for $f(\sigma,\tau)$: it in no way implies that $f$
depends only on the spatial worldsheet coordinate.

\para
We see that $g_{\alpha\beta}$ isn't quite the same as the pull-back metric $\gamma_{\alpha\beta}$ defined in equation \eqn{gametric}; the two differ by the conformal factor $f$. However, this doesn't matter because, rather remarkably,  $f$ drops out of the equation of motion \eqn{polyeom}. This is because the $\sqrt{-g}$ term scales as $f$, while the inverse metric $\gabi$ scales as $f^{-1}$ and the two pieces cancel. We therefore see that Nambu-Goto and the Polyakov actions result in the same equation of motion for $X$.

\para
In fact, we can see more directly that the Nambu-Goto and Polyakov actions coincide. We may replace $g_{\alpha\beta}$ in the Polyakov action \eqn{poly} with its equation of motion  $g_{\alpha\beta} = 2f\, \gamma_{\alpha\beta}$. The factor of $f$ also drops out of the action for the same reason that it dropped out of the equation of motion. In this manner, we recover the Nambu-Goto action \eqn{ng}.

\subsubsection{Symmetries of the Polyakov Action}

The fact that the presence of the factor $f(\sigma,\tau)$ in \eqn{ggam} didn't actually affect the equations of motion for $X^\mu$ reflects the existence of an extra symmetry which the Polyakov action
enjoys. Let's look more closely at this. Firstly, the Polyakov action still has the two symmetries of the Nambu-Goto action,
\begin{itemize}
\item Poincar\'e invariance. This is a global symmetry on the worldsheet.
\be X^\mu\rightarrow \Lambda^\mu_{\ \nu}X^\nu + c^\mu\ .\nn\ee
\item Reparameterization invariance, also known as diffeomorphisms. This is a gauge symmetry on the worldsheet. We may redefine
the worldsheet coordinates as $\sigma^\alpha\rightarrow \tilde{\sigma}^\alpha(\sigma)$. The
fields $X^\mu$ transform as worldsheet scalars, while $g_{\alpha\beta}$ transforms in the manner appropriate for a 2d metric.
\be  X^\mu(\sigma) &\rightarrow& \tilde{X}^\mu(\tilde{\sigma}) = X^\mu(\sigma)\nn\\\
\gab(\sigma) &\rightarrow& \tilde{g}_{\alpha\beta}(\tilde{\sigma})=
\frac{\partial\sigma^\gamma}{\partial\tilde{\sigma}^\alpha}\,
\frac{\partial\sigma^\delta}{\partial\tilde{\sigma}^\beta}\,g_{\gamma\delta}(\sigma)\nn\ee
It will sometimes be useful to work infinitesimally. If we make the coordinate change $\sigma^\alpha\rightarrow \tilde{\sigma}^\alpha = \sigma^\alpha - \eta^\alpha(\sigma)$, for
some small $\eta$. The transformations of the fields then become,
\be \delta X^\mu(\sigma) &=& \eta^\alpha \partial_\alpha X^\mu\nn\\
\delta g_{\alpha\beta}(\sigma) &=& \nabla_\alpha \eta_\beta + \nabla_\beta \eta_\alpha\nn
\ee
where the covariant derivative is defined by $\nabla_\alpha \eta_\beta = \partial_\alpha\eta_\beta - \Gamma^\sigma_{\alpha\beta}\eta_\sigma$ with the Levi-Civita connection associated to the worldsheet metric given by the usual expression,
\be \Gamma^\sigma_{\alpha\beta} = \ft12\, g^{\sigma\rho}(\partial_\alpha g_{\beta\rho}+\partial_\beta g_{\rho\alpha} - \partial_\rho \gab)\nn\ee
\end{itemize}
Together with these familiar symmetries, there is also a new symmetry which is novel to the Polyakov action. It is called {\it Weyl invariance}.
\begin{itemize}
\item Weyl Invariance. Under this symmetry, $X^\mu(\sigma)\rightarrow X^\mu(\sigma)$, while the metric changes as
\be \gab(\sigma) \rightarrow \Omega^2(\sigma)\,\gab(\sigma)\ .\label{weyl}\ee
Or, infinitesimally, we can write $\Omega^2(\sigma) = e^{2\phi(\sigma)}$ for small $\phi$ so that
\be \delta \gab(\sigma) = 2\phi(\sigma)\,\gab(\sigma)\ .\nn\ee
It is simple to see that the Polyakov action is invariant under this transformation: the factor of $\Omega^2$ drops out just as the factor of $f$ did in equation \eqn{ggam}, canceling between  $\sqrt{-g}$ and the inverse metric $\gabi$. This is a gauge symmetry of the string, as seen by the fact that the parameter $\Omega$ depends on the worldsheet coordinates $\sigma$. This means that two metrics which are related by a Weyl transformation
\eqn{weyl} are to be considered as the same physical state.
\end{itemize}
\begin{figure}[htb]
\begin{center}
\epsfxsize=4.7in\leavevmode\epsfbox{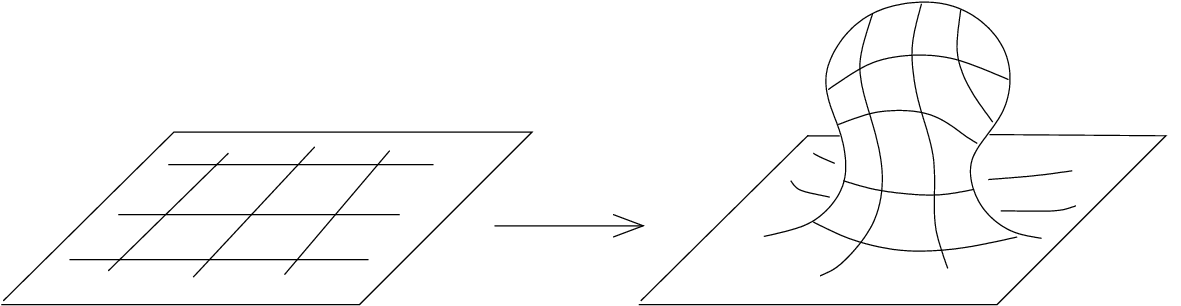}
\end{center}
\caption{An example of a Weyl transformation}
\end{figure}
How should we think of Weyl invariance? It is not a coordinate change. Instead it is the invariance of the theory under a local change of scale which preserves the angles between all lines. For example the two worldsheet metrics shown in the figure are viewed by the Polyakov string as equivalent. This is rather surprising! And, as you might imagine, theories with this property are extremely rare.  It should be clear from the discussion above that the property of Weyl invariance is special to two dimensions, for only there does the scaling factor coming from the determinant $\sqrt{-g}$ cancel that coming from the inverse metric. But even in two dimensions,
if we wish to keep Weyl invariance then we are strictly limited in the kind of interactions that can
be added to the action. For example, we would not be allowed a potential term for the worldsheet scalars of the form,
\be \int d^2\sigma \sqrt{-g}\, V(X) \ .\nn\ee
These break Weyl invariance. Nor can we add a worldsheet cosmological constant term,
\be \mu \int d^2\sigma \sqrt{-g} \ . \nn\ee
This too breaks Weyl invariance. We will see later in this course that the requirement of Weyl
invariance becomes even more
stringent in the quantum theory. We will also see what kind of interactions terms can be
added to the worldsheet. Indeed, much of this course can be thought of as the study of
theories with Weyl invariance.

\subsubsection{Fixing a Gauge}
\label{gaugefixing}

As we have seen, the equation of motion \eqn{polyeom} looks pretty nasty. However, we can use
the redundancy inherent in the gauge symmetry to choose coordinates in which they simplify.  Let's think about what we can do with the gauge symmetry.

\para
Firstly, we have two reparameterizations to play with. The worldsheet metric has three independent components. This means that we expect to be able to set any two of the metric components to a value of our choosing. We will choose to make the metric locally conformally flat, meaning
\be \gab = e^{2\phi} \eta_{\alpha\beta}\ ,\label{gconf}\ee
where $\phi(\sigma,\tau)$ is some function on the worldsheet. You can check that this is possible by writing down the change of the metric under a coordinate transformation and seeing that the differential equations which result from the condition \eqn{gconf} have solutions, at least locally. Choosing a  metric of the form \eqn{gconf} is known as {\it conformal gauge}.

\para
We have only used reparameterization invariance to get to the metric \eqn{gconf}. We still have Weyl transformations to play with. Clearly, we can use these to remove the last independent component of the metric and set $\phi=0$ such that,
\be \gab=\eta_{\alpha\beta} \ .\label{flat}\ee
We end up with the flat metric on the worldsheet in Minkowski coordinates.

\subsubsection*{A Diversion: How to make a metric flat}

The fact that we can use Weyl invariance to make any two-dimensional metric flat is an important
result. Let's take a quick diversion from our main discussion to see a different proof that
isn't tied to the choice of Minkowski coordinates on the worldsheet.
We'll work in 2d Euclidean space to avoid annoying minus signs. Consider two metrics related by a Weyl transformation, $g^\prime_{\alpha\beta} = e^{2\phi}\gab$. One can check that the Ricci scalars of the two metrics are related by,
\be \sqrt{g^\prime}R^\prime  = \sqrt{g}(R-2\nabla^2\phi)\ .\label{riccirel}\ee
We can therefore pick a $\phi$ such that the new metric has vanishing Ricci scalar, $R^\prime = 0$,
simply by solving this differential equation for $\phi$. However, in two dimensions (but not in higher dimensions) a vanishing Ricci scalar implies a flat metric.
The reason is simply that there aren't too many indices to play with. In particular, symmetry of the Riemann tensor in two dimensions means that it must take the form,
\be R_{\alpha\beta\gamma\delta} = \frac{R}{2}(g_{\alpha\gamma}g_{\beta\delta}-g_{\alpha\delta}g_{\beta\gamma})\ .\nn\ee
So $R^\prime =0$ is enough to ensure that $R^\prime_{\alpha\beta\gamma\delta} =0$, which means that the manifold is flat. In equation \eqn{flat}, we've further used reparameterization invariance to pick coordinates in which the flat metric is the Minkowski metric.

\subsubsection*{The equations of motion and the stress-energy tensor}

With the choice of the flat metric \eqn{flat}, the Polyakov action simplifies tremendously and
becomes the theory of $D$ free scalar fields. (In fact, this simplification happens in any
conformal gauge).
\be S= -\frac{1}{4\pi\ap} \int d^2\sigma \ \partial_\alpha X \cdot \partial^\alpha X \ ,\label{thatseasy}\ee
and the equations of motion for $X^\mu$ reduce to the free wave equation,
\be \partial_\alpha\partial^\alpha X^\mu = 0\ .\label{wave}\ee
Now that looks too good to be true! Are the horrible equations \eqn{polyeom} really equivalent to a free wave equation? Well, not quite. There is something that we've forgotten: we picked a choice of
gauge for the metric $\gab$. But we must still make sure that the equation of motion for $\gab$ is satisfied. In fact, the variation of the action with respect to the metric gives rise to a rather special quantity: it is the stress-energy tensor, $T_{\alpha\beta}$. With a particular choice of normalization convention, we define the stress-energy tensor to be
\be \tab = -\frac{2}{T}\,\frac{1}{\sg}\,\frac{\partial S}{\partial \gabi}\ .\nn\ee
We varied the Polyakov action with respect to $\gab$ in \eqn{varyg}. When we set
$\gab=\eta_{\alpha\beta}$ we get
\be \tab = \pa X\cdot \partial_\beta X - \ft12\eta_{\alpha\beta}\,\eta^{\rho\sigma}\partial_\rho X\cdot \partial_\sigma X\ .\ee
The equation of motion associated to the metric $\gab$ is simply $\tab=0$. Or, more explicitly,
\be T_{01} &=& \dot{X}\cdot X^\prime = 0 \nn\\
T_{00}=T_{11} &=& \ft12(\dot{X}^2+X^{\prime\,2})=0\ .\label{cons}\ee
We therefore learn that the equations of motion of the string are the free wave equations \eqn{wave}  subject to the two constraints \eqn{cons} arising from the equation of motion $\tab=0$.

\newpage
\subsubsection*{Getting a feel for the constraints}
\EPSFIGURE{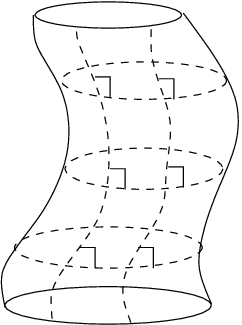,height=90pt}{}

\noindent Let's try to get some intuition for these constraints. There is a simple meaning of the first constraint in \eqn{cons}: we must choose our parameterization such that lines of constant $\sigma$ are  perpendicular to the lines of constant $\tau$, as shown in the figure.

\para
But we can do better. To gain more physical insight, we need to make use of the fact
that we haven't quite exhausted our gauge symmetry. We will discuss this more in Section
\ref{lightconeq}, but for now one can check that there is enough remnant gauge symmetry
to allow us to go to static gauge,
\be X^0\equiv t = R\tau\ , \nn\ee
so that $(X^0)^\prime = 0$ and $\dot{X}^0=R$, where $R$ is a constant that is needed on dimensional grounds. The interpretation of this constant will become clear shortly. Then, writing $X^\mu=(t,\vec{x})$, the equation of motion for spatial components is the free wave equation,
\be \ddot{\vec{x}}-\vec{x}^{\,\prime\prime} = 0\nn\ee
while the constraints become
\be
\dot{\vec{x}}\cdot\vec{x}^{\,\prime} &=& 0 \nn\\
\dot{\vec{x}}{}^{\ \!2}+\vec{x}^{\,\prime}{}^{\,2} &=& R^2 \label{haveafeel}\ee
The first constraint tells us that the motion of the string must be perpendicular to the
string itself. In other words, the physical modes of the string are transverse oscillations.
There is no longitudinal mode. We'll also see this again in Section \ref{lightconeq}.

\para
From the second constraint, we can understand the meaning of the constant $R$: it is related to the length of the string when $\dot{\vec{x}}=0$,
\be \int d\sigma \,\sqrt{(d\vec{x}/d\sigma)^2} = 2\pi R\ .\nn\ee
Of course, if we have a stretched string with $\dot{\vec{x}}=0$ at one moment of time, then it
won't stay like that for long. It will contract under its own tension. As this happens, the second constraint equation relates the length of the string to the instantaneous velocity of the string.

\subsection{Mode Expansions}

Let's look at the equations of motion and constraints more closely. The equations of motion
\eqn{wave} are easily solved. We introduce lightcone coordinates on the worldsheet,
\be \sigma^\pm = \tau \pm \sigma \ ,\nn\ee
in terms of which the equations of motion simply read
\be \p_+\p_- X^\mu = 0\nn\ee
The most general solution is,
\be X^\mu(\sigma,\tau) = X^\mu_L(\sigma^+) + X^\mu_R(\sigma^-)\nn\ee
for arbitrary functions $X^\mu_L$ and $X^\mu_R$. These describe left-moving and right-moving waves respectively. Of course the solution must still obey both the constraints \eqn{cons} as well as the periodicity condition,
\be X^\mu(\sigma,\tau)=X^\mu(\sigma+2\pi,\tau)\ .\label{periodicity}\ee
The most general, periodic solution can be expanded in Fourier modes,
\be
X^\mu_L(\sigma^+) &=& \ft12 x^\mu + \ft12 \ap p^\mu \,\sigma^+ + i\sfap\sum_{n\neq 0}
\frac{1}{n}\,\tilde{\alpha}_n^\mu\, e^{-in\sigma^+} \ ,\nn\\
X^\mu_R(\sigma^-) &=& \ft12 x^\mu + \ft12 \ap p^\mu \,\sigma^- + i\sfap\sum_{n\neq 0}
\frac{1}{n}\,{\alpha}_n^\mu\, e^{-in\sigma^-} \ .\label{mode}\ee
This mode expansion will be very important when we come to the quantum theory. Let's make a few simple comments here.
\begin{itemize}
\item Various normalizations in this expression, such as the $\ap$ and factor of $1/n$ have been chosen for later convenience.
\item $X_L$ and $X_R$ do not individually satisfy the periodicity condition \eqn{periodicity} due to the terms linear in $\sigma^\pm$. However, the sum of them is invariant under $\sigma\rightarrow \sigma +2\pi$ as required.
\item The variables $x^\mu$ and $p^\mu$ are the position and momentum of the center of mass of the string. This can be checked, for example, by studying the Noether currents arising from the spacetime translation symmetry $X^\mu\rightarrow X^\mu+c^\mu$. One finds that the conserved charge is indeed $p^\mu$.
\item Reality of $X^\mu$ requires that the coefficients of the Fourier modes, $\an$ and $\tilan$, obey
\be \an= (\alpha^\mu_{-n})^\star\ \ \ ,\ \ \ \tilan=(\tilde{\alpha}^\mu_{-n})^\star\ .\label{realalpha}\ee
\end{itemize}

\subsubsection{The Constraints Revisited}

We still have to impose the two constraints \eqn{cons}. In the worldsheet lightcone coordinates $\sigma^\pm$, these become,
\be (\partial_+X)^2 = (\partial_- X)^2 = 0 \ .\label{cons+-}\ee
These equations give constraints on the momenta $p^\mu$ and the Fourier modes $\an$ and $\tilan$. To see what these are, let's look at
\be \partial_-X^\mu = \partial_-X^\mu_R &=& \frac{\ap}{2}\,p^\mu + \sfap\sum_{n\neq 0}\an\,e^{-in\sigma^-} \nn \\ &=& \sfap \sum_n \an e^{-in\sigma^-}\nn\ee
where in the second line the sum is over all $n\in {\bf Z}$, and we have defined $\alpha_0^\mu$ to be
\be \alpha^\mu_0 \equiv \sfap p^\mu\ .\nn\ee
The constraint \eqn{cons+-} can then be written as
\be (\partial_-X)^2 &=& \frac{\ap}{2} \sum_{m,p} \alpha_m\cdot \alpha_p\,e^{-i(m+p)\sigma^-} \nn\\ &=& \frac{\ap}{2} \sum_{m,n} \alpha_m\cdot \alpha_{n-m}\,e^{-in\sigma^-} \nn\\
&\equiv& \ap \sum_n L_n\, e^{-in\sigma^-} =0  \ .\nn\ee
where we have defined the sum of oscillator modes,
\be L_n = \frac{1}{2} \sum_m\,\alpha_{n-m}\cdot \alpha_m \ .\label{ln}\ee
We can also do the same for the left-moving modes, where we again define an analogous sum of operator modes,
\be \tilde{L}_n = \frac{1}{2} \sum_m\,\tilde{\alpha}_{n-m}\cdot\tilde{\alpha}_m \ . \label{tiln}\ee
with the zero mode defined to be,
\be \tilde{\alpha}^\mu_0 \equiv \sfap p^\mu\ .\nn\ee
The fact that $\tilde{\alpha}_0^\mu= \alpha_0^\mu$ looks innocuous but is a key point to remember
when we come to quantize the string. The $L_n$ and $\tilde{L}_n$ are the Fourier modes of the constraints. Any classical solution of the string of the form \eqn{mode} must further obey the infinite number of constraints,
\be L_n = \tilde{L}_n = 0 \ \ \ \ n\in {\bf Z}\ .\nn \ee
We'll meet these objects $L_n$ and $\tilde{L}_n$ again in a more general context when we come to discuss conformal field theory.

\para
The constraints arising from $L_0$ and $\tilde{L}_0$ have a rather special interpretation. This
is because they include the square of the spacetime momentum $p^\mu$. But, the square of the spacetime momentum is an important quantity in Minkowski space: it is the square of the rest mass of a particle,
\be p_\mu p^\mu = - M^2 \ .\nn\ee
So the $L_0$ and $\tilde{L}_0$ constraints tell us the effective mass of a string in terms of the excited oscillator modes, namely
\be M^2 = \frac{4}{\ap} \sum_{n > 0} \alpha_n\cdot\alpha_{-n} = \frac{4}{\ap} \sum_{n> 0} \tilde{\alpha}_n\cdot\tilde{\alpha}_{-n} \label{levelm}\ee
Because both  $\alpha_0^\mu$ and $\tilde{\alpha}_0^\mu$ are equal to $\sqrt{\ap/2}\, p^\mu$, we have two expressions for the invariant mass: one in terms of right-moving oscillators $\an$ and one in terms of left-moving oscillators $\tilan$. And these two terms must be equal to each other. This is known as {\it level matching}. It will play an important role in the next section where we turn to the quantum
theory.

%\subsubsection*{The Hamiltonian}
%
%Let's also consider the Hamiltonian of the system. Starting from the gauge fixed Lagrangian %\eqn{thatseasy}, we can easily compute the Hamiltonian,
%%
%\be H = \frac{T}{2}\int d\sigma \ \dot{X}^2 + X^{\prime\,2} \ee
%%
%Note that this is proportional to the constraint \eqn{gcons} and is therefore zero on any %physical solution.  This shouldn't surprise us. We saw the same thing for the relativistic %point particle.

\newpage
\section{The Quantum String}
\label{quantum}

Our goal in this section is to quantize the string. We have seen that the string action involves a
gauge symmetry and whenever we wish to quantize a gauge theory we're presented with a
number of different ways in which we can proceed. If we're working in the canonical formalism,
this usually boils down to one of two choices:
\begin{itemize}
\item We could first quantize the system and then subsequently impose the constraints that arise from gauge fixing as operator equations on the physical states of the system. For example, in QED this is the Gupta-Bleuler method of quantization that we use in Lorentz gauge. In string theory it consists of treating all fields $X^\mu$, including time $X^0$, as operators and imposing the constraint equations  \eqn{cons} on the states. This is usually called covariant quantization.
\item The alternative method is to first solve all of the constraints of the system to determine the space of physically distinct classical solutions. We then quantize these physical solutions. For example, in QED, this is the way we proceed in Coulomb gauge. Later in this chapter, we will see a simple way to solve the constraints of the free string.
\end{itemize}
Of course, if we do everything correctly, the two methods should agree. Usually, each presents a slightly different challenge and offers a different viewpoint.

\para
In these lectures, we'll take a brief look at the first method of covariant quantization. However, at the slightest sign of difficulties, we'll bail! It will be useful enough to see where the problems lie.
We'll then push forward with the second method described above which is known as lightcone quantization
in string theory. Although we'll succeed in pushing quantization through to the end, our derivations will be a little cheap and unsatisfactory in places. In Section \ref{polyakov} we'll return to all these
issues, armed with more sophisticated techniques from conformal field theory.

\subsection{A Lightning Look at Covariant Quantization}
\label{covquansec}

We wish to quantize $D$ free scalar fields $X^\mu$ whose dynamics is governed by the action \eqn{thatseasy}. We subsequently wish to impose the constraints
\be \dot{X}\cdot X^\prime = \dot{X}^2 + {X}^{\prime\,2}=0\ .\label{consagain}
\ee
The first step is easy. We promote $X^\mu$ and their conjugate momenta $\Pi_\mu = (1/2\pi\ap) \dot{X}_\mu$ to operator valued fields obeying the canonical equal-time commutation relations,
\be &[X^\mu(\sigma,\tau),\Pi_\nu(\sigma^\prime,\tau)]=i\delta(\sigma-\sigma^\prime)
\,\delta^\mu_{\ \nu} &\ , \nn\\\ &[X^\mu(\sigma,\tau),X^\nu(\sigma^\prime,\tau)] = [\Pi_\mu
(\sigma,\tau),\Pi_\nu(\sigma^\prime,\tau)] = 0&\ .\nn\ee
We translate these into commutation relations for the Fourier modes $x^\mu$, $p^\mu$, $\an$ and $\tilan$. Using the mode expansion \eqn{mode} we find
\be [x^\mu,p_\nu]=i\delta^\mu_{\ \nu} \ \ \ {\rm and}\ \ \
{[}\an,\alpha_m^\nu{]}=
{[}\tilan,\tilde{\alpha}_m^\nu{]}
=n\, \eta^{\mu\nu}\delta_{n+m,\,0}\ ,\label{acom}\ee
with all others zero. The commutation relations for $x^\mu$ and $p^\mu$ are expected for operators governing the position and momentum of the center of mass of the string. The commutation relations of $\an$ and $\tilan$ are those of harmonic oscillator creation and annihilation operators in disguise. And the disguise
isn't that good. We just need to define (ignoring the $\mu$ index for now)
\be a_n = \frac{\alpha_n}{\sqrt{n}} \ \ \ ,\ \ \ a_n^\dagger =
\frac{\alpha_{-n}}{\sqrt{n}}\ \ \  \ \ \ \ \ \ \ \ \ \ \ n>0\ \label{whatais}\ee
Then \eqn{acom} gives the familiar $[a_n,a^\dagger_m]=\delta_{mn}$. So each scalar field gives rise to
two infinite towers of creation and annihilation operators, with $\alpha_n$ acting as a rescaled annihilation operator for $n>0$ and as a creation operator for $n<0$. There are two towers
because we have right-moving modes $\alpha_n$ and left-moving modes $\tilde{\alpha}_n$.

\para
With these commutation relations in hand we can now start building the Fock space of our theory. We introduce a vacuum state of the string $|0\rangle$, defined to obey
\be  \an\,\vac=\tilan\,\vac=0\ \ \ \ \ \ \ \ \ \ \ {\rm for} \ \ n>0 \label{vac}\ee
%
%\EPSFIGURE{ground.eps,height=20pt}{A cartoon of the ground state of the string}
%
%
The vacuum state of string theory has a different interpretation from the analogous object in field theory. This is not the vacuum state of spacetime. It is instead the vacuum state of a single string.
This is reflected in the fact that the operators $x^\mu$ and $p^\mu$ give extra structure to the
vacuum. The true ground state of the string is $|0\rangle$, tensored with a spatial wavefunction $\Psi(x)$. Alternatively, if we work in momentum space, the vacuum carries another quantum number,
$p^\mu$, which is the eigenvalue of the momentum operator. We should therefore write the vacuum as $|0;p\rangle$, which still obeys \eqn{vac}, but now also
\be \hat{p}^\mu\,|0;p\rangle = p^\mu|0;p\rangle \ee
where (for the only time in these lecture notes) we've put a hat on the momentum operator $\hat{p}^\mu$
on the left-hand side of this equation to distinguish it from the eigenvalue $p^\mu$ on the right-hand side.
%
%\EPSFIGURE{excited.eps,height=40pt}{A cartoon of a typical excited state of the string.}
%
%
\para
We can now start to build up the Fock space by acting with creation operators $\an$ and $\tilan$ with $n<0$. A generic state comes from acting with any number of these creation
operators on the vacuum,
\be (\alpha^{\mu_1}_{-1})^{n_{\mu_1}}(\alpha_{-2}^{\mu_2})^{n_{\mu_2}}\ldots (\tilde{\alpha}^{\nu_1}_{-1})^{n_{\nu_1}}(\tilde{\alpha}_{-2}^{\nu_2})^{n_{\nu_2}}\ldots|0;p\rangle \nn\ee
Each state in the Fock space is a different excited state of the string. Each has the interpretation of a different species of particle in spacetime. We'll see exactly what particles they are shortly. But for now, notice that because there's an infinite number of ways to excite a string there are an infinite
number of different species of particles in this theory.

\subsubsection{Ghosts}

There's a problem with the Fock space that we've constructed: it doesn't have positive norm. The reason for this is that one of the scalar fields, $X^0$, comes with the wrong sign kinetic term in the action \eqn{thatseasy}. From the perspective of the commutation relations, this issue raises its head in
presence of the spacetime Minkowski metric in the expression
\be [\alpha^\mu_n,\alpha^{\nu\,\dagger}_m]=n\,\eta^{\mu\nu}\,\delta_{n,m}\ .\nn\ee
This gives rise to the offending negative norm states, which come with an odd number of timelike oscillators excited, for example
\be \langle p^\prime; 0|\alpha_1^0\alpha_{-1}^0|0;p\rangle \sim - \delta^{D}(p-p^\prime) \nn\ee
This is the first problem that arises in the covariant approach to quantization. States with negative norm are referred to as {\it ghosts}. To make sense of the theory, we have to make sure that they
can't be produced in any physical processes.
Of course, this problem is familiar from attempts to quantize QED in Lorentz gauge. In that case,
gauge symmetry rides to the rescue since the ghosts are removed by imposing the gauge fixing
constraint. We must hope that the same happens in string theory.

\subsubsection{Constraints}

Although we won't push through with this programme at the present time, let us briefly look at what kind of constraints we have in string theory. In terms of Fourier modes, the classical constraints can be written as $L_n=\tilde{L}_n=0$, where
\be L_n=\frac{1}{2}\sum_m\alpha_{n-m}\cdot \alpha_m \nn\ee
and similar for $\tilde{L}_n$. As in the Gupta-Bleuler quantization of QED, we don't impose all of these as operator equations on the Hilbert space. Instead we only require that the operators $L_n$
and $\tilde{L}_n$ have vanishing matrix elements when sandwiched between two physical states $|{\rm phys}\rangle$ and $|{\rm phys}^\prime\rangle$,
\be \langle{\rm phys}^\prime| L_n|{\rm phys}\rangle = \langle{\rm phys}^\prime| \tilde{L}_n|{\rm phys}\rangle = 0\nn\ee
Because $L_n^\dagger=L_{-n}$, it is therefore sufficient to require
\be L_n|{\rm phys}\rangle = \tilde{L}_n |{\rm phys}\rangle = 0 \ \ \ \ {\rm for}\ \ n>0 \label{itsl}\ee
However, we still haven't explained how to impose the constraints $L_0$ and $\tilde{L}_0$. And these present a problem that doesn't arise in the case of QED. The problem is that, unlike for $L_n$ with $n\neq 0$, the operator $L_0$ is not uniquely defined when we pass to the quantum theory. There is an operator ordering ambiguity arising from the commutation relations \eqn{acom}. Commuting the $\an$ operators past each other in $L_0$ gives rise to extra constant terms.
\\ {} \\
{\bf Question:} How do we know what order to put the $\an$ operators in the quantum operator $L_0$? Or the $\tilan$ operators in $\tilde{L}_0$?
\\ {} \\
{\bf Answer:} We don't! Yet. Naively it looks as if each different choice will define a different theory when we impose the constraints. To make this ambiguity manifest, for now let's just pick a choice of ordering. We define the quantum operators to be normal ordered, with the annihilation operators
$\alpha^i_n$, $n>0$, moved to the right,
\be L_0=\sum_{m=1}^\infty \alpha_{-m}\cdot \alpha_m + \frac{1}{2}\,\alpha_0^2 \ \ \ ,\ \ \
\tilde{L}_0=\sum_{m=1}^\infty \tilde{\alpha}_{-m}\cdot \tilde{\alpha}_m + \frac{1}{2}\,\tilde{\alpha}_0^2 \nn\ee
Then the ambiguity rears its head in the different constraint equations that we could impose, namely
\be (L_0-a)|{\rm phys}\rangle = (\tilde{L}_0-a)|{\rm phys}\rangle =0 \label{itslzero}\ee
for some constant $a$.

\para
As we saw classically, the operators $L_0$ and $\tilde{L}_0$ play an important role in determining the spectrum of the string because they include a term quadratic in the momentum $\alpha_0^\mu=\tilde{\alpha}_0^\mu = \sqrt{\ap /2}\,p^\mu$. Combining the expression \eqn{levelm} with our constraint equation for $L_0$ and $\tilde{L}_0$, we find the spectrum of the string is given by,
\be M^2 = \frac{4}{\ap}\left(-a + \sum_{m=1}^\infty \alpha_{-m}\cdot\alpha_m\right) =
\frac{4}{\ap}\left(-a + \sum_{m=1}^\infty \tilde{\alpha}_{-m}\cdot\tilde{\alpha}_m\right) \nn\ee
We learn therefore that the undetermined constant $a$ has a direct physical effect: it changes the mass spectrum of the string. In the quantum theory, the sums over $\alpha^\mu_n$ modes are related to the number operators for the harmonic oscillator: they count the number of excited modes of the string. The level matching in the quantum theory tells us that the number of left-moving modes must equal the number of right-moving modes.

\para
Ultimately, we will find that the need to decouple the ghosts forces us to make a unique choice for the constant $a$. (Spoiler alert: it turns out to be $a=1$). In fact, the requirement that there are no ghosts is much stronger than this. It also restricts the number of scalar fields that we have in the theory. (Another spoiler: $D=26)$. If you're interested in how this works in covariant formulation then you can read about it in the book by Green, Schwarz and Witten. Instead, we'll show how to quantize the
string and derive these values for $a$ and $D$ in lightcone gauge. However, after a trip through the world of conformal field theory, we'll come back to these ideas in a context which is closer to the covariant approach.

\subsection{Lightcone Quantization}
\label{lightconeq}

We will now take the second path described at the beginning of this section. We will try to find a parameterization of all classical solutions of the string. This is equivalent to finding the classical phase space of the theory. We do this by solving the constraints \eqn{consagain} in the classical theory, leaving behind only the physical degrees of freedom.

\para
Recall that we fixed the gauge to set the worldsheet metric to
\be \gab = \eta_{\alpha\beta} \ .\nn
\ee
However, this isn't the end of our gauge freedom. There still remain gauge transformations which preserve this choice of metric. In particular, any coordinate transformation $\sigma \rightarrow \tilde{\sigma}(\sigma)$ which changes the metric by
\be \eta_{\alpha\beta} \rightarrow \Omega^2(\sigma)\eta_{\alpha\beta} \ ,\label{factor}\ee
can be undone by a Weyl transformation. What are these coordinate transformations? It's simplest to answer this using lightcone coordinates on the worldsheet,
\be \sigma^\pm = \tau \pm \sigma\ ,\label{wl}\ee
where the flat metric on the worldsheet takes the form,
\be ds^2 = -d\sigma^+ d\sigma^-\nn\ee
In these coordinates, it's clear that any transformation of the form
\be \sigma^+\rightarrow \tilde{\sigma}^+(\sigma^+)\ \ \ \ ,\ \ \ \sigma^-\rightarrow \tilde{\sigma}^-(\sigma^-)\ ,\label{whatsleft}\ee
simply multiplies the flat metric by an overall factor \eqn{factor} and so can be undone by
a compensating Weyl transformation. Some quick comments on this surviving gauge symmetry:
\begin{itemize}
\item Recall that in Section \ref{gaugefixing} we used the argument that 3 gauge invariances (2 reparameterizations + 1 Weyl) could be used to fix 3 components of the worldsheet metric $\gab$. What happened to this argument? Why do we still have some gauge symmetry left? The reason is that $\tilde{\sigma}^\pm$ are functions of just a single variable, not two. So we did fix nearly all our gauge symmetries. What is left is a set of measure zero amongst the full gauge symmetry that we started with.
\item The remaining reparameterization invariance \eqn{whatsleft} has an important physical implication. Recall that the solutions to the equations of motion are of the form $X_L^\mu(\sigma^+)+ X^\mu_R(\sigma^-)$ which looks like $2D$ functions worth of solutions. Of course, we still have the constraints which, in terms of $\sigma^\pm$, read
\be (\partial_+ X)^2 = (\partial_-X)^2 = 0\ , \label{consag}\ee
   which seems to bring the number down to $2(D-1)$ functions. But the reparameterization invariance \eqn{whatsleft} tells us that even some of these are fake since we can always change what we mean by $\sigma^\pm$. The physical solutions of the string are therefore actually described by $2(D-2)$ functions. But this counting has a nice interpretation: the degrees of freedom describe the {\it transverse} fluctuations of the string.
\item The above comment reaches the same conclusion as the discussion in Section \ref{gaugefixing}. There, in an attempt to get some feel for the constraints, we claimed that we could go to static gauge $X^0=R\tau$ for some dimensionful parameter $R$. It is easy to check that this is simple to do using reparameterizations of the form \eqn{whatsleft}. However, to solve the string constraints in full, it turns out that static gauge is not that useful. Rather we will use something called ``lightcone gauge".
\end{itemize}

\subsubsection{Lightcone Gauge}

We would like to gauge fix the remaining reparameterization invariance \eqn{whatsleft}. The best way to do this is called lightcone gauge. In counterpoint to the worldsheet lightcone coordinates \eqn{wl}, we introduce the spacetime lightcone coordinates,
\be X^\pm = \sqrt{\frac{1}{2}}(X^0 \pm X^{D-1}) \ .\label{stlc}\ee
Note that this choice picks out a particular time direction and a particular spatial direction. It means that any calculations that we do involving $X^\pm$ will not be manifestly Lorentz invariant.
You might think that we needn't really worry about this. We could try to make the following argument:
``The equations may not {\it look}
Lorentz invariant but, since we started from a Lorentz invariant theory, at the end of the
day any  physical process
is guaranteed to obey this symmetry". Right?! Well, unfortunately not. One of the more
interesting and subtle aspects of quantum field theory is the possibility of anomalies: these
are symmetries of the classical theory that do not survive the journey of quantization.
When we come to the quantum theory, if our equations don't look Lorentz invariant then there's a
real possibility that it's because the underlying physics actually isn't Lorentz invariant.
Later we will need to spend some time figuring out under what circumstances our quantum
theory keeps the classical Lorentz symmetry.

\para
In lightcone coordinates, the spacetime Minkowski metric reads
\be ds^2 = -2dX^+ dX^- + \sum_{i=1}^{D-2} dX^i dX^i\nn\ee
This means that indices are raised and lowered with $A_+=-A^-$ and $A_-=-A^+$ and $A_i=A^i$. The product of spacetime vectors reads $A\cdot B = - A^+B^--A^-B^++A^iB^i$.

\para
Let's look at the solution to the equation of motion for $X^+$. It reads,
\be X^+ = X^+_L(\sigma^+) + X_R^+(\sigma^-)\ .\nn\ee
We now gauge fix. We use our freedom of reparameterization invariance to choose coordinates such
 that
\be X^+_L = \ft12\,x^+ + \ft12\, \ap p^+ \sigma^+ \ \ \ \ ,\ \ \ \
X^+_R = \ft12\,{x^+} + \ft12\,\ap p^+ \sigma^- \ . \nn\ee
\EPSFIGURE{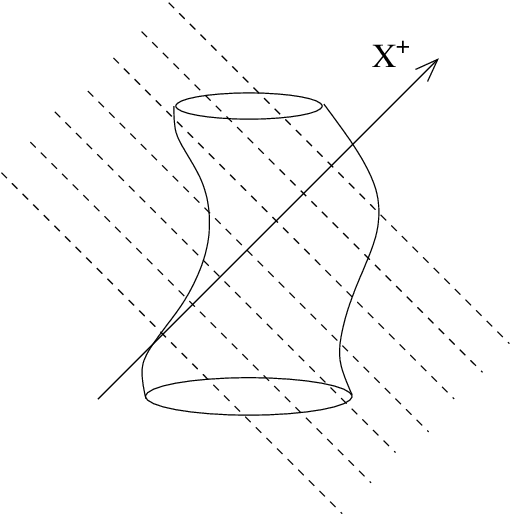,height=120pt}{}
\noindent You might think that we could go further and eliminate $p^+$ and $x^+$ but this isn't possible because we don't quite have the full freedom of reparameterization invariance since all functions should remain periodic in $\sigma$. The upshot of this choice of gauge is that
\be X^+ = x^+ + \ap p^+\,\tau  \ . \label{lcg}\ee

\noindent This is {\it lightcone gauge}. Notice that, as long as $p^+\neq 0$, we can always shift $x^+$
by a shift in $\tau$.

\para
There's something a little disconcerting about the choice \eqn{lcg}. We've identified a timelike worldsheet coordinate with a null spacetime coordinate. Nonetheless, as you can see from the
figure, it seems to be a good parameterization of the worldsheet. One could imagine that the
parameterization might break if the string is actually massless and travels in the $X^-$ direction,
with $p^+=0$. But otherwise, all should be fine.

\subsubsection*{Solving for $X^-$}

The choice \eqn{lcg} does the job of fixing the reparameterization invariance \eqn{whatsleft}.
As we will now see, it also renders the constraint equations trivial.
The first thing that we have to worry about is the possibility of  extra constraints arising
from this new choice of gauge fixing. This can be checked by looking at the equation of motion
for $X^+$,
\be \partial_+\partial_-X^-=0\nn\ee
But we can solve this by the usual ansatz,
\be X^- = X^-_L(\sigma^+) + X_R^-(\sigma^-)\ .\nn\ee
We're still left with all the other constraints \eqn{consag}. Here we see the real benefit of working
in lightcone gauge (which is actually what makes  quantization possible at all): $X^-$ is completely determined by these constraints. For example, the first of these reads
\be 2\partial_+X^-\partial_+X^+=\sum_{i=1}^{D-2}\partial_+X^i\partial_+ X^i \ee
which, using \eqn{lcg}, simply becomes
\be \partial_+X^-_L = \frac{1}{\ap p^+} \sum_{i=1}^{D-2}\,\partial_+X^i\partial_+X^i\ .\label{minus1}\ee
Similarly,
\be \partial_-X^-_R = \frac{1}{\ap p^+} \sum_{i=1}^{D-2}\,\partial_-X^i\partial_-X^i\ .\label{minus2}\ee
So, up to an integration constant, the function $X^-(\sigma^+,\sigma^-)$ is completely determined in terms of the other fields. If we write the usual mode expansion for $X^-_{L/R}$
\be
X^-_L(\sigma^+) &=& \ft12 x^- + \ft12\, \ap p^- \,\sigma^+ + i\sfap\sum_{n\neq 0}
\frac{1}{n}\,\tilde{\alpha}_n^-\, e^{-in\sigma^+} \ ,\nn\\
X^-_R(\sigma^-) &=& \ft12 x^- + \ft12\, \ap p^- \,\sigma^- + i\sfap\sum_{n\neq 0}
\frac{1}{n}\,{\alpha}_n^-\, e^{-in\sigma^-} \ .\nn\ee
then $x^-$ is the undetermined integration constant, while $p^-$, $\alpha_n^-$ and $\tilde{\alpha}_n^-$ are all fixed by the constraints \eqn{minus1} and \eqn{minus2}. For example, the oscillator modes $\alpha^-_n$ are given by,
\be \alpha^-_n = \sqrt{\frac{1}{2\ap}}\frac{1}{p^+}\sum_{m=-\infty}^{+\infty}\sum_{i=1}^{D-2} \alpha_{n-m}^i\alpha^i_m \ ,\label{iknowalpha}\ee
A special case of this is the $\alpha_0^-=\sqrt{\ap/2}\,p^-$ equation, which reads
\be \frac{\ap p^-}{2} = \frac{1}{2 p^+} \sum_{i=1}^{D-2}\left(\ft12\,\ap p^i p^i + \sum_{n\neq 0}\alpha_n^i\alpha_{-n}^i\right)\ .\label{p-1}\ee
We also get another equation for $p^-$ from the $\tilde{\alpha}^-_0$ equation arising
from \eqn{minus1}
\be \frac{\ap p^-}{2} = \frac{1}{2 p^+} \sum_{i=1}^{D-2}\left(\ft12\,\ap p^i p^i + \sum_{n\neq 0}\tilde{\alpha}_n^i\tilde{\alpha}_{-n}^i\right)\ .\label{p-2}\ee
From these two equations, we can reconstruct the old, classical, level matching conditions \eqn{levelm}.
But now with a difference:
\be M^2 = 2p^+p^- - \sum_{i=1}^{D-2}p^ip^i = \frac{4}{\ap}\sum_{i=1}^{D-2}\sum_{n>0} \alpha^i_{-n}\alpha_n^i  = \frac{4}{\ap}\sum_{i=1}^{D-2}\sum_{n>0} \tilde{\alpha}^i_{-n}\tilde{\alpha}_n^i \ .\label{lcmass}\ee
The difference is that now the sum is over oscillators $\alpha^i$ and $\tilde{\alpha}^i$ only,
with $i=1,\ldots,D-2$.
We'll refer to these as {\it transverse} oscillators. Note that the string isn't necessarily living
in the $X^0$-$X^{D-1}$ plane, so these aren't literally the transverse excitations of
the string. Nonetheless, if we specify the $\alpha^i$ then all other oscillator modes
are determined. In this sense, they are the physical excitation of the string.

\para
Let's summarize the state of play so far.  The most general classical solution is described in terms of $2(D-2)$ transverse oscillator modes $\alpha_n^i$ and $\tilde{\alpha}_n^i$, together with a number of zero modes describing the center of mass and momentum of the string: $x^i,p^i,p^+$ and $x^-$. But $x^+$ can be absorbed by a shift of $\tau$ in \eqn{lcg} and $p^-$ is constrained to obey \eqn{p-1} and \eqn{p-2}. In fact, $p^-$ can be thought of as (proportional to) the lightcone Hamiltonian. Indeed, we know that $p^-$ generates translations in $x^+$, but this is equivalent to shifts in $\tau$.

\subsubsection{Quantization}
\label{casimirsec}

Having identified the physical degrees of freedom, let's now quantize. We want to impose commutation relations.  Some of these are easy:
\be [x^i,p^j]=i\delta^{ij}\ \ \ ,\ \ \ [x^-,p^+]=-i  \nn\\
{[}\alpha^i_n,\alpha_m^j{]}={[}\tilde{\alpha}_n^i,\tilde{\alpha}^j_m{]} = n\delta^{ij}\delta_{n+m,0}
\ . \label{lccom}\ee
all of which follow from the commutation relations \eqn{acom} that we saw in covariant quantization\footnote{{\bf Mea Culpa:} We're not really supposed to
do this. The whole point of the approach that we're taking is to quantize just the physical degrees of freedom. The resulting commutation relations are not, in general, inherited from the larger theory that we started with simply by closing our eyes and forgetting about all the other fields that we've gauge fixed. We can see the problem by looking at \eqn{iknowalpha}, where $\alpha^-_n$ is determined in terms of $\alpha^i_n$. This means that the commutation relations for $\alpha_n^i$ might be infected by those of $\alpha_n^-$ which could potentially give rise to extra terms.  The correct procedure to deal with this is to figure out the Poisson bracket structure of the physical degrees of freedom in the classical theory. Or, in fancier language, the symplectic form on the phase space which
schematically looks like
\be \omega \sim \int d\sigma\ -d\dot{X}^+\wedge dX^- - d\dot{X}^-\wedge dX^++2d\dot{X}^i
\wedge dX^i\ , \nn\ee
The reason that the commutation relations \eqn{lccom} do not get infected is because the $\alpha^-$ terms in the symplectic form come multiplying $X^+$. Yet $X^+$ is given in \eqn{lcg}. It has no oscillator modes. That means that the symplectic form doesn't pick up the Fourier modes of $X^-$, and so doesn't receive any corrections from $\alpha^-_n$. The upshot of this is that the naive commutation relations \eqn{lccom} are actually right.}.

\para
What to do with $x^+$ and $p^-$? We could implement $p^-$ as the Hamiltonian acting on states. In fact,
it will prove slightly more elegant (but equivalent) if we promote both $x^+$ and $p^-$ to operators with the expected commutation relation,
\be [x^+,p^-]=-i \ .\ee
This is morally equivalent to writing $[t,H]=-i$ in non-relativistic quantum mechanics, which is true on a formal level. In the present context, it means that we can once again choose states to be eigenstates
of $p^\mu$, with $\mu=0,\ldots,D$, but the constraints \eqn{p-1} and \eqn{p-2} must still be imposed as  operator equations on the physical states. We'll come to this shortly.

\para
The Hilbert space of states is very similar to that described in covariant quantization: we define a
vacuum state, $|0;p\rangle$ such that
\be \hat{p}^\mu|0;p\rangle = p^\mu|0;p^\mu\rangle \ \ \ \ ,\ \ \ \
\alpha_n^i|0;p\rangle = \tilde{\alpha}_n^i|0;p\rangle = 0\ \ \ \ {\rm for}\ \ n>0 \label{lcground}\ee
and we build a Fock space by acting with the creation operators $\alpha^i_{-n}$ and $\tilde{\alpha}^i_{-n}$ with $n>0$. The difference with the covariant quantization is that we
only act with transverse oscillators which carry a spatial index $i=1,\ldots, D-2$.
For this reason, the Hilbert space
is, by construction, positive definite. We don't have to worry about ghosts.

\subsubsection*{The Constraints}

Because $p^-$ is not an independent variable in our theory, we must impose the constraints \eqn{p-1} and \eqn{p-2} by hand as operator equations which define the physical states. In the classical theory, we
saw that these constraints are equivalent to mass-shell conditions \eqn{lcmass}.

\para
But there's a problem when we go to the quantum theory. It's the same problem that we saw in covariant quantization: there's an ordering ambiguity in the sum over oscillator modes on the right-hand side of \eqn{lcmass}. If we choose all operators to be normal ordered then this ambiguity reveals itself in an overall constant, $a$, which we have not yet determined. The final result for the mass of states in lightcone gauge is:
\be M^2 = \frac{4}{\ap}\left(\sum_{i=1}^{D-2}\sum_{n>0} \alpha_{-n}^i\alpha_{n}^i - a\right)
= \frac{4}{\ap}\left(\sum_{i=1}^{D-2}\sum_{n>0} \tilde{\alpha}_{-n}^i\tilde{\alpha}_{n}^i - a\right)
\ \nn\ee
Since we'll use this formula quite a lot in what follows, it's useful to introduce quantities related to
the number operators of the harmonic oscillator,
\be N = \sum_{i=1}^{D-2}\sum_{n>0} \alpha^i_{-n}\alpha_n^i\ \ \  ,\ \ \ \tilde{N} = \sum_{i=1}^{D-2}\sum_{n>0} \tilde{\alpha}^i_{-n}\tilde{\alpha}_n^i \ .\label{nntilde}\ee
These are not quite number operators because of the factor of $1/\sqrt{n}$ in \eqn{whatais}. The value of $N$ and $\tilde{N}$ is often called the level. Which, if nothing else, means that the name ``level matching" makes sense. We now have
\be M^2 = \frac{4}{\ap}(N-a) = \frac{4}{\ap}(\tilde{N}-a)\ .\label{mlc}\ee
How are we going to fix $a$? Later in the course we'll see the correct way to do it. For now, I'm just going to give you a quick and dirty derivation.

\subsubsection*{The Casimir Energy}

What follows is a heuristic derivation of the normal ordering constant $a$. Suppose that we didn't notice that there was any ordering ambiguity and instead took the naive classical result directly over to the quantum theory, that is
\be
\frac{1}{2}\sum_{n\neq 0} \alpha_{-n}^i\alpha_n^i = \frac{1}{2}\sum_{n<0} \alpha_{-n}^i\alpha_n^i
+\frac{1}{2}\sum_{n>0} \alpha_{-n}^i\alpha_n^i\ .\nn\ee
where we've left the sum over $i=1,\ldots, D-2$ implicit.
We'll now try to put this in normal ordered form, with the annihilation operators $\alpha_n^i$ with $n>0$ on the right-hand side. It's the first term that needs changing. We get
\be && \frac{1}{2}\sum_{n<0}\left[\alpha_n^i\alpha_{-n}^i - n(D-2)\right]
+\frac{1}{2}\sum_{n>0} \alpha_{-n}^i\alpha_n^i = \sum_{n>0} \alpha_{-n}^i\alpha_n^i + \frac{D-2}{2}\sum_{n>0} n \nn\ \ \ .
\ee
The final term clearly diverges. But it at least seems to have a physical interpretation: it is the
sum of zero point energies of an infinite number of harmonic oscillators. In fact, we came across
exactly the same type of term in the course on quantum field theory where we learnt that, despite
the divergence, one can still extract interesting physics from this.  This is the physics of the
Casimir force.

\para
Let's recall the steps that we took to derive the Casimir force. Firstly, we introduced an ultra-violet cut-off $\epsilon \ll 1$, probably muttering some words about no physical plates being able to withstand very high energy quanta. Unfortunately, those words are no longer available to us in string theory, but let's proceed regardless. We replace the divergent sum over integers by the expression,
\be \sum_{n=1}^\infty n\  \longrightarrow \ \sum_{n=1}^\infty n e^{-\epsilon n} &=& -\frac{\partial}{\partial \epsilon}\sum_{n=1}^\infty e^{-\epsilon n} \nn\\ &=& -\frac{\partial}{\partial\epsilon} (1-e^{-\epsilon})^{-1} \nn\\
% &=& \frac{e^\epsilon}{(e^\epsilon -1)^2} \nn\\
&=& \frac{1}{\epsilon^2}-\frac{1}{12}+{\cal O}(\epsilon)\nn\ee
Obviously the $1/\epsilon$ piece diverges as $\epsilon\rightarrow 0$. This term should be renormalized
away. In fact, this is necessary to preserve the Weyl invariance of the Polyakov action since it
contributes to a cosmological constant on the worldsheet. After this renormalization,
we're left with the answer
\be \sum_{n=1}^\infty n = -\frac{1}{12} \ .\nn\ee
While heuristic, this argument does predict the correct physical Casimir energy measured in one-dimensional systems. For example, this effect is seen in simulations of quantum spin chains.

\para
What does this mean for our string? It means that we should take the unknown constant $a$ in the mass formula \eqn{mlc} to be,
\be M^2 = \frac{4}{\ap}\left(N-  \frac{D-2}{24} \right)
= \frac{4}{\ap}\left(\tilde{N} - \frac{D-2}{24} \right)
\ .\label{lcmassa}\ee
This is the formula that we will use to determine the spectrum of the string.

\subsubsection*{Zeta Function Regularization}

I appreciate that the preceding argument is not totally convincing. We could spend some time making it more robust at this stage, but it's best if we wait until later in the course when we will have the
tools of conformal field theory at our disposal. We will eventually revisit this issue and
provide a respectable derivation of the Casimir energy in Section \ref{cassec}. For now I merely offer an
even less convincing argument, known as zeta-function regularization.

\para
The zeta-function is defined, for ${\rm Re}(s)>1$, by the sum
\be \zeta(s) = \sum_{n=1}^{\infty} n^{-s} \ .\nn\ee
But $\zeta(s)$ has a unique analytic continuation to all values of $s$. In particular,
\be \zeta(-1) = -\frac{1}{12} \ .\nn\ee
Good? Good. This argument is famously unconvincing the first time you meet it! But it's actually
a very useful trick for getting the right answer.

\subsection{The String Spectrum}

Finally, we're in a position to analyze the spectrum of a single, free string.

\subsubsection{The Tachyon}

Let's start with the ground state $|0;p\rangle$ defined in \eqn{lcground}. With no oscillators
excited, the mass formula \eqn{lcmassa} gives
\be M^2 = -\frac{1}{\ap}\frac{D-2}{6}\ .\label{tmass}\ee
But that's a little odd. It's a negative mass-squared. Such particles are called {\it tachyons}.
\EPSFIGURE{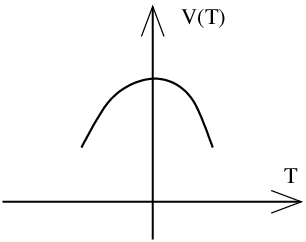,height=80pt}{}

\para
\noindent In fact, tachyons aren't quite as pathological as you might think. If you've heard of these objects
before, it's probably in the context of special relativity where they're strange beasts which
always travel faster than the speed of light. But that's not the right interpretation. Rather we should think more in the language of quantum field theory. Suppose that we have a field in spacetime --- let's call it $T(X)$ --- whose quanta will give rise to this particle. The mass-squared of the particle is
simply the quadratic term in the action, or
\be M^2 = \left.\frac{\partial ^2 V(T)}{\partial T^2}\right|_{T=0}\nn
\ee
So the negative mass-squared in \eqn{tmass} is telling us that we're expanding around a maximum of the potential for the tachyon field as shown in the figure. Note that from this perspective, the Higgs field in the standard model at $H=0$ is also a tachyon.
\EPSFIGURE{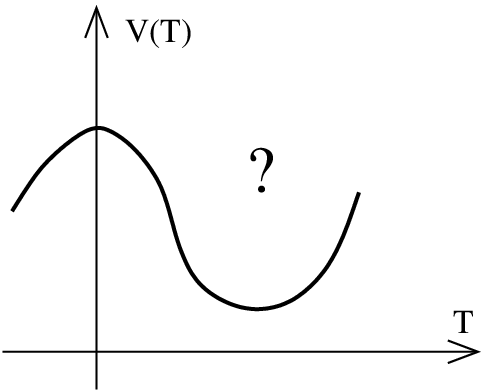,height=90pt}{}

\para
\noindent The fact that string theory turns out to sit at an unstable point in the tachyon field is unfortunate. The natural question is whether the potential has a good minimum elsewhere, as shown in the figure to the right. No one knows the answer to this! Naive attempts to understand this don't work.
We know that around $T=0$, the leading order contribution to the potential is negative and
quadratic. But there are further terms that we can compute using techniques that we'll describe
in Section \ref{scattering}. An expansion of the tachyon potential around $T=0$ looks like
\be V(T)= \frac{1}{2}M^2 T^2 + c_3T^3 + c_4 T^4 + \ldots
\nn\ee
It turns out that the $T^3$ term in the potential does give rise to a minimum. But the $T^4$ term destabilizes it again. Moreover, the $T$ field starts to mix with other scalar fields in the theory that we will come across soon. The ultimate fate of the tachyon in the bosonic string is not yet understood.

\para
The tachyon is a problem for the bosonic string. It may well be that this theory makes
no sense --- or, at the very least, has no time-independent stable solutions. Or perhaps we
just haven't worked out how to correctly deal with the tachyon. Either way,
the problem does not arise when we introduce fermions on the worldsheet and study the superstring.
This will involve several further technicalities which we won't get into in
this course. Instead, our time will be put to better use if we continue to study the bosonic string since all the lessons that we learn will carry over directly to the superstring. However, one should be
aware that the problem of the unstable vacuum will continue to haunt us throughout this course.

\para
Although we won't describe it in detail, at several times along our journey we'll make
an aside about how calculations work out for the superstring.

\subsubsection{The First Excited States}

We now look at the first excited states. If we act with a creation operator $\alpha^j_{-1}$, then
the level matching condition \eqn{mlc} tells us that we also need to act with a
$\tilde{\alpha}_{-1}^i$ operator. This gives us $(D-2)^2$ particle states,
\be \tilde{\alpha}_{-1}^i\alpha_{-1}^j\,|0;p\rangle \ ,\label{first}\ee
each of which has mass
\be M^2 = \frac{4}{\ap}\left(1-\frac{D-2}{24}\right)\ .\nn\ee
But now we seem to have a problem. Our states have space indices $i,j=1,\ldots D-2$. The operators
$\alpha^i$ and $\tilde{\alpha}^i$ each transform in the vector representation of   $SO(D-2)\subset SO(1,D-1)$ which is manifest in lightcone gauge. But ultimately we want these states
to fit into some representation of the full Lorentz $SO(1,D-1)$ group. That looks as if it's going to be hard to arrange. This is the first manifestation of the comment that we made after equation \eqn{stlc}: it's tricky to see Lorentz invariance in lightcone gauge.

\para
To proceed, let's recall Wigner's classification of representations of the Poincar\'e group.
We start by looking at massive particles in ${\bf R}^{1,D-1}$. After going to the rest frame of
the particle by setting $p^\mu=(p,0\ldots,0)$, we can watch how any internal indices transform
under the little group $SO(D-1)$ of spatial rotations. The upshot of this is that any massive
particle must form a representation of $SO(D-1)$.
But the particles described by \eqn{first} have $(D-2)^2$ states. There's no way to package
these states into a representation of $SO(D-1)$ and this means that there's no way that the first excited
states of the string can form a massive representation of the $D$-dimensional Poincar\'e group.

\para
It looks like we're in trouble. Thankfully, there's a way out. If the states are massless, then we can't go to the rest frame. The best that we can do is choose a spacetime momentum for the particle of the form $p^\mu =(p,0,\ldots,0,p)$. In this case, the particles fill out a representation of the little group $SO(D-2)$. This means that massless particles get away with having fewer internal states than
massive particles. For example, in four dimensions
the photon has two polarization states, but a massive spin-1 particle
must have three.

\para
The first excited states \eqn{first} happily sit in a representation of $SO(D-2)$. We learn
that if we want the quantum theory to preserve
the $SO(1,D-1)$ Lorentz symmetry that we started with, then these states will
have to be massless. And this is only the case if the dimension of spacetime is
\be D=26\ .\nn\ee
This is our first derivation of the critical dimension of the bosonic string.

\para
Moreover, we've found that our theory contains a bunch of massless particles.
And massless particles are interesting
because they give rise to long range forces. Let's look more closely at what massless particles the string has given us. The states \eqn{first} transform in the ${\bf 24}\otimes {\bf 24}$ representation of $SO(24)$.  These decompose into three irreducible representations:
\be \mbox{traceless symmetric} \ \oplus\ \mbox{anti-symmetric}\ \oplus\ \mbox{singlet (=trace)}\nn\ee
To each of these modes, we associate a massless field in spacetime such that the string oscillation can be identified with a quantum of these fields. The fields are:
\be G_{\mu\nu}(X)\ \ \ \ ,\ \ \ \ \ B_{\mu\nu}(X)\ \ \ \ \ ,\ \ \ \ \ \Phi(X)\ee
Of these, the first is the most interesting and we shall have more to say momentarily.
The second is an anti-symmetric tensor field which is usually
called the anti-symmetric tensor field. It also goes by the names of the ``Kalb-Ramond field" or,
in the language of differential geometry, the ``2-form".
The scalar field is called the {\it dilaton}.  These three massless fields are common to all string theories. We'll
learn more about the role these fields play later in the course.

\para
The particle in the symmetric traceless representation of $SO(24)$ is particularly interesting.
This is a massless spin 2 particle.  However, there are general arguments, due originally to
Feynman and Weinberg, that {\it any} theory of interacting
massless spin two particles must be equivalent to general relativity\footnote{A very readable
description of this can be found in the first few chapters of the Feynman Lectures on Gravitation.}.
We should therefore identify the field $G_{\mu\nu}(X)$ with the metric of spacetime.
Let's pause briefly to review the thrust of these arguments.

\subsubsection*{Why Massless Spin 2 = General Relativity}

Let's call the spacetime metric $G_{\mu\nu}(X)$. We can expand around flat space by writing
\be G_{\mu\nu}=\eta_{\mu\nu} + h_{\mu\nu}(X)\ .\nn\ee
Then the Einstein-Hilbert action has an expansion in powers of $h$. If we truncate to quadratic
order, we simply have a free theory which we may merrily quantize in the usual canonical fashion:
we promote $h_{\mu\nu}$ to an operator and introduce the associated creation and annihilation
operators $a_{\mu\nu}$ and $a_{\mu\nu}^\dagger$. This way of looking at gravity is anathema to those
raised in the geometrical world of general relativity. But from a particle physics language
it is very standard: it is simply the quantization of a massless spin 2 field, $h_{\mu\nu}$.

\para
However, even on this simple level, there is a problem
due to the indefinite signature of the spacetime Minkowski metric.
The canonical quantization relations of the creation
and annihilation operators are schematically of the form,
\be [a_{\mu\nu},a^\dagger_{\rho\sigma}] \sim \eta_{\mu\rho}\eta_{\nu\sigma} + \eta_{\mu\sigma} \eta_{\nu\rho}\nn\ee
But this will lead to a Hilbert space with negative norm states coming from acting with time-like
creation operators. For example, the one-graviton state of the form,
\be a^\dagger_{0i}|0\rangle \ee
suffers from a negative norm. This should be becoming familiar by now: it is the usual
problem that we run into if we try to covariantly quantize a gauge theory. And, indeed,
general relativity is a gauge theory. The gauge transformations are diffeomorphisms.
We would hope that this saves the theory of quantum gravity from these
negative norm states.

\para
Let's look a little more closely at what the gauge symmetry looks like for small fluctuations
$h_{\mu\nu}$.
We've butchered the Einstein-Hilbert action and left only terms quadratic in $h$.
Including all the index contractions, we find
\be S_{EH} = \frac{M_{pl}^2}{2} \int d^4x\ \left[\partial_\mu h^\rho_{\ \rho} \partial_\nu h^{\mu\nu} - \partial^\rho
h^{\mu\nu} \partial_\mu h_{\rho\nu} + \frac{1}{2} \partial_\rho h_{\mu\nu} \partial^\rho h^{\mu\nu}
-\frac{1}{2} \partial_\mu h^\nu_{\ \nu} \partial^\mu h^{\rho}_{\ \rho}\right] + \ldots\nn\ee
One can check that this truncated action is invariant under the gauge symmetry,
\be h_{\mu\nu} \ \longrightarrow \ h_{\mu\nu} + \partial_\mu\xi_\nu + \partial_\nu \xi_\mu \label{smallgauge}\ee
for any function $\xi_\mu(X)$. The gauge symmetry is the remnant of diffeomorphism invariance,
restricted to small deviations away from flat space. With this gauge invariance in hand one
can show that, just like QED, the negative norm states decouple from all physical processes.

\para
To summarize, theories of massless spin 2 fields only make sense if there is a gauge symmetry
to remove the negative norm states. In general relativity, this gauge symmetry descends
from diffeomorphism invariance. The argument of Feynman and Weinberg now runs this logic in reverse.
It goes as follows: suppose that we have a
massless, spin 2 particle. Then, at the linearized level, it must be invariant under the gauge
symmetry \eqn{smallgauge} in order to eliminate the negative norm states. Moreover,
this symmetry must survive when interaction terms are introduced. But the only way to do
this is to ensure that the resulting theory obeys diffeomorpism invariance. That means the
theory of any interacting, massless spin 2 particle is Einstein gravity,
perhaps supplemented by higher derivative terms.

\para
We haven't yet shown that string theory includes interactions for $h_{\mu\nu}$
but we will come to this later in the course. More importantly, we will also explicitly see how
Einstein's field equations arise directly in string theory.

\subsubsection*{A Comment on Spacetime Gauge Invariance}

We've surreptitiously put $\mu,\nu=0,\ldots, 25$ indices on the spacetime fields, rather than
$i,j=1,\ldots,24$. The reason we're allowed to do this is because both $G_{\mu\nu}$ and $B_{\mu\nu}$
enjoy a spacetime gauge symmetry which allows us to eliminate appropriate modes. Indeed, this is
exactly the gauge symmetry \eqn{smallgauge} that entered the discussion above.
It isn't possible to see these
spacetime gauge symmetries from the lightcone formalism of the string since, by construction,
we find only the physical states (although, by consistency alone, the gauge symmetries must be there).
One of the main
advantages of pushing through with the covariant calculation is that it does allow us to
see how the spacetime gauge symmetry emerges from the string worldsheet. Details can be found
in Green, Schwarz and Witten. We'll also briefly return to this issue in Section \ref{polyakov}.

\subsubsection{Higher Excited States}

We rescued the Lorentz invariance of the first excited states by choosing $D=26$ to ensure that they
are massless. But now we've used this trick once, we still have to worry about all the other
excited states. These also carry indices that take the range $i,j=1,\ldots, D-2=24$ and, from the mass formula \eqn{lcmassa}, they will all be massive and so must form representations of
$SO(D-1)$. It looks like we're in trouble again.

\para
Let's examine the string at  level $N=\tilde{N}=2$. In the right-moving sector, we now have two different states: $\alpha_{-1}^i\alpha_{-1}^j |0\rangle$ and $\alpha_{-2}^i|0\rangle$. The same is true for the left-moving sector, meaning that the total set of states at level 2 is (in notation that is hopefully
obvious, but probably technically wrong)
\be (\alpha_{-1}^i\alpha_{-1}^j  \oplus \alpha_{-2}^i) \otimes (\tilde{\alpha}_{-1}^i\tilde{\alpha}_{-1}^j  \oplus \tilde{\alpha}_{-2}^i)|0;p\rangle\ .\nn \ee
These states have mass $M^2=4/\ap$. How many states do we have? In the left-moving sector, we have,
\be \ft12(D-2)(D-1)+(D-2)=\ft12D(D-1)-1\ .\nn\ee
But, remarkably, that does fit nicely into a representation of $SO(D-1)$, namely the traceless symmetric tensor representation.

\para
In fact, one can show that all excited states of the string fit nicely into $SO(D-1)$ representations. The only consistency requirement that we need for Lorentz invariance is to fix up the first excited states: $D=26$.

\para
Note that if we are interested in a fundamental theory of quantum gravity, then all these excited states will have masses close to the Planck scale so are unlikely to be observable in particle physics
experiments.  Nonetheless, as we shall see when we come to discuss scattering amplitudes, it is
the presence of this infinite tower of states that tames the ultra-violet behaviour of gravity.

\subsection{Lorentz Invariance Revisited}

The previous discussion allowed to us to derive both the critical dimension and the spectrum
of string theory in the quickest fashion. But the derivation creaks a little in places. The calculation
of the Casimir energy is unsatisfactory the first time one sees it. Similarly, the explanation of the
need for massless particles at the first excited level is correct, but seems rather cheap considering
the huge importance that we're placing on the result.

\para
As I've mentioned a few times already, we'll shortly do better and gain some physical insight into these issues, in particular the critical dimension. But here I would just like to briefly sketch how one can
be a little more rigorous within the framework of lightcone quantization. The question, as
we've seen, is whether one preserves spacetime Lorentz symmetry when we quantize in lightcone gauge. We can examine this more closely.

\para
Firstly, let's go back to the action for free scalar fields \eqn{thatseasy} before we imposed  lightcone gauge fixing. Here the full Poincar\'e symmetry was manifest: it appears as a global symmetry on the worldsheet,
\be X^\mu \rightarrow \Lambda^\mu_{\ \nu}X^\nu + c^\mu\ee
But recall that in field theory, global symmetries give rise to Noether currents and their associated conserved charges. What are the Noether currents associated to this Poincar\'e transformation? We can start with the translations $X^\mu\rightarrow X^\mu + c^\mu$. A quick computation shows that the
current is,
\be P^\alpha_\mu = T \partial^\alpha X_\mu \label{transnoeth}\ee
which is indeed a conserved current since  $\partial_\alpha P^\alpha_\mu=0$ is simply the
equation of motion. Similarly, we can compute the $\ft12 D(D-1)$ currents associated to Lorentz transformations.
They are,
\be J^\alpha_{\mu\nu} = P_\mu^\alpha X_\nu - P_\nu^\alpha X_\mu\nn\ee
It's not hard to check that $\partial_\alpha J^\alpha_{\mu\nu}=0$ when the equations of motion are
obeyed.

\para
The conserved charges arising from this current are given by $M_{\mu\nu}=\int d\sigma J_{\mu\nu}^\tau$. Using the mode expansion \eqn{mode} for $X^\mu$, these can be written as
\be {\cal M}^{\mu\nu} & =& (p^\mu x^\nu-p^\nu x^\mu) -i\sum_{n=1}^\infty \frac{1}{n}\left(\alpha_{-n}^\nu\alpha_{n}^\mu - \alpha_{-n}^\mu\alpha_{n}^\nu\right) -i\sum_{n=1}^\infty \frac{1}{n}\left(
\tilde{\alpha}_{-n}^\nu \tilde{\alpha}_{n}^\mu - \tilde{\alpha}_{-n}^\mu\tilde{\alpha}_{n}^\nu\right)
\nn\\ &\equiv& l^{\mu\nu} + S^{\mu\nu} + \tilde{S}^{\mu\nu}
\nn\ee
The first piece, $l^{\mu\nu}$, is the orbital angular momentum of the string while the remaining pieces
$S^{\mu\nu}$ and $\tilde{S}^{\mu\nu}$ tell us the angular momentum  due to excited
oscillator modes. Classically, these obey the Poisson brackets of the Lorentz algebra. Moreover,
if we quantize in the covariant approach, the corresponding operators obey the commutation relations
of the Lorentz Lie algebra, namely
\be [{\cal M}^{\rho\sigma},{\cal
M}^{\tau\nu}]=\eta^{\sigma\tau}{\cal M}^{\rho\nu}-\eta^{\rho\tau}
{\cal M}^{\sigma\nu}+\eta^{\rho\nu}{\cal M}^{\sigma\tau} -
\eta^{\sigma\nu}{\cal M}^{\rho\tau} \nn\label{lorcom}\ee
However, things aren't so easy in lightcone gauge. Lorentz invariance is not guaranteed and, in
general, is not there. The right way to go about looking for it is to make sure that the
Lorentz algebra above is reproduced by the generators ${\cal M}^{\mu\nu}$. It turns out that the smoking
gun lies in the commutation relation,
\be [{\cal M}^{i-},{\cal M}^{j-}]=0 \nn\ee
Does this equation hold in lightcone gauge? The problem is that it involves the operators $p^-$ and $\alpha_n^-$, both of which are fixed by \eqn{iknowalpha} and \eqn{p-1} in terms of the other operators. So the task
is to compute this commutation relation $[{\cal M}^{i-},{\cal M}^{j-}]$, given the commutation relations \eqn{lccom} for the physical degrees of freedom, and check that it vanishes. To do this, we re-instate
the ordering ambiguity $a$ and the number of spacetime dimension $D$ as arbitrary variables and proceed.

\para
The part involving orbital
angular momenta $l^{i-}$ is fairly straightforward. (Actually, there's a small subtlety because we
must first make sure that the operator $l^{\mu\nu}$ is Hermitian by replacing $x^\mu p^\nu$ with $\ft12(x^\mu p^\nu + p^\nu x^\mu)$). The real difficulty comes from computing the commutation
relations $[S^{i-},S^{j-}]$.
This is messy\footnote{The original, classic, paper where lightcone quantization was first implemented is Goddard, Goldstone, Rebbi and Thorn ``{\it Quantum Dynamics of a Massless Relativistic String}",
Nucl. Phys. B56 (1973). A pedestrian walkthrough of this calculation can be found in the
lecture notes by Gleb Arutyunov. A link is given on the course webpage.}.
After a tedious computation, one finds,
\be
[{\cal M}^{i-},{\cal M}^{j-}]= \frac{2}{(p^+)^2} \sum_{n>0}\left(\left[\frac{D-2}{24}-1\right]n
+\frac{1}{n}\left[a-\frac{D-2}{24}\right]\right)(\alpha_{-n}^i\alpha_n^j-\alpha_{-n}^j\alpha_n^i)
+(\alpha\leftrightarrow\tilde{\alpha})
\nn\ee
The right-hand side does not, in general, vanish. We learn that the relativistic string can only be quantized in flat Minkowski space if we pick,
\be D=26\ \ \ {\rm and}\ \ \ a=1\ . \nn\ee

\subsection{A Nod to the Superstring}
\label{supernodsec}

We won't provide details of the superstring in this course, but will pause occasionally to
make some pertinent comments. Although what follows is nothing more than a list of facts, it
will hopefully be helpful in orienting you when you do come to study this material.

\para
The key difference between the bosonic string and the superstring is the addition of
fermionic modes on its worldsheet. The resulting worldsheet theory is supersymmetric. (At
least in the so-called Neveu-Schwarz-Ramond formalism). Hence the name ``superstring".
Applying the kind of quantization procedure we've discussed in this section, one finds the following results:
\begin{itemize}
\item The critical dimension of the superstring is $D=10$.
\item There is no tachyon in the spectrum.
\item The massless bosonic fields $G_{\mu\nu}$, $B_{\mu\nu}$ and $\Phi$ are all part of the spectrum
of the superstring. In this context, $B_{\mu\nu}$ is sometimes
referred to as the Neveu-Schwarz  2-form. There are also massless spacetime fermions, as well as
further massless bosonic fields. As we now discuss, the exact form of these extra bosonic fields depends on exactly what superstring theory we consider.
\end{itemize}
While the bosonic string is unique, there are a number of  discrete choices that one
can make when adding fermions to the worldsheet. This gives rise to a handful of different
perturbative superstring theories. (Although later developments reveal that they are actually all
part of the same framework which sometimes goes by the name of {\it M-theory}). The most important
of these discrete options is whether we add fermions in both the left-moving and right-moving sectors
of the string, or whether we choose  the fermions to move only in one direction, usually taken to
be right-moving. This gives  rise to two different classes of string theory.
\begin{itemize}
\item Type II strings have both left and right-moving worldsheet fermions. The resulting spacetime
theory in $D=10$ dimensions has ${\cal N}=2$ supersymmetry, which means 32 supercharges.
\item Heterotic strings have just right-moving fermions. The resulting spacetime theory has ${\cal N}=1$
supersymmetry, or 16 supercharges.
\end{itemize}
In each of these cases, there is then one further discrete choice that we can make. This leaves us with
four superstring theories. In each case, the massless bosonic fields include $G_{\mu\nu}$, $B_{\mu\nu}$ and $\Phi$
together with a number of extra fields. These are:
\begin{itemize}
\item {\bf Type IIA}:  In the type II theories, the extra massless bosonic excitations of the string are referred to as
{\it Ramond-Ramond}  fields. For Type IIA, they are a 1-form $C_\mu$ and a 3-form
$C_{\mu\nu\rho}$.
Each of these is to be thought of as a gauge field. The gauge invariant information lies in the field
strengths which take the form $F=dC$.
\item {\bf Type IIB}: The Ramond-Ramond gauge fields consist of a scalar $C$, a 2-form $C_{\mu\nu}$ and a 4-form $C_{\mu\nu\rho\sigma}$. The 4-form is restricted to have a  self-dual field strength: $F_{5}={}^\star F_{5}$. (Actually, this statement is almost true...we'll look a little
    closer at this in Section \ref{noddysec}).
\item {\bf Heterotic $SO(32)$}: The heterotic strings do not have Ramond-Ramond fields. Instead, each comes with a non-Abelian gauge field in spacetime. The heterotic strings are named after the gauge group. For example, the Heterotic $SO(32)$ string gives rise to an $SO(32)$ Yang-Mills theory in ten dimensions.
\item {\bf Heterotic $E_8\times E_8$}: The clue is in the name. This string gives rise to an $E_8\times E_8$ Yang-Mills
field in ten-dimensions.
\end{itemize}
It is sometimes said that there are five perturbative superstring theories in ten dimensions. Here we've only
mentioned four. The remaining theory is called Type I and includes open strings moving in flat
ten dimensional space as well as closed strings. We'll mention it in passing in the following section.

\newpage
\section{Open Strings and D-Branes}
\label{open}

\EPSFIGURE{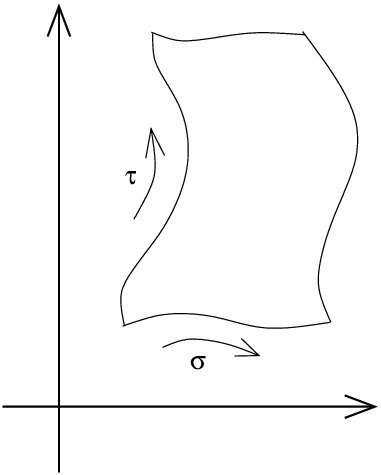,height=120pt}{}
\noindent In this section we discuss the dynamics of open strings. Clearly their distinguishing
feature is the existence of two end points. Our goal is to understand the effect of these
end points.
The spatial coordinate of the string is parameterized by
\be \sigma \in [0,\pi]\ .\nn\ee
The dynamics of a generic point on a string is governed by local physics. This means that a
generic point has no idea if it is part of a closed string or an open string. The dynamics
of an open string must therefore still be described by the Polyakov action.
But this must now be supplemented by something else: boundary conditions to tell
us how the end points move. To see this, let's look at the Polyakov action in conformal gauge
\be S = -\frac{1}{4\pi\ap}\int d^2\sigma \ \partial_\alpha X \cdot \partial^\alpha X \ .\nn\ee
As usual, we derive the equations of motion by finding the extrema of the action. This involves an integration by parts. Let's consider the string evolving from some initial configuration
at $\tau=\tau_i$ to some final configuration at $\tau=\tau_f$:
\be \delta S &=& -\frac{1}{2\pi\ap}\int_{\tau_i}^{\tau_f} d\tau \int_0^\pi d\sigma \ \, \partial_\alpha X\cdot \partial^\alpha \delta X \nn\\ &=& \frac{1}{2\pi\ap} \int d^2\sigma \  (\partial^\alpha\partial_\alpha X)\cdot \delta X + \mbox{total derivative}\nn\ee
For an open string the total derivative picks up the boundary contributions
\be \frac{1}{2\pi\ap}\left[ \int_0^\pi d\sigma \ \dot{X}\cdot
\delta X \right]^{\tau=\tau_f}_{\tau=\tau_i}\ - \frac{1}{2\pi\ap}\left[\int^{\tau_f}_{\tau_i}d\tau\
X^\prime\cdot \delta X \right]_{\sigma=0}^{\sigma=\pi}\nn\ee
The first term is the kind that we always get when using the principle of least action. The equations of motion are derived by requiring that $\delta X^\mu= 0$ at $\tau=\tau_i$ and $\tau_f$ and so it vanishes. However, the second term is novel. In order for it too to vanish, we require
\be \partial_\sigma X^\mu \, \delta X_\mu =0 \ \ \ \ \ {\rm at}\ \sigma =0,\pi\nn\ee
There are two different types of boundary conditions that we can impose to satisfy this:
\begin{itemize}
\item Neumann boundary conditions.
\be \partial_\sigma X^\mu = 0 \ \ \ \ {\rm at}\ \sigma=0,\pi \label{neumann}\ee
Because there is no restriction on $\delta X^\mu$, this condition allows the end of the string to move freely. To see the consequences of this, it's useful to repeat what we did for the closed string and work in static gauge with $X^0\equiv t= R\tau$, for some dimensionful constant $R$. Then, as in equations \eqn{haveafeel}, the constraints read
\be
\dot{\vec{x}}\cdot\vec{x}^{\,\prime} &=& 0 \ \ \ {\rm and}\ \ \
\dot{\vec{x}}{}^{\ \!2}+\vec{x}^{\,\prime}{}^{\,2} =R^2 \nn\ee
But at the end points of the string, $\vec{x}^{\,\prime}=0$. So the second equation
tells us that $|d\vec{x}/dt| =1$. Or, in other words, the end point of the string moves
at the speed of light.
\item Dirichlet boundary conditions
\be \delta X^\mu = 0\ \ \ \ \ {\rm at} \ \sigma=0,\pi\label{dirichlet}\ee
This means that the end points of the string lie at some constant position, $X^\mu = c^\mu$, in
space.
\end{itemize}
\EPSFIGURE{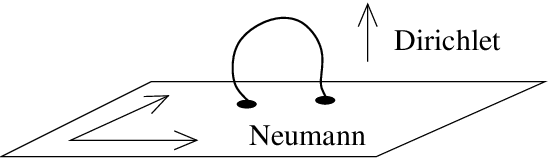,height=50pt}{}
\noindent At first sight, Dirichlet boundary conditions may seem a little odd. Why on earth would the strings be fixed at some point $c^\mu$? What is special about that point? Historically people were pretty hung up about this and Dirichlet boundary conditions were
rarely considered until the mid-1990s. Then everything changed due to an insight of Polchinski...

\para
Let's consider Dirichlet boundary conditions for some coordinates, and
Neumann for the others. This means that at both end points of the string, we have
\be \partial_\sigma X^a  &=& 0 \ \ \ \ \ \ \ \ {\rm for}\ \ a =0,\ldots,p \nn\\
X^I &=& c^I \ \ \ \ \ \ \ \ {\rm for}\ \ I = p+1,\ldots,D-1 \label{pbrane}\ee
This fixes the end-points of the string to lie in a $(p+1)$-dimensional hypersurface in spacetime such that the  $SO(1,D-1)$ Lorentz group is broken to,
\be SO(1,D-1) \rightarrow SO(1,p) \times SO(D-p-1) \ .\nn\ee
This hypersurface is called a {\it D-brane} or, when we want to specify its
dimension, a D$p$-brane. Here D stands for Dirichlet, while $p$ is the number of spatial
dimensions of the brane. So, in this language, a D$0$-brane is a particle; a $D1$-brane
is itself a string; a D$2$-brane a membrane and so on. The brane sits at specific
positions $c^I$ in the transverse space.
But what is the interpretation of this hypersurface?

\para
It turns out that the D-brane hypersurface should be thought of as a new, dynamical object in its own right. This is a conceptual leap that is far from obvious. Indeed, it took decades for people to
fully appreciate this fact. String theory is not just a theory of strings: it also contains higher dimensional branes. In Section \ref{dbranesec} we will see how these D-branes develop a life of their own. Some
comments:

\begin{itemize}
\item We've defined D-branes that are  infinite in space. However, we could just
as well define finite D-branes by specifying closed surfaces on which the string can end.
\item There are many situations where we want to describe  strings that have Neumann boundary conditions in all directions,
meaning that the string is free to move throughout spacetime. It's best to understand this in
terms of a space-filling D-brane. No Dirichlet conditions means D-branes are everywhere!
\item The D$p$-brane described above always has Neumann boundary conditions in the $X^0$
direction. What would it mean to have  Dirichlet conditions
for $X^0$? Obviously this is a little weird since the object is now localized at
a fixed point in time. But there is an interpretation of such an object: it is an {\it instanton}.
This ``D-instanton" is usually referred to as a D$(-1)$-brane. It is related to tunneling effects in the quantum theory.
\end{itemize}

\subsubsection*{Mode Expansion}

We take the usual mode expansion for the string, with $X^\mu = X^\mu_L(\sigma^+)
+X^\mu_R(\sigma^-)$, and
\be
X^\mu_L(\sigma^+) &=& \ft12 x^\mu +  \ap p^\mu \,\sigma^+ + i\sfap\sum_{n\neq 0}
\frac{1}{n}\,\tilde{\alpha}_n^\mu\, e^{-in\sigma^+} \ ,\nn\\
X^\mu_R(\sigma^-) &=& \ft12 x^\mu +  \ap p^\mu \,\sigma^- + i\sfap\sum_{n\neq 0}
\frac{1}{n}\,{\alpha}_n^\mu\, e^{-in\sigma^-} \ .\label{openmode}\ee
The boundary conditions impose relations on the modes of the string.
They are easily checked to be:
\begin{itemize}
\item Neumann boundary conditions, $\partial _\sigma X^a=0$, at the end points require that
\be \alpha_n^a = \tilde{\alpha}_n^a\label{neualpha}\ee
\item Dirichlet boundary conditions, $X^I=c^I$, at the end points require that
\be x^I=c^I\ \ \ \ ,\ \ \ \ p^I=0\ \ \ \ \ ,\ \ \ \ \ \alpha^I_n = -\tilde{\alpha}_n^I
\nn\ee
\end{itemize}
So for both boundary conditions, we only have one set of oscillators, say $\alpha_n$. The $\tilde{\alpha}_n$ are then determined by the boundary conditions.

\para
It's worth pointing out that there is a factor of 2 difference in the $p^\mu$ term between
the open string \eqn{openmode} and the closed string \eqn{mode}. This is to ensure that $p^\mu$ for
the open string retains the interpretation of the
spacetime momentum of the string when $\sigma \in [0,\pi]$. To see this, one needs to
check the Noether current associated to translations of $X^\mu$ on the worldsheet: it was given in \eqn{transnoeth}. The conserved charge is then
\be P^\mu = \int_0^\pi d\sigma\, (P^\tau)^\mu = \frac{1}{2\pi\ap}\int_0^\pi d\sigma\,\dot{X}^\mu
=p^\mu\nn\ee
as advertised. Note that we've needed to use the Neumann conditions \eqn{neualpha}
to ensure that the Fourier modes don't contribute to this integral.

\subsection{Quantization}

To quantize, we promote the fields $x^a$ and $p^a$ and $\alpha_n^\mu$
to operators. The other elements in the mode expansion are fixed by the boundary conditions.
An obvious, but important, point is that the position and momentum degrees of freedom, $x^a$ and $p^a$, have  a spacetime index that takes values
$a=0,\ldots p$. This means that the spatial wavefunctions
only depend on the coordinates of the brane not the whole spacetime. Said another, quantizing an open string gives rise to states which are restricted to lie on the brane.

\para
To determine the spectrum, it is again simplest to work in lightcone gauge. The
spacetime lightcone coordinate is chosen to lie within the brane,
\be X^\pm = \sqrt{\frac{1}{2}}\,(X^0\pm X^p)\nn\ee
Quantization now proceeds in the same manner as for the closed string until we arrive at the
mass formula for states which is a sum over the transverse  modes of the string.
\be M^2 = \frac{1}{\ap}\left(\sum_{i=1}^{p-1}\sum_{n>0}\alpha^i_{-n}\alpha^i_n +
\sum_{i=p+1}^{D-1} \sum_{n>0}\alpha^i_{-n}\alpha^i_n -a\right) \nn\ee
The first sum is over modes parallel to the brane, the second over modes perpendicular
to the brane.  It's worth commenting on the differences with the closed string formula. Firstly, there is an
overall factor of 4 difference. This can be traced to the lack of the factor of $1/2$ in front of $p^\mu$
in the mode expansion that we discussed above. Secondly, there is a sum only over $\alpha$ modes. The $\tilde{\alpha}$ modes are not independent because of the boundary conditions.

\subsubsection*{Open and Closed}

In the mass formula, we have once again left the normal ordering  constant $a$ ambiguous.
As in the closed string case, requiring the Lorentz symmetry of the quantum theory --- this
time  the reduced symmetry $SO(1,p)\times SO(D-p-1)$ ---  forces us to choose
\be D=26\ \ \ {\rm and}\ \ \ a=1\ . \nn\ee
These are the same values that we found for the closed string. This reflects an important
fact: the open string and closed string are not different theories. They are both different
states inside the same theory.

\para
\EPSFIGURE{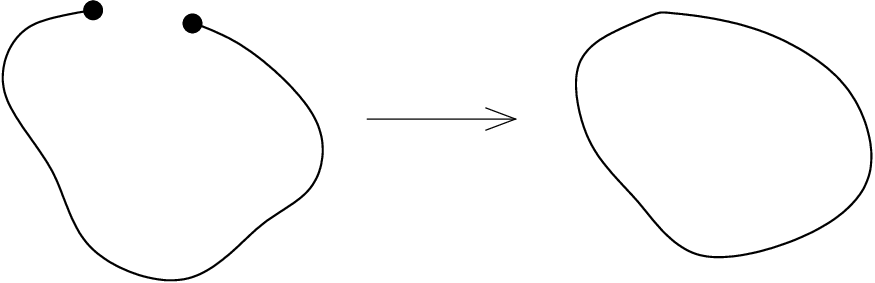,height=50pt}{}
More precisely, theories of open strings
necessarily contain closed strings. This is because, once we consider interactions, an
open string can join to form a closed string as shown in the figure. We'll look
at interactions in Section \ref{scattering}.  The question of whether this works
the other way --- meaning whether closed string
theories require open strings --- is a little more involved and is cleanest to
state in the context of the superstring. For type II superstrings,
the open strings and D-branes are necessary ingredients. For heterotic superstrings,
there appear to be no open strings and no D-branes. For the bosonic theory, it seems
likely that the open strings are a necessary ingredient although I don't know of a
killer argument. But since we're not sure whether the theory exists due to the presence
of the tachyon, the point is probably moot. In the remainder of these lectures, we'll
view the bosonic string in the same manner as the type II string and assume that the
theory includes both closed strings and open strings with their associated D-branes.

\subsubsection{The Ground State}

The ground state is defined by
\be \alpha_n^i|0;p\rangle = 0 \ \ \ \ \ \ \ \ \ \ n>0\nn\ee
The spatial index now runs over $i=1,\ldots,p-1,p+1,\ldots,D-1$. The ground state has mass
\be M^2 = - \frac{1}{\ap}\nn\ee
It is again tachyonic. Its mass is half that of the closed string tachyon. As we commented
above, this time the tachyon is confined to the brane. In contrast to the
closed string tachyon, the
open  string tachyon is now fairly well understood and its potential
\EPSFIGURE{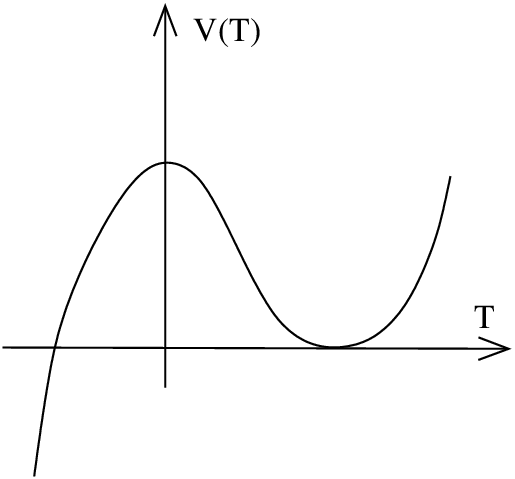,height=100pt}{}
\noindent
is of the form
shown in the
figure.
The interpretation is that the brane is unstable. It will decay, much like a
resonance state in field theory. It does this by dissolving into closed
string modes. The end point of this process -- corresponding
to the minimum at $T>0$ in the figure -- is simply a state with no D-brane.
The difference between the value of the potential at the minimum and at $T=0$
is the tension of the D-brane.

\para
Notice that although there is a minimum of the potential at $T>0$, it is not
 a global minimum. The potential seems to drop off without bound
to the left. This is still not well understood. There are suggestions that it
is related in some way to the closed string tachyon.

\subsubsection{First Excited States: A World of Light}

The first excited states are massless. They fall into two classes:
\begin{itemize}
\item Oscillators longitudinal to the brane,
\be \alpha_{-1}^a |0;p\rangle\ \ \ \ \ a=1,\ldots,p-1\nn\ee
The spacetime indices $a$ lie within the brane so this state transforms under the $SO(1,p)$
Lorentz group. It is a spin 1 particle on the brane or, in other words, it is a photon.
We introduce a gauge field $A_a$ with $a=0,\ldots, p$ lying on the brane whose
quanta are identified with this photon.
\item Oscillators transverse to the brane,
\be \alpha^I_{-1}|0;p\rangle \ \ \ \ \ \ I=p+1,\ldots,D-1\nn\ee
These states are scalars under the $SO(1,p)$ Lorentz group of the brane. They can
be thought of as arising from scalar fields $\phi^I$ living on the brane. These scalars
have a nice interpretation: they are fluctuations of the brane in the transverse directions. This is our first hint that the D-brane is a dynamical object. Note
that although the $\phi^I$ are  scalar fields under the $SO(1,p)$ Lorentz group of the brane, they
do transform as a vector under the $SO(D-p-1)$ rotation group transverse to the
brane. This appears as a global symmetry on the brane worldvolume.
\end{itemize}

\newpage
\subsubsection{Higher Excited States and Regge Trajectories}

\EPSFIGURE{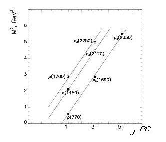,height=160pt}{}
\noindent  At level $N$, the mass of the string state is
\be M^2 = \frac{1}{\ap}(N-1)\nn\ee
The maximal spin of these states arises from the symmetric tensor. It is
\be J_{max} = N = \ap M^2 +1 \nn\ee
Plotting the spin vs. the mass-squared, we find straight lines. These are usually
called {\it Regge trajectories}. (Or sometimes Chew-Fraschuti trajectories). They are
seen in Nature in both the spectrum of mesons and baryons. Some examples involving
$\rho$-mesons are shown in the figure. These stringy Regge trajectories suggest a
naive cartoon picture of mesons as two rotating quarks connected by a
confining flux tube.

\para
The value of the string tension required to match the hadron spectrum of QCD is
$T\sim 1$ GeV. This relationship between the strong interaction
and the open string was one of the original motivations for the development of
string theory and it is from here that the parameter $\ap$ gets its (admittedly rarely
used) name ``Regge slope". In these enlightened modern times, the connection between the
open string and quarks lives on in the AdS/CFT correspondence.
{} \\ {}

\subsubsection{Another Nod to the Superstring}

Just as supersymmetry eliminates the closed string tachyon, so it removes the
open string tachyon. Open strings are an ingredient of the type II string theories.
The possible D-branes are
\begin{itemize}
\item Type IIA string theory has stable D$p$-branes with $p$ even.
\item Type IIB string theory has stable D$p$-branes with $p$ odd.
\end{itemize}
The most important reason that D-branes are stable in the type II string
theories is that they are charged under the Ramond-Ramond fields.
(This was actually Polchinski's insight that made people take D-branes
seriously). However, type II string theories also
contain unstable branes, with $p$ odd in type IIA and $p$ even in type IIB.

\para
The fifth string theory (which was actually the first to be discovered) is  called Type I. Unlike
the other string theories, it contains both open and closed strings moving in flat ten-dimensional
Lorentz-invariant spacetime. It can be thought of as the Type IIB theory with a bunch of space-filling
D9-branes, together with something called an orientifold plane. You can read about this in Polchinski.

\para
As we mentioned above, the heterotic string doesn't have (finite energy)
D-branes. This is due to an inconsistency
in any attempt to reflect left-moving modes into right-moving modes.

\subsection{Brane Dynamics: The Dirac Action}
\label{diracsec}

We have introduced D-branes as fixed boundary conditions for the open string. However,
we've already seen a hint that these objects are dynamical in their own right, since
the massless scalar excitations $\phi^I$ have a natural interpretation as transverse
fluctuations of the brane. Indeed, if a theory includes both open strings and closed strings,
then the D-branes have to be dynamical because there can be no rigid objects in a theory
of gravity. The dynamical nature of D-branes will become clearer as the course progresses.

\para
But any dynamical object should have an action which describes
how it moves. Moreover, after our discussion in Section \ref{classical}, we already
know what this is! On grounds of Lorentz invariance and reparameterization invariance alone,
the action must be a higher dimensional extension of the Nambu-Goto action. This is
\be S_{Dp} = -T_p\,\int d^{p+1}\xi\ \sqrt{-\det\gamma} \label{dric}\ee
where $T_p$ is the tension of the D$p$-brane which we will determine later, while $\xi^a$, $
a=0,\ldots p$, are the worldvolume coordinates of the brane.  $\gamma_{ab}$ is the pull back of the spacetime metric onto the worldvolume,
\be
\gamma_{ab} = \frac{\partial X^\mu}{\partial \xi^a}\,\frac{\partial X^\nu}{\partial \xi^b} \eta_{\mu\nu}\ .\nn\ee
This is called the {\it Dirac action}. It was first written down by Dirac for a membrane
some time before Nambu and Goto rediscovered it in the context of the string.

\para
To make contact with the fields $\phi^I$, we can use the reparameterization invariance
of the Dirac action to go to static gauge. For an infinite, flat D$p$-brane we can choose
\be X^a = \xi^a \ \ \ \ \ a=0,\ldots,p \ .\nn\ee
The dynamical transverse coordinates are then identified with the fluctuations $\phi^I$
through
\be X^I(\xi) = 2\pi\ap\,\phi^I(\xi)\ \ \ \ \ \ I=p+1,\ldots, D-1\nn\ee
%
%Then expanding in powers of $\partial_a\phi^I$, we find the action
%
%\be S_{Dp}= -T_p\int d^{p+1}\xi \ \left(1 + \alpha^{\prime\,2}(\partial_a\phi^I)^2 + \ldots %\right) \ee
%%
However, the Dirac action can't be the whole story. It describes the transverse fluctuations
of  the D-brane, but has nothing to say about the $U(1)$ gauge field $A_\mu$ which lives on the
D-brane. There must be some action which describes how this gauge field moves as well.
We will return to this in Section \ref{background}.

\subsubsection*{What's Special About Strings?}

We could try to quantize the Dirac action \eqn{dric} for a D-brane in the same manner that we
quantized the action for the string. Is this possible? The answer, at present, is no. There appear to be
both technical and conceptual obstacles . The technical issue is just that it's hard.
Weyl invariance was one of our chief weapons in attacking the string, but it doesn't
hold for higher dimensional objects.

\para
The conceptual issue is that quantizing a membrane, or higher dimensional object, would
not give rise to a discrete spectrum of states which have the interpretation of
particles. In this way, they appear to be fundamentally different from the string.

\para
\EPSFIGURE{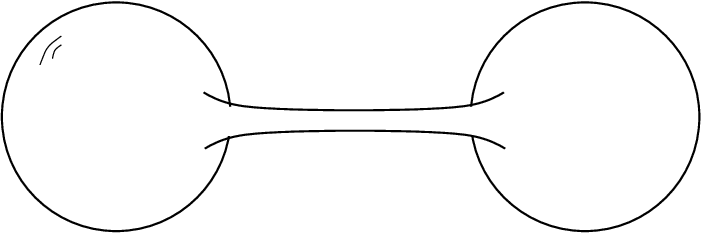,height=50pt}{}
Let's get some intuition for why this is the case. The energy of a string is proportional
to its length. This ensures that strings behave more or less like familiar elastic bands.
What about D2-branes? Now the energy is proportional to the area. In the back of your mind,
you might be thinking of a rubber-like sheet. But membranes, and higher dimensional objects,
governed by the Dirac action don't behave as household rubber sheets. They are more flexible.
This is because a membrane can form many different shapes with the
same area. For example, a tubular membrane of length $L$ and radius $1/L$ has
the same area for all values of $L$; short and stubby, or long and thin. This means
that long thin spikes can develop on a membrane at no extra cost of energy. In
particular, objects connected by long thin tubes have the same energy, regardless of
their separation. After quantization, this property gives rise to a continuous spectrum
of states. A quantum membrane, or higher dimensional object, does not have the single particle
interpretation that we saw for the string. The expectation is that the quantum membrane
should describe multi-particle states.

\subsection{Multiple Branes: A World of Glue}
\label{gluesec}

\EPSFIGURE{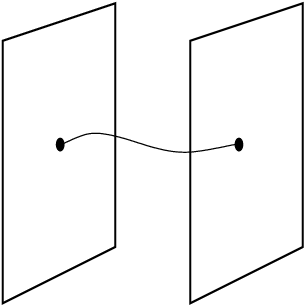,height=100pt}{}
Consider two parallel D$p$-branes. An open string now has options. It could either
end on the same brane, or stretch between the two branes. Let's consider the
string that stretches between the two. It obeys
\be
X^I(0,\tau) = c^I\ \ \ \ {\rm and}\ \ \ \ X^I(\pi,\tau)=d^I\nn\ee
where $c^I$ and $d^I$ are the positions of the two branes. In terms of the
mode expansion, this requires

\be X^I=c^I+\frac{(d^I-c^I)\sigma}{\pi} + \mbox{oscillator modes}\nn\ee
The classical constraints then read
\be \partial_+X\cdot \partial_+X = \alpha^{\prime\,2}p^2 + \frac{|\vec{d}-\vec{c}|^2}{4\pi^2}
+ \mbox{oscillator modes} =0\nn\ee
which means the classical mass-shell condition is
\be M^2 = \frac{|\vec{d}-\vec{c}|^2}{(2\pi\ap)^2} + \mbox{oscillator modes}\nn\ee
The extra term has an obvious interpretation: it is the mass of a classical string
stretched between the two branes. The quantization of this string proceeds as before.
After we include the normal ordering constant, the ground state of this string is only tachyonic if $|\vec{d}-\vec{c}|^2 < 4\pi^2\alpha^{\prime}$. Or in other words, the ground state is tachyonic
if the branes approach to a sub-stringy distance.

\para
There is an obvious generalization of this to the case of $N$ parallel branes.
Each end point of the string has $N$ possible places on which to end. We can label
each end point with a number $m,n=1,\ldots,N$ which tell us which brane it ends on.
This label is sometimes referred to as a {\it Chan-Paton factor}.

\para
Consider now the situation where all branes lie at the same position in spacetime. Each
end point can lie on one of $N$ different branes, giving $N^2$ possibilities in total. Each
of these strings has the mass spectrum of an open string, meaning that there are now $N^2$
different particles of each type. It's natural to arrange the associated fields to
 sit inside  $N\times N$ Hermitian matrices. We then have the open string tachyon $T^m_{\ n}$
and the massless fields
\be (\phi^I)^m_{\ n}\ \ \ ,\ \ \ \ (A_a)^m_{\ n}\ee
Here the components of the matrix tell us which string the field came from. Diagonal components arise from strings which have both ends on the same brane.

\para
The gauge field $A_a$ is particularly interesting. Written in this way, it looks like a $U(N)$ gauge connection. We will later see that this is indeed the case.
One can show that as $N$ branes coincide, the $U(1)^N$ gauge symmetry of the branes is
enhanced to $U(N)$. The scalar fields $\phi^I$ transform in the adjoint of this symmetry.

\newpage
\section{Introducing Conformal Field Theory}

The purpose of this section is to get comfortable with the basic language of two
dimensional conformal field theory\footnote{Much of the material covered in this
section was first described in the ground breaking paper by Belavin, Polyakov and
Zamalodchikov, ``{\it Infinite Conformal Symmetry in Two-Dimensional Quantum Field
Theory}", Nucl. Phys. B241 (1984). The application to string theory was explained
by Friedan, Martinec and Shenker in  ``{\it Conformal Invariance, Supersymmetry and
String Theory}", Nucl. Phys. B271 (1986). The canonical reference for learning conformal
field theory is the excellent review by Ginsparg. A link can be found on the course webpage.}.
This topic which has many applications outside
of string theory, most notably in statistical physics where it offers a description of
critical phenomena. Moreover, it turns out that conformal
field theories in two dimensions provide rare examples of interacting, yet  exactly solvable,
 quantum field theories.  In
recent years, attention has focussed on conformal field theories in higher dimensions
due to their role in the AdS/CFT correspondence.

\para
A {\it conformal transformation} is a change of coordinates
$\sigma^\alpha\rightarrow \tilde{\sigma}^\alpha(\sigma)$ such that the metric changes by
\be \gab(\sigma) \rightarrow \Omega^2(\sigma)\gab(\sigma)\label{cft}\ee
 A {\it
conformal field theory} (CFT) is a field theory which is invariant under these transformations.
This means that the physics of the theory looks the same at
all length scales. Conformal field theories cares about angles, but not about distances.

\para
A transformation of the form \eqn{cft} has a different interpretation depending on whether we are considering a
fixed background metric $\gab$, or a dynamical background metric. When the metric is
dynamical, the transformation is a diffeomorphism; this is a gauge symmetry. When the
background is fixed, the transformation should be thought of as an honest, physical symmetry,
taking the point $\sigma^\alpha$ to point $\tilde{\sigma}^\alpha$. This is now a global
symmetry with the corresponding conserved currents.

\para
In the context of string theory in the Polyakov formalism, the metric is dynamical and the
transformations \eqn{cft} are residual gauge transformations: diffeomorphisms which can be
undone by a Weyl transformation.

\para
In contrast, in this section we will be primarily interested in theories defined on fixed backgrounds.
Apart from a few noticeable exceptions, we will usually take this background to be flat.
This is the situation that we are used to when studying quantum field theory.

\para
Of course, we can alternate between thinking of theories as defined on fixed or fluctuating backgrounds.
Any theory of 2d gravity which enjoys both diffeomorphism and Weyl invariance will reduce to a conformally
invariant theory when the background metric is fixed. Similarly, any conformally invariant theory can be
coupled to 2d gravity where it will give rise to a classical theory which enjoys both diffeomorphism and Weyl
invariance. Notice the caveat ``classical"! In some sense, the whole point of this course is to
understand when this last statement also holds at the quantum level.

\para
Even though conformal field theories are a subset of quantum field theories, the language used
to describe them is a little different. This is partly out of necessity. Invariance under the
transformation \eqn{cft} can only hold if the theory has no preferred length scale. But this
means that there can be nothing in the theory like a mass or a Compton wavelength. In other words,
conformal field theories only support massless excitations. The questions that we ask are not
those of particles and S-matrices. Instead we will be concerned with correlation functions and
the behaviour of different operators under conformal transformations.

\subsubsection{Euclidean Space}
\label{complexsection}

Although we're ultimately interested in Minkowski signature worldsheets, it will be much
simpler and elegant if we work instead with Euclidean worldsheets. There's no funny business
here --- everything we do could also be formulated in Minkowski space.

\para
The Euclidean worldsheet coordinates are $(\sigma^1,\sigma^2) = (\sigma^1, i\sigma^0)$ and
it will prove useful to form the complex coordinates,
\be z=\sigma^1+i\sigma^2\ \ \ \ {\rm and}\ \ \ \ \bar{z}=\sigma^1-i\sigma^2\nn\ee
which are the Euclidean analogue of the lightcone coordinates.
Motivated by this analogy, it is common to refer to holomorphic functions as ``left-moving"
and anti-holomorphic functions as ``right-moving".

\para
The holomorphic derivatives are
\be \partial_z\equiv \partial = \frac{1}{2}(\partial_1-i\partial_2)\ \ \ \ {\rm and}\ \ \ \
\partial_{\bar{z}}\equiv \bar{\partial} = \frac{1}{2}(\partial_1+i\partial_2)\nn\ee
These obey $\partial z=\bar{\partial}\bar{z}=1$ and $\partial\bar{z}=\bar{\partial}z=0$.
We will usually work in flat Euclidean space, with metric
\be ds^2=(d\sigma^1)^2+ (d\sigma^2)^2 = dz\,d\bar{z}\ee
In components, this flat metric reads
\be g_{zz}= g_{\bz\bz} = 0 \ \ \ {\rm and}\ \ \ g_{z\bz} = \frac{1}{2}\nn\ee
With this convention, the measure factor is $dzd\bz=2d\sigma^1d\sigma^2$. We define
the delta-function such that $\int d^2z\, \delta(z,\bz)=1$.  Notice that because we
also have $\int d^2\sigma\,\delta(\sigma)=1$, this means that there
is a factor of $2$ difference between
the two delta functions. Vectors
naturally have their indices up: $v^z=(v^1+iv^2)$ and $v^{\bar{z}}=(v^1-iv^2)$. When indices are
down, the vectors are $v_z=\ft12(v^1-iv^2)$ and $v_{\bar{z}}=\ft12(v^1+iv^2)$.

\subsubsection{The Holomorphy of Conformal Transformations}

In the complex Euclidean coordinates $z$ and $\bz$, conformal transformations of
flat space are simple: they are any holomorphic change of coordinates,
\be z\rightarrow z^\prime  = f(z) \ \ \ \ {\rm and}\ \ \ \ \bz\rightarrow\bz^\prime = \bar{f}(\bz)\nn\ee
Under this transformation,
$ds^2 = dzd\bz \ \rightarrow\ |df/dz|^2\,dzd\bz$,
which indeed takes the form \eqn{cft}. Note that we have an infinite number of conformal transformations ---
in fact, a whole functions worth $f(z)$. This is special to conformal field theories in two
dimensions. In higher dimensions, the space of conformal transformations is a finite dimensional group.
For theories defined on ${\bf R}^{p,q}$, the conformal group is $SO(p+1,q+1)$ when $p+q>2$.

\para
A couple of particularly simple and important examples of 2d conformal transformations are
\begin{itemize}
\item $z\rightarrow z+a$: This is a translation.
\item $z\rightarrow \zeta z$: This is a rotation for $|\zeta|=1$ and a scale
transformation (also known as a {\it dilatation}) for real $\zeta \neq 1$.
\end{itemize}
For many purposes, it's simplest to treat $z$ and $\bz$ as independent variables. In doing this, we're
really extending the worldsheet from ${\bf R}^2$ to ${\bf C}^2$. This will allow us to make use of various
theorems from complex methods. However, at the end of the day we should remember that we're really sitting on
the real slice ${\bf R}^2\subset {\bf C}^2$ defined by $\bz=z^\star$.

\subsection{Classical Aspects}

We start by deriving some properties of classical theories which are invariant
under conformal transformations \eqn{cft}.

\subsubsection{The Stress-Energy Tensor}

One of the most important objects in any field theory is the {\it stress-energy tensor} (also known
as the energy-momentum tensor). This is defined in the usual way as the matrix of conserved
currents which arise from translational invariance,
\be \delta\sigma^\alpha=\epsilon^\alpha\ .\nn\ee
In flat spacetime, a translation is a special case of a conformal transformation.

\para
There's a cute way to derive the stress-energy tensor in any theory. Suppose for the moment
that we are in flat space $\gab=\eab$. Recall that we can usually
derive conserved currents by promoting the constant parameter $\epsilon$ that appears in the
symmetry to a function of the spacetime coordinates. The change in the action must then be
of the form,
\be \delta S = \int d^2\sigma \ J^\alpha\,\partial_\alpha\epsilon\label{actchange}\ee
for some function of the fields, $J^\alpha$. This ensures that the variation of the
action vanishes when $\epsilon$
is  constant, which is of course the definition of a symmetry. But when the equations
of motion are satisfied, we must have $\delta S=0$ for all variations $\epsilon(\sigma)$, not
just constant $\e$. This
means that when the equations of motion are obeyed, $J^\alpha$ must satisfy
\be \partial_\alpha J^\alpha = 0\nn\ee
The function $J^\alpha$ is our conserved current.

\para
Let's see how this works for translational invariance. If we promote $\epsilon$ to a function
of the worldsheet variables, the change of the action must be of the form \eqn{actchange}. But
what is $J^\alpha$? At this point we do the cute thing. Consider the same theory, but now
coupled to a dynamical background metric $\gab(\sigma)$. In other words, coupled to gravity.
Then we could view the transformation
\be \delta \sigma^\alpha = \epsilon^\alpha(\sigma)\nn\ee
as a diffeomorphism, and we know that the theory is invariant as long as we make the
corresponding change to the metric
\be \delta \gab = \partial_\alpha \epsilon_\beta + \partial_\beta \epsilon_\alpha\ . \nn\ee
This means that if we just make the transformation of the coordinates in our original theory, then
the change in the action  must  be the opposite of  what we get if we just transform the metric.
(Because doing both together leaves the action invariant). So we have
\be \delta S = -\int d^2\sigma\ \frac{\partial S}{\partial \gab}\,\delta\gab = -2\int d^2\sigma \
\frac{\partial S}{\partial \gab}\,\partial_\alpha\epsilon_\beta \nn\ee
Note that $\partial S/\partial \gab$ in this expression is really a functional derivatives
but we won't be careful about using notation to indicate this.
We now have the conserved current arising from translational invariance. We will add a normalization
constant which is standard in string theory (although not necessarily in other areas) and define the stress-energy
tensor to be
\be T_{\alpha\beta} = -\frac{4\pi}{\sqrt{g}}\,\frac{\partial S}{\partial g^{\alpha\beta}}\label{se}\ee
If we have a flat worldsheet, we evaluate $\tab$ on $\gab=\delta_{\alpha\beta}$ and the resulting
expression obeys $\partial^\alpha \tab=0$. If we're working on a curved worldsheet, then the
energy-momentum tensor is covariantly conserved, $\nabla^\alpha\tab=0$.

\subsubsection*{The Stress-Energy Tensor is Traceless}

In conformal theories, $\tab$ has a very important property: its trace vanishes. To see this,
let's vary the action with respect to a scale transformation which is a special case of a
conformal transformation,
\be \delta \gab = \epsilon \gab\ee
Then we have
\be \delta S = \int d^2\sigma\ \frac{\partial S}{\partial g_{\alpha\beta}}\,\delta \gab = -\frac{1}{4\pi} \int d^2\sigma
\sqrt{g}\,\epsilon\, T^\alpha_{\ \alpha}\nn\ee
But this must vanish in a conformal theory because scaling transformations are a symmetry. So
\be T^\alpha_{\ \alpha} = 0 \nn\ee
This is the key feature of a conformal field theory in any dimension.
Many theories have this feature at the classical level,
including Maxwell theory and Yang-Mills theory in four-dimensions. However, it
is much harder to preserve at the quantum level. (The weight of the world rests on the
fact that Yang-Mills theory fails to
be conformal at the quantum level). Technically the difficulty arises due to the need to
introduce a scale when regulating the theories.  Here we will be interested in two-dimensional
theories which succeed in preserving the conformal symmetry at the quantum level.
\\{}\\
{\bf Looking Ahead:} Even when the conformal invariance survives in a 2d quantum theory, the vanishing
trace $T^\alpha_{\ \alpha}=0$ will only turn out to hold in flat space. We will derive this result
in section \ref{weylanom}.

\subsubsection*{The Stress-Tensor in Complex Coordinates}

In complex coordinates, $z=\sigma^1+i\sigma^2$, the vanishing of the trace
$T^\alpha_{\ \alpha}=0$ becomes
\be T_{z\bz}=0 \nn\ee
Meanwhile, the conservation equation $\partial_\alpha T^{\alpha\beta}=0$ becomes $\partial T^{zz}=
\pb T^{\bz\bz}=0$. Or, lowering the indices on $T$,
\be \pb\, T_{zz} =0\ \ \ \ {\rm and}\ \ \ \ \p \, T_{\bz\bz} = 0\nn\ee
In other words, $T_{zz}=T_{zz}(z)$ is a holomorphic function while $T_{\bz\bz}=T_{\bz\bz}(\bz)$ is an
anti-holomorphic function. We will often use the simplified notation
\be T_{zz}(z) \equiv T(z)\ \ \ {\rm and}\ \ \  T_{\bz\bz}(\bz)\equiv \bar{T}(\bz)\nn\ee

\subsubsection{Noether Currents}

The stress-energy tensor $\tab$ provides the Noether currents for translations.
What are
the currents associated to the other conformal transformations? Consider the infinitesimal change,
\be z^\prime  = z + \epsilon(z)\ \ \ ,\ \ \ \bz^\prime  = \bz + \bar{\epsilon}(\bz)\nn\ee
where, making contact with the two examples above, constant $\epsilon$ corresponds to a
translation while $\epsilon(z)\sim z$ corresponds to a rotation and dilatation.
To compute the current, we'll use the same trick that we saw before: we promote the parameter $\epsilon$ to depend on the worldsheet coordinates. But it's already a function of half of the worldsheet
coordinates, so this now means $\epsilon(z) \rightarrow \epsilon(z,\bz)$. Then we can compute the change
in the action, again using the fact that we can make a compensating change in the metric,
\be \delta S &=& -\int d^2\sigma \ \ppp{S}{\gabi}\,\delta \gabi
\nn\\ &=& \frac{1}{2\pi}\int d^2\sigma\ \tab\, (\partial^\alpha\delta\sigma^\beta) \nn\\
&=& \frac{1}{2\pi}\int d^2z\ \frac{1}{2}\left[T_{zz}\,(\partial^z\delta z) + T_{\bz\bz}\,
(\partial^{\bz}\delta\bz)\right]
\nn\\ &=& \frac{1}{2\pi} \int d^2z\ \left[T_{zz}\,\partial_{\bz}\epsilon + T_{\bz\bz}\,\partial_z\bar{\epsilon}\right]
\label{actchang}\ee
Firstly note that if $\epsilon$ is holomorphic and $\bar{\epsilon}$ is anti-holomorphic, then we immediately have
$\delta S=0$. This, of course, is the statement that we have a symmetry on our hands. (You may wonder where in the above  derivation we used the fact that the theory was conformal. It lies in the transition to the third line where we needed $T_{z\bz}=0$).

\para
At this stage, let's use the trick of treating $z$ and $\bz$ as independent variables.
We look at separate currents that come from shifts in $z$ and shifts $\bz$.
Let's first look at the symmetry
\be \delta z = \epsilon(z)\ \ \ ,\ \ \ \delta\bz=0\nn\ee
We can read off the conserved current from \eqn{actchang} by using the standard trick of
letting the small parameter depend on position. Since $\epsilon(z)$ already depends on position,
this means promoting $\epsilon \rightarrow \epsilon(z) f (\bar{z})$ for some function $f$, and
then looking at the $\bar{\partial}f$ terms in \eqn{actchang}. This gives us the current
\be J^z=0\ \ \ {\rm and}\ \ \ J^{\bz} = T_{zz}(z)\,\epsilon(z) \equiv T(z)\,\epsilon(z)\label{confcur}\ee
Importantly, we find that the current itself is also holomorphic. We can check that this is indeed a conserved current: it should satisfy $\partial_\alpha J^\alpha = \partial_z J^z + \partial_{\bz}J^{\bz}=0$. But in fact it does so
with room to spare: it satisfies the much stronger condition $\partial_{\bz} J^{\bz}=0$.

\para
Similarly, we can look at transformations $\delta \bz = \bar{\epsilon}(\bz)$ with $\delta z=0$. We get the anti-holomorphic current $\bar{J}$,
\be \bar{J}^z=\bt(\bz)\,\bar{\epsilon}(\bz) \ \ \  {\rm and}\ \ \ \bar{J}^{\bz} = 0\label{anticc}\ee

\subsubsection{An Example: The Free Scalar Field}

Let's illustrate some of these ideas about classical conformal theories with the free
scalar field,
\be S=\frac{1}{4\pi\ap}\int d^2\sigma\ \partial_\alpha X\,\partial^\alpha X\nn\ee
Notice that there's no overall minus sign, in contrast to our earlier action \eqn{thatseasy}.
That's because we're now working with a Euclidean worldsheet metric. The theory of a free
scalar field is, of course, dead easy. We can compute anything we like in this theory. Nonetheless,
it will still exhibit enough structure to provide an example of all the abstract
concepts that we will come across in CFT. For this reason, the free scalar field will prove
a good companion throughout this part of the lectures.

\para
Firstly, let's just check that this free scalar field is actually conformal. In particular,
we can look at rescaling $\sigma^\alpha\rightarrow \lambda\sigma^\alpha$.
If we view this in the sense of an active transformation, the coordinates remain fixed
but the value of the field at point $\sigma$ gets moved to point $\lambda\sigma$. This
means,
\be X(\sigma)\ \rightarrow\ X(\lambda^{-1}\sigma) \ \ \ \ {\rm and} \ \ \ \
\ppp{X(\sigma)}{\sigma^\alpha}\ \rightarrow \ \ppp{X(\lambda^{-1}\sigma)}{\sigma^\alpha}
= \frac{1}{\lambda}\ppp{X(\tilde{\sigma})}{\tilde{\sigma}} \nn\ee
where we've defined $\tilde{\sigma}=\lambda^{-1}\sigma$. The factor of $\lambda^{-2}$ coming
from the two derivatives in the Lagrangian then cancels the Jacobian factor from the measure
$d^2\sigma = \lambda^2\,
d^2\tilde{\sigma}$, leaving the action invariant. Note that any polynomial interaction term for
$X$ would break conformal invariance.

\para
The stress-energy tensor for this theory is defined using \eqn{se},
\be \tab = - \frac{1}{\ap}\left(\partial_\alpha X \partial_\beta X - \frac{1}{2}\,\delta_{\alpha\beta}
(\partial X)^2\right)\ ,\label{classicalt}\ee
which indeed satisfies $T^\alpha_{\ \alpha}=0$ as it should. The stress-energy tensor looks
much simpler in complex coordinates. It is simple to check that $T_{z\bz}=0$ while
\be T = -\frac{1}{\ap}\,\partial X\partial X\ \ \ \ {\rm and}\ \ \ \ \bt=-\frac{1}{\ap}
\,\bp X\bp X\nn\ee
The equation of motion for $X$ is $\p\bp X=0$. The general classical solution decomposes as,
\be X(z,\bz) = X(z) + \bar{X}(\bz)\nn\ee
When evaluated on this solution, $T$ and $\bt$ become holomorphic and anti-holomorphic
functions respectively.

\subsection{Quantum Aspects}

So far our discussion has been entirely classical. We now turn to the quantum theory. The
first concept that we want to discuss is actually a feature of any quantum field theory. But
it really comes into its own in the context of CFT: it is the {\it operator product expansion}.

\subsubsection{Operator Product Expansion}
\label{opesection}

Let's first describe what we mean by a {\it local} operator in a CFT. We will also refer
to these objects as {\it fields}.  There is a slight difference in terminology between
CFTs and more general quantum field theories. Usually in quantum
field theory, one reserves the term ``field" for the objects $\phi$ which sit in the action
and are integrated over in the path integral. In contrast, in CFT the term ``field" refers to
any local expression that we can write down. This includes $\phi$, but also includes
derivatives $\p^n\phi$ or composite operators such as $e^{i\phi}$. All of these are thought of as different fields in a CFT. It should be clear from this that the set of all ``fields" in a
CFT is always infinite even though, if you were used to working with quantum field theory, you would
talk about  only a finite number of fundamental objects $\phi$. Obviously, this is nothing to
be scared about. It's just a change of language: it doesn't mean that our theory got harder.

%\para
%This difference in terminology between CFT and other quantum field theories reflects what we
%can do with a CFT. The fields $\phi$ that appears in the path integral are only fundamental
%objects if the field theory is weakly coupled, since only then does their quantization lead
%directly to the particle spectrum. However, in general the relevant degrees of freedom may
%have nothing to do with the fields that we integrate over in the path integral. The techniques
%of CFT often allow us to move away from weakly coupled field theories, and this is perhaps
%the reason that the terminology is a little different.

\para
We now define the {\it operator product expansion} (OPE). It is a statement about what happens as
local operators approach each other.
The idea is that two local operators inserted at nearby points can be closely approximated
by a string of operators at one of these points. Let's denote all the local operators of the CFT by
${\cal O}_i$, where $i$ runs over the set of all operators. Then the OPE is
\be {\cal O}_i(z,\bz)\,{\cal O}_j(w,\bw) = \sum_{k}C_{ij}^k(z-w,\bz-\bw)\,{\cal O}_k(w,\bw)
\label{ope}\ee
\EPSFIGURE{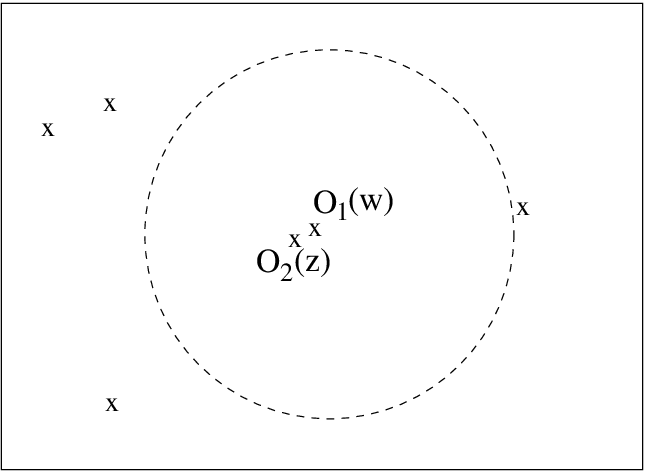,height=100pt}{}
\noindent Here $C_{ij}^k(z-w,\bz-\bw)$ are a set of functions which, on grounds of
translational invariance,
depend only on the separation between the two operators. We will write a lot of operator
equations of the form \eqn{ope} and it's important to clarify exactly what they mean:
they are always to be understood as statements which hold as operator insertions inside
time-ordered correlation functions,
\be \langle {\cal O}_i(z,\bz)\,{\cal O}_j(w,\bw)\ldots\ \rangle = \sum_{k}C_{ij}^k(z-w,\bz-\bw)\
\langle {\cal O}_k(w,\bw)\ldots\ \rangle \nn\ee
where the $\ldots$ can be any other operator insertions that we choose. Obviously it would be tedious
to continually write $\langle\ldots \rangle$. So we don't. But it's always implicitly there.
There are further caveats about the OPE that are worth stressing
\begin{itemize}
\item The correlation functions are always assumed to be time-ordered. (Or something similar
that we will discuss in Section \ref{radial}). This means that as far as the OPE is
concerned, everything commutes since the ordering of operators
is determined inside the correlation function anyway. So we must have
${\cal O}_i(z,\bz)\,{\cal O}_j(w,\bw)= {\cal O}_j(w,\bw)\,{\cal O}_i(z,\bz)$.
(There is a caveat here: if the operators are Grassmann objects, then they pick
up an extra minus sign when commuted, even inside time-ordered products).
\item The other operator insertions in the correlation function (denoted $\ldots$ above) are arbitrary. {\it Except} they should be at a distance large compared to $|z-w|$. It turns
    out --- rather remarkably --- that in a CFT the OPEs are exact statements, and have a radius
    of convergence equal to the distance to the nearest other insertion. We will return to this in Section \ref{stateopmap}. The radius of convergence is denoted in the figure by the dotted line.
\item The OPEs have singular behaviour as $z\rightarrow w$. In fact, this singular behaviour will
really be the only thing we care about! It will turn out to contain the same information as
commutation relations, as well as telling us how operators transform under symmetries. Indeed, in many equations we will simply write the singular terms in the OPE and denote the non-singular terms as $+\ldots$.
\end{itemize}

\subsubsection{Ward Identities}

 The spirit of Noether's theorem in quantum field theories is captured by operator equations known as {\it Ward Identities}. Here we derive the Ward identities associated to conformal invariance. We start
 by considering a general theory with a symmetry. Later we will restrict to conformal symmetries.

\subsubsection*{Games with Path Integrals}

We'll take this opportunity to get comfortable with some basic techniques using path
integrals. Schematically, the path integral takes the form
 \be Z = \int {\cal D}\phi\ e^{-S[\phi]}\nn\ee
where $\phi$ collectively denote all the fields (in the path integral sense...not the CFT sense!).
A symmetry of the quantum theory is such that an infinitesimal transformation
\be \phi^\prime  = \phi + \epsilon\delta \phi\nn\ee
leaves both the action {\it and} the measure invariant,
\be S[\phi^\prime] = S[\phi]\ \ \ \ {\rm and} \ \ \ \ {\cal D}\phi^\prime = {\cal D}\phi\nn
\ee
(In fact, we only really need the combination ${\cal D}\phi\,e^{-S[\phi]}$ to be invariant but
this subtlety won't matter in this course). We use the same trick that we employed
earlier in the classical
theory and promote $\epsilon\rightarrow \epsilon(\sigma)$. Then, typically, neither the action
nor the measure are invariant but, to leading order in $\epsilon$, the change has to be proportional
to $\p \e$. We have
\be Z &\longrightarrow& \int {\cal D}\phi^\prime\ \exp\left(-S[\phi^\prime]\right) \nn\\ &=&
\int {\cal D}\phi\ \exp\left(-S[\phi] -\frac{1}{2\pi}\int J^\alpha\,\partial_\alpha
\epsilon\right) \nn\\ &=& \int {\cal D}\phi\ e^{-S[\phi]}\left(1 -\frac{1}{2\pi}\int  J^\alpha\,\partial_\alpha\epsilon \right)\nn\ee
where the factor of $1/2\pi$ is merely a convention and $\int$ is shorthand for
$\int d^2\sigma \sqrt{g}$. Notice that the current $J^\alpha$ may now also have contributions
from the measure transformation as well as the action.

\para
Now comes the clever step. Although the integrand has changed, the actual value of the partition
function can't have changed at all. After all, we just redefined a dummy integration variable $\phi$.
So the expression above must be equal to the original $Z$. Or, in other words,
\be \int {\cal D}\phi\,e^{-S[\phi]}\ \left(\int J^\alpha\,\partial_\alpha\epsilon\right) = 0
\nn\ee
Moreover, this must hold for all $\epsilon$. This gives us the quantum version of Noether's
theorem: the vacuum expectation value of the divergence of the current vanishes:
\be \langle \partial_\alpha J^\alpha\rangle =0\ . \nn\ee
We can repeat these tricks of this sort to derive some stronger statements. Let's
see what happens when we have
other insertions in the path integral. The time-ordered correlation function is given by
\be \langle {\cal O}_1(\sigma_1)\ldots{\cal O}_n(\sigma_n)\rangle = \frac{1}{Z}\int {\cal D}\phi\,e^{-S[\phi]}
\,{\cal O}_1(\sigma_1)\ldots{\cal O}_n(\sigma_n)\nn\ee
We can think of these as operators inserted at particular points on the plane as shown in the
figure. As we described above, the operators $\calo_i$ are any general expressions that we can form
from the $\phi$ fields. Under the symmetry of interest, the operator will change in some way, say
\be {\cal O}_i\rightarrow {\cal O}_i+\epsilon\,\delta{\cal O}_i\nn\ee
We once again promote $\epsilon\rightarrow \e(\sigma)$. As our first pass, let's pick a choice of $\epsilon(\sigma)$ which only has support away
from the operator insertions as shown in the Figure 21. Then,
\be \delta{\cal O}_i(\sigma_i) = 0\nn\ee
\EPSFIGURE{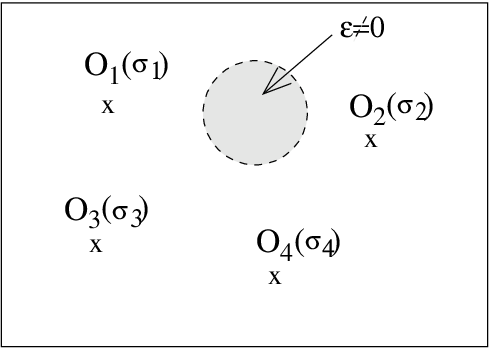,height=90pt}{}
\noindent
and the above derivation goes through in exactly the same way to give
\be \langle \partial_\alpha J^\alpha(\sigma)\,{\cal O}_1(\sigma_1)\ldots {\cal O}_n(\sigma_n)\rangle
=0 \ \ \ \ \ {\rm for}\ \sigma\neq\sigma_i \nn\ee
Because this holds for any operator insertions away from $\sigma$,
from the discussion in
Section \ref{opesection} we are entitled to write the operator equation
\be \partial_\alpha J^\alpha =0\ \ \ \ \ \ \ \ \ \ \ \ \ \ \ \ \ \ \ \ \ \ \nn\ee
\EPSFIGURE{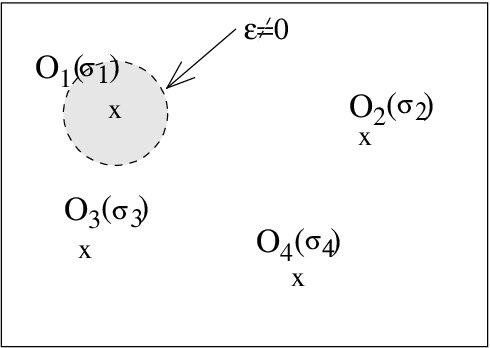,height=90pt}{}
%\noindent
But what if there are operator insertions that lie at the same point as $J^\alpha$? In
other words, what
happens as $\sigma$ approaches one of the insertion points? The resulting formulae are
called Ward identities. To derive these, let's take $\epsilon(\sigma)$ to have
support in some region that includes the point $\sigma_1$, but not the other points
as shown in Figure 22. The simplest choice is just to take $\epsilon(\sigma)$ to be constant
inside the shaded region, and zero outside.
Now using the same procedure as before, we find that the
original correlation function is equal to,
\be \frac{1}{Z}\int {\cal D}\phi\,e^{-S[\phi]}\ \left(1-\frac{1}{2\pi} \int
J^\alpha\,\partial_\alpha\epsilon\right) \ ({\cal O}_1+\epsilon\,\delta{\cal O}_1)\,{\cal O}_2
\ldots {\cal O}_n
\nn\ee
Working to leading order in $\epsilon$, this gives
\be
-\frac{1}{2\pi} \int_{\epsilon}\ \partial_\alpha
\langle J^\alpha(\sigma)\,{\cal O}_1(\sigma_1)\ldots\rangle = \langle\delta{\cal O}_1(\sigma_1)
\ldots\rangle\label{ward}\ee
where the integral on the left-hand-side is only over the region of non-zero $\epsilon$. This
is the {\it Ward Identity}.

\subsubsection*{Ward Identities for Conformal Transformations}

Ward identities \eqn{ward} hold for any symmetries. Let's now see what they give when applied
to conformal transformations. There are two further
steps needed in the derivation. The first simply comes from the fact that we're working
in two dimensions and we can use Stokes' theorem to convert the integral on the left-hand-side
of \eqn{ward} to a line integral around the boundary.
Let $\hat{n}^\alpha$ be the unit vector normal to the
boundary. For any vector $J^\alpha$, we have
\be \int_\epsilon\ \partial_\alpha J^\alpha = \oint_{\partial\epsilon}\, J_\alpha\hat{n}^\alpha
&=& \oint_{\partial\epsilon}\, (J_1\,d\sigma^2-J_2\,d\sigma^1) =
-i\oint_{\partial\epsilon}\, (J_z\,dz-J_{\bz}\,d\bz)\nn\ee
where we have written the expression both in Cartesian coordinates $\sigma^\alpha$ and complex
coordinates on the plane. As described in Section \ref{complexsection}, the complex components
of the vector with indices down are defined as $J_z=\ft12(J_1-iJ_2)$ and $J_{\bz}=\ft12(J_1+iJ_2)$.
So, applying this to the Ward identity \eqn{ward}, we find for two dimensional theories
\be \frac{i}{2\pi}\oint_{\partial\epsilon}dz\ \langle J_z(z,\bz)\,{\cal O}_1(\sigma_1)\ldots\rangle
-\frac{i}{2\pi} \oint_{\partial\epsilon}d\bz\ \langle J_{\bz}(z,\bz)\,{\cal O}_1(\sigma_1)\ldots\rangle
=\langle\delta{\cal O}_1(\sigma_1)\ldots\rangle\nn\ee
So far our derivation holds for any conserved current $J$ in two dimensions. At this stage
we specialize to the currents that arise from conformal transformations \eqn{confcur} and
\eqn{anticc}. Here something nice happens because $J_z$ is holomorphic while $J_{\bz}$ is
anti-holomorphic. This means that the contour integral simply picks up the residue,
\be \frac{i}{2\pi}\oint_{\partial\epsilon}dz\, J_z(z){\cal O}_1(\sigma_1)  = -\,\Res[J_z{\cal O}_1]\nn\ee
where this means the residue in the OPE between the two operators,
\be J_z(z)\,{\cal O}_1(w,\bw) = \ldots + \frac{\Res[J_z{\cal O}_1(w,\bw)]}{z-w}+\ldots\nn\ee
So we find a rather nice way of writing the Ward identities for conformal transformations. If we again
view $z$ and $\bar{z}$ as independent variables, the Ward identities split into two pieces. From
the change $\delta z = \epsilon(z)$, we get
\be \delta {\cal O}_1(\sigma_1) = -\Res[J_z(z){\cal O}_1(\sigma_1)] = -\Res[\e(z)T(z){\cal O}_1(\sigma_1)]
\label{wardope}\ee
where, in the second equality, we have used the expression for the conformal current
\eqn{confcur}.
Meanwhile, from the change $\delta \bar{z} = \bar{\e}(\bar{z})$, we have
\be \delta {\cal O}_1(\sigma_1) = - \Res[\bar{J}_{\bz}(\bz){\cal O}_1(\sigma_1)]
=-\Res[\bar{\e}(\bz)\bt(\bz){\cal O}_1(\sigma_1)]\nn\ee
where the minus sign comes from the fact that the $\oint d\bz$
boundary integral is taken in the opposite direction.

\para
This result means that if we know the OPE between an operator and the stress-tensors $T(z)$
and $\bt(\bz)$, then we immediately know how the operator transforms
under conformal symmetry. Or, standing this on its head, if we know how an operator transforms
then we know at least some part of its OPE with $T$ and $\bt$.

\subsubsection{Primary Operators}
\label{primarysec}

The Ward identity allows us to start piecing together some OPEs by looking at how operators
transform under conformal symmetries.  Although we don't yet know
the action of general conformal symmetries, we can start to make progress by looking at the
two simplest examples.
\\{}\\
{\bf Translations:} If $\delta z=\epsilon$, a constant, then all operators transform as
\be {\cal O}(z-\epsilon) = {\cal O}(z) - \epsilon\,\partial{\cal O}(z) + \ldots \nn\ee
The Noether current for translations is the stress-energy tensor $T$. The Ward identity in
the form \eqn{wardope} tells us that the OPE of $T$ with any operator ${\cal O}$ must be
of the form,
\be T(z)\,{\cal O}(w,\bw) = \ldots + \frac{\partial{\cal O}(w,\bw)}{z-w} + \ldots \label{tope}
\ee
Similarly, the OPE with $\bt$ is
\be \bt(\bz)\,{\cal O}(w,\bw) = \ldots + \frac{\bp{\cal O}(w,\bw)}{\bz-\bw} + \ldots \label{btope}
\ee
\\
{\bf Rotations and Scaling:} The transformation
\be z\rightarrow z+\epsilon z\ \ \ \ {\rm and}\ \ \ \bz\rightarrow \bz+\bar{\e}\bz\label{rsz}\ee
describes rotation for $\e$ purely imaginary, and scaling (dilatation) for $\e$ real. Not
all operators have good transformation
properties under these actions. This is entirely analogous to the statement in
quantum mechanics that not all states transform nicely under the Hamiltonian $H$ and
angular momentum operator $L$. However, in quantum mechanics we know that the eigenstates of
$H$ and $L$ can be chosen as a basis of the Hilbert space provided, of
course, that $[H,L]=0$.

\para
The same statement holds for operators in a CFT:
we can choose a basis of local operators that have good transformation properties under rotations and dilatations.
In fact, we will see in Section \ref{stateopmap} that the statement about local operators actually follows
from the statement about states. 
\\{}\\
{\bf Definition:}  An operator ${\cal O}$ is said
to have {\it weight} $(h,\tilde{h})$ if, under $\delta z=\e z$ and $\delta\bz=\bar{\e}\bz$, $\calo$ transforms as
\be \delta {\cal O} = -\e(h{\cal O}+z\,\partial {\cal O})-\bar{\e}(\tilde{h}{\cal O}+\bz\,\bp {\cal O})\label{quasip}\ee
The terms  $\partial{\cal O}$ in this expression would be there for any operator. They simply
come from expanding ${\cal O}(z-\e z, \bz-\bar{\e}\bz)$. The terms $h{\cal O}$ and
$\tilde{h}{\cal O}$ are  special to operators which 
are eigenstates of dilatations and rotations.  Some comments:
\begin{itemize}
\item Both $h$ and $\tilde{h}$ are real numbers.  In a unitary CFT, all operators have
$h,\tilde{h} \geq 0$. We will prove this is Section \ref{unitsec}.
\item The weights are not as unfamiliar as they appear. They simply tell us how operators
transform under rotations and scalings. But we already have names for these concepts from
undergraduate days. The eigenvalue under rotation is usually called the {\it spin}, $s$,
and is given in terms of the weights as
\be s=h-\tilde{h}\nn\ee
Meanwhile, the {\it scaling dimension} $\Delta$ of an operator is
\be \Delta = h+\tilde{h}\nn\ee
\item To motivate these definitions, it's worth recalling how rotations and scale transformations
act on the underlying coordinates.
Rotations are implemented by the operator
\be L=-i(\sigma^1\partial_2-\sigma^2\partial_1)=z\partial-\bz\bp\nn\ee
while the dilation operator $D$ which gives rise to scalings is
\be D=\sigma^\alpha \partial_\alpha = z\partial + \bz\bp\nn\ee
\item The scaling dimension is nothing more than the familiar ``dimension" that we usually associate to fields and
operators by dimensional analysis. For example, worldsheet derivatives always
increase the dimension of an operator by
one: $\Delta[\partial]=+1$.
The tricky part is that the naive dimension that fields have in the classical theory is
not necessarily the same as the dimension in the quantum theory.

\end{itemize}
Let's compare the transformation law  \eqn{quasip} with the Ward identity
\eqn{wardope}. The Noether current arising from rotations and scaling $\delta z = \e z$ was given in
\eqn{confcur}: it is $J(z)=zT(z)$. This means that
the residue of the $J{\cal O}$ OPE will determine the $1/z^2$ term in the $T{\cal O}$ OPE. Similar
arguments hold, of course, for $\delta \bz=\bar{\e}\bz$ and $\bt$. So, the upshot of this is
that, for an operator $\calo$ with weight $(h,\tilde{h})$, the OPE with $T$ and $\bt$ takes the form
\be T(z)\,{\cal O}(w,\bw) &=& \ldots + h\frac{{\cal O}(w,\bw)}{(z-w)^2} + \frac{\partial{\cal O}(w,\bw)}{z-w} +
\ldots \nn\\
\bt(\bz)\,{\cal O}(w,\bw) &=& \ldots + \tilde{h}\frac{{\cal O}(w,\bw)}{(\bz-\bw)^2} + \frac{\bp{\cal O}(w,\bw)}{\bz-\bw}
+\ldots \nn\ee

\subsubsection*{Primary Operators}

A {\it primary} operator is one whose OPE with $T$ and $\bt$ truncates at order
$(z-w)^{-2}$ or order $(\bz-\bw)^{-2}$ respectively.
There are no higher singularities:
\be T(z)\,{\cal O}(w,\bw) &=&  h\frac{{\cal O}(w,\bw)}{(z-w)^2} + \frac{\partial{\cal O}(w,\bw)}{z-w} +
\mbox{non-singular} \nn\\
\bt(\bz)\,{\cal O}(w,\bw) &=& \tilde{h}\frac{{\cal O}(w,\bw)}{(\bz-\bw)^2} + \frac{\bp{\cal O}(w,\bw)}{\bz-\bw}
+ \mbox{non-singular} \nn\ee
Since we now know all singularities in the $T\calo$
OPE, we can reconstruct the transformation under all conformal transformations.
The importance of primary operators
is that they have particularly simple transformation properties.
Focussing on $\delta z=\e(z)$, we have
\be \delta {\cal O}(w,\bw)&=&-\Res[\e(z)\,T(z)\,{\cal O}(w,\bw)]
\nn\\ &=& -\Res\left[\e(z)\left(
h\frac{{\cal O}(w,\bw)}{(z-w)^2}+\frac{\partial{\cal O}(w,\bw)}{z-w}+\ldots \right)\right]\nn\ee
We want to look at smooth conformal transformations, and so require that $\e(z)$ itself has no singularities at $z=w$. We can then Taylor expand
\be \e(z)=\e(w)+\e^\prime(w)\,(z-w) + \ldots \nn\ee
We learn that the infinitesimal change of a primary operator under a general
conformal transformation $\delta z = \e(z)$ is
\be \delta{\cal O}(w,\bw) = -h\e^\prime(w)\,{\cal O}(w,\bw) - \e(w)\,\p{\cal O}(w,\bw)\label{primary}\ee
There is a similar expression for the anti-holomorphic transformations $\delta\bz = \bar{\e}(\bz)$.

\para
Equation \eqn{primary} holds for infinitesimal conformal transformations. It is a simple matter to integrate up to find how primary operators change under a finite conformal transformation,
\be z \rightarrow \tilde{z}(z)\ \ \ \ {\rm and}\ \ \ \ \bz\rightarrow \bar{\tilde{z}}(\bz)\nn\ee
The general
transformation of a primary operator is given by
\be \calo(z,\bz) \ \rightarrow\ \tilde{\calo}(\tilde{z},\bar{\tilde{z}}) = \left(\ppp{\tilde{z}}{z}\right)^{-h}\,\left(
\ppp{\bar{\tilde{z}}}{\bz}\right)^{-\tilde{h}}\,\calo(z,\bz)\label{goprimary}\ee
It will turn out that one of the main objects of interest in a CFT is the spectrum of weights $(h,\tilde{h})$
of primary fields. This will be equivalent to computing the particle mass spectrum
in a quantum field theory. In the context of statistical mechanics,
the weights of primary operators are the critical exponents.

\subsection{An Example: The Free Scalar Field}
\label{freesec}

Let's look at how all of this works for the free scalar field. We'll start by familiarizing
ourselves with some techniques using the path integral. The action is,
\be S=\frac{1}{4\pi\ap}\is\,\p_\alpha X\,\p^\alpha X\label{xact}\ee
The classical equation of motion is $\partial^2 X=0$. Let's start by seeing how to derive the
analogous statement in the quantum theory using the path integral. The key fact that we'll need is
that the integral of a total derivative vanishes in the path integral just as it does in an
ordinary integral. From this we have,
\be 0 = \int {\cal D}X\ \frac{\delta}{\delta X(\sigma)}\,e^{-S} = \int {\cal D}X\ e^{-S}\,\left[
\frac{1}{2\pi\ap}\,\partial^2 X(\sigma)\right]\nn\ee
But this is nothing more than the Ehrenfest theorem which states that expectation values
of operators obey the classical equations of motion,
\be \langle \partial^2 X(\sigma)\rangle =0\nn\ee

\subsubsection{The Propagator}

The next thing that we want to do is compute the propagator for $X$. We could do this using
canonical quantization, but it will be useful to again see how it works using the path
integral. This time we look at,
\be 0 = \int {\cal D}X\ \frac{\delta}{\delta X(\sigma)}\,\left[e^{-S}\,X(\sigma^\prime)\right]
  &= \int {\cal D}X\ e^{-S}\,\left[
\frac{1}{2\pi\ap}\,\partial^2 X(\sigma)\,X(\sigma^\prime) + \delta(\sigma-\sigma^\prime)\right]\nn\ee
So this time we learn that
\be \langle \partial^2 X(\sigma)\,X(\sigma^\prime)\rangle = -2\pi\ap\,\delta(\sigma-\sigma^\prime)
\label{letsolve}\ee
Note that if we'd computed this in the canonical approach, we would have found the same answer: the
$\delta$-function arises in this calculation because all correlation functions are time-ordered.

\para
We can now treat \eqn{letsolve} as a differential equation for the propagator $\langle X(\sigma)X(\sigma^\prime)\rangle$.  To solve this equation, we need the following standard result
\be {\partial^2} \ln(\sigma-\sigma^\prime)^2
%= 2\p\pb\ln|z|^2
= 4\pi \delta(\sigma-\sigma^\prime)\label{logeqn}\ee
Since this is important, let's just quickly check that it's true. It's a simple
application of Stokes' theorem. Set $\sigma^\prime=0$ and integrate over $\int d^2\sigma$.
We obviously get $4\pi$ from the right-hand-side. The left-hand-side
gives
\be \int d^2\sigma\ \partial^2\ln (\sigma_1^2+\sigma_2^2) = \int d^2\sigma\ \partial^\alpha\left(\frac{2\sigma_\alpha}{\sigma_1^2+\sigma_2^2}\right)
= 2\oint \frac{(\sigma_1\,d\sigma^2-\sigma_2d\sigma^1)}{\sigma_1^2+\sigma_2^2}
\nn\ee
Switching to polar coordinates $\sigma_1+i\sigma_2=re^{i\theta}$, we can rewrite
this expression as
\be 2\int \frac{r^2d\theta}{r^2} = 4\pi\nn\ee
confirming \eqn{logeqn}. Applying this result to our equation \eqn{letsolve}, we get the
propagator of a free scalar in two-dimensions,
\be \langle X(\sigma)X(\sigma^\prime)\rangle = -\frac{\ap}{2}\,\ln(\sigma-\sigma^\prime)^2  \nn\ee
The propagator has a singularity as $\sigma\rightarrow \sigma^\prime$. This is an ultra-violet divergence and is common to all field theories. It also has a singularity
as $|\sigma-\sigma^\prime|\rightarrow \infty$. This is telling us something important that
we mention below in Section \ref{goldsec}.

\para
Finally, we could repeat our trick of looking at total derivatives in the
path integral, now with other operator insertions $\calo_1(\sigma_1),\ldots
\calo_n(\sigma_n)$ in the path integral. As long as $\sigma,\sigma^\prime\neq \sigma_i$,
then the whole analysis goes through as before. But this is exactly our criterion to
write the operator product equation,
\be X(\sigma)X(\sigma^\prime) = -\frac{\ap}{2}\,\ln(\sigma-\sigma^\prime)^2 + \ldots
\label{xprob}\ee
We can also write this in complex coordinates. The classical equation of motion
$\p\bp X=0$ allows us to split the operator $X$ into left-moving and right-moving
pieces,
\be X(z,\bz) = X(z) + \bar{X}(\bz)\nn\ee
We'll focus just on the left-moving piece. This has the operator product expansion,
\be X(z)X(w) = -\frac{\ap}{2}\,\ln(z-w)+\ldots
\nn\ee
The logarithm means that $X(z)$ doesn't have any nice properties under the conformal
transformations. For this reason,  the ``fundamental field" $X$ is not really
the object of interest in this theory! However, we can look at the derivative
of $X$. This has a rather nice looking OPE,
\be \partial X(z)\,\partial X(w) = -\frac{\ap}{2}\,\frac{1}{(z-w)^2} + \mbox{non-singular}
\label{pxpx}\ee

\subsubsection{An Aside: No Goldstone Bosons in Two Dimensions}
\label{goldsec}

The infra-red divergence in the propagator has an important physical implication.
Let's start by pointing out one of the big differences between quantum mechanics and
quantum field theory in $d=3+1$ dimensions. Since the language used to describe
these two theories is rather different, you may not even be aware that this
difference exists.

\para
Consider the quantum mechanics of a particle on a line. This is a $d=0+1$ dimensional
theory of a free scalar field $X$. Let's prepare the particle in some localized
state -- say a Gaussian wavefunction $\Psi(X) \sim \exp(-X^2/L^2)$. What then
happens? The wavefunction starts to spread out. And the spreading doesn't stop. In fact,
the would-be ground state of the system is a uniform  wavefunction of infinite width, which
isn't a state in the Hilbert space because it is non-normalizable.

\para
Let's now compare this to the situation of a free scalar field $X$ in a $d=3+1$ dimensional
field theory. Now we think of this as a scalar without potential. The physics is
very different: the theory has an infinite number of ground states, determined
by the expectation value $\vev{X}$. Small fluctuations around this vacuum are massless: they
are Goldstone bosons for broken translational invariance $X\rightarrow X +c$.

\para
We see that the physics is very different in field theories in $d=0+1$ and $d=3+1$ dimensions.
The wavefunction spreads along flat directions in quantum mechanics, but not in higher dimensional
field theories. But what happens in $d=1+1$ and $d=2+1$ dimensions? It turns out that field theories
in $d=1+1$ dimensions are more like quantum mechanics: the wavefunction spreads. Theories in
$d=2+1$ dimensions and higher exhibit the opposite behaviour: they have Goldstone bosons. The
place to see this is the propagator. In $d$ spacetime dimensions, it takes the form
\be \langle X(r)\,X(0)\rangle \sim \left\{\begin{array}{lr}1/{r^{d-2}}\ \ \ \ \ \ &\ \ \ \ \ d\neq 2 \\ \ln r & d=2\end{array}\right.\nn\ee
which diverges at large $r$ only for $d=1$ and $d=2$. If we perturb the vacuum slightly by
inserting the operator $X(0)$, this correlation function tells us how this perturbation falls off
with distance. The infra-red divergence in low dimensions is telling us that the wavefunction
wants to spread.

\para
The spreading of the wavefunction in low dimensions means that there is no spontaneous symmetry
breaking and no Goldstone bosons. It is usually referred to as the Coleman-Mermin-Wagner theorem.
Note, however, that it certainly doesn't prohibit massless excitations in two dimensions: it
only prohibits
Goldstone-like massless excitations.

\subsubsection{The Stress-Energy Tensor and Primary Operators}
\label{fsfnowsec}

We want to compute the OPE of $T$ with other operators. Firstly, what is $T$? We computed it
in the classical theory in \eqn{classicalt}. It is,
\be T=-\frac{1}{\ap}\,\partial X\partial X\ee
But we need to be careful about what this means in the quantum theory. It involves the
product of two operators defined at the same point, and this is bound to mean
divergences if we just treat it naively. In canonical quantization, we would be
tempted to  normal order  by putting all annihilation operators to the
right. This guarantees that the vacuum has zero energy. Here we do something
that is basically equivalent, but without reference to creation and annihilation
operators. We write
\be T = -\frac{1}{\ap}\,:\partial X\partial X:\ \  \equiv\
-\frac{1}{\ap}\,\mathop{\mbox{limit}}_{z\rightarrow w}\left(\partial X(z)\partial X(w)
-\langle\partial X(z)\partial X(w) \rangle\right)\label{normalt}\ee
which, by construction, has $\langle T\rangle =0$.

\para
With this definition of $T$, let's start to compute the OPEs to determine the
primary fields in the theory.
\\{}\\
{\bf Claim 1:} $\partial X$ is a primary field with weight $h=1$ and $\tilde{h}=0$.
\\{}\\
{\bf Proof:} We need to figure out how to take products of normal ordered operators
\be
T(z)\,\partial X(w) = -\frac{1}{\ap}: \partial X(z)\partial X(z): \partial X(w)\nn\ee
The operators on the left-hand side are time-ordered (because all operator expressions of
this type are taken to live inside time-ordered correlation functions). In contrast, the
right-hand side is a product of normal-ordered operators. But we know how to change normal
ordered products into time ordered products: this is the content of Wick's theorem. Although
we have defined normal ordering in \eqn{normalt} without reference to creation and
annihilation operators, Wick's theorem still holds. We must sum over all possible contractions
of pairs of operators, where the term ``contraction" means that we replace the pair
by the propagator,
\be
\overbrace{\partial X(z)\,\partial X(w)}
= -\frac{\ap}{2}\,\frac{1}{(z-w)^2}
\nn\ee
Using this, we have
\be T(z)\partial X(w) = -\frac{2}{\ap}\,\partial X(z)\,\left(
-\frac{\ap}{2}\frac{1}{(z-w)^2}+\mbox{non-singular}\right) \nn\ee
Here the ``non-singular'' piece includes the totally normal ordered term $:T(z)\partial X(w):$.
It is only the singular part that interests us. Continuing, we have
\be T(z)\partial X(w) =\frac{\partial X(z)}{(z-w)^2} +\ldots = \frac{\partial X(w)}{(z-w)^2}
+\frac{\partial^2 X(w)}{z-w} + \ldots\nn\ee
This is indeed the OPE for a primary operator of weight $h=1$. \hfill$\Box$

\para
Note that higher derivatives $\p^n X$ are not primary for $n>1$. For example, $\p^2X$ is an operator with 
weight $(h,\tilde{h})=(2,0)$, but is
not a primary operator, as we see from the OPE,
\be T(z)\,\p^2 X(w) = \p_w \left[ \frac{\p X(w)}{(z-w)^2} +\ldots\right]
= \frac{2\p X(w)}{(z-w)^3}
+ \frac{2\p^2X(w)}{(z-w)^2}+\ldots\nn\ee
The fact that the field $\p^n X$ has weight $(h,\tilde{h})=(n,0)$ fits our natural intuition:
each derivative provides spin $s=1$ and dimension $\Delta=1$, while the field $X$ does not appear
to be contributing, presumably reflecting the fact that it has naive, classical dimension zero.
However,  in the quantum theory, it is not correct to say that $X$ has vanishing dimension:
it has an ill-defined dimension due to the logarithmic behaviour of its OPE \eqn{xprob}.
This is responsible for the following, more surprising, result
\\ {\ } \\
{\bf Claim 2:} The field $:e^{ikX}:$ is primary with weight $h=\tilde{h}=\ap k^2/4$.

\para
This result is not what we would guess from the classical theory. Indeed, it's obvious
that it has a quantum origin because the weight is proportional to $\ap$, which sits outside the
action in the same place that $\hbar$ would (if we hadn't set it to one).
Note also that
this means that the spectrum of the free scalar field is continuous. This is related
to the fact that the range of $X$ is non-compact. Generally, CFTs will have a
discrete spectrum.
\\{}\\
{\bf Proof:} Let's first compute the OPE with $\p X$. We have
\be \p X(z)\,:e^{ikX(w)}: &=& \sum_{n=0}^\infty\,\frac{(ik)^n}{n!}\,\p X(z)\,:X(w)^n:\nn
\\ &=& \sum_{n=1}^\infty\,\frac{(ik)^n}{(n-1)!}\,:X(w)^{n-1}:\,\left(-\frac{\ap}{2}\,\frac{1}{z-w}
\right) +\ldots \nn\\ &=& -\frac{i\ap k}{2}\,\frac{:e^{ikX(w)}:}{z-w}+\ldots\label{willneedthis}\ee
From this, we can compute the OPE with $T$.
\be T(z)\,:e^{ikX(w)}: &=& -\frac{1}{\ap}\,:\p X(z)\p X(z):\ :e^{ikX(w)}: \nn\\
&=& \frac{\ap k^2}{4}\,\frac{:e^{ikX(w)}:}{(z-w)^2}+ ik\,\frac{:\p X(z) e^{ikX(w)}:}{z-w}+\ldots
\nn\ee
where the first term comes from two contractions, while the second term comes from
a single contraction. Replacing $\p_z$ by $\p_w$ in the final term we get
\be T(z)\,:e^{ikX(w)}:
= \frac{\ap k^2}{4}\,\frac{:e^{ikX(w)}:}{(z-w)^2}+ \frac{\p_w :e^{ikX(w)}:}{z-w} +\ldots
\ee
showing that $:e^{ikX(w)}:$ is indeed primary. We will encounter this operator
frequently later, but will choose to simplify notation and drop the normal ordering colons.
Normal ordering will just be assumed from now on. \hfill$\Box$.

\para
Finally, lets check to see the OPE of $T$ with itself. This is again just an
exercise in Wick contractions.
\be T(z)\,T(w) &=& \frac{1}{\alpha^{\prime\,2}}\ :\p X(z)\,\p X(z):\ :\p X(w)\,\p X(w): \nn\\
&=& \frac{2}{\alpha^{\prime\,2}}\left(-\frac{\ap}{2}\,\frac{1}{(z-w)^2}\right)^2 -
\frac{4}{\alpha^{\prime\,2}}
\, \frac{\ap}{2}\,\frac{:\p X(z)\,\p X(w):}{(z-w)^2}+\ldots\nn\ee
The factor of 2 in front of the first term comes from the two ways of performing
two contractions; the factor of 4 in the second term comes from the number of
ways of performing a single contraction. Continuing,
\be  T(z)\,T(w) &=& \frac{1/2}{(z-w)^4} + \frac{2T(w)}{(z-w)^2} - \frac{2}{\ap}
\frac{\p^2X(w)\,\p X(w)}{z-w} +\ldots \nn\\
&=& \frac{1/2}{(z-w)^4} + \frac{2T(w)}{(z-w)^2} +
\frac{\p T(w)}{z-w} + \ldots \label{scalartt}\ee
We learn that $T$ is {\it not} a primary operator in the theory of a
single free scalar field. It is an operator of weight $(h,\tilde{h})=(2,0)$,
but it fails the primary test on account of the $(z-w)^{-4}$ term. In fact,
this property of the stress energy tensor a general feature of
all CFTs which we now explore in more detail.

\subsection{The Central Charge}

In any CFT, the most prominent example of an operator which is not primary is the
stress-energy tensor itself.

\para
For the free scalar field, we have already seen that $T$ has 
weight $(h,\tilde{h})=(2,0)$. This remains true in any CFT. The reason for this is simple:
 $\tab$ has dimension  $\Delta =2$ because we obtain the energy by integrating over space.
It has spin $s=2$ because it is a symmetric 2-tensor. But these two pieces of information
are equivalent to the statement that $T$ is an 
operator of weight $(2,0)$. Similarly, $\bt$ has weight $(0,2)$. This means that the
$TT$ OPE takes the form,
\be T(z)\,T(w) = \ldots + \frac{2T(w)}{(z-w)^2} + \frac{\p T(w)}{z-w} + \ldots \nn\ee
and similar for $\bt\bt$. What other terms could we have in this expansion? Since
each term has dimension $\Delta=4$, any operators that appear on the right-hand-side must
be of the form
\be \frac{\calo_n}{(z-w)^n}
\ee
where $\Delta[\calo_n]=4-n$. But, in a unitary CFT there are no operators with $h$, $\tilde{h}<0$. (We will
prove this shortly). So the most singular term that we can have is of order $(z-w)^{-4}$.
Such a term must be multiplied by a constant. We write,
\be T(z)\,T(w) = \frac{c/2}{(z-w)^4}+ \frac{2T(w)}{(z-w)^2} + \frac{\p T (w)}{z-w}
+\ldots \nn\ee
and, similarly,
\be \bt(\bz)\,\bt(\bw) = \frac{\tilde{c}/2}{(\bz-\bw)^4}+ \frac{2\bt(\bw)}{(\bz-\bw)^2} +
\frac{\bp \bt (\bw)}{\bz-\bw}+\ldots \nn\ee
The constants $c$ and $\tilde{c}$ are called the {\it central charges}. (Sometimes they are
referred to as left-moving and right-moving central charges). They are perhaps
the most important numbers characterizing the CFT. We can already get some
intuition for the information contained in these two numbers.
Looking back at the free scalar field \eqn{scalartt} we see that it has
$c=\tilde{c}=1$. If we instead considered $D$ non-interacting free scalar fields,
we would get $c=\tilde{c}=D$. This gives us a hint: $c$ and $\tilde{c}$ are
somehow measuring the number of degrees of freedom in the CFT. This is true
in a deep sense! However, be warned: $c$ is not necessarily an integer.

\para
Before moving on, it's worth pausing to explain why we didn't include a $(z-w)^{-3}$
term in the $TT$ OPE. The reason is that the OPE must obey $T(z)T(w) = T(w)T(z)$
because, as explained previously, these operator equations are all taken to hold
inside time-ordered correlation functions. So the
quick answer is that a $(z-w)^{-3}$ term would not be invariant under $z\leftrightarrow w$.
However, you may wonder how the $(z-w)^{-1}$ term manages to satisfy this property. Let's
see how this works:
\be T(w)\, T(z)=\frac{c/2}{(z-w)^4} + \frac{2T(z)}{(z-w)^2} + \frac{\p T(z)}{w-z} +\ldots\nn\ee
Now we can Taylor expand $T(z) = T(w) + (z-w)\p T(w)+\ldots$ and $\p T(z) = \p T(w) +\ldots$.
Using this in the above expression, we find
\be T(w)\, T(z)=\frac{c/2}{(z-w)^4} + \frac{2T(w)+2(z-w)\p T(w)}{(z-w)^2} - \frac{\p T(w)}{z-w} +\ldots = T(z)\,T(w)\nn\ee
This trick of Taylor expanding saves the $(z-w)^{-1}$ term. It wouldn't work for the
$(z-w)^{-3}$ term.

\subsubsection*{The Transformation of Energy}

So $T$ is not primary unless $c=0$. And we will see shortly that all theories have $c>0$.
What does this mean for the transformation of $T$?
\be \delta T(w) &=& -\Res [ \epsilon(z)\,T(z)\,T(w)]
\nn\\ &=& -\Res \left[\epsilon(z)\left(\frac{c/2}{(z-w)^4} + \frac{2T(w)}{(z-w)^2}
+ \frac{\p T(w)}{z-w}+\ldots\right)\right]\nn\ee
If $\epsilon(z)$ contains no singular terms, we can expand
\be \e (z) = \e(w) + \e^\prime(w)(z-w) + \frac{1}{2}\e^{\prime\prime}
(z-w)^2 + \frac{1}{6}\e^{\prime\prime\prime}(w)(z-w)^3+\ldots\nn\ee
from which we find
\be \delta T(w) = -\e(w)\,\partial T(w) -2\e^\prime(w)\,T(w) -\frac{c}{12}\e^{\prime\prime\prime}
(w)\label{tinf}\ee
This is the infinitesimal version. We would like to know what becomes of $T$ under the
finite conformal transformation $z\rightarrow \tilde{z}(z)$. The answer turns out to be
\be\tilde{T}(\tilde{z}) = \left(\ppp{\tilde{z}}{z}\right)^{-2}\,\left[T(z)-\frac{c}{12}\, S(\tilde{z},z)\right]\label{ttrans}\ee
where $S(\tilde{z},z)$ is known as the {\it Schwarzian} and is defined by
\be S(\tilde{z},z) = \left(\frac{\p^3\tilde{z}}{\p z^3}\right)\,\left(\ppp{\tilde{z}}{z}\right)^{-1} - \frac{3}{2}\left(\frac{\p^2 \tilde{z}}
{\p z^2}\right)^2\left(\ppp{\tilde{z}}{z}\right)^{-2}\label{schwarz}\ee
It is simple to check that the Schwarzian has the right infinitesimal form to give \eqn{tinf}.
Its key property is that it preserves the  group structure of successive conformal transformations.

\subsubsection{c is for Casimir}
\label{cassec}

Note that the extra term in the transformation \eqn{ttrans} of $T$ does not depend
on $T$ itself. In particular, it will be the same evaluated on all states. It only
affects the constant term --- or zero mode --- in the energy. In other words, it
is the Casimir energy of the system.

\para
Let's look at an example that will prove to be useful later for the string. Consider
the Euclidean cylinder, parameterized by
\be w=\sigma+i\tau\ \ \ \ \ ,\ \ \ \ \sigma\in [0,2\pi)\nn\ee
\EPSFIGURE{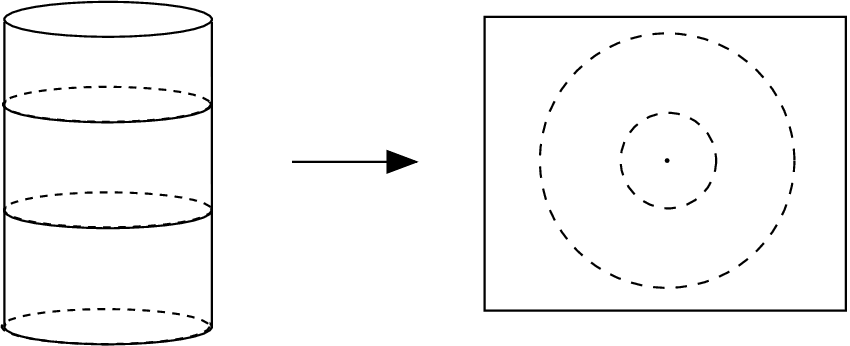,height=90pt}{}
\noindent
We can make a conformal transformation from the cylinder to the complex plane by
\be z=e^{-iw}\nn\ee
The fact that the cylinder and the plane are related by a conformal map means that
if we understand a given CFT on the cylinder, then we immediately understand it on
the plane. And vice-versa. Notice that constant time slices on the cylinder are
mapped to circles of constant radius. The origin, $z=0$, is the distant past, $\tau\rightarrow -\infty$.

\para
What becomes of $T$ under this transformation? The Schwarzian can be easily calculated to be
$S(z,w)=1/2$. So we find,
\be T_{\rm cylinder}(w) = -z^2\, T_{\rm plane}(z) + \frac{c}{24}\label{weuse}\ee
Suppose that the ground state energy vanishes when the theory is defined
on the plane: $\langle T_{\rm plane} \rangle=0$.
What happens on the cylinder? We want to look at the
Hamiltonian, which is defined by
\be H \equiv \int d\sigma\ T_{\tau\tau} = -\int d\sigma\, (T_{ww} + \bar{T}_{\bw\bw})\nn\ee
The conformal transformation then tells us that the ground state energy on the cylinder is
\be E = -\frac{2\pi(c+\tilde{c})}{24}\nn\ee
This is indeed the (negative) Casimir energy on a cylinder. For a free scalar field, we have
$c=\tilde{c}=1$ and the energy density $E/2\pi = -1/12.$ This is the same result that we
got in Section \ref{casimirsec}, but this time with no funny business where we throw out
infinities.

\subsubsection*{An Application: The L\"uscher Term}

If we're looking at a physical system, the cylinder will have a radius $L$. In this
case, the Casimir energy is given by $E = -2\pi(c+\tilde{c})/24L$. There is an
application of this to QCD-like theories. Consider two quarks in a confining theory,
separated by a distance $L$. If the tension of the confining flux tube is $T$, then
the string will be stable as long as $TL \lesssim m$, the mass of the lightest quark.
The energy of the stretched string as a function of $L$ is given by
\be E(L) = TL + a - \frac{\pi c}{24 L} + \ldots \nn\ee
Here $a$ is an undetermined constant, while $c$ counts the number of degrees of freedom of
the QCD flux tube. (There is no analog of $\tilde{c}$ here because of the reflecting boundary
conditions at the end of the string). If the string has no internal degrees of freedom, then $c=2$ for the two transverse fluctuations. This contribution to the string energy is
known as the {\it L\"uscher term}.

\subsubsection{The Weyl Anomaly}
\label{weylanom}

There is another  way in which the central charge affects the stress-energy
tensor. Recall that in the classical theory, one of the defining features of a CFT
was the vanishing of the trace of the stress tensor,
\be T^\alpha_{\ \alpha}=0\nn\ee
However, things are more subtle in the quantum theory. While $\langle T^\alpha_{\ \alpha}\rangle$ indeed vanishes in flat space, it will not longer be true if we place the theory on a
curved background. The purpose of this section is to show that
\be \vev{T^\alpha_{\ \alpha}}=-\frac{c}{12} R\label{cr}\ee
where $R$ is the Ricci scalar of the 2d worldsheet. Before we derive this formula, some quick
comments:
\begin{itemize}
\item Equation \eqn{cr} holds for any state in the theory --- not just the vacuum. This reflects the fact that it comes from regulating short distant divergences in the theory. But, at short distances all finite energy states look basically the same.

\item Because $\vev{T^\alpha_{\ \alpha}}$ is the same for any state it must be equal to something that depends only on the background metric. This something should be local and must be dimension 2. The only candidate is the Ricci scalar $R$. For this reason, the formula $\vev{T^\alpha_{\ \alpha}}\sim R$ is the most general possibility. The only question is: what is the coefficient. And, in particular, is it non-zero?

\item By a suitable choice of coordinates,  we can always put any 2d metric in the form $\gab=e^{2\omega}\delta_{\alpha\beta}$. In these coordinates, the Ricci scalar is given by
\be R = -2e^{-2\omega}\partial^2\omega\label{rforweyl}\ee
which depends explicitly on the function $\omega$. Equation \eqn{cr} is then telling us
that any conformal theory with $c\neq 0$ has at least one physical observable,
$\langle T^\alpha_{\ \alpha}\rangle$, which takes different values on backgrounds
related by a Weyl transformation $\omega$. This result is referred to as the
{\it Weyl anomaly}, or sometimes as the trace anomaly.

\item There is also a Weyl anomaly for conformal field theories in higher dimensions. For
example, $4d$ CFTs are characterized by two numbers, $a$ and $c$, which appear as
coefficients in the Weyl anomaly,
\be \langle T^\mu_{\ \mu}\rangle_{4d} = \frac{c}{16\pi^2}\,C_{\rho\sigma\kappa\lambda}C^{\rho\sigma\kappa\lambda}
-\frac{a}{16\pi^2}\tilde{R}_{\rho\sigma\kappa\lambda}\tilde{R}^{\rho\sigma\kappa\lambda}\nn\ee
where $C$ is the Weyl tensor and $\tilde{R}$ is the dual of the Riemann tensor.

\item  Equation \eqn{cr} involves only the left-moving central charge
$c$. You might wonder what's special about the left-moving sector. The answer, of course,
is nothing. We also have
\be \vev{T^\alpha_{\ \alpha}}=-\frac{\tilde{c}}{12}\,R\nn\ee
In flat space, conformal field theories with different $c$ and $\tilde{c}$ are perfectly
acceptable. However, if we wish these theories to be consistent in fixed, curved
backgrounds, then we require $c=\tilde{c}$. This is an example of a {\it gravitational anomaly}.

\item The fact that Weyl invariance requires $c=0$ will prove crucial in string theory. We
shall return to this in Chapter \ref{polyakov}.
\end{itemize}

%
%\be
%\delta T(z) = -\e(z)\,\partial T(z) -2\,\e^{\,\prime}(z)\,T(z) %-\frac{c}{12}\e^{\,\prime\prime\prime}
%(z)\nn\ee
%%
%But we know that a conformal transformation can be thought of as a coordinate
%transformation, accompanied by a compensating Weyl transformation. The first two
%terms above reflect the way that a 2-tensor transforms under the transformation
%%
%\be z\rightarrow \tilde{z}=z+\epsilon(z)\nn\ee
%%
%Therefore the last term, $-c\epsilon^{\,\prime\prime\prime}/12$, must be telling us how
%$T$ changes under the compensating Weyl transformation,
%%
%\be d\tilde{z}d\bar{\tilde{z}} = (1+\e^{\,\prime}+\bar{\e}^{\,\prime})\,dzd\bz
%= \exp\left({\e^{\,\prime}+\bar{\e}^{\,\prime}}\right)\,dzd\bz\nn\ee
%%
%So the compensating Weyl transformation is
%%
%\be 2\phi = -(\e^{\,\prime}+\bar{\e}^{\,\prime})\nn\ee
%%
%Let's set $\bar{\e}=0$ for now, and just focus on the holomorphic sector and $T(z)$.
%Under a Weyl transformation,
%%
%\be \delta_{Weyl} T(z) = -\frac{c}{12}\,\e^{\,\prime\prime\prime} = \frac{c}{6}\,\partial^2\, %\phi
%\nn\ee
%%

We will now prove the Weyl anomaly formula \eqn{cr}. Firstly, we need to derive an intermediate
formula: the $T_{z\bz}\,T_{w\bw}$ OPE. Of course, in the classical theory we found that
conformal invariance requires $T_{z\bz}=0$. We will now show that it's a little more subtle in
the quantum theory.

\para
Our starting point is the equation for energy conservation,
\be \p T_{z\bz} = -\pb\,T_{zz}\nn\ee
Using this, we can express our desired OPE in terms of the familiar $TT$ OPE,
\be \p_zT_{z\bz}(z,\bz)\ \p_wT_{w\bw}(w,\bw) = \pb_{\bz} T_{zz}(z,\bz) \ \pb_{\bw}T_{ww}(w,\bw)
=
\pb_{\bz}\pb_{\bw}\left[\frac{c/2}{(z-w)^4}+\ldots\right]\ \ \ \ \ \ \label{odt}\ee
Now you might think that the right-hand-side just vanishes: after all, it is an
anti-holomorphic derivative $\pb$ of a holomorphic quantity. But we shouldn't
be so cavalier because there is a singularity at $z=w$. For example,  consider
the following equation,
\be \bp_{\bz}\p_z\,\ln|z-w|^2 = \bp_{\bz} \,\frac{1}{z-w} = 2\pi \delta(z-w,\bz-\bw)
\label{clogeqn}\ee
We proved this statement after equation \eqn{logeqn}. (The factor of 2 difference from
\eqn{logeqn} can be traced to the conventions we
defined for complex coordinates in Section \ref{complexsection}). Looking at the
intermediate step in \eqn{clogeqn}, we again have an anti-holomorphic derivative of a
holomorphic function and you might be tempted to say that this also  vanishes.
But you'd be wrong: subtle things happen because of the singularity and equation
\eqn{clogeqn} tells us that the function $1/z$ secretly depends on $\bar{z}$. (This should
really be understood as a statement about distributions, with the delta function integrated
against arbitrary test functions). Using this result, we can write
\be \bp_{\bz}\bp_{\bw}\,\frac{1}{(z-w)^4} = \frac{1}{6}\,\pb_{\bz}\pb_{\bw}\,\left( \p^2_z\,
\p_w\frac{1}{z-w}\right) = \frac{\pi}{3}\,\p^2_z\,\p_w\bp_{\bw}\,\delta(z-w,\bz-\bw)\nn\ee
Inserting this into the correlation function \eqn{odt}, and stripping off the
$\p_z\p_w$ derivatives on both sides, we end up with what we want,
\be T_{z\bz}(z,\bz)\ T_{w\bw}(w,\bw) = \frac{c\pi}{6}\,\p_z\bp_{\bw}\,\delta(z-w,\bz-\bw)
\label{tzbzope}\ee
So the OPE of $T_{z\bz}$ and $T_{w\bw}$ almost vanishes, but there's some strange singular
behaviour going on as $z\rightarrow w$. This is usually referred to as a contact term
between operators and, as we have shown,  it is needed to ensure the conservation of  energy-momentum. We will now see that this contact term is responsible for
the Weyl anomaly.

\para
We assume that $\vev{T^\alpha_{\ \alpha}}=0$ in flat
space. Our goal is to derive an expression for $\vev{T^\alpha_{\ \alpha}}$ close to
flat space. Firstly,
consider the
change of $\vev{T^\alpha_{\ \alpha}}$
under a general shift of the metric $\delta\gab$. Using the definition of the
energy-momentum tensor \eqn{se}, we have
\be \delta\vev{T^\alpha_{\ \alpha}(\sigma)} &=&
\delta\int {\cal D}\phi\ e^{-S}\,T^{\alpha}_{\ \alpha}(\sigma)
\nn\\ &=& \frac{1}{4\pi}\int {\cal D}\phi\ e^{-S}\,\left(T^{\alpha}_{\ \alpha}(\sigma)\,\int d^2\sigma^\prime \sqrt{g}\ \delta g^{\beta\gamma}\, T_{\beta\gamma}(\sigma^\prime)\, \right)\nn\ee
If we now restrict to a Weyl transformation, the change to a flat metric is
$\delta \gab = 2\omega\delta_{\alpha \beta}$, so the change in the inverse
metric is $\delta \gabi = -2\omega\delta^{\alpha\beta}$. This gives
\be \delta\vev{T^\alpha_{\ \alpha}(\sigma)} = -
\frac{1}{2\pi}\int {\cal D}\phi\ e^{-S}\,\left(T^{\alpha}_{\ \alpha}(\sigma)\,\int d^2\sigma^\prime
\ \omega(\sigma^\prime)\, T^{\beta}_{\ \beta}(\sigma^\prime)\, \right)\label{itsthis}\ee
Now we see why the OPE \eqn{tzbzope} determines the Weyl anomaly. We need to change
between complex coordinates and Cartesian coordinates, keeping track of factors of 2.
We have
\be T^\alpha_{\ \alpha}(\sigma)\,T^{\beta}_{\ \beta}(\sigma^\prime) = 16\,T_{z\bz}(z,\bz)\ T_{w\bw}(w,\bw)
\nn\ee
Meanwhile, using the conventions laid down in \ref{complexsection},
we have $8\p_z\pb_{\bw}\delta(z-w,\bz-\bw) = -\p^2 \,\delta(\sigma-\sigma^\prime)$. This gives
us the OPE in Cartesian coordinates
\be T^\alpha_{\ \alpha}(\sigma)\,T^{\beta}_{\ \beta}(\sigma^\prime) = -\frac{c\pi}{3}\,\p^2
\,\delta(\sigma-\sigma^\prime)\nn\ee
We now plug this into \eqn{itsthis} and integrate by parts to move the
two derivatives onto the conformal factor $\omega$. We're left with,
\be \delta \vev{T^\alpha_{\ \alpha}}=\frac{c}{6}\,\p^2\omega \ \ \ \Rightarrow\ \ \
\vev{T^\alpha_{\ \alpha}} = -\frac{c}{12}R\nn\ee
where, to get to the final step, we've used \eqn{rforweyl} and, since we're working
infinitesimally, we can replace $e^{-2\omega}\approx 1$. This completes the
proof of the Weyl anomaly, at least for spaces infinitesimally close to flat space.
The fact that $R$ remains on the right-hand-side for general 2d surfaces follows
simply from the comments after equation \eqn{cr}, most pertinently the
need for the expression to be reparameterization invariant.

\subsubsection{c is for Cardy}

The Casimir effect and the Weyl anomaly have a similar smell. In both, the central
charge provides an extra contribution to the energy. We now demonstrate a different avatar
of the central charge: it tells us the density of high energy states.

\para
We will study  conformal field theory on a Euclidean torus. We'll keep our
normalization $\sigma \in [0,2\pi)$, but now we also take $\tau$ to be periodic,
lying in the range
\be \tau \in [0,\beta)\nn\ee
The partition function of a theory with periodic Euclidean time has a very natural
interpretation: it is related to the free energy of the theory at temperature $T=1/\beta$.
\be Z[\beta] = \Tr\,e^{-\beta H} = e^{-\beta F}\ee
At very low temperatures, $\beta \rightarrow \infty$, the free energy is dominated by the
lowest energy state. All other states are exponentially suppressed. But we saw in
\ref{cassec} that the vacuum state on the cylinder has Casimir energy $H=-c/12$. In
the limit of low temperature, the partition function is therefore approximated by
\be Z \ \rightarrow\  e^{c\beta/12}\ \ \ \ {\rm as}\ \beta\rightarrow \infty\ee
\EPSFIGURE{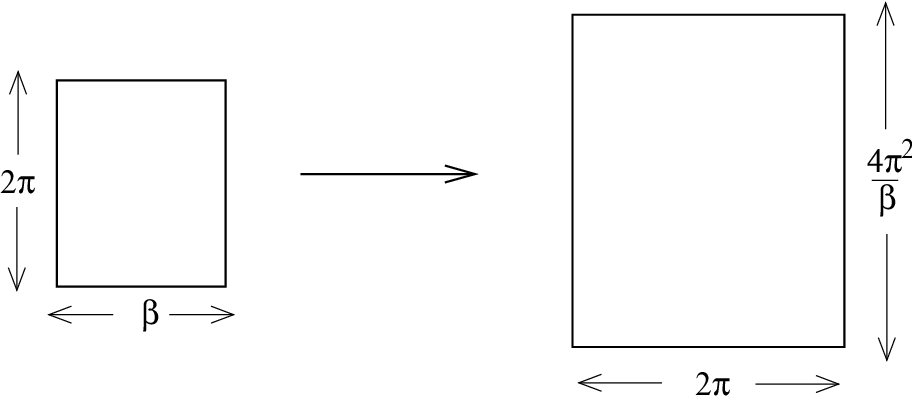,height=95pt}{}
\noindent Now comes the trick. In Euclidean space, both directions of the torus are on
equal footing. We're perfectly at liberty to decide that $\sigma$ is ``time"
and $\tau$ is ``space". This can't change the value of the partition function.
So let's make the swap. To compare to our original partition function, we want
the spatial direction to have range $[0,2\pi)$. Happily, due to the conformal nature
of our theory, we arrange this through the scaling
\be \tau \rightarrow \frac{2\pi}{\beta}\,\tau\ \ \  \ ,\ \ \ \ \sigma\rightarrow
\frac{2\pi}{\beta}\,\sigma\nn\ee
Now we're back where we started, but with the temporal direction taking values
in $\sigma\in [0,4\pi^2/\beta$. This tells us that the high-temperature and low-temperature
partition functions are related,
\be Z[{4\pi^2}/{\beta}]= Z[\beta]\nn\ee
This is called modular invariance. We'll come across it again in Section \ref{oneloopsec}.
Writing $\beta^\prime = 4\pi^2/\beta$, this tells us the very high temperature behaviour of the
partition function
\be Z[\beta^\prime]\ \rightarrow\ e^{c\pi^2/3\beta^\prime} \ \ \ \ {\rm as}\ \ \ \beta^\prime
\rightarrow 0\nn\ee
But the very high temperature limit of the partition function is sampling all states in the
theory. On entropic grounds, this sampling is dominated by the high energy states. So
this computation is telling us how many high energy states there are.

\para
To see this more explicitly, let's do some elementary manipulations
in statistical mechanics.
Any system has a density of states $\rho(E)=e^{\,S(E)}$, where $S(E)$ is the entropy.
The free energy is given by
\be e^{-\beta F} = \int dE\ \rho(E)\,e^{-\beta E} = \int dE \ e^{S(E)-\beta E}\nn\ee
In two dimensions, all systems have an entropy which scales at large energy as
\be S(E)\rightarrow N \sqrt{E}\label{seee}\ee
The coefficient $N$  counts the number of degrees of freedom. The fact
that $S\sim \sqrt{E}$ is equivalent to the fact that $F\sim T^2$, as befits
an energy density in a theory with one spatial dimension. To see this, we need only
approximate the integral by the saddle point $S^\prime (E_\star)=\beta$. From
\eqn{seee}, this gives us the free energy
\be F \sim N^2 T^2 \nn\ee
We can now make the statement about the central charge more explicit. In a conformal
field theory, the entropy of high energy states is given by
\be S(E) \sim \sqrt{cE}\nn\ee
This is Cardy's formula.

\subsubsection{c has a Theorem}

The connection between the central charge and the degrees of freedom in a theory is given
further weight by  a result of Zamalodchikov, known as the {\it c-theorem}. The idea of the
c-theorem is to stand back and look at the space of all theories and the renormalization
group (RG) flows between them.

\para
Conformal field theories are special. They are the fixed points of the renormalization
group, looking the same at all length scales. One can consider perturbing a conformal
field theory by adding an extra term to the action,
\be S\rightarrow S + \alpha\int d^2\sigma\ \calo(\sigma)\nn\ee
Here $\calo$ is a local operator of the theory, while $\alpha$ is some coefficient. These
perturbations fall into three classes, depending on the dimension $\Delta$ of $\calo$.
\begin{itemize}
\item $\Delta < 2$: In this case, $\alpha$ has positive dimension: $[\alpha]=2-\delta$.
Such deformations are called {\it relevant} because they are important in the infra-red.
RG flow takes us away from our original CFT. We only stop flowing when we
hit a new CFT (which could be trivial with $c=0$).
\item $\Delta = 2$: The constant $\alpha$ is dimensionless. Such deformations are
called {\it marginal}. The deformed theory defines a new CFT.
\item $\Delta > 2$: The constant $\alpha$ has negative dimension. These deformations
are irrelevant. The infra-red physics is still described by the original CFT. But the
ultra-violet physics is altered.
\end{itemize}
We expect information is lost as we flow from an ultra-violet theory to the infra-red.
The c-theorem makes this intuition precise. The theorem exhibits a function $c$ on the space
of all theories which monotonically decreases along RG flows. At the fixed
points, $c$ coincides with the central charge of the CFT.

\subsection{The Virasoro Algebra}

So far our discussion has been limited to the operators of the CFT. We haven't said
anything about states. We now remedy this. We start by taking a closer look
at the map between the cylinder and the plane.

\subsubsection{Radial Quantization}
\label{radial}

\EPSFIGURE{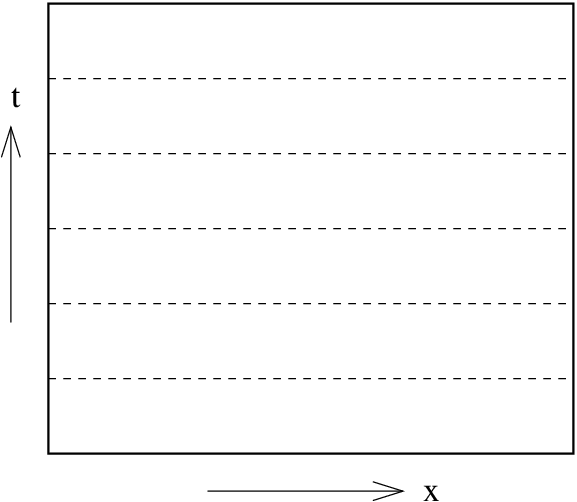,height=90pt}{}
\noindent To discuss states in a quantum field theory we need to think about where they
live and how they evolve. For example, consider a two dimensional quantum field
theory defined on the plane. Traditionally, when quantizing this theory, we parameterize
the plane by Cartesian coordinates $(t,x)$ which we'll call ``time" and ``space".
The states live on spatial slices. The Hamiltonian generates time translations and
hence governs the evolution of states.

\para
However, the map between the cylinder and the plane suggests a different
way to quantize a CFT on the plane.  The
complex coordinate on the cylinder is taken to be $\omega$, while the coordinate on
the plane is $z$. They are related by,
\be \omega = \sigma + i\tau\ \ \ \ ,\ \ \ \ z=e^{-i\omega}\nn\ee
On the cylinder, states live on spatial slices of constant $\sigma$ and evolve by the
Hamiltonian,
\be H = \partial_\tau\nn\ee
After the map to the plane, the Hamiltonian becomes the dilatation operator
\be D = z\p + \bz\pb\nn\ee
If we want the states on the plane to remember their cylindrical roots, they should live on
circles of constant radius. Their evolution is governed by the dilatation operator $D$.
This approach to a theory is known as {\it radial quantization}.

\para
Usually in a quantum field theory, we're interested in time-ordered correlation
functions. Time ordering on the cylinder becomes radial ordering on the plane. Operators
in correlation functions are ordered so that those inserted at larger radial distance are
moved to the left.

\begin{figure}[htb]
\begin{center}
\epsfxsize=3.7in\leavevmode\epsfbox{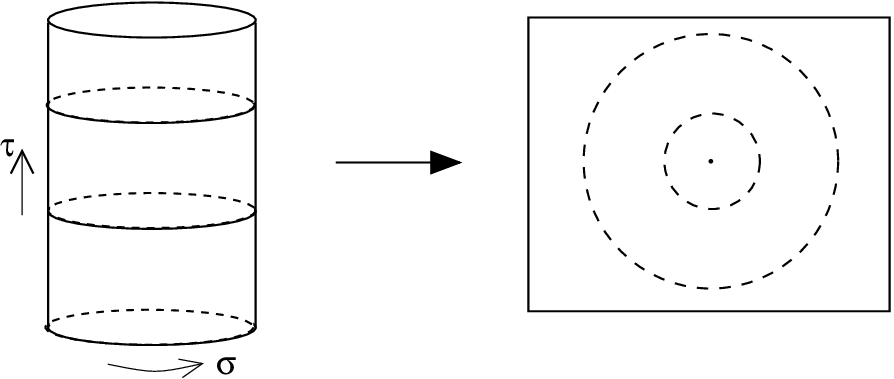}
\end{center}
\caption{The map from the cylinder to the plane.}
%\label{mir}
\end{figure}

\subsubsection*{Virasoro Generators}

Let's look at what becomes of the stress tensor $T(z)$ evaluated on the plane. On the cylinder,
we would decompose $T$ in a Fourier expansion.
\be
T_{\rm cylinder}(w) = -\sum_{m=-\infty}^\infty\ L_me^{imw} + \frac{c}{24}\nn\ee
After the transformation \eqn{weuse} to the plane, this becomes the Laurent expansion
\be T(z) =\sum_{m=-\infty}^{\infty}\,\frac{L_m}{z^{m+2}}\nn\ee
As always, a similar statement holds for the right-moving sector
\be \bt(\bz) = \sum_{m=-\infty}^\infty\,\frac{\tilde{L}_m}{\bz^{m+2}}\nn\ee
We can invert these expressions to get $L_m$ in terms of $T(z)$. We need to take a
suitable contour integral
\be L_n = \frac{1}{2\pi i}\oint dz \ z^{n+1}\,T(z)\ \ \ \ \ ,\ \ \ \ \
\tilde{L}_n = \frac{1}{2\pi i}\oint d\bz \ \bz^{n+1}\,\bt(\bz) \label{qln}\ee
where, if we just want $L_n$ or $\tilde{L}_n$, we must make sure that there are no other insertions
inside the contour.

\para
In radial quantization,
$L_n$ is the conserved charge associated to the conformal transformation $\delta z = z^{n+1}$.
To see this, recall that the corresponding Noether current, given in \eqn{confcur}, is
$J(z)=z^{n+1}T(z)$. Moreover, the contour integral $\oint dz$
maps to the integral around spatial slices on the cylinder. This tells us that $L_n$ is
the conserved charge where ``conserved"  means
that it is constant under time evolution on the cylinder, or under radial evolution on the
plane. Similarly,
$\tilde{L}_n$ is the conserved charge associated to the conformal transformation
$\delta \bz=\bz^{n+1}$.

\para
When we go to the quantum theory, conserved charges become generators for the transformation.
Thus the operators $L_n$ and $\tilde{L}_n$ generate the conformal transformations
$\delta z = z^{n+1}$ and $\delta \bz = \bz^{n+1}$.
They are known as the {\it Virasoro}
generators. In particular, our two favorite conformal transformations are
\begin{itemize}
 \item $L_{-1}$ and $\tilde{L}_{-1}$ generate translations in the plane.
 \item $L_0$ and $\tilde{L}_0$ generate scaling and rotations.
\end{itemize}
The Hamiltonian of the system --- which measures the energy of states on the cylinder ---
is mapped into the dilatation operator on the plane. When acting on states of the
theory, this operator is represented as
\be D = L_0+\tilde{L}_0\nn\ee

\subsubsection{The Virasoro Algebra}

If we have some number of conserved charges, the first thing that we should do is compute
their algebra. Representations of this algebra then classify the states of the theory.
(For example, think angular momentum in the hydrogen atom). For conformal symmetry, we want
to determine the algebra obeyed by the $L_n$ generators. It's a nice fact that the commutation
relations are actually encoded $TT$ OPE. Let's see how this works.

\para
We want to compute $[L_m,L_n]$. Let's write $L_m$ as a contour integral over $\oint dz$ and
$L_n$ as a contour integral over $\oint dw$. (Note: both $z$ and $w$ denote coordinates
on the complex plane now). The commutator is
\be [L_m,L_n] = \left(\oint\frac{dz}{2\pi i}\oint\frac{dw}{2\pi i}\ -\ \oint\frac{dw}{2\pi i}\oint\frac{dz}{2\pi i}\right)\,z^{m+1}w^{n+1}\,T(z)\,T(w)\nn\ee
What does this actually mean?! We need to remember that all operator equations
are to be viewed as living inside time-ordered correlation functions. Except, now
we're working on the z-plane, this statement has transmuted into radially ordered
correlation functions: outies to the left, innies to the right.

\para
So $L_mL_n$ means $\raisebox{-8.9ex}{\epsfxsize=1.4in\epsfbox{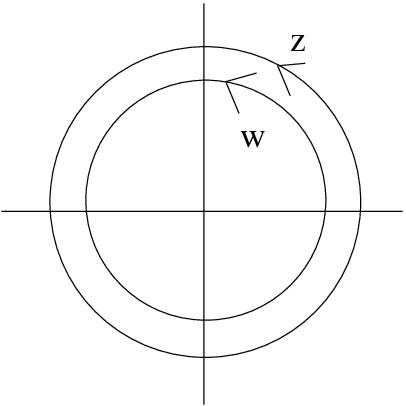}}$
while $L_n L_m$ means $\raisebox{-8.9ex}{\epsfxsize=1.4in\epsfbox{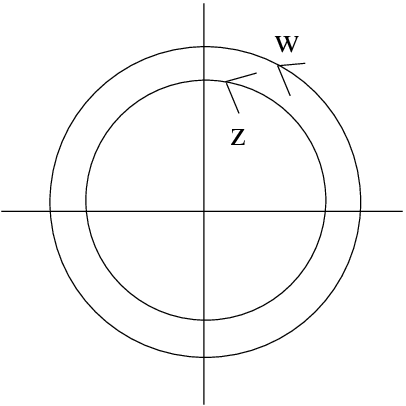}}$.

\para
The trick to computing the commutator is to first fix $w$ and do the  $\oint dz$
integrations. The resulting contour is,
\be \raisebox{-8.9ex}{\epsfxsize=4.4in\epsfbox{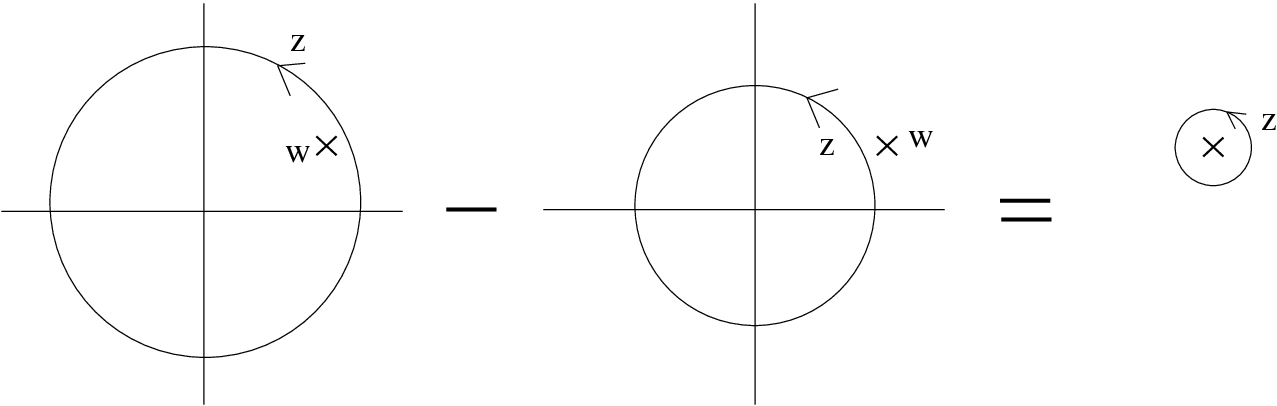}}\nn\ee

\para
In other words, we do the $z$-integration around a fixed point $w$, to get
\be
[L_m,L_n] &=& \oint \frac{dw}{2\pi i}\oint_w\frac{dz}{2\pi i} \ z^{m+1}w^{n+1}\,T(z)\,T(w) \nn\\
&=& \oint\frac{dw}{2\pi i}\,\Res\left[z^{m+1}w^{n+1}\left(\frac{c/2}{(z-w)^4}+\frac{2T(w)}{(z-w)^2}
+\frac{\p T(w)}{z-w} + \ldots\right)\right]
\nn\ee
To compute the residue at $z=w$, we first need to Taylor expand $z^{m+1}$ about the point $w$,
\be z^{m+1} &=& w^{m+1} + (m+1)w^m(z-w) + \frac{1}{2}m(m+1)w^{m-1}(z-w)^2
\nn\\ &&\ \ \ \ \ +\frac{1}{6}m(m^2-1)w^{m-2}(z-w)^3 + \ldots
\nn\ee
The residue then picks up a contribution from each of the three terms,
\be [L_m,L_n] = \oint \frac{dw}{2\pi i}\ w^{n+1}\left[ w^{m+1}\p T(w) + 2(m+1)w^m\, T(w)
+\frac{c}{12}m(m^2-1)w^{m-2}\right]\nn\ee
To proceed, it is simplest to integrate the first term by parts. Then we do the $w$-integral.
But for both the first two terms, the resulting integral is of the form  \eqn{qln} and gives
us $L_{m+n}$. For the third term, we pick up the pole. The end result is
\be [L_m,L_n] = (m-n)L_{m+n} + \frac{c}{12}m(m^2-1)\delta_{m+n,0}\nn\ee
This is the {\it Virasoro algebra}. It's quite famous. The $\tilde{L}_n$'s satisfy exactly
the same algebra, but with $c$ replaced by $\tilde{c}$. Of course, $[L_n,\tilde{L}_m]=0$.
The appearance of $c$ as an extra term in the Virasoro algebra is the reason it is
called the ``central charge". In general, a central charge is an extra term in an
algebra that commutes with everything else.

\subsubsection*{Conformal = Diffeo + Weyl}

We can build some intuition for the Virasoro algebra. We know that the $L_n$'s generate
conformal transformations $\delta z = z^{n+1}$. Let's consider something closely related:
a coordinate transformation $\delta z  = z^{n+1}$.
These are generated by the vector fields
\be l_n = z^{n+1}\partial_z\label{littleln}\ee
But it's a simple matter to compute their commutation relations:
\be [l_n,l_m]=(m-n)l_{m+n}\nn\ee
So this is giving us the first part of the Virasoro algebra. But what about the
central term? The key point to remember is that, as we stressed at the
beginning of this chapter, a conformal transformation is not just a reparameterization
of the coordinates: it is a reparameterization, followed by a compensating Weyl rescaling.
The central term in the Virasoro algebra is due to the Weyl rescaling.

\subsubsection{Representations of the Virasoro Algebra}

With the algebra of conserved charges at hand, we can now start to see how the conformal
symmetry classifies the states into representations.

\para
Suppose that we have some state $\ket{\psi}$ that is an eigenstate of $L_0$ and $\tilde{L}_0$.
\be L_0\ket{\psi} = h\ket{\psi} \ \ \ ,\ \ \ \tilde{L}_0\ket{\psi} = \tilde{h}\ket{\psi}\nn\ee
Back on the cylinder, this corresponds to some state with energy
\be \frac{E}{2\pi} = h+\tilde{h} - \frac{c+\tilde{c}}{24}\nn\ee
For this reason, we'll refer to the eigenvalues $h$ and $\tilde{h}$ as the energy of
the state. By acting with the $L_n$ operators, we can get further states with eigenvalues
\be L_0L_n\ket{\psi} = (L_nL_0-nL_n)\ket{\psi} = (h-n)L_n\ket{\psi}\nn\ee
This tells us that $L_n$ are raising and lowering operators depending on the sign of $n$. When $n>0$, $L_n$ lowers the energy of the state and $L_{-n}$ raises the energy of the state. If the spectrum is
to be bounded below, there must be some states which are annihilated by all $L_n$ and
$\tilde{L}_n$ for $n>0$. Such states are called {\it primary}. They obey
\be L_n\ket{\psi} = \tilde{L}_n\ket{\psi}= 0\ \ \ \ \mbox{for all} \ n>0\nn\ee
In the language of representation theory, they are also called highest weight states. They are
the states of lowest energy.

\para
Representations of the Virasoro algebra can now be built by acting on the primary states with
raising operators $L_{-n}$ with $n>0$. Obviously this results in an infinite tower of
states. All states obtained in this way are called {\it descendants}. From an initial
primary state $\ket{\psi}$, the tower fans out...
\be &\ket{\psi}& \nn\\ &L_{-1}\ket{\psi}& \nn\\ &L_{-1}^2\ket{\psi}\ ,\ L_{-2}\ket{\psi}&
 \nn\\ &L_{-1}^3\ket{\psi}\ ,\ L_{-1}L_{-2}\ket{\psi}\ ,\ L_{-3}\ket{\psi}& \nn\ee
The whole set of states is called a {\it Verma} module. They are the irreducible representations of the Virasoro algebra. This means that if we know the spectrum of primary states, then
we know the spectrum of the whole theory.

\para
Some comments:
\begin{itemize}
\item The vacuum state $\ket{0}$ has $h=0$. This state obeys
\be L_n\ket{0}=0\ \ \ \ \mbox{for all}\ n\geq 0\label{lnand0}\ee
Note that this state preserves the maximum number of symmetries: like all primary states,
it is annihilated by $L_n$ with $n>0$, but it is also annihilated by $L_0$. This fits
with our intuition that the vacuum state should be invariant under as many symmetries as
possible. You might think that we could go further and require that the vacuum state
obeys $L_n\ket{\,0}=0$ for all $n$. But that isn't consistent with the central charge
term in Virasoro algebra. The requirements \eqn{lnand0} are the best we can do.
\item This discussion should be ringing bells. We saw something very similar in the covariant quantization of the string, where we imposed conditions \eqn{itsl} as
    constraints. We will see the connection between the primary states and the spectrum of the string in Section \ref{polyakov}.
\item There's a subtlety that you should be aware of: the states in the Verma
module are not necessarily all  independent. It could be that some linear
combination of the states vanishes. This linear combination is known as a null
state. The existence of null states depends on the values of $h$ and $c$. For example,
suppose that we are in a theory in which the central charge is $c=2h(5-8h)/(2h+1)$,
where $h$ is the energy of a primary state $\ket{\psi}$. Then it is simple to check that
the following combination has vanishing norm:
\be L_{-2}\ket{\psi} -\frac{3}{2(2h+1)}\,L_{-1}^2\ket{\psi} \ee
\item There is a close relationship between the primary states and the primary operators defined in Section \ref{primarysec}. In fact, the energies $h$ and $\tilde{h}$ of primary states will turn out to be exactly the weights of primary operators in the theory. This connection will be described in Section \ref{stateopmap}.
\end{itemize}

\subsubsection{Consequences of Unitarity}
\label{unitsec}

There is one physical requirement that a theory must obey which we have so far neglected
to mention: {\it unitarity}. This is the statement that
probabilities are conserved when we are in Minkowski signature
spacetime. Unitarity follows immediately if we have a Hermitian Hamiltonian which governs
time evolution. But so far our discussion has been somewhat algebraic, and we've not
enforced this condition. Let's do so now.

\para
We retrace our footsteps back to the Euclidean cylinder, and then back again to the Minkowski
cylinder where we can ask questions about time evolution. Here the Hamiltonian density takes
the form
\be {\cal H} = T_{ww} + T_{\bw\bw} = \sum_n L_ne^{-in\sigma^+} + \tilde{L}_ne^{-in\sigma^-} \nn\ee
So for the Hamiltonian to be Hermitian, we require
\be L_n=L_{-n}^\dagger\nn\ee
This requirement imposes some strong constraints on the structure of
CFTs.  Here we look at a couple of trivial, but important,
constraints that arise due to unitarity and the requirement that the physical
Hilbert space does not contain negative norm states.

\begin{itemize}
\item $h\geq 0$: This fact follows from looking at the norm,
\be |L_{-1}\ket{\psi}\!|^{\,2} = \bra{\psi}L_{+1}L_{-1}\ket{\psi} = \bra{\psi}[L_{+1},L_{-1}]\ket{\psi} = 2h\bra{\psi}\psi\rangle \geq 0\nn\ee
The only state with $h=0$ is the vacuum state $\ket{\,0}$.
\item $c>0$: To see this, we can look at
\be |L_{-n}\ket{0}\!|^{\,2} = \bra{0}[L_n,L_{-n}]\ket{0} = \frac{c}{12}n(n^2-1) \geq 0
\ee
So $c \geq 0$. If $c=0$, the only state in the vacuum module is the vacuum itself. It turns out that, in fact, the only state in the whole theory is the vacuum itself. Any non-trivial
CFT has $c>0$.
\end{itemize}
There are many more requirements of this kind that constrain the theory. In fact, it
turns out that for CFTs with $c<1$ these requirements are enough to classify and
solve all theories.

\subsection{The State-Operator Map}
\label{stateopmap}

In this section we describe one particularly important aspect of conformal field
theories: a map between states and local operators.

\para
Firstly, let's get some perspective. In a typical quantum field theory, the states and
local operators are very different objects. While local operators live at a point in
spacetime, the states live over an entire spatial slice. This is most clear if we
write down a Schr\"odinger-style wavefunction. In field theory, this object is actually
a wave-functional, $\Psi[\phi(\sigma)]$, describing the probability for every field
configuration $\phi(\sigma)$ at each point $\sigma$ in space (but at a fixed time).

\para
Given that states and local operators are such very different beasts, it's a little
surprising that in a CFT there is an isomorphism between them: it's
called the state-operator map. The key point is that the distant past in the
cylinder gets mapped to a single point $z=0$ in the complex plane. So specifying a
state on the cylinder in the far past is equivalent to specifying a local disturbance
at the origin.

\para
To make this precise, we need to recall how to write down wavefunctions using
path integrals. Different states are computed by putting different boundary
conditions on the functional integral. Let's start by returning to quantum
mechanics and reviewing a few simple facts. The
propagator for a particle to move from position $x_i$ at time $\tau_i$ to position
$x_f$ at time $\tau_f$ is given by
\be G(x_f,x_i) = \int_{x(\tau_i)=x_i}^{x(\tau_f)=x_f}{\cal D}x\ e^{iS}\nn\ee
This means that if our system starts off in some state described by the wavefunction
$\psi_i(x_i)$ at time $\tau_i$ then (ignoring the overall normalization)
it evolves to the state
\be \psi_f(x_f,\tau_f) = \int dx_i\,G(x_f,x_i)\,\psi_i(x_i,\tau_i)\nn\ee
There are two lessons to take from this. Firstly, to determine the value of the
wavefunction at a given point $x_f$, we evaluate the path integral restricting to
paths which satisfy $x(\tau_f) =x_f$. Secondly, the initial state $\psi(x_i)$ acts
as a weighting factor for the integral over initial boundary conditions.

\para
Let's now write down the same formula in a field theory, where we're dealing with
wavefunctionals. We'll work with the Euclidean path integral on the cylinder.
If we start with some state $\Psi_i[\phi_i(\sigma)]$ at
time $\tau_i$, then it will evolve to the state
\be \Psi_f[\phi_f(\sigma),\tau_f] =
\int {\cal D}\phi_i\int_{\phi(\tau_i)=\phi_i}^{\phi(\tau_f)=\phi_f}
{\cal D}\phi\ e^{-S[\phi]}\ \Psi_i[\phi_i(\sigma),\tau_i]\nn\ee
\EPSFIGURE{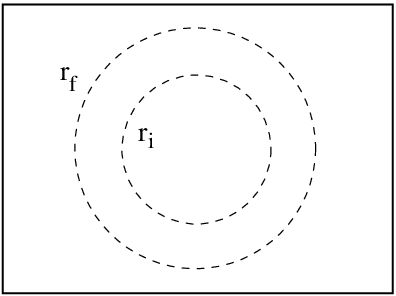,height=90pt}{}
\noindent How do we write a similar expression for states after the map to the complex plane?
Now the states are defined on circles of constant radius, say $|z|=r$, and
evolution is governed by the dilatation operator. Suppose the initial
state is defined at $|z|=r_i$.
In the path integral, we integrate over all fields with fixed boundary
conditions $\phi(r_i)=\phi_i$ and $\phi(r_f)=\phi_f$ on the two edges of the annulus
shown in the figure,
\be \Psi_f[\phi_f(\sigma),r_f] =
\int {\cal D}\phi_i\int_{\phi(r_i)=\phi_i}^{\phi(r_f)=\phi_f}
{\cal D}\phi\ e^{-S[\phi]}\ \Psi_i[\phi_i(\sigma),r_i]\nn\ee
\EPSFIGURE{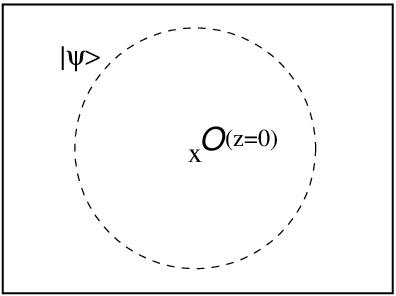,height=90pt}{}
\noindent
This is the traditional way to define a state in field theory, albeit with a
slight twist because we're working in radial quantization. We see that the effect of
the initial state is to change the weighting of the path integral over the inner ring at $|z|=r_i$.

\para
Let's now see what happens as we take the initial state back to the far past and, ultimately,
to $z=0$? We must now integrate over the whole disc $|z|\leq r_f$, rather
than the annulus. The only effect of the initial state is now to change the
weighting of the path integral at the point $z=0$.
But that's exactly what we mean by a local operator inserted at that point. This means
that each local operator $\calo(z=0)$ defines a different state in the theory,
\be \Psi[{\phi}_f;r] = \int^{\phi(r) =\phi_f}{\cal D}\phi\ e^{-S[\phi]}\,\calo(z=0)\nn\ee
We're now integrating over all field configurations within the disc, including all
possible values of the field at $z=0$, which is analogous to integrating over the
boundary conditions $\int {\cal D}\phi_i$ on the inner circle.

\begin{itemize}
\item The state-operator map is only true in conformal field theories where
we can map the cylinder to the plane. It also holds in conformal field theories in
higher dimensions (where ${\bf R}\times {\bf S}^{D-1}$ can be mapped to the plane ${\bf R}^D$).
In non-conformal field theories, a typical local operator creates many different states.
\item The state-operator map does not say that the number of states in the theory is
equal to the number of operators: this is never true. It does say that the states are in one-to-one correspondence with the {\it local} operators.
\item You might think that you've seen something like this before. In the canonical quantization of free fields, we create states in a Fock space by acting with creation operators.  That's {\it not} what's going on here! The creation operators are just about as far from local operators as you can get. They are the Fourier transforms of local operators.
\item There's a special state that we can create this way: the vacuum. This arises by inserting the identity operator ${\bf 1}$ into the path integral. Back in the cylinder picture, this just means that we propagate the state back to time $\tau=-\infty$ which is a standard trick used in the Euclidean path integral to project out all but the ground state. For this reason the vacuum
    is sometimes referred to, in operator notation, as $\ket{{\bf 1}}$.
\end{itemize}

\subsubsection{Some Simple Consequences}

Let's use the state-operator map to wrap up a few loose ends that have arisen in
our study of conformal field theory.

\para
Firstly, we've defined two objects that we've called ``primary": states and operators.
The state-operator map relates the two.
Consider the state $\ket{\calo}$, built from inserting a primary operator
$\calo$ into the path integral at $z=0$. We can look at,
\be L_n\ket{\calo}&=&\oint\frac{dz}{2\pi i}\ z^{n+1}\,T(z)\,\calo(z=0) \nn\\
&=& \oint\frac{dz}{2\pi i}\ z^{n+1}\left(\frac{h\calo}{z^2} + \frac{\p\calo}{z}+\ldots
\right)\label{lgoes}\ee
You may wonder what became of the path integral $\int {\cal D}\phi\,e^{-S[\phi]}$ in
this expression. The answer is that it's still implicitly there. Remember that operator expressions such as \eqn{qln} are always taken to hold inside correlation functions.
But putting an operator in the correlation function is the same thing as putting it
in the path integral, weighted with $e^{-S[\phi]}$.

\para

From \eqn{lgoes} we can see the effect of various generators on states
\begin{itemize}
\item $L_{-1}\ket{\calo} = \ket{\p\calo}$: In fact, this is true for all operators, not just primary ones. It is expected since $L_{-1}$ is the translation generator.
\item $L_0\ket{\calo}=h\ket{\calo}$: This is true of any operator with a well defined transformation under scaling.
\item $L_n\ket{\calo}=0$ for all $n>0$. This is true only of primary operators $\calo$. Moreover, it is our requirement for $\ket{\calo}$ to be a primary state.
\end{itemize}
This has an important consequence. We stated earlier that one of the most important things to
compute in a CFT is the spectrum of weights of primary operators. This seems like a slightly
obscure thing to do. But now we see that it has a much more direct, physical meaning. It
is the spectrum of energy and angular momentum of states of the theory defined on the cylinder.

\para
Another loose end: when defining operators which carry specific weight, we made the statement
that we could always work in a basis of operators which have specified eigenvalues
under $D$ and $L$. This follows immediately from the statement that we can always
find a basis of eigenstates of $H$ and $L$ on the cylinder.

\para
\begin{figure}[htb]
\begin{center}
\epsfxsize=4.2in\leavevmode\epsfbox{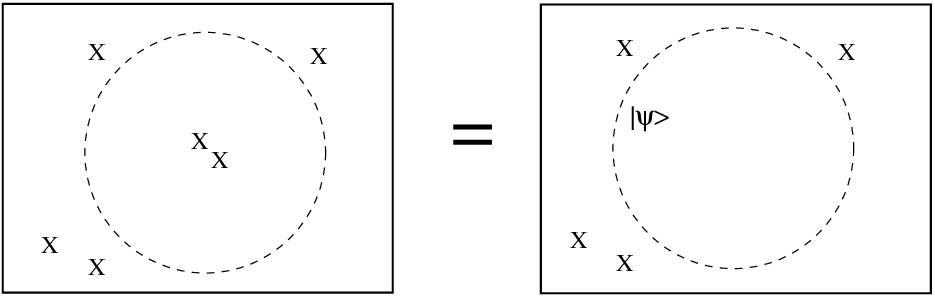}
\end{center}
\caption{}
%\label{mir}
\end{figure}

Finally, we can use this idea of the state-operator map to understand why the
OPE works so well in conformal field theories. Suppose that we're interested
in some correlation function, with operator insertions as shown in the figure.
The statement of the OPE is that we can replace the two inner operators by a
sum of operators at $z=0$, {\it independent} of what's going on outside of the dotted
line. As an operator statement, that sounds rather surprising. But this follows by computing
the path integral up to the dotted line, by which point the only effect of the two
operators is to determine what state we have.  This provides us a way of
understanding why the OPE is exact in CFTs, with a radius of convergence equal to the next-nearest insertion.

\subsubsection{Our Favourite Example: The Free Scalar Field}
\label{fsfsec}

Let's illustrate the state-operator map by returning yet again to the free scalar field.
On a Euclidean cylinder, we have the mode expansion
\be X(w,\bw) = x +  \ap p\,\tau + i\sfap\sum_{n\neq 0}
\frac{1}{n}\,\left(\alpha_n\,e^{inw} + \tilde{\alpha}_n\, e^{in\bw}\right) \nn\ee
where we retain the requirement of reality in Minkowski space, which gave us
$\alpha^\star_n = \alpha_{-n}$ and $\tilde{\alpha}^\star_n = \tilde{\alpha}_{-n}$.
We saw in Section \ref{freesec} that $X$ does not have good conformal properties. Before
transforming to the $z=e^{-iw}$ plane, we should work with the primary field on the cylinder,
\be \partial_wX(w,\bw) = - \sfap \sum_n \alpha_n\,e^{inw}\ \ \ \ \ \ {\rm with}\ \alpha_0
\equiv i\sqrt{\frac{\ap}{2}}p\nn\ee
Since $\p X$ is a primary field of weight $h=1$, its transformation to the plane is given by
\eqn{goprimary} and reads
\be \p_z X(z) = \left(\ppp{z}{w}\right)^{-1}\p_wX(w) = -i\sfap\sum_{n} \frac{\alpha_n}{z^{n+1}}\ \nn\ee
and similar for $\pb X$. Inverting this gives an equation for $\alpha_n$ as a
contour integral,
\be \alpha_n = i\sqrt\frac{2}{\ap}\oint \frac{dz}{2\pi i} \ z^n\,\p X(z)\label{qam}\ee
Just as the $TT$ OPE allowed us to determine the $[L_m,L_n]$ commutation relations
in the previous section, so the $\p X \p X$ OPE contains the information about the
$[\alpha_m,\alpha_n]$ commutation relations. The calculation is straightforward,
\be [\alpha_m,\alpha_n] &=& -\frac{2}{\ap}
\left(\oint\frac{dz}{2\pi i}\oint\frac{dw}{2\pi i}\ -\ \oint\frac{dw}{2\pi i}\oint\frac{dz}{2\pi i}\right)\ z^mw^n\,\p X(z)\,\p X(w)\nn\\\ &=&
-\frac{2}{\ap}\oint\frac{dw}{2\pi i}\ \Res_{z=w}\left[z^m w^n\left(\frac{-\ap/2}{(z-w)^2}
+\ldots\right)\right] \nn\\ &=& m\oint \frac{dw}{2\pi i}\ w^{m+n-1} = m\delta_{m+n,0}\nn\ee
where, in going from the second to third line, we have Taylor expanded $z$ around $w$.
Hearteningly, the final result agrees with the commutation relation \eqn{acom}
that we derived in string theory using canonical quantization.

\subsubsection*{The State-Operator Map for the Free Scalar Field}

Let's now look at the map between states and local operators. We know from canonical
quantization that the Fock space is defined by acting with creation operators
$\alpha_{-m}$ with $m>0$ on the vacuum $\vac$. The vacuum state itself obeys
$\alpha_m\vac=0$ for $m>0$. Finally, there is also the zero mode $\alpha_0\sim p$
which provides all states with another quantum number. A general state is given by
\be \prod_{m=1}^\infty\,\alpha_{-m}^{k_m}\,\ket{0;p}\nn\ee
Let's try and recover these states by inserting operators into the path integral.
Our first task is to check whether the vacuum state is indeed equivalent to the insertion
of the the identity operator. In other words, is the ground state
wavefunctional of the theory on the circle $|z|=r$ really given by
\be \Psi_0[X_f] = \int^{X_f(r)} {\cal D}X\ e^{-S[X]}\ \ \ \ \ \ \ ?
\label{isthisvac}\ee
We want to check that this satisfies the definition of the vacuum state, namely
$\alpha_m\vac=0$ for $m>0$. How do we act on the wavefunctional with an
operator? We should still integrate over all field configurations $X(z,\bz)$,
subject to the boundary conditions at $X(|z|=r) = X_f$. But now we
should insert the  contour integral \eqn{qam} at some $|w|<r$ (because, after all,
the state is only going to vanish after
we've hit it with $\alpha_m$, not before!). So we look at
\be \alpha_m\Psi_0[X_f] = \int^{X_f} {\cal D}X \ e^{-S[X]}
\oint \frac{dw}{2\pi i} \, w^m\p X(w)\nn\ee
The path integral is weighted by the action \eqn{xact} for a free scalar field.
If a given configuration diverges somewhere inside the disc $|z|<r$, then the action also
diverges. This ensures that only smooth functions $\p X(z)$, which have no singularity
inside the disc, contribute. But for such
functions we have
\be
\oint \frac{dw}{2\pi i} \, w^m\p X(w) = 0\ \ \ \ \ \mbox{for all}\ m\geq 0\nn\ee
So the state \eqn{isthisvac} is indeed the vacuum state. In fact, since $\alpha_0$ also
annihilates this state, it is identified as the vacuum state with vanishing momentum.

\para
What about the excited states of the theory?
\\{}\\
{\bf Claim:} $\alpha_{-m}\vac = \ket{\p^mX}$. By which we mean that the state $\alpha_{-m}\vac$ can be built from the path integral,
\be
\alpha_{-m}\vac = \int {\cal D}X\ e^{-S[X]}\ \p^m X(z=0)\label{xcite}\ee
\\
{\bf Proof:} We can check this by acting on $|\p^mX\rangle$ with the annihilation
operators $\alpha_{n}$.
\be
\alpha_n\ket{\p^mX} &\sim&
\int^{X_f(r)} {\cal D}X\ e^{-S[X]}\
\oint\frac{dw}{2\pi i} \ w^n\,\p X(w)\,\p^m X(z=0)\nn\ee
We can focus on the operator insertions and use the OPE \eqn{pxpx}. We drop the path
integral and just focus on the operator equation (because, after all, operator equations
only make sense in correlation functions which is the same thing as in path integrals). We
have
\be  \oint\frac{dw}{2\pi i}\ \left.w^n\,\partial_z^{m-1}\frac{1}{(w-z)^2}\right|_{z=0}
= m!\oint \frac{dw}{2\pi i}\ w^{n-m-1} = 0 \ \ \ \ \ \ \ \mbox{unless $m=n$}
\nn\ee
This confirms that the state \eqn{xcite} has the right properties. \hfill$\Box$

\para
Finally, we should worry about the zero mode, or momentum $\alpha_0\sim p$. It is
simple to show using the techniques above (together with the OPE \eqn{willneedthis})
that the momentum of a state arises by the insertion of the primary operator $e^{ipX}$. For example,
\be |0;p\rangle \sim \int {\cal D}X\ e^{-S[X]}\ e^{ipX(z=0)} \ .\nn\ee

\subsection{Brief Comments on Conformal Field Theories with Boundaries}
\label{opencftsec}

\EPSFIGURE{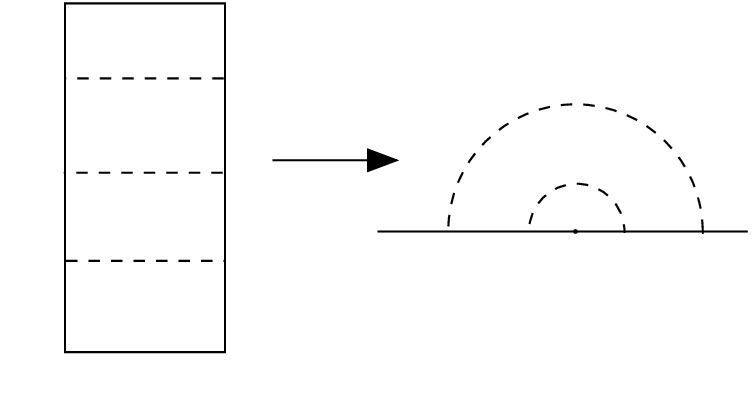,height=100pt}{}
The open string lives on the infinite strip with spatial coordinate $\sigma \in [0,\pi]$.
Here we make just a few brief comments on the corresponding conformal field theories.

\para
As before, we can define the complex coordinate $w=\sigma + i\tau$ and make the
conformal map
\be z = e^{-iw}\nn\ee
This time the map takes us to the upper-half plane: ${\rm Im} z \geq 0$. The end
points of the string are mapped to the real axis, ${\rm Im} z =0$.

\para
Much of our previous discussion goes through as before. But now we need to take care of
boundary conditions at ${\rm Im}z=0$. Let's first look at $\tab$. Recall that the stress-energy
tensor exists because of translational invariance. We still have translational invariance
in the direction
parallel to the boundary --- let's call the associated tangent vector $t^\alpha$. But
translational invariance is broken perpendicular to the boundary --- we call the normal
vector $n^\alpha$. The upshot of this is that $T_{\alpha\beta}t^\beta$ remains a conserved
current.

\para
To implement Neumann boundary conditions, we insist that none of the current flows out of the
boundary. The condition is
\be \tab n^\alpha t^\beta =0\ \ \ \ \ {\rm at}\ {\rm Im}z=0\nn\ee
In complex coordinates, this becomes
\be T_{zz} = T_{\bz\bz} \ \ \ \ \ \ {\rm at} \ {\rm Im}z=0\nn\ee
There's a simple way to implement this: we extend the definition of $T_{zz}$ from the
upper-half plane to the whole complex plane by defining
\be T_{zz}(z) = T_{\bz\bz}(\bz)\nn\ee
For the closed string we had both functions $T$ and $\bt$ in the whole plane.
But for the open string, we have just
one of these -- say, $T$, ---  in the whole plane. This contains the same information as
both $T$ and $\bt$ in the
upper-half plane. It's simpler to work in the whole plane and focus just on $T$.
Correspondingly, we now have just a single set of
Virasoro generators,
\be L_n= \oint \frac{dz}{2\pi i}\,z^{n+1}\,T_{zz}(z)\nn\ee
There is no independent $\tilde{L}_n$ for the open string.

\para
A similar doubling trick works when computing the propagator for the free scalar field. The
scalar field $X(z,\bz)$ is only defined in the upper-half plane. Suppose we want to
implement Neumann boundary conditions. Then the propagator is defined by
\be \langle X(z,\bz)\,X(w,\bw)\rangle = G(z,\bz;w,\bw)\nn\ee
which obeys $\p^2G = -2\pi \ap\,\delta(z-w,\bz-\bw)$ subject to the boundary condition
\be \left.\partial_\sigma\,G(z,\bz;w,\bw)\right|_{\sigma=0}=0\nn\ee
But we solve problems like this in our electrodynamics courses. A useful way of proceeding
is to introduce an ``image charge" in the lower-half plane. We now let $X(z,\bz)$
vary over the whole complex plane with its dynamics governed by the propagator
\be G(z,\bz;w,\bw) = -\frac{\ap}{2}\,\ln|z-w|^2-\frac{\ap}{2}\,\ln|z-\bw|^2\label{image}\ee
%
%\para
%\EPSFIGURE{openso.eps,height=110pt}{}
%
Much of the remaining discussion of CFTs carries forward with only minor differences. However,
there is one point that is simple but worth stressing because it will be of importance later.
This concerns the state-operator map. Recall the logic that leads us to this idea: we consider
a state at fixed time on the strip, and propagate it back to past infinity
$\tau\rightarrow -\infty$. After the map to the half-plane, past infinity is again the origin.
But now the origin lies on the boundary. We learn that the state-operator map relates
states to local operators defined on the boundary.

\para This fact ensures that theories on a strip have fewer states than those on the cylinder.
For example, for a free scalar field, Neumann boundary conditions require $\p X=\bp X$ at
${\rm Im} z= 0$. (This follows from the requirement that $\p_\sigma X=0$ at $\sigma =0,\pi$ on the
strip). On the cylinder, the operators $\p X$ and $\bp X$ give rise to different states; on the
strip they give rise to the same state. This, of course, mirrors what we've seen for the quantization
of the open string where boundary conditions mean that we have only half the oscillator modes to play
with.

%\para
%We won't make any further explicit
%comments about CFTs with boundaries,
% but will instead mention some relevant points as and when we need them ---
%see, for example, Sections \ref{openspecsec} and \ref{venezsec}.

\newpage
\section{The Polyakov Path Integral and Ghosts}
\label{polyakov}

At the beginning of the last chapter, we stressed that there are two very different interpretations of conformal symmetry depending on whether we're thinking of a fixed
2d background or a dynamical
2d background. In applications to statistical physics, the background is fixed and conformal
symmetry is a global symmetry. In contrast, in string theory the background is dynamical.
Conformal symmetry is a gauge symmetry, a remnant of diffeomorphism invariance and Weyl
invariance.

\para
But gauge symmetries are not symmetries at all. They are redundancies in
our description of the system. As such, we can't afford to lose them and it is imperative
that they don't suffer  an anomaly in the quantum theory. At worst, theories with gauge
anomalies make no sense. (For example, Yang-Mills theory coupled to only left-handed fundamental
fermions is a nonsensical theory for this reason). At best, it may be possible to recover
the quantum theory, but it almost certainly has nothing to do with the theory that you
started with.

\para
Piecing together some results from the previous chapter, it looks like we're in
trouble. We saw that the Weyl symmetry is anomalous since the expectation value
of the stress-energy tensor takes different values on backgrounds related by a Weyl symmetry:
\be \vev{T^\alpha_{\ \alpha}} = -\frac{c}{12}\,R\nn\ee
On fixed backgrounds, that's merely interesting. On dynamical backgrounds, it's fatal. What
can we do? It seems that the only way out is to ensure that our theory has $c=0$. But
we've already seen that $c>0$ for all non-trivial, unitary CFTs. We seem to have reached
an impasse.
In this section we will discover the loophole. It turns out that we do indeed require
 $c=0$, but there's a way to achieve this that makes sense.

\subsection{The Path Integral}

In Euclidean space the Polyakov action is given by,
\be S_{\rm Poly}=\frac{1}{4\pi \ap}\int d^2\sigma\sqrt{g}\ \gabi\,\p_\alpha X^\mu\,\p_\beta X^\nu\ \delta_{\mu\nu}
\nn\ee
From now on, our analysis of the string will be in terms of the path integral\footnote{The analysis of the string path integral was first performed by Polyakov in
``{\it Quantum geometry of bosonic strings},", Phys.\ Lett.\  B {\bf 103}, 207 (1981). The
paper weighs in at a whopping 4 pages. As a follow-up, he took another 2.5 pages to analyze
the superstring in ``{\it Quantum geometry of fermionic strings},''
Phys.\ Lett.\  B {\bf 103}, 211 (1981).}. We
integrate over all embedding coordinates $X^\mu$ and all
worldsheet metrics $\gab$.  Schematically, the path integral is given by,
\be Z = \frac{1}{{\rm Vol}}\,\int {\cal D}g{\cal D}X\ e^{-S_{\rm Poly}[X,g]}\nn\ee
\EPSFIGURE{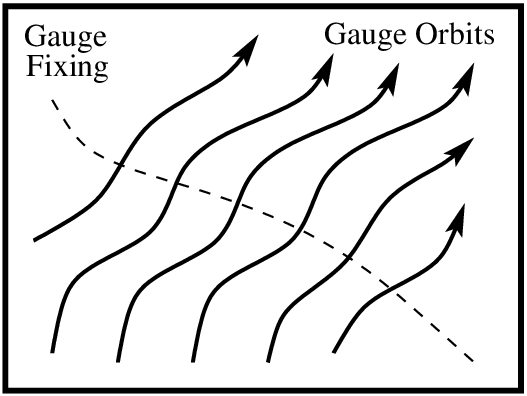,height=100pt}{}
The ``{\rm Vol}" term is all-important. It refers to the fact that we shouldn't be
integrating over all field configurations, but only those physically distinct
configurations not related by diffeomorphisms and Weyl symmetries. Since the
path integral, as written, sums over all fields, the ``Vol" term means
that we need to divide out by the volume of the gauge action on field space.

\para
To make the situation more explicit, we need to split the integration over all
field configurations into two pieces:
those corresponding to physically distinct configurations --- schematically depicted as
the dotted line in the figure --- and those corresponding to
gauge transformations --- which are shown  as solid lines. Dividing by ``Vol" simply
removes the piece of the partition function which comes from integrating along the solid-line
gauge orbits.

\para
In an ordinary integral, if we change coordinates then we pick up a Jacobian factor
for our troubles. The path integral is no different. We want to decompose our integration
variables into physical fields and gauge orbits. The tricky part is to figure out what
Jacobian we get. Thankfully, there is a
standard method to determine the Jacobian, first introduced by Faddeev and Popov. This method
works for all gauge symmetries, including Yang-Mills, and you will also learn about it
in the ``Advanced Quantum Field Theory" course.

\subsubsection{The Faddeev-Popov Method}

We have two gauge symmetries: diffeomorphisms and Weyl transformations. We will schematically
denote both of these by $\zeta$. The change of the metric under a general gauge transformation
is  $g\rightarrow g^{\,\zeta}$. This is shorthand for,
\be \gab(\sigma)\ \longrightarrow\  g^{\,\zeta}_{\alpha\beta}(\sigma^\prime) = e^{2\omega(\sigma)}\,\ppp{\sigma^\gamma}{\sigma^{\prime\,\alpha}}
\ppp{\sigma^\delta}{\sigma^{\prime\,\beta}}\,g_{\gamma\delta}(\sigma)\nn\ee
In two dimensions these gauge symmetries allow us to put the metric into any form
that we like --- say, $\hat{g}$. This is called the fiducial metric and will represent
our choice of gauge fixing. Two caveats:
\begin{itemize}
\item Firstly, it's not true that we can put any 2d metric into the form $\hat{g}$ of
our choosing. This is only true locally. Globally, it remains true if the worldsheet has
the topology of a cylinder or a sphere, but not for higher genus surfaces. We'll revisit
this issue in Section \ref{scattering}.
\item Secondly, fixing the metric locally to $\hat{g}$ does not fix all the gauge
symmetries. We still have the conformal symmetries to deal with. We'll revisit this
in the Section \ref{scattering} as well.
\end{itemize}

Our goal is to only integrate over physically inequivalent configurations. To achieve this,
first consider the integral over the gauge orbit of $\hat{g}$. For some value of the gauge
transformation $\zeta$, the configuration $g^{\,\zeta}$ will coincide with our
original metric $g$. We can put a delta-function in the integral to get
\be \int {\cal D}\zeta\ \delta(g-\hat{g}{}^{\,\zeta}) = \Delta^{-1}_{FP}[g]\label{dfp}\ee
This integral isn't equal to one because we need to take into account the Jacobian
factor. This is analogous to the statement that $\int dx\,\delta(f(x)) = 1/|f^{\prime}|$,
evaluated at points where
$f(x)=0$. In the above equation, we have written this Jacobian factor as $\DFPI$.
The inverse of this, namely $\DFP$, is called the {\it Faddeev-Popov determinant}. We
will evaluate it explicitly shortly. Some comments:
\begin{itemize}
\item This whole procedure is rather formal and runs into the usual difficulties with trying
to define the path integral. Just as for Yang-Mills theory, we will find that it results
in sensible answers.
\item We will assume that our gauge fixing is good, meaning that the dotted line in the
previous figure cuts through each
physically distinct configuration exactly once.
Equivalently, the
 integral over gauge transformations ${\cal D}\zeta$ clicks
exactly once with the delta-function   and we don't have to worry about discrete
ambiguities (known as Gribov copies in QCD).
\item The measure is taken to be the analogue of the Haar measure for Lie groups, invariant
under left and right actions
\be {\cal D}\zeta = {\cal D}(\zeta^\prime\zeta) = {\cal D}(\zeta\zeta^\prime)\nn\ee
\end{itemize}
Before proceeding, it will be useful to prove a quick lemma:
\\{}\\
{\bf Lemma:} $\DFP[g]$ is gauge invariant. This means that
\be \DFP[g]=\DFP[g^{\,\zeta}]\nn\ee
{\bf Proof:}
\be \DFPI[g^{\,\zeta}] &=& \int{\cal D}\zeta^\prime\ \delta(g^{\,\zeta}-\hat{g}^{\,\zeta^\prime}) \nn\\
&=& \int {\cal D}\zeta^\prime\ \delta(g-\hat{g}^{\,\zeta^{-1}\zeta^\prime})\nn\\
&=& \int {\cal D}\zeta^{\prime\prime}\ \delta(g-\hat{g}^{\,\zeta^{\prime\prime}})
= \DFPI[g]\nn\ee
where, in the second line, we have used the fact that the measure is invariant.\hfill$\Box$

\para
Now we can employ the Faddeev-Popov procedure. We start by inserting a factor of unity into
the path integral, in the guise of
\be 1 = \DFP[g]\ \int{\cal D}\zeta\ \delta(g-\hat{g}^{\,\zeta})\nn\ee
We'll call the resulting path integral expression $Z[\hat{g}]$ since it depends
on the choice of fiducial metric $\hat{g}$. The
first thing we do is use the $\delta(g-\hat{g}^{\,\zeta})$ delta-function to do the
integral over metrics,
\be Z[\hat{g}]&=&\frac{1}{{\rm Vol}}\int{\cal D}\zeta{\cal D}X{\cal D}g\ \DFP[g]
\,\delta(g-\hat{g}^{\,\zeta})\,e^{-S_{\rm Poly}[X,g]} \nn\\ &=&
\frac{1}{{\rm Vol}}\int{\cal D}\zeta{\cal D}X\ \DFP[\hat{g}^{\,\zeta}]
\,e^{-S_{\rm Poly}[X,\hat{g}^{\,\zeta}]}
\nn\\ &=& \frac{1}{{\rm Vol}}\int{\cal D}\zeta{\cal D}X\ \DFP[\hat{g}]
\,e^{-S_{\rm Poly}[X,\hat{g}]} \nn\ee
where to go to the last line, we've used the fact that the action is gauge invariant, so
$S[X,g]=S[X,g^{\,\zeta}]$, and, as we proved in the lemma above, the Faddeev-Popov
determinant is also gauge invariant.

\para
But now, nothing depends on the gauge transformation $\zeta$. Indeed, this is precisely the
integration over the gauge orbits that we wanted to isolate and it cancels the ``Vol"
factor sitting outside. We're left with
\be Z[\hat{g}] = \int {\cal D}X\ \DFP[\hat{g}]\,e^{-S_{\rm Poly}[X,\hat{g}]}\label{fullpf}\ee
This is the integral over physically distinct configurations --- the dotted line in
the previous figure. We see that the Faddeev-Popov determinant is precisely the Jacobian
factor that we need.

\subsubsection{The Faddeev-Popov Determinant}
\label{fpdsec}

We still need to compute $\DFP[\hat{g}]$. It's defined in \eqn{dfp}. Let's look at
gauge transformations $\zeta$ which are close to the identity. In this case, the
delta-function $\delta(g-\hat{g}^{\,\zeta})$ is going to be non-zero when the
metric $g$ is close to the fiducial metric $\hat{g}$. In fact, it will be sufficient
to look at the delta-function $\delta(\hat{g}-\hat{g}^{\,\zeta})$, which is only
non-zero when $\zeta=0$. We take an infinitesimal
Weyl transformation parameterized by $\omega(\sigma)$ and an
infinitesimal diffeomorphism $\delta\sigma^\alpha = v^\alpha(\sigma)$. The change in the metric is
\be \delta \hat{g}_{\alpha\beta}  = 2\omega\hat{g}_{\alpha\beta} + \nabla_\alpha v_\beta + \nabla_\beta v_\alpha\nn\ee
Plugging this into the delta-function, the expression for the Faddeev-Popov determinant
becomes
\be \DFPI[\hat{g}]=\int {\cal D}\omega{\cal D}v\ \delta(2\omega\hat{g}_{\alpha\beta}
+\nabla_{\alpha}v_\beta + \nabla_\beta v_\alpha)\label{yeswecan}\ee
where we've replaced the integral ${\cal D}\zeta$ over the gauge group with the integral
${\cal D}\omega{\cal D}v$ over the Lie algebra of group since we're near the identity.
(We also suppress the subscript on $v_\alpha$ in the measure factor to keep things
looking tidy).

\para
At this stage it's useful to  represent the delta-function in its integral, Fourier form. For
a single delta-function, this is $\delta(x) = \int dp\ \exp(2\pi ip\,x)$. But
the delta-function in \eqn{yeswecan} is actually a delta-functional: it restricts a whole
function. Correspondingly, the integral representation is in terms of a functional integral,
\be \DFPI[\hat{g}] = \int {\cal D}\omega{\cal D}v{\cal D}\beta
\ \exp\left(2\pi i\int d^2\sigma\ \sqrt{\hat{g}}\ \beta^{\alpha\beta}
[2\omega\hat{g}_{\alpha\beta}
+\nabla_{\alpha}v_\beta + \nabla_\beta v_\alpha]\right)\nn\ee
where $\beta^{\alpha\beta}$ is a symmetric 2-tensor on the worldsheet.

\para
We now simply do the $\int {\cal D}\omega$ integral. It doesn't come with any
derivatives, so it merely acts as a Lagrange multiplier, setting
\be \beta^{\alpha\beta} \hat{g}_{\alpha\beta} =0\nn\ee
In other words, after performing the $\omega$ integral, $\beta^{\alpha\beta}$ is symmetric
and traceless. We'll take this to be the definition of $\beta^{\alpha\beta}$ from now on. So, finally
we have
\be  \DFPI[\hat{g}] = \int {\cal D}v{\cal D}\beta
\ \exp\left(4\pi i\int d^2\sigma\ \sqrt{\hat{g}}\ \beta^{\alpha\beta}
\nabla_{\alpha}v_\beta\right)\nn\ee

\subsubsection{Ghosts}

The previous manipulations  give us an expression for $\DFPI$. But we
want to invert it to get $\DFP$. Thankfully, there's a simple
way to achieve this. Because the integrand is quadratic in $v$ and $\beta$, we know that
the integral computes the inverse determinant of the operator $\nabla_\alpha$. (Strictly
speaking, it computes the inverse determinant of the projection of $\nabla_\alpha$ onto
symmetric, traceless tensors. This observation is important because it means
the relevant operator is a square matrix which is necessary to talk about a determinant).
But we also
know how to write down an expression for the determinant $\DFP$, instead of its inverse,
in terms of
path integrals: we
simply need to replace the commuting integration variables with anti-commuting fields,
\be \beta_{\alpha\beta} &\longrightarrow& b_{\alpha\beta} \nn\\ v^\alpha & \longrightarrow
& c^\alpha\nn\ee
where $b$ and $c$ are both Grassmann-valued fields (i.e. anti-commuting). They are known as
{\it ghost fields}. This gives us our final expression for the Faddeev-Popov determinant,
\be \DFP[g] = \int{\cal D}b{\cal D}c\ \exp[iS_{\rm ghost}]\nn\ee
where the ghost action is defined to be
\be S_{\rm ghost} = \frac{1}{2\pi}\int d^2\sigma \sqrt{g}\ b_{\alpha\beta}\nabla^\alpha c^\beta\label{ghost1}\ee
and we have chosen to rescale the $b$ and $c$ fields at this last step to get a
factor of $1/2\pi$
sitting in front of the action. (This only changes the normalization of the partition function
which doesn't matter). Rotating back to Euclidean space, the factor of $i$ disappears.
The expression for the full partition function \eqn{fullpf} is
\be Z[\hat{g}]=\int {\cal D}X{\cal D}b{\cal D}c\ \exp\left(-S_{\rm Poly}[X,\hat{g}] -
S_{\rm ghost}[b,c,\hat{g}]\right)\nn\ee
Something lovely has happened. Although the ghost fields were introduced as some auxiliary
constructs, they now appear on the same footing as the dynamical fields $X$. We learn
that gauge
fixing comes with a price: our theory has extra ghost fields.

\para
The role of these ghost fields is to cancel the unphysical gauge degrees of freedom, leaving
only the $D-2$ transverse modes of $X^\mu$. Unlike lightcone quantization, they achieve
this in a way which preserves Lorentz invariance.

\subsubsection*{Simplifying the Ghost Action}

The ghost action \eqn{ghost1} looks fairly simple. But it looks even simpler if we
work in conformal gauge,
\be \hat{g}_{\alpha\beta} = e^{2\omega}\delta_{\alpha\beta}\nn\ee
The determinant is $\sqrt{\hat{g}}=e^{2\omega}$. Recall that in complex coordinates,
the measure is $d^2\sigma =\ft12 d^2z$, while we can lower the index on
the covariant derivative using $\nabla^z=g^{z\bar{z}}\nabla_{\bz}=2e^{-2\omega}\nabla_{\bz}$.
We have
\be S_{\rm ghost} = \frac{1}{2\pi}\int d^2z\ \left( b_{zz}\nabla_{\bz}c^z + b_{\bz\bz}\nabla_z c^{\bz}\right)
\nn\ee
In deriving this, remember that there is no field $b_{z\bz}$ because $b_{\alpha\beta}$
is traceless. Now comes the nice part: the covariant derivatives are actually just
ordinary derivatives. To see why this is the case, look at
\be \nabla_{\bz} c^z = \p_{\bz}c^z + \Gamma^z_{\bz\alpha}c^\alpha\nn\ee
But the Christoffel symbols are given by
\be \Gamma^z_{\bz\alpha} = \frac{1}{2}g^{z\bz}\left(\p_{\bz}g_{\alpha\bz}
+\p_\alpha g_{\bz\bz} - \p_{\bz}g_{\bz\alpha}\right)=0\ \ \ \ \ {\rm for}\ \alpha=z,\bz
\nn\ee
So in conformal gauge, the ghost action factorizes into two  free theories,
\be S_{\rm ghost} = \frac{1}{2\pi} \int d^2z\ b_{zz}\,\p_{\bz}c^z +
b_{\bz\bz}\,\p_zc^{\bz}\nn\ee
The action doesn't depend on the conformal factor $\omega$.
In other words, it is Weyl invariant without any need to change $b$ and $c$: these
are therefore both neutral under Weyl transformations.

\para
(It's worth pointing out that $b_{\alpha\beta}$ and $c^{\alpha}$ are neutral
under Weyl transformations. But if we raise or lower these indices, then the
fields pick up factors of the metric. So $b^{\alpha\beta}$ and $c_{\alpha}$ would
not be neutral under Weyl transformations).

\subsection{The Ghost CFT}

Fixing the Weyl and diffeomorphism gauge symmetries has left us with two
new dynamical ghost fields, $b$ and $c$. Both are Grassmann (i.e. anti-commuting)
variables. Their dynamics is governed by a CFT. Define
\be b= b_{zz}\ \ \ \ &,&\ \ \ \ \bar{b}=b_{\bz\bz} \nn\\
c=c^z\ \ \ \ \ &,&\ \ \ \ \bar{c}=c^{\bz}\nn\ee
The ghost action is given by
\be S_{\rm ghost} = \frac{1}{2\pi}\int d^2z\ \left(b\,\pb c + \bar{b}\,\p\bar{c}
\right)\nn\ee
Which gives the equations of motion
\be \pb b = \p\bar{b} = \pb c = \p\bar{c}=0\nn\ee
So we see that $b$ and $c$ are holomorphic fields, while $\bar{b}$ and $\bar{c}$ are
anti-holomorphic.

\para
Before moving onto quantization, there's one last bit of information we need from the
classical theory: the stress tensor for the $bc$ ghosts. The calculation
is a little bit fiddly. We use the general definition of the stress tensor \eqn{se},
which requires us to return to the theory \eqn{ghost1} on a general background and vary the
metric $g^{\alpha\beta}$. The complications are twofold. Firstly, we pick up a contribution
from the Christoffel symbol that is lurking inside the covariant derivative $\nabla^\alpha$.
Secondly, we must also remember that $b_{\alpha\beta}$ is traceless. But this is a condition
which itself depends on the metric: $b_{\alpha\beta}g^{\alpha\beta}=0$. To account for
this we should add a Lagrange multiplier to the action imposing tracelessness. After correctly
varying the metric, we may safely retreat back to flat space where the end result is
rather simple. We have $T_{z\bz}=0$, as we must for any conformal
theory. Meanwhile, the holomorphic and anti-holomorphic parts of the stress tensor are given by,
\be T = 2(\partial c)\,b  + c\, \partial b\ \ \ \ ,\ \ \ \ \bar{T}=2(\bar{\partial}\bar{c})\,
\bar{b} + \bar{c}\,\bar{\partial}\bar{b}.
\ee

\subsubsection*{Operator Product Expansions}

We can compute the OPEs of these fields using the standard path integral techniques
that we employed in the last chapter. In what follows, we'll just focus on the
holomorphic piece of the CFT.  We have, for example,
\be 0 = \int {\cal D}b{\cal D}c\ \frac{\delta}{\delta b(\sigma)}\ \left[e^{-S_{\rm ghost}}
\,b(\sigma^\prime)\right ] = \int {\cal D}b{\cal D}c\ s^{-S_{\rm ghost}}\left[
-\frac{1}{2\pi}\,\pb c(\sigma)\,b(\sigma^\prime)+\delta(\sigma-\sigma^\prime)\right]\nn\ee
which tells us that
\be \pb c(\sigma)\,b(\sigma^\prime) = 2\pi \,\delta(\sigma-\sigma^\prime)\nn\ee
Similarly, looking at $\delta/\delta c(\sigma)$ gives
\be \pb b(\sigma)\,c(\sigma^\prime) = 2\pi\,\delta(\sigma-\sigma^\prime)\nn\ee
We can integrate both of these equations using our favorite formula
$\bp(1/z) = 2\pi \delta(z,\bz)$.  We learn that the OPEs between fields are given by
\be b(z)\,c(w) &=& \frac{1}{z-w} + \ldots\nn\\ c(w)\,b(z) &=& \frac{1}{w-z} + \ldots\nn\ee
In fact the second equation follows from the first equation and Fermi statistics. The
OPEs of $b(z)\,b(w)$ and $c(z)\,c(w)$ have no singular parts. They vanish as $z\rightarrow
w$.

\para
Finally, we need the stress tensor of the theory. After normal ordering, it is given by
\be T(z) = 2:\p c(z)\,b(z): + :c(z)\p b(z):\nn\ee
We will shortly see that with this choice, $b$ and $c$ carry appropriate weights for tensor
fields which are neutral under Weyl rescaling.

\subsubsection*{Primary Fields}

We will now show that both $b$ and $c$ are primary fields, with weights $h=2$ and $h=-1$ respectively. Let's
start by looking at $c$. The OPE with the stress tensor is
\be T(z)\,c(w) &=& 2:\p c(z)\,b(z):c(w) + :c(z)\,\p b(z): c(w) \nn\\
&=& \frac{2\p c(z)}{z-w}-\frac{c(z)}{(z-w)^2} +\ldots= -\frac{c(w)}{(z-w)^2}+
\frac{\p c(w)}{z-w}+\ldots\nn\ee
confirming that $c$ has weight $-1$. When taking the OPE with $b$, we need to be a little
more careful with minus signs. We get
\be T(z)\,b(w) &=& 2:\p c(z)\,b(z):b(w) + :c(z)\,\p b(z): b(w) \nn\\
&=& -2b(z)\left(\frac{-1}{(z-w)^2}\right)-\frac{\p b(z)}{z-w} = \frac{2b(w)}{(z-w)^2}+
\frac{\p b(w)}{z-w}+\ldots\nn\ee
showing that $b$ has weight 2. As we've pointed out a number of times,
conformal = diffeo + Weyl. We mentioned
earlier that the fields $b$ and $c$ are neutral under Weyl transformations. This
is reflected in their weights, which are due solely to diffeomorphisms as dictated
by their index structure: $b_{zz}$ and $c^z$.

\subsubsection*{The Central Charge}

Finally, we can compute the $TT$ OPE to determine the central charge of the $bc$ ghost system.
\be T(z)\,T(w)  &=& 4:\p c(z)b(z)\!:\,:\p c(w)b(w)\!: + 2:\p c(z)b(z)\!:\,:
c(w)\p b(w)\!: \nn\\ &&\ \ + 2:c(z)\p b(z)\!:\,:\p c(w)b(w)\!: +
:c(z)\p b(z)\!:\,:c(w)\p b(w)\!: \nn\ee
For each of these terms, making two contractions gives a $(z-w)^{-4}$ contribution to the OPE.
There are also two ways to make a single contraction. These give $(z-w)^{-1}$ or $(z-w)^{-2}$ or
$(z-w)^{-3}$ contributions depending on what the derivatives hit. The end result is
\be T(z)\,T(w)  &=& \frac{-4}{(z-w)^4} + \frac{4:\p c(z)b(w)\!:}{(z-w)^2}
-\frac{4:b(z)\p c(w)\!:}{(z-w)^2} \nn\\ &&\ \ -\frac{4}{(z-w)^4} +\frac{2:\p c(z)\,\p b(w)\!:}{z-w}
-\frac{4:b(z)c(w)\!:}{(z-w)^3} \nn\\ &&\ \ -\frac{4}{(z-w)^4} -\frac{4:c(z)b(w)\!:}{(z-w)^3}
+\frac{2:\p b(z)\p c(w)\!:}{z-w} \nn\\ && \ \ -\frac{1}{(z-w)^4}
-\frac{:c(z)\p b(w)\!:}{(z-w)^2} + \frac{\p b(z) c(w)\!:}{(z-w)^2} +\ldots\nn
\ee
After some Taylor expansions to turn $f(z)$ functions into $f(w)$ functions, together with a little collecting of terms, this can be written as,
\be T(z)\,T(w) = \frac{-13}{(z-w)^4}+\frac{2T(w)}{(z-w)^2} + \frac{\p T(w)}{z-w} +\ldots\nn\ee
The first thing to notice is that it indeed has the form expected of $TT$ OPE. The second,
and most important, thing to notice is the  central charge of the $bc$ ghost system: it is
\be c=-26 \nn\ee

\subsection{The Critical ``Dimension" of String Theory}

Let's put the pieces together. We've learnt that gauge fixing the diffeomorphisms and
Weyl gauge symmetries results in the introduction of ghosts which contribute central
charge $c=-26$. We've also learnt that the Weyl symmetry is anomalous unless $c=0$.
Since the Weyl symmetry is a gauge symmetry, it's crucial that we keep it. We're forced
to add exactly the right degrees of freedom to the string to cancel the contribution
from the ghosts.

\para
The simplest possibility is to add $D$ free scalar fields. Each of these contributes $c=1$
to the central charge, so the whole procedure is only consistent if we pick
\be D=26 \nn\ee
This agrees with the result we found in Chapter 2: it is the critical dimension of string theory.
\para
However, there's no reason that we have to work with free scalar fields. The consistency requirement is merely that the degrees of freedom of the string are described by a
CFT with $c=26$. Any CFT will do. Each such CFT describes a different background in which a string
can propagate. If you like, the space of CFTs with $c=26$ can be thought of as the space
of classical solutions of string theory.

\para
We learn that the ``critical dimension" of string theory is something of a misnomer:
it is really a ``critical central charge". Only
for rather special CFTs can this central charge be thought of as a spacetime dimension.

\para
For example, if we wish to describe strings moving in 4d Minkowski space, we can take $D=4$
free scalars (one of which will be timelike) together with some other $c=22$ CFT. This CFT
may have a geometrical interpretation, or it may be something more abstract.
The CFT with $c=22$ is sometimes called the ``internal sector" of the theory. It
is what we really mean when we talk about the ``extra hidden dimensions of string theory".
We'll see some examples of CFTs describing curved spaces in Section \ref{background}.

\para
There's one final subtlety: we need to be careful with the transition back to
Minkowski space. After all, we want one of the directions of the CFT, $X^0$, to
have the wrong sign kinetic term. One safe way to do this is to keep $X^0$ as a
free scalar field, with the remaining degrees of freedom described by some $c=25$ CFT.
This doesn't seem quite satisfactory
though since it doesn't allow for spacetimes which evolve in time --- and, of course,
 these are certainly necessary if we wish to understand early universe cosmology. There are still
some technical obstacles to understanding the worldsheet of the string in time-dependent backgrounds.
To make progress, and discuss string cosmology, we usually bi-pass this issue by working with
the low-energy effective action which we will derive in Section \ref{background}.

\subsubsection{The Usual Nod to the Superstring}

The superstring has another gauge symmetry on the worldsheet: supersymmetry. This gives
rise to more ghosts, the so-called $\beta\gamma$ system, which turns out to have
central charge $+11$.
Consistency then requires that the degrees of freedom of the string have central
charge $c=26-11=15$.

\para
However, now the CFTs must themselves be invariant under supersymmetry,
which means that bosons come matched with fermions. If we add $D$ bosons, then
we also need to add $D$ fermions. A free boson has $c=1$,
while a free fermion has $c=1/2$. So, the total number of free bosons that we
should add is $D(1+1/2) = 15$, giving us the critical dimension of the superstring:
\be D=10
\nn\ee

\subsubsection{An Aside: Non-Critical Strings}
\label{polyncsec}

Although it's a slight departure from the our main narrative, it's worth pausing
to mention what Polyakov actually did in his four page paper. His main focus was
not critical strings, with $D=26$, but rather {\it non-critical} strings with $D\neq 26$.
From the discussion above, we know that these suffer from a Weyl anomaly. But it turns
out that there is a way to make sense of the situation.

\para
The starting point is to abandon Weyl invariance from the beginning. We
start with $D$ free scalar fields coupled to a dynamical worldsheet metric
$g_{\alpha\beta}$. (More generally, we could have any CFT). We still
want to keep reparameterization invariance, but now we ignore the constraints
of Weyl invariance. Of course, it seems likely that this isn't going to have too
much to do with the Nambu-Goto string, but let's proceed anyway. Without Weyl
invariance, there is one extra term that it is natural to add to the 2d theory:
a worldsheet cosmological constant $\mu$,
\be S_{\rm non-critical} = \frac{1}{4\pi\ap} \int d^2\sigma \sqrt{g}\left(
\gabi\p_\alpha X^\mu \p_\beta X_\mu  + \mu\right)\nn\ee
Our goal will be to understand how the partition function changes under a
Weyl rescaling. There will be two contributions: one from the explicit $\mu$ dependence and one from the Weyl anomaly. Consider two metrics related by a Weyl transformation
\be \hat{g}_{\alpha\beta} = e^{2\omega} {g}_{\alpha\beta}\nn\ee
As we vary $\omega$, the partition function $Z[\hat{g}]$ changes as
\be \frac{1}{Z}\,\ppp{Z}{\omega} &=& \frac{1}{Z} \int {\cal D}\phi \ e^{-S}\,\left(-\ppp{S}{\hat{g}_{\alpha\beta}}\,\ppp{\hat{g}_{\alpha\beta}}{\omega}
\right)\nn\\ &=& \frac{1}{Z}\int {\cal D}\phi\ e^{-S}\,\left(-\frac{1}{2\pi}\,\sqrt{\hat{g}}\,T^{\alpha}_{\ \alpha}
\right) \nn\\ &=& \frac{c}{24\pi}\,\sqrt{\hat{g}}\,\hat{R} - \frac{1}{2\pi\ap} \,\mu e^{2\omega}\nn\\ &=& \frac{c}{24\pi}\,\sqrt{{g}}({R}-2\nabla^2 \omega) - \frac{1}{2\pi\ap}
\,\mu e^{2\omega}
\nn\ee
where, in the last two lines, we used the Weyl anomaly \eqn{cr} and the relationship
between Ricci curvatures \eqn{riccirel}. The central charge appearing in these
formulae includes the contribution from the ghosts,
\be c=D-26\nn\ee
We can now just treat this as a differential
equation for the partition function $Z$ and solve.
This allows us to express the partition function $Z[\hat{g}]$, defined on one worldsheet
metric, in terms of $Z[g]$, defined on another. The relationship is,
\be Z[\hat{g}] = Z[{g}] \exp\left[-\frac{1}{4\pi\ap}\int d^2\sigma\ \sqrt{{g}}\left(2\mu e^{2\omega} - \frac{c\ap}{6}\left({g}^{\alpha\beta}\,\partial_\alpha\omega\,\partial^\beta
\omega + {R}\omega\right)\right)\right]\nn\ee
We see that the scaling mode $\omega$ inherits a kinetic term. It now appears as a new
dynamical scalar field in the theory. It is often called the Liouville field on
account of the exponential potential term multiplying $\mu$. Solving this theory
is quite hard\footnote{A good review can be found Seiberg's article ``{\it
Notes on Quantum Liouville Theory and Quantum Gravity}", Prog. Theor. Phys.
Supl. 102 (1990) 319.}. Notice also that our
new scalar field $\omega$ appears in the final term multiplying the Ricci scalar $R$. We
will describe the significance of this in Section \ref{bsec}. We'll also see another
derivation of this kind of Lagrangian in Section \ref{ncsec}.

\subsection{States and Vertex Operators}

In Chapter 2 we determined the spectrum of the string in flat space. What is the
spectrum for a general string background? The theory consists of the $b$ and $c$ ghosts, together
with a $c=26$ CFT.
At first glance, it seems that we have a greatly enlarged Hilbert space since
we can act with creation operators from all fields, including the ghosts.
However, as you might expect, not all of these states will be physical. After correctly
accounting for the gauge symmetry, only some subset survives.

\para
The elegant method to determine the physical Hilbert space in a gauge fixed action with
ghosts is known as {\it BRST quantization}. You will learn about it in the ``Advanced Quantum
Field Theory" course where you will apply it to Yang-Mills theory.
Although a correct construction of the string spectrum employs the BRST method, we won't describe
it here for lack of time. A very clear description of the general method and its application
to the string can be found in Section 4.2 of Polchinski's book.

\para
Instead, we will make do with a poor man's
attempt to determine the spectrum of the string. Our strategy is to simply pretend that
the ghosts aren't there and focus
on the states created by the fields of the matter CFT (i.e. the $X^\mu$ fields if we're
talking about flat space). As we'll explain in the next section, if we're only interested in tree-level
scattering amplitudes then this will suffice.

\para
To illustrate how to compute the spectrum of the string, let's go back to flat
$D=26$ dimensional Minkowski space and the discussion of covariant quantization
in Section \ref{covquansec}. We found that physical states $\ket{\Psi}$ are subject
to the Virasoro constraints \eqn{itsl} and \eqn{itslzero} which read
\be L_n\ket{\Psi} &=& 0\ \ \  \ \ \ \ \ \ \ \ \ {\rm for}\ n>0\nn\\ L_0\ket{\Psi}&=&a\ket{\Psi} \nn\ee
and similar for $\tilde{L}_n$,
\be \tilde{L}_n\ket{\Psi} &=& 0\ \ \  \ \ \ \ \ \ \ \ \ {\rm for}\ n>0\nn\\ \tilde{L}_0\ket{\Psi}&=&\tilde{a}\ket{\Psi} \nn\ee
where we have, just briefly, allowed for the possibility of different normal
ordering coefficients  $a$ and $\tilde{a}$ for the left- and right-moving sectors.
But there's a name for states in a conformal field theory obeying these requirements: they are
primary states of weight $(a,\tilde{a})$.

\para
So how do we fix the normal ordering ambiguities $a$ and $\tilde{a}$?
A simple way is to first replace the states with operator insertions on the worldsheet using the
state-operator map: $\ket{\Psi}\rightarrow {\cal O}$. But we have a further requirement on
the operators $\calo$:
gauge invariance. There are two gauge symmetries: reparameterization invariance, and Weyl symmetry.
Both restrict the possible states.

\para
Let's start by considering reparameterization invariance. In the last section, we happily placed
operators at specific points on the worldsheet. But in a theory with a dynamical metric,
this doesn't give rise to a diffeomorphism invariant operator. To make an object that is
invariant under reparameterizations of the worldsheet coordinates, we should integrate over
the whole worldsheet. Our operator insertions (in conformal gauge) are therefore of the form,
\be V \sim \int d^2z\ \calo\label{vop}\ee
Here the $\sim$ sign reflects the fact that we've dropped an overall normalization
constant which we'll return to in the next section.

\para
Integrating over the worldsheet takes care of diffeomorphisms. But what about Weyl
symmetries? The measure $d^2z$ has weight $(-1,-1)$ under rescaling. To compensate,
the operator $\calo$ must have weight $(+1,+1)$. This is how we fix the normal ordering ambiguity: we require $a=\tilde{a}=1$. Note that this agrees with the
normal ordering coefficient $a=1$ that we derived in lightcone quantization in Chapter 2.

\para
This, then, is the rather rough derivation of the string spectrum. The physical states are
the primary states of the CFT with weight $(+1,+1)$. The operators \eqn{vop} associated to
these states are called {\it vertex operators}.

\subsubsection{An Example: Closed Strings in Flat Space}
\label{closedvsec}

Let's use this new language to rederive the spectrum of the closed string in flat space. We
start with the ground state of the string, which was previously identified as a tachyon.
As we saw
in Section 4, the vacuum
of a CFT is associated to the identity operator. But we also have the zero modes. We can give
the string momentum $p^\mu$ by acting with the operator
$e^{ip\cdot X}$. The vertex operator associated to the ground state of the string is
therefore
\be V_{\rm tachyon} \sim \int d^2z\ :e^{ip\cdot X}:\label{tvertex}\ee
In Section \ref{fsfnowsec}, we showed that the operator $e^{ip\cdot X}$ is primary with
weight $h=\tilde{h} = \ap p^2/4$. But Weyl invariance requires that the operator has
weight $(+1,+1)$. This is only true if the mass of the state is
\be M^2 \equiv -p^2 = -\frac{4}{\ap}\nn\ee
This is precisely the mass of the tachyon that we saw in Section 2.

\para
Let's now look at the first excited states. In covariant quantization, these are of the
form $\zeta_{\mu\nu}\,\alpha_{-1}^\mu\,\tilde{\alpha}_{-1}^\nu\,\ket{0;p}$, where
$\zeta_{\mu\nu}$ is a constant tensor that
determines the type of state, together with its polarization. (Recall: traceless
symmetric $\zeta_{\mu\nu}$ corresponds to the graviton, anti-symmetric $\zeta_{\mu\nu}$
corresponds to the $B_{\mu\nu}$ field, and the trace of $\zeta_{\mu\nu}$ corresponds
to the scalar known as the dilaton). From \eqn{xcite}, the vertex operator
associated to this state is,
\be V_{\rm excited} \sim \int d^2z\ :e^{ip\cdot X}\,\p X^\mu\pb X^\nu:\,\zeta_{\mu\nu}
\label{1vertex}\ee
where $\p X^\mu$ gives us a $\alpha_{-1}^\mu$ excitation, while $\bp X^\mu$ gives a
$\tilde{\alpha}_{-1}^\mu$ excitation. It's easy to check that the weight of
this operator is $h=\tilde{h}=1+\ap p^2/4$. Weyl invariance therefore requires that
\be p^2=0\nn\ee
confirming that the first excited states of the string are indeed massless. However,
we still need to check that the operator in \eqn{1vertex} is actually primary. We know
that $\p X$ is primary, and we know that $e^{ip\cdot X}$ is primary, but now we want
to consider them both sitting together inside the normal ordering.  This means that there 
are extra terms in the Wick contraction which give rise to $1/(z-w)^3$ terms in the OPE, 
potentially ruining the primacy of our operator. One such term arises from a double contraction, 
one of which includes the $e^{ip\cdot X}$ operator. This gives rise to an offending 
term proportional to $p^\mu\zeta_{\mu\nu}$. The same kind of contraction with $\bar{T}$ gives 
rise to a term proportional to $p^\nu\zeta_{\nu\mu}$. In order for these terms to vanish, 
the polarization tensor must satisfy
\be p^\mu \zeta_{\mu\nu}=p^\nu\zeta_{\mu\nu}=0\nn\ee
which is precisely the transverse polarization condition expected for a massless particle.

\subsubsection{An Example: Open Strings in Flat Space}
\label{openspecsec}

As explained in Section \ref{opencftsec}, vertex operators for the open-string
are inserted on the boundary $\p{\cal M}$ of the worldsheet. We still need to ensure that
these operators are diffeomorphism invariant which is achieved by integrating
over $\p{\cal M}$. The vertex operator for the open
string tachyon is
\be V_{\rm tachyon} \sim \int_{\p {\cal M}}ds\ :e^{ip\cdot X}:\nn\ee
We need to figure out the dimension of the boundary operator $:e^{ip\cdot X}:$. It's
not the same as for the closed string. The reason is due to presence of the
image charge in the propagator \eqn{image} for a free scalar field on a space with
boundary. This propagator appears in the Wick contractions in the OPEs and affects
the weights. Let's see why this is the case. Firstly, we look at a single scalar field $X$,
\be \p X(z)\,:e^{ipX(w,\bw)}: &=& \sum_{n=1}^\infty\frac{(ip)^n}{(n-1)!}:X(w,\bw)^{n-1}:
\left(-\frac{\ap}{2}\frac{1}{z-w}-\frac{\ap}{2}\frac{1}{z-\bw}\right) +\ldots
\nn\\ &=& -\frac{i\ap p}{2}\,:e^{ipX(w,\bw)}:\,\left(\frac{1}{z-w}+\frac{1}{z-\bw}\right)
+\ldots \nn\ee
With this result, we can now compute the OPE with $T$,
\be T(z)\,:e^{ipX(w,\bw)}:&=& \frac{\ap p^2}{4}\,:e^{ipX}:\left(\frac{1}{z-w}
+\frac{1}{z-\bw}\right)^2 + \ldots\nn\ee
When the operator $:e^{ipX(w,\bw)}:$ is placed on the boundary $w=\bw$, this becomes
\be T(z):e^{ipX(w,\bw)}: = \frac{\ap p^2\,:e^{ipX(w,\bw)}:}{(z-w)^2} +\ldots\nn\ee
This tells us that the boundary operator $:e^{ip\cdot X}:$ is indeed primary, with weight
$\ap p^2$.

\para
For the open string, Weyl invariance requires that operators have weight $+1$ in order
to cancel the scaling dimension of $-1$ coming from the boundary integral $\int ds$. So the
mass of the open string ground state is
\be M^2 \equiv -p^2 = -\frac{1}{\ap}\nn\ee
in agreement with the mass of the open string tachyon computed in Section 3.

\para
The vertex operator for the photon is
\be V_{\rm photon} \sim \int_{\p {\cal M}}ds\ \zeta_a\,:\p X^a\,e^{ip\cdot X}:\label{photonv}\ee
where the index $a=0,\ldots p$ now runs only over those directions with Neumann boundary
conditions that lie parallel to the brane worldvolume.
The requirement that this is a primary operator gives $p^a\zeta_{a}=0$, while Weyl
invariance tells us that $p^2=0$. This is the expected behaviour for the momentum and polarization
of a photon.

\subsubsection{More General CFTs}

Let's now consider a string propagating in four-dimensional Minkowski space ${\cal M}_4$,
together with some internal CFT with $c=22$. Then any primary operator of the internal
CFT with weight $(h,h)$ can be assigned momentum $p^\mu$, for $\mu=0,1,2,3$ by
dressing the operator with  $e^{ip\cdot X}$. In order to get a primary operator of weight $(+1,+1)$ as required, we must have
\be \frac{\ap p^2}{4} = 1-h\nn\ee
We see that the mass spectrum of closed string states is given by
\be M^2 = \frac{4}{\ap}(h-1)\nn\ee
where $h$ runs over the spectrum of primary operators of the internal CFT.
Some comments:
\begin{itemize}
\item
Relevant operators in the internal CFT have $h<1$ and give rise to tachyons
in the spectrum. Marginal operators, with $h=1$, give massless particles. And irrelevant
operators result in massive states.
\item Notice that requiring the vertex operators to be Weyl invariant determines
the mass formula for the state. We say that the vertex operators are ``on-shell",
in the same sense that external legs of Feynman diagrams are on-shell. We will have
more to say about this in the next section.
\end{itemize}

\newpage
\section{String Interactions}
\label{scattering}

So far, despite considerable effort, we've only discussed the free string. We now wish to
consider interactions. If we take the analogy with quantum
field theory as our guide, then we might be led to think that interactions require us to add
various non-linear terms to the action. However, this isn't the case. Any attempt to add extra
non-linear terms for the string won't be consistent with our precious gauge symmetries.
Instead, rather remarkably, all the information about interacting strings is already contained in the free theory described by the Polyakov action. (Actually, this statement is almost true).

\para
\EPSFIGURE{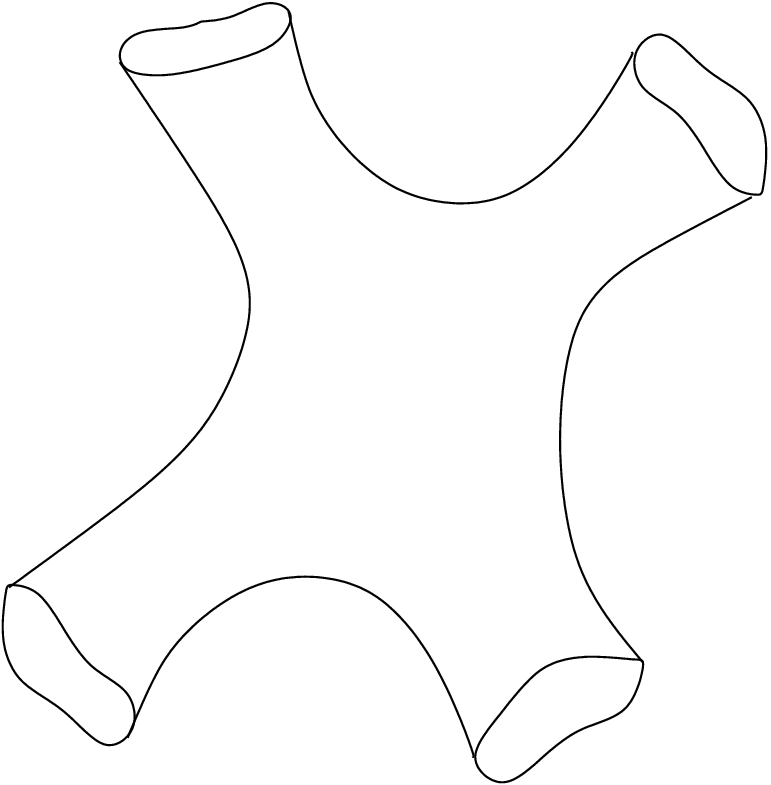,height=120pt}{}
\noindent
To see that this is at least feasible, try to draw a cartoon picture of two strings
interacting. It looks something like the worldsheet shown in the figure. The worldsheet
is smooth.
In Feynman diagrams in quantum field theory, information about  interactions is
inserted at vertices, where different lines meet. Here there are no such points.
Locally, every part of
the diagram looks like a free propagating string. Only globally do we see that the diagram
describes interactions.

\subsection{What to Compute?}

If the information about string interactions is already contained in
the Polyakov action, let's go ahead and compute something! But what should we compute? One obvious
thing to try is the probability for a particular configuration of strings at an
early time to evolve into a new configuration at some later time. For example,
we could try to compute the amplitude associated to the diagram above, stipulating fixed
curves for the string ends.

\para
No one knows how to do this. Moreover, there are words that we can drape around
this failure that suggests this  isn't really a  sensible thing to compute.
I'll now try to explain these words. Let's start by returning
to the familiar framework of quantum field theory in a fixed background.
There the basic objects that we can compute are correlation functions,
\be \langle \phi(x_1)\ldots \phi(x_n)\rangle\label{notgi}\ee
After a Fourier transform, these describe Feynman diagrams in which the external legs
carry arbitrary momenta. For this reason, they are referred to as {\it off-shell}. To get
the scattering amplitudes, we simply need to put the external legs on-shell (and perform
a few other little tricks captured in the LSZ reduction formula).

\para
The discussion above needs amendment if we turn on gravity. Gravity is a gauge theory and the
gauge symmetries are diffeomorphisms. In a gauge theory, only gauge invariant observables
make sense. But the correlation function \eqn{notgi} is not gauge invariant because its value
changes under a diffeomorphism which maps the points $x_i$ to another point. This
emphasizes an important fact: there are no local off-shell gauge invariant observables in a theory of
gravity.

%\para
%There is another, more physical, way to say this. What use is a Feynman diagram in which the external %legs are off-shell? The simplest answer is that you can think of the diagram as
%part of a larger diagram. For example, in deep inelastic scattering, the incoming photon is far from %off-shell and this provides a useful probe of QCD. But, in order to isolate this
%diagram, it's necessary that the photon probe arises from a sector external to QCD.
%But in a theory of gravity there is no external
%sector which can provide off-shell states. Gravity couples to everything.

\para
There is another way to say this. We know, by causality, that space-like separated operators should
commute in a quantum field theory. But in gravity the question of whether operators are space-like
separated becomes a dynamical issue and the causal structure can fluctuate due to quantum effects.
This provides another reason why we are unable to define local gauge invariant observables in any
theory of quantum gravity.

\para
Let's now return to string theory. Computing the evolution of string configurations for a finite
time is analogous to computing off-shell correlation functions in QFT. But string theory
is a theory of gravity so such things probably don't make sense. For this reason, we retreat
from attempting to compute correlation functions, back to the S-matrix.

%This argument sounds like
%a ``deep truth". Like all deep truths, it would disappear in a puff of smoke if anyone came
%along and showed how to compute these things

\subsubsection*{The String S-Matrix}

\EPSFIGURE{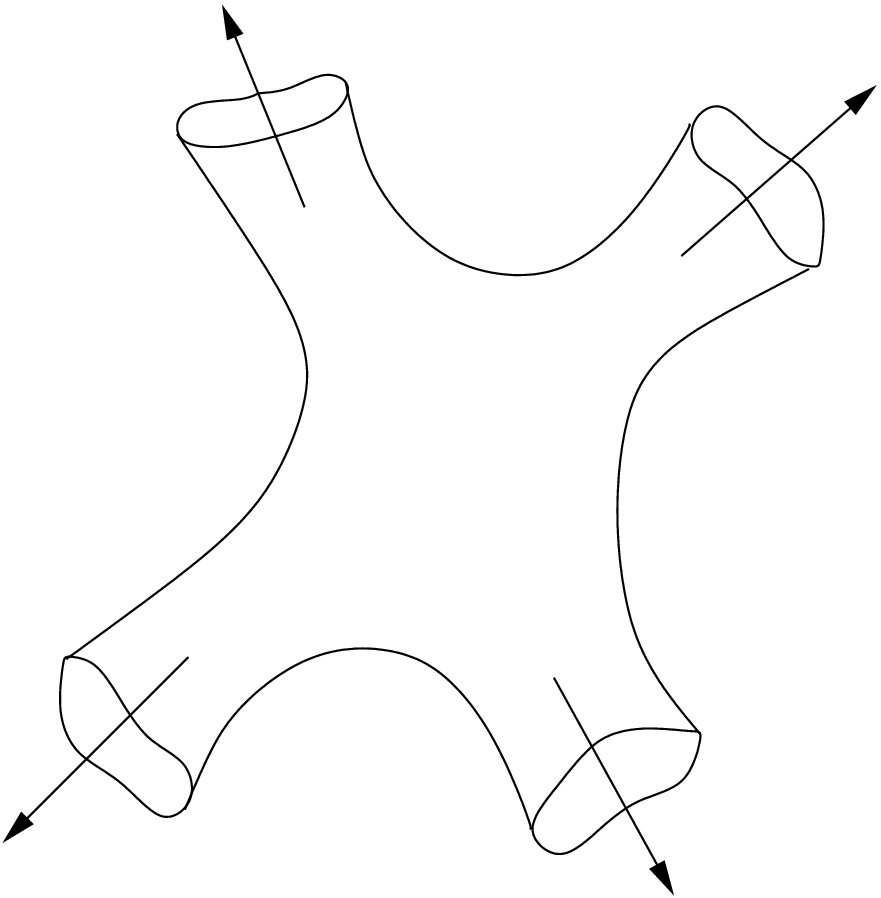,height=130pt}{}
\noindent
The object that we can compute in string theory is the S-matrix. This is obtained by taking
the points in the correlation function to infinity: $x_i\rightarrow \infty$. This is
acceptable because, just like in the case of QED,  the redundancy of the system consists
of those gauge transformations which die off  asymptotically. Said another way, points on the
boundary don't fluctuate in quantum gravity.
(Such fluctuations would be over an infinite volume of space and
are suppressed due to their infinite action).

\para
So what we're really going to calculate is a diagram of the type shown in the
figure, where all external legs are taken to infinity. Each of these legs can
be placed in a different state of the free string and assigned some spacetime
momentum $p_i$. The resulting expression is the string {\it S-matrix}.

\para
Using the state-operator map, we  know that each of these states at infinity is
equivalent to the insertion of an appropriate vertex operator on the worldsheet.
Therefore, to  compute this S-matrix element we use a conformal transformation to bring each
of these infinite legs to a finite distance. The end result is a worldsheet with
the topology

\EPSFIGURE{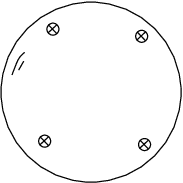,height=80pt}{}
\noindent
 of the sphere,  dotted with vertex operators where the legs used to be.
However, we already saw in the previous section that the constraint of Weyl
invariance meant that vertex operators are necessarily on-shell. Technically,
this is the reason that we can only compute on-shell correlation functions in
string theory.

\subsubsection{Summing Over Topologies}

The Polyakov path integral instructs us to sum over all metrics. But what about
worldsheets of different topologies? In fact, we should also sum over these. It
is this sum that gives the perturbative expansion of string theory. The scattering of
two strings receives contributions from worldsheets of the form
\be \raisebox{-8.1ex}{\epsfxsize=4.5in\epsfbox{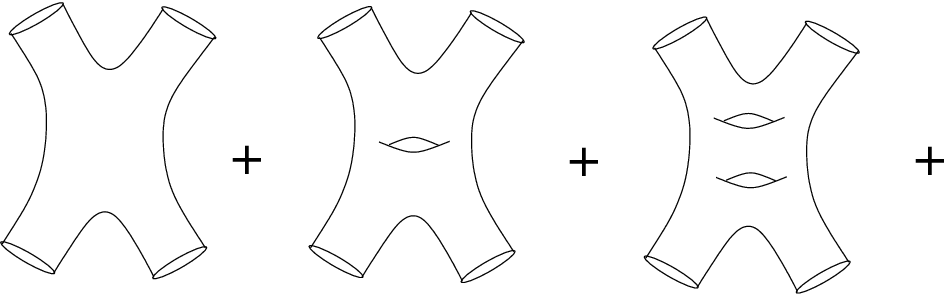}}\label{wssum}\ee
The only thing that we need to know is how to weight these different worldsheets.
Thankfully, there is a
very natural coupling on the string that we have yet to consider and this will do
the job. We augment the Polyakov action by
\be S_{\rm string} = S_{\rm Poly} + \lambda \chi
\label{lchi}\ee
Here $\lambda$ is simply a real number, while $\chi$ is given by an integral over the
(Euclidean) worldsheet
\be \chi = \frac{1}{4\pi} \int d^2\sigma\ \sqrt{g} R \label{chi}\ee
where $R$ is the Ricci scalar of the worldsheet metric.
This looks like the Einstein-Hilbert term for gravity on the worldsheet. It is simple
to check that it is invariant under reparameterizations and Weyl transformations.

\para
In four-dimensions, the Einstein-Hilbert term makes gravity dynamical. But life is very different in 2d. Indeed, we've already seen that all the components of the metric can be gauged away so there are no propagating degrees of freedom associated to $\gab$. So, in two-dimensions, the term \eqn{chi} doesn't make gravity dynamical: in fact, classically, it doesn't do anything at all!

\para
The reason for this is that $\chi$ is a topological invariant. This means that it doesn't actually depend on the metric $\gab$ at all -- it depends only on the topology of the worldsheet. (More precisely, $\chi$ only depends on those global properties of the metric which themselves depend on the topology of the worldsheet). This is the content of the  Gauss-Bonnet theorem: the integral of the Ricci scalar $R$ over the worldsheet gives an integer, $\chi$, known as the Euler number of the worldsheet. For a worldsheet without boundary (i.e. for the closed string) $\chi$ counts the number of handles $h$ on the worldsheet. It is given by,
\be \chi = 2-2h = 2(1-g) \ee
where $g$ is called the {\it genus} of the surface. The simplest examples are shown in the figure. The sphere has $g=0$ and $\chi =2$; the torus has $g=1$ and $\chi=0$. For higher $g>1$, the
Euler character $\chi$ is negative.
\begin{figure}[htb]
\begin{center}
\epsfxsize=4.2in\leavevmode\epsfbox{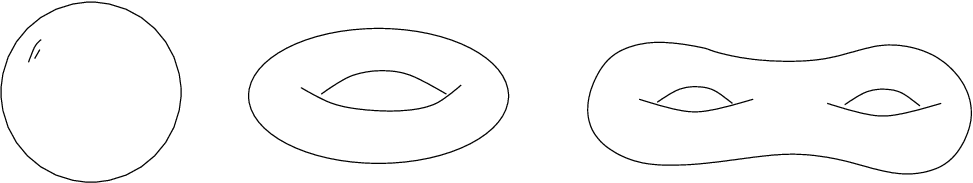}
\end{center}
\caption{Examples of increasingly poorly drawn Riemann surfaces with $\chi=2,0$ and $-2$.}
\end{figure}

\para
Now we see that the number $\lambda$ --- or, more precisely, $e^\lambda$ ---  plays  the role
of the string coupling. The integral
over worldsheets is weighted by,
\be \mathop{\sum_{\rm topologies}}_{\rm metrics}\ e^{-S_{\rm string}}
\sim\sum_{\rm topologies}\ e^{-2\lambda(1-g)}\int {\cal D}X{\cal D}g \ e^{-S_{\rm Poly}}\nn\ee
For $e^\lambda\ll 1$, we have a good perturbative expansion in which we sum over all
topologies. (In fact, it is an asymptotic
expansion, just as in quantum field theory). It is standard to define the string
coupling constant as
\be g_s=e^\lambda\nn\ee
After a conformal map, tree-level scattering corresponds to a worldsheet with the topology of a
sphere: the amplitudes are proportional to $1/g_s^2$. One-loop scattering corresponds to
toroidal worldsheets and, with our normalization, have no power of $g_s$. (Although, obviously,
these are suppressed by $g_s^2$ relative to tree-level processes). The end result is that
the sum over worldsheets in \eqn{wssum} becomes a sum over Riemann surfaces of increasing
genus, with vertex operators inserted for the initial and final states,
\be \raisebox{-10.1ex}{\epsfxsize=4.5in\epsfbox{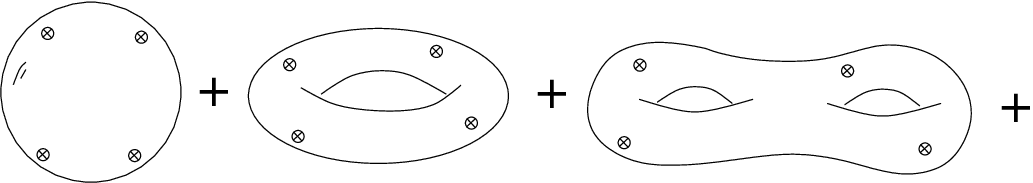}}\nn\ee

\noindent
The Riemann surface of genus $g$ is weighted by
\be (g_s^{2})^{g-1}\nn\ee
While it may look like we've introduced a new parameter $g_s$ into the theory and
added the coupling \eqn{lchi} by hand, we will later see why this coupling is
a necessary part of the theory and provide an interpretation for $g_s$.

\subsubsection*{Scattering Amplitudes}

We now have all the information that we need to explain how to compute string
scattering amplitudes. Suppose that we want to
compute the S-matrix for $m$ states: we will label them as $\Lambda_i$ and assign them
spacetime momenta $p_i$. Each has a corresponding vertex operator $V_{\Lambda_i}(p_i)$.
The S-matrix element is then computed by evaluating the correlation function in the
2d conformal field theory, with insertions of the vertex operators.
\be {\cal A}^{(m)}(\Lambda_i,p_i) = \sum_{\rm topologies} g_s^{-\chi}\ \frac{1}{\rm Vol}
\int {\cal D}X{\cal D}g\ e^{-S_{\rm Poly}}\ \prod_{i=1}^m\ V_{\Lambda_i}(p_i)\nn\ee
This is a rather peculiar equation. We are interpreting the correlation functions of a
two-dimensional theory as the S-matrix for a theory in $D=26$ dimensions!

\para
To properly compute the correlation function, we should introduce  the $b$ and $c$ ghosts that
we saw in the last chapter and treat them carefully. However, if we're only interested
in tree-level amplitudes, then we can proceed naively and ignore the ghosts.
The reason can be seen in the ghost action \eqn{ghost1} where we see that
the ghosts couple only to the worldsheet metric,
not to the other worldsheet fields. This means that if our gauge fixing procedure fixes
the worldsheet metric completely --- which it does for worldsheets with the topology of
a sphere --- then we can forget about the ghosts. (At least, we can forget about them as soon
as we've made sure that the Weyl anomaly cancels). However, as we'll explain in \ref{oneloopsec},
for higher genus worldsheets,
the gauge fixing does not fix the metric completely and there are residual dynamical modes of the
metric, known as moduli,
which couple the ghosts and matter fields. This is analogous
to the statement in field theory that we only need to worry about ghosts
running in loops.

\subsection{Closed String Amplitudes at Tree Level}

The tree-level scattering amplitude is given by the correlation function of the 2d theory,
evaluated on the sphere,
\be {\cal A}^{(m)} = \frac{1}{g^2_s}\,\frac{1}{\rm Vol}\int {\cal D}X{\cal D}g\
e^{-S_{\rm Poly}}\ \prod_{i=1}^m\ V_{\Lambda_i}(p_i)\nn\ee
where $V_{\Lambda_i}(p_i)$ are the vertex operators associated to the states.

\para
\EPSFIGURE{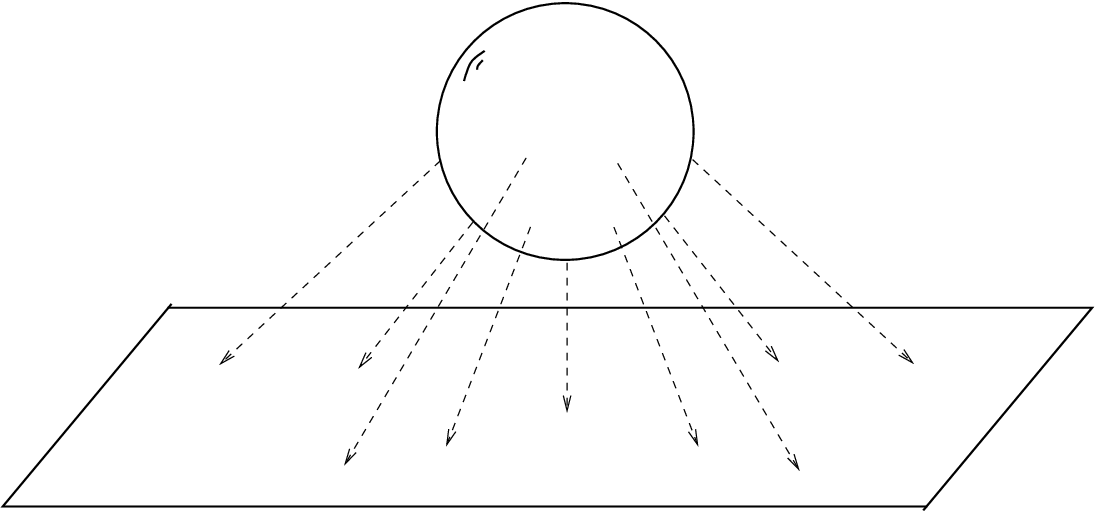,height=80pt}{}
\noindent
We want to integrate over all metrics on the sphere. At first glance that sounds
rather daunting
but, of course, we have the gauge symmetries of diffeomorphisms and Weyl transformations
 at our disposal. Any metric on the sphere is
conformally
equivalent to the flat metric on the plane. For example, the round metric on the sphere
of radius $R$ can be written as
\be ds^2 = \frac{4R^2}{(1+|z|^2)^2}\,dzd\bz\nn\ee
which is manifestly conformally equivalent to the plane, supplemented by the point at infinity.
The conformal map from the sphere to the plane is the stereographic projection depicted in the
diagram. The south pole of the sphere is mapped to the origin; the north pole is mapped to the
point at infinity.
Therefore, instead of integrating over all metrics, we may gauge fix diffeomorphisms and
Weyl transformations to leave ourselves with the seemingly easier task of computing
correlation functions on the plane.

\subsubsection{Remnant Gauge Symmetry: SL(2,C)}

There's a subtlety. And it's  a subtlety that we've seen before: there is a
residual gauge symmetry. It is the conformal group, arising from diffeomorphisms
which can be undone by Weyl transformations. As we saw in Section 4, there are an
infinite number of such conformal transformations. It looks like we have a whole lot
of gauge fixing still to do.

\para
However, global issues actually mean that
there's less remnant gauge symmetry than you might think.
In Section 4, we only looked at infinitesimal conformal transformations, generated by
the Virasoro operators $L_n$, $n\in {\bf Z}$. We did not examine whether these
transformations are well-defined and invertible over all of space. Let's take a look at this.
Recall that the coordinate changes associated to $L_n$ are generated by the vector
fields \eqn{littleln},
\be l_n = z^{n+1}\p_z\nn\ee
which result in the shift $\delta z = \epsilon z^{n+1}$. This is non-singular at $z=0$
only for $n\geq -1$. If we restrict to smooth maps, that gets rid of half the
transformations right away. But, since we're ultimately interested in the sphere, we
now also need to worry about the point at $z=\infty$ which, in stereographic projection,
is just the north pole of the sphere. To do this, it's useful to work with the coordinate
\be u=\frac{1}{z}\nn\ee
The generators of coordinate transformations for the $u$ coordinate are
\be l_n=z^{n+1}\p_z = \frac{1}{u^{n+1}}\ppp{u}{z}\p_u=-u^{1-n}\p_u\nn\ee
which is non-singular at $u=0$ only for $n\leq 1$.

\para
Combining these two results, the only generators of the conformal group that are
non-singular over the whole Riemann sphere are $l_{-1}$, $l_0$ and $l_1$ which
act infinitesimally as
\be l_{-1}&:&\ \  z\ \rightarrow\ z+\epsilon \nn\\ l_0&:&\ \  z\ \rightarrow\ (1+\epsilon)z
\nn\\ l_1 &:&\ \  z\ \rightarrow\ (1+\e z)z\nn\ee
The global version of these transformations is
\be
l_{-1}&:&\ \  z\ \rightarrow\ z+\alpha \nn\\ l_0&:&\ \  z\ \rightarrow\ \lambda z
\nn\\ l_1 &:&\ \  z\ \rightarrow\ \frac{z}{1-\beta z}\nn\ee
which can be combined to give the general transformation
\be z\ \rightarrow\ \frac{az+b}{cz+d}\label{sl2c}\ee
with $a,b,c$ and $d\in {\bf C}$. We have four complex parameters, but we've only got
three transformations. What happened? Well, one transformation is fake because
an overall scaling of the
parameters doesn't change $z$. By such a rescaling, we can always
insist that the parameters obey
\be ad-bc=1\nn\ee
The transformations \eqn{sl2c} subject to this constraint have the group structure
$SL(2;{\bf C})$, which is the group of $2\times 2$ complex matrices with unit determinant.
In fact, since the transformation is blind to a flip in sign of all the parameters, the actual
group of global conformal transformations is $SL(2;{\bf C})/{\bf Z}_2$, which
is sometimes written as $PSL(2;{\bf C})$. (This ${\bf Z}_2$ subtlety won't be important
for us in what follows).

\para
The remnant global transformations on the sphere are known as {\it conformal
Killing vectors} and the group $SL(2;{\bf C})/{\bf Z}_2$ is the {\it conformal
Killing group}. This group allows us to take any three points on the plane and
move them to three other points of our choosing. We will shortly make use
of this fact to gauge fix, but for now we leave the $SL(2;{\bf C})$ symmetry intact.

\subsubsection{The Virasoro-Shapiro Amplitude}

We will now compute the S-matrix for closed string tachyons. You might think that this is
the least interesting thing to compute: after all, we're ultimately interested in the
superstring which doesn't have tachyons. This is true, but it turns out that tachyon
scattering is much simpler than everything else, mainly because we don't have a plethora of
extra indices on the states to worry about. Moreover, the lessons that we will learn
from tachyon scattering hold for the scattering of other states as well.

\para
The $m$-point tachyon scattering amplitude is given by the flat
space correlation function
\be {\cal A}^{(m)}(p_1,\ldots,p_m)=\frac{1}{g_s^2}\frac{1}{{\rm Vol}(SL(2;{\bf C}))}
\int {\cal D}X\ e^{-S_{\rm Poly}}\ \prod_{i=1}^m V(p_i)\nn\ee
where the tachyon vertex operator is given by,
\be V(p_i) = g_s\int d^2z\ e^{ip_i\cdot X} \equiv g_s\int d^2z\ \hat{V}(z,p_i)
\label{tvo}\ee
Note that, in contrast to \eqn{tvertex}, we've added an appropriate normalization factor to the vertex operator. Heuristically, this reflects the fact that the operator is associated to the addition of a
closed string mode. A rigorous derivation of this normalization can be found in Polchinski.

\para
The amplitude can therefore be written as,
\be {\cal A}^{(m)}(p_1,\ldots,p_m)=\frac{g_s^{m-2}}{{\rm Vol}(SL(2;{\bf C}))} \int\prod_{i=1}^m d^2z_i\
\langle \hat{V}(z_1,p_1)\ldots \hat{V}(z_m,p_m)\rangle\nn\ee
where the expectation value $\langle\dots\rangle$ is computed using the gauge fixed Polyakov
action. But the gauge fixed Polyakov action is simply a free theory and our correlation function
is something eminently computable: a Gaussian integral,
\be
\langle \hat{V}(z_1,p_1)\ldots \hat{V}(z_m,p_m)\rangle = \int {\cal D}X
\exp\left(-\frac{1}{2\pi\ap}\int d^2z\ \p X\cdot \pb X\right)\ \exp\left(
i\sum_{i=1}^mp_i\cdot X(z_i,\bz_i)\right)
\nn\ee
The normalization in front of the Polyakov action is now $1/2\pi\ap$ instead of $1/4\pi\ap$
because we're working with complex coordinates and we need to remember that $\p_\alpha\p^\alpha = 4\p\bp$
and $d^2z=2d^2\sigma$.

\subsubsection*{The Gaussian Integral}

We certainly know how to compute Gaussian integrals. Let's go slow. Consider the following
general integral,
\be \int {\cal D}X\ \exp\left(\int d^2z\ \frac{1}{2\pi\ap} X\cdot \p\bp X +i J\cdot X\right)
\sim \exp\left(\frac{\pi\ap}{2}\int d^2zd^2z^\prime\ J(z,\bz)\,\frac{1}{\p\bp}\,J(z^\prime,\bar{z}^\prime)\right)\nn\ee
Here the $\sim$ symbol reflects the fact that we've dropped a whole lot of irrelevant
normalization terms, including $\det^{-1/2}(-\p\bp)$. The inverse operator $1/\p\bp$ on
the right-hand-side of this equation is shorthand for the propagator $G(z,z^\prime)$ which solves
\be \p\bp G(z,\bz;z^\prime,\bar{z}^\prime) = \delta(z-z^\prime,\bz-\bar{z}^\prime)\nn\ee
As we've seen several times before, in two dimensions this propagator is given by
\be G(z,\bz;z^\prime,\bar{z}^\prime) = \frac{1}{2\pi}\ln|z-z^\prime|^2\nn\ee

\subsubsection*{Back to the Scattering Amplitude}

Comparing our scattering amplitude with this general expression, we need to take the source $J$
to be
\be J(z,\bz) = \sum_{i=1}^mp_i\ \delta(z-z_i,\bz-\bar{z}_i)\nn\ee
Inserting this into the Gaussian integral gives us an expression for the amplitude
\be {\cal A}^{(m)} \sim \frac{g_s^{m-2}}{{\rm Vol}(SL(2;{\bf C}))}\int \prod_{i=1}^m
d^2z_i\ \exp\left(\frac{\ap}{2}\sum_{j,l}p_j\cdot p_l\ \ln|z_j-z_l|\right)\nn\ee
The terms with $j=l$ seem to be problematic. In fact, they should just be left out.
This follows from correctly implementing normal ordering and leaves us with
\be {\cal A}^{(m)} \sim \frac{g_s^{m-2}}{{\rm Vol}(SL(2;{\bf C}))}\int \prod_{i=1}^m
d^2z_i\ \prod_{j<l} |z_j-z_l|^{\,\ap p_j\cdot p_l}\label{halfamp}\ee
Actually, there's something that we missed. (Isn't there always!). We certainly expect scattering in flat space to obey momentum conservation, so there should be a $\delta^{(26)}(\sum_{i=1}^m p_i)$
in the amplitude. But where is it? We missed it because we were a little too quick
in computing the Gaussian integral. The operator $\p\bar{\p}$ annihilates the zero mode,
$x^\mu$, in the mode expansion. This means that its inverse, $1/\p\bar{\p}$, is not
well-defined. But it's easy to  deal with this by treating the zero mode separately.
The derivatives $\p^2$ don't see $x^\mu$, but the source $J$ does. Integrating over the zero mode in
the path integral gives us our delta function
\be \int dx \ \exp(i\sum_{i=1}^mp_i\cdot x) \sim \delta^{26}(\sum_{i=1}^m
p_i)\nn\ee
So, our final result for the amplitude is
\be {\cal A}^{(m)} \sim \frac{g_s^{m-2}}{{\rm Vol}(SL(2;{\bf C}))}\ \delta^{26}(\sum_ip_i)\ \int \prod_{i=1}^m
d^2z_i\ \prod_{j<l} |z_j-z_l|^{\,\ap p_j\cdot p_l}\label{tamp}\ee

\subsubsection*{The Four-Point Amplitude}

We will compute only the four-point amplitude for two-to-two scattering of tachyons.
The ${\rm Vol}(SL(2;{\bf C}))$ factor is there to remind us that we still have a remnant
gauge symmetry floating around. Let's now fix this. As we mentioned before, it provides
enough freedom for us to take any three points on the plane
and move them to any other three points. We will make use of this to set
\be z_1=\infty \ \ \ ,\ \ \ z_2=0\ \ \ ,\ \ \ z_3=z\ \ \ ,\ \ \ z_4=1\nn\ee
Inserting this into the amplitude \eqn{tamp}, we find ourselves with just a single integral
to evaluate,
\be {\cal A}^{(4)} \sim g_s^2\ \delta^{26}(\sum_ip_i)\ \int d^2z\ |z|^{\,\ap p_2\cdot p_3}\,|1-z|^{\,\ap p_3\cdot p_4}\label{aint4}\ee
(There is also an overall factor of $|z_1|^4$, but this just gets absorbed into an
overall normalization constant).  We still need to do the integral. It can be evaluated
exactly in terms of gamma functions. We relegate the proof  to Appendix
\ref{appendix}, where we show that
\be  \int d^2z \ |z|^{2a-2}|1-z|^{2b-2} = \frac{2\pi\Gamma(a)\Gamma(b)
\Gamma(c)}{\Gamma(1-a)\Gamma(1-b)\Gamma(1-c)}\label{dotheint}\ee
where $a+b+c=1$.

\para
\EPSFIGURE{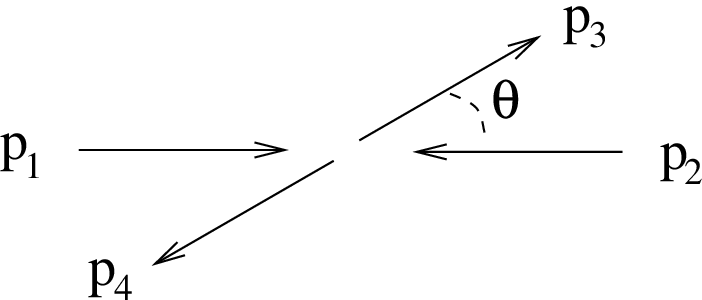,height=50pt}{}
\noindent
Four-point scattering amplitudes are typically expressed in terms of Mandelstam variables.
We choose $p_1$ and $p_2$ to be incoming momenta, and $p_3$ and $p_4$ to be outgoing momenta,
as shown in the figure. We then define
\be s=-(p_1+p_2)^2\ \ \ \ ,\ \ \ \ t=-(p_1+p_3)^2\ \ \ \ ,\ \ \ \ u=-(p_1+p_4)^2\nn\ee
These obey
\be s+t+u=-\sum_i p_i^2 = \sum_i M_i^2 = -\frac{16}{\ap}\nn\ee
where, in the last equality, we've inserted the value of the tachyon mass \eqn{tmass}. Writing
the scattering amplitude \eqn{aint4} in terms of Mandelstam variables, we have our final answer
\be
{\cal A}^{(4)}\sim g_s^2\ \delta^{26}(\sum_ip_i)\ \frac{\Gamma(-1-\ap s/4)\Gamma(-1-\ap t/4)\Gamma(-1-\ap u/4)}
{\Gamma(2+\ap s/4)\Gamma(2+\ap t/4)\Gamma(2+\ap u/4)}\label{vsa}\ee
This is the {\it Virasoro-Shapiro amplitude} governing tachyon scattering in the closed bosonic
string.

\para
Remarkably, the Visasoro-Shapiro amplitude was almost the first equation of string theory!
(That honour actually goes to the Veneziano amplitude which is the analogous expression
for open string tachyons and will be derived in Section \ref{venicesec}).
These amplitudes were written down long before people knew that
they had anything to do with strings: they simply exhibited some interesting and
surprising properties. It took several years of work to realise that they actually describe the
scattering of strings. We will now start to tease apart the Virasoro-Shapiro amplitude to see
some of the properties that got people hooked many years ago.

\subsubsection{Lessons to Learn}

So what's the physics lying behind the scattering amplitude \eqn{vsa}? Obviously it is symmetric
in $s$, $t$ and $u$. That is already surprising and we'll return to it shortly. But we'll
start by fixing $t$ and looking at  the properties of the amplitude as we vary $s$.

\para
The first thing to notice is that $\amp$ has poles. Lots of poles. They come from the factor
of $\Gamma(-1-\ap s/4)$ in the numerator. The first of these poles appears when
\be -1-\frac{\ap s}{4}=0\ \ \ \ \Rightarrow \ \ \ \ s=-\frac{4}{\ap}\nn\ee
But that's the mass of the tachyon! It means that, for $s$ close to
$-4/\ap$, the amplitude has the form of
a familiar scattering amplitude in quantum field theory with a cubic vertex,
\be \raisebox{-3.1ex}{\epsfxsize=1.0in\epsfbox{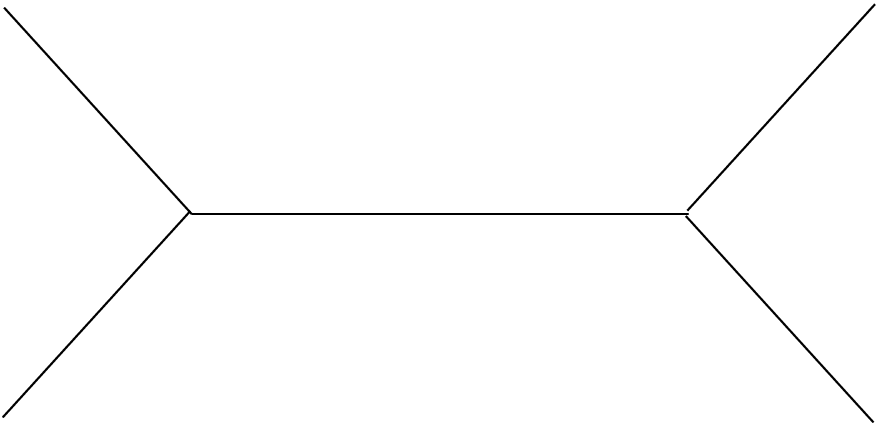}}\  \sim\  \frac{1}{s-M^2}\nn\ee
where $M$ is the mass of the exchanged particle, in this case the tachyon.

\para
Other poles in the amplitude occur at $s=4(n-1)/\ap$ with $n\in {\bf Z}^+$. This is precisely
the mass formula for the higher states of the closed string. What we're learning is that the
string amplitude is summing up an infinite number of tree-level field theory diagrams,
\be \raisebox{-3.1ex}{\epsfxsize=3.2in\epsfbox{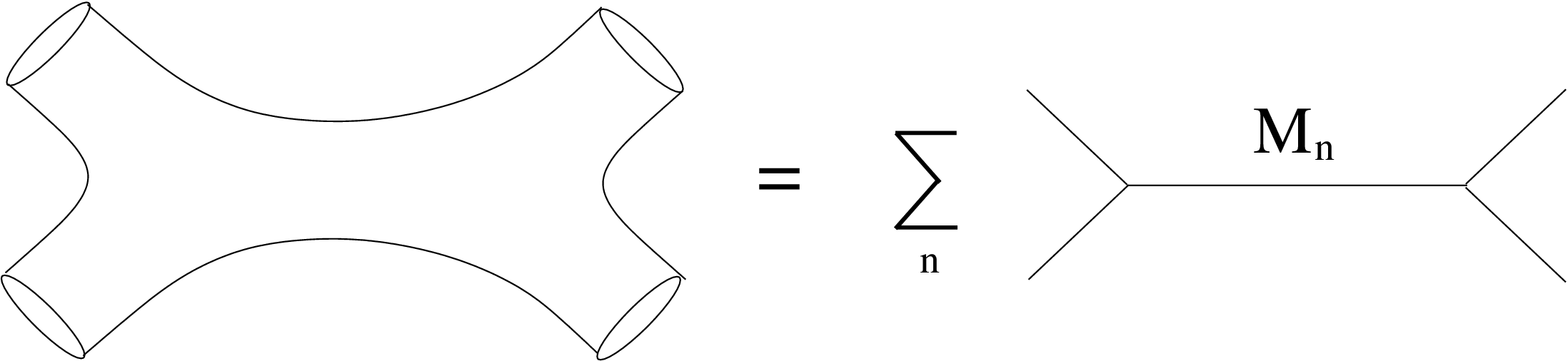}}\nn\ee
{}\\
\noindent
where the exchanged particles are all the different states of the free string.

\para
In fact, there's more information about the spectrum of states hidden within these
amplitudes. We can look at the residues of the poles at $s=4(n-1)/\ap$, for $n=0,1,\ldots$.
These residues are rather complicated functions of $t$, but the highest power of
momentum that appears for each  pole is
\be \amp \sim \sum_{n=0}^\infty \frac{t^{2n}}{s-M_n^2}\label{spinscat}\ee
The power of the momentum is telling us the highest spin of the particle states at level $n$. To
see why this is, consider a field corresponding to a spin $J$ particle. It has a whole bunch
of Lorentz indices, $\chi_{\mu_1\ldots\mu_J}$. In a cubic interaction, each of these must be
soaked up by derivatives. So we have $J$ derivatives at each vertex, contributing powers of
$({\rm momentum})^{2J}$ to the numerator of the Feynman diagram. Comparing with the string scattering
amplitude, we see that the highest spin particle at level $n$ has $J=2n$. This is indeed the result that we
saw from the canonical quantization of the string in Section 2.

\para
Finally, the amplitude \eqn{vsa} has a property that is very different from amplitudes
in field theory. Above, we framed our
discussion by keeping $t$ fixed and expanding in $s$. We could just have well done the opposite:
fix $s$ and look at poles in $t$. Now the string amplitude has the interpretation of an infinite
number of $t$-channel scattering amplitudes, one for each state of the string
\be \raisebox{-3.1ex}{\epsfxsize=2.0in\epsfbox{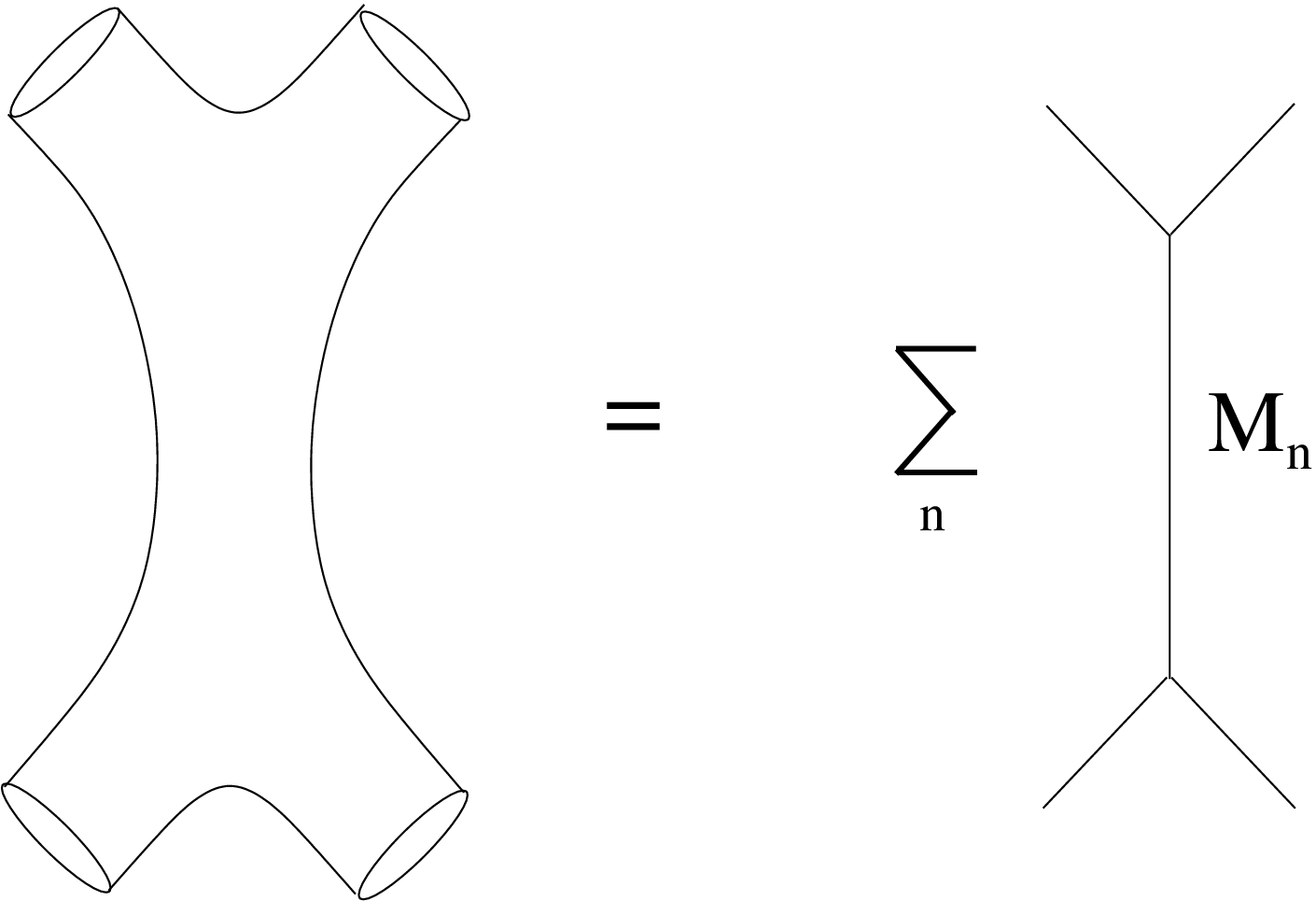}}\nn\ee
Usually in field theory, we sum up both $s$-channel and $t$-channel scattering amplitudes. Not
so in string theory. The sum over an infinite number of $s$-channel amplitudes can be reinterpreted
as an infinite sum of $t$-channel amplitudes. We don't include both: that would be overcounting.
(Similar statements hold for $u$). The fact that the same amplitude can be written as a
sum over $s$-channel poles {\it or} a sum over $t$-channel poles is sometimes referred to as ``duality".
(A much overused word). In the early days, before it was known that string theory was a theory
of strings, the subject inherited its name from this duality property of amplitudes:
it was called the {\it dual resonance model}.

\subsubsection*{High Energy Scattering}

Let's use this amplitude to see what happens when we collide strings at high energies. There are
different regimes that we could look at. The most illuminating is $s$, $t\rightarrow \infty$, with
$s/t$ held fixed. In this limit, all the exchanged momenta become large. It corresponds
to high-energy scattering with the angle $\theta$ between incoming and outgoing particles kept fixed. To see
this consider, for example, massless particles (our amplitude is really for tachyons, but the
same considerations hold). We take the incoming and outgoing momenta to be
\be p_1=\frac{\sqrt{s}}{2}(1,1,0,\ldots)\ \ \ &,&\ \ \ p_2=\frac{\sqrt{s}}{2}(1,-1,0,\ldots)\nn\\
p_3=\frac{\sqrt{s}}{2}(1,\cos\theta,\sin\theta,\ldots)\ \ \ &,&\ \ \ p_4=\frac{\sqrt{s}}{2}(1,-
\cos\theta,-\sin\theta,\ldots)\nn\ee
Then we see explicitly that $s\rightarrow \infty$ and $t\rightarrow \infty$ with the ratio
$s/t$ fixed also keeps the scattering angle $\theta$ fixed.

\para
We can evaluate the scattering amplitude $\amp$ in this limit by using $\Gamma(x)\sim \exp(x\ln x)$.
We send $s\rightarrow \infty$ avoiding the poles. (We can achieve this by sending $s\rightarrow
\infty$ in a slightly imaginary direction. Ultimately this is valid because all the higher string
states are actually unstable in the interacting theory which will shift their poles off the real
axis once taken into account). It is simple to check that the amplitude drops off exponentially quickly at
high energies,
\be \amp \sim g_s^2\,\delta^{26}(\sum_ip_i)\,\exp\left(-\frac{\ap}{2}(s\ln s + t \ln t
+u \ln u)\right)
%\exp\left(-\ap  s\ln s\right)
\ \ \ \ {\rm as}\ s\rightarrow \infty\label{highscat}\ee
The exponential fall-off seen in \eqn{highscat} is
much faster than the amplitude of any field theory which, at best, fall off with
power-law decay at high energies and, at worse, diverge.
For example, consider the individual terms \eqn{spinscat} corresponding to the amplitude for
$s$-channel processes involving the exchange of particles with spin $2n$. We see that the exchange
of a spin 2 particle results in a divergence in this limit. This is reflecting something you already
know about gravity: the dimensionless coupling is $G_NE^2$ (in four-dimensions) which becomes large
for large energies. The exchange of higher spin particles gives rise to even worse
divergences. If we were to truncate the infinite sum \eqn{spinscat} at any finite $n$,
the whole thing would diverge.  But infinite sums can do
things that finite sums can't and the final behaviour of the amplitude \eqn{highscat} is much
softer than any of the individual terms. The infinite number of particles in string theory
conspire to render finite any divergence arising from an individual particle species.

\para
Phrased in terms of the $s$-channel exchange of particles, the high-energy behaviour of string theory
seems somewhat miraculous. But there is another viewpoint where it's all very obvious. The power-law
behaviour of scattering amplitudes is characteristic of point-like charges. But, of course,
the string isn't a point-like object. It is extended and fuzzy at length scales comparable to
$\sqrt{\ap}$. This is the reason the amplitude has such soft high-energy behaviour. Indeed, this idea
that smooth extended objects give rise to scattering amplitudes that decay exponentially at high
energies is something that you've seen before in non-relativistic quantum mechanics. Consider, for
example, the scattering of a particle off a Gaussian potential. In the Born approximation, the
differential cross-section is just given by the Fourier transform which is again a Gaussian, now
decaying exponentially for large momentum.

%\para
%Historically, the idea that amplitudes, or cross-sections, should decay exponentially at high
%energies is tied up with one of the most famous experiments of the 20th century: this was the
%behaviour that Geiger, Marsden and Rutherford expected to see in their alpha particle scattering
%experiment. This was based on the ``plum-pudding'' model of the atom as an extended object.
%The fact that the cross-section didn't die off exponentially showed that the nucleus of
%the atom is (effectively) point-like. Since that time, all high-energy scattering experiments
%have continued to reveal power-law fall-off, as expected from a description of physics based on
%quantum field theory. However, if string theory does turn out to describe Nature, Geiger
%and Marsden would have eventually seen their desired exponential fall-off...they would just need
%to crank up the energy of their experiment by $N$ orders of magnitude. (where, as we'll see
%in the next Section, $N$ lies somewhere
%between 6 and 20, depending on your favourite string model!).

\para
It's often said that theories of quantum gravity should have a ``minimum length", sometimes
taken to be the Planck scale. This is roughly true in string theory, although not in
any crude simple manner. Rather, the minimum length  reveals
itself in different ways depending on which question is being asked. The above discussion
highlights one example of this: strings can't probe distance scales shorter than $l_s = \sqrt{\ap}$ simply because they are themselves fuzzy at this scale.
It  turns out that D-branes are much better probes of sub-stringy physics and provide
a different view on the short distance structure of spacetime.
We will also see another manifestation of the minimal length scale of string theory
in Section \ref{ttsec}.

%\para
%So far, we've only discussed tree-level scattering. We will turn to loops in Section \ref{oneloopsec}
%where we will find another manifestation of the good UV behaviour
%of string theory. However, already
%we can see that things are looking good. All scattering amplitudes come with exponential suppressions
%at high-energies and even if there were divergences at one-loop (there aren't!) we would expect
%them to be at most power-law. Any potential divergences that arise in string theory would have to work very %hard to overcome the high-energy suppression \eqn{highscat}

\subsubsection*{Graviton Scattering}

Although we've derived the result \eqn{highscat} for tachyons, all tree-level amplitudes have this
soft fall-off at high-energies. Most notably, this includes graviton scattering. As we noted
above, this is in sharp contrast to general relativity for which tree-level scattering amplitudes
diverge at high-energies. This is the first place to see that UV problems of general relativity
might have a good chance of being cured in string theory.

\para
Using the techniques described in this section, one can compute
$m$-point tree-level amplitudes for graviton scattering.
If we restrict attention to low-energies (i.e. much smaller than $1/\sqrt{\ap}$), one can show
that these coincide with the amplitudes derived from  the Einstein-Hilbert action in $D=26$
dimensions
\be S = \frac{1}{2\kappa^2}\int d^{26}X\sqrt{-G}\ {\cal R}
\nn\ee
where ${\cal R}$ is the $D=26$ Ricci scalar (not to be confused with the worldsheet Ricci scalar
which we call $R$). The gravitational coupling, $\kappa^2$ is related to Newton's constant in
26 dimensions. It plays no role for pure gravity, but is important when we couple to matter. We'll
see shortly that it's given by
\be \kappa^2 \approx g_s^2 (\ap)^{12}\nn\ee
We won't explicitly compute graviton scattering amplitudes in this course, partly because
they're fairly messy and partly because building up the Einstein-Hilbert action from
$m$-particle scattering is hardly the best way to look at general relativity. Instead, we shall
derive the Einstein-Hilbert action in a much better fashion in Section \ref{background}.

\subsection{Open String Scattering}
\label{venezsec}

So far our discussion has been entirely about closed strings. There is a very similar
story for open strings. We again compute S-matrix elements. Conformal symmetry now maps tree-level scattering to the disc, with vertex operators inserted  on the boundary of the disc.
\begin{figure}[htb]
\begin{center}
\epsfxsize=3.2in\leavevmode\epsfbox{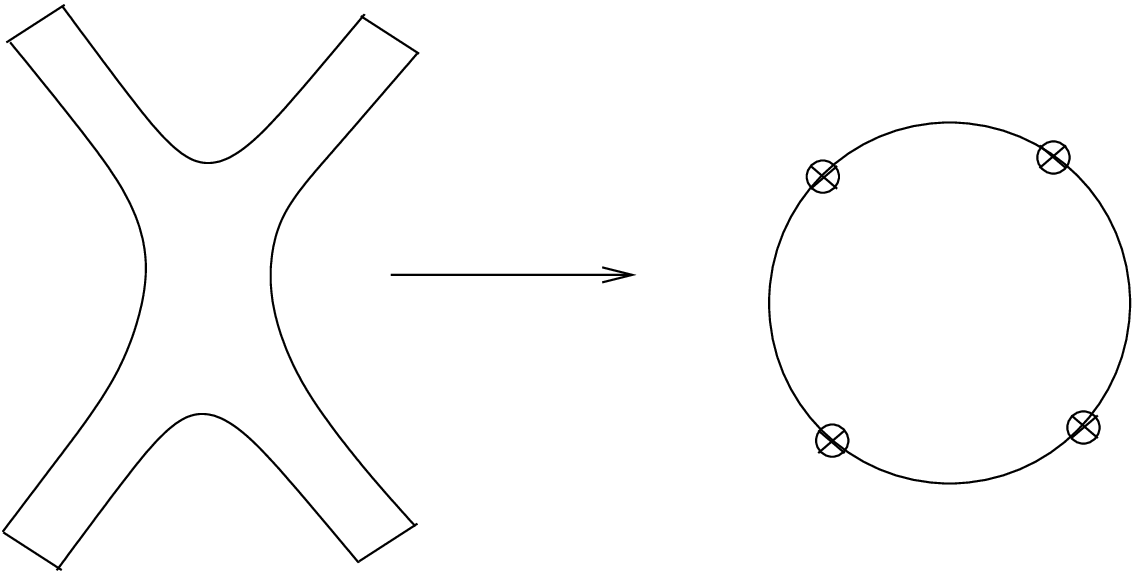}
\end{center}
\caption{The conformal map from the open string worldsheet to the disc.}
\end{figure}

\para
For the open string, the string coupling constant
that we add to the Polyakov action requires the addition of a
boundary term to make it well defined,
\be \chi = \frac{1}{4\pi} \int_{\cal M} d^2\sigma\ \sqrt{g} R + \frac{1}{2\pi}\int_{\p {\cal M}} ds\,k
\label{openchi}\ee
where $k$ is the geodesic curvature of the boundary. To define it, we introduce two unit vectors
on the worldsheet: $t^\alpha$ is tangential to the boundary, while $n^\alpha$ is normal and points
outward from the boundary. The geodesic curvature is the defined as
\be k = -t^\alpha n_\beta \nabla_\alpha t^\beta\nn\ee
Boundary terms of the type seen in \eqn{openchi} are also needed in general relativity for
manifolds with boundaries: in that context, they are referred to as Gibbons-Hawking terms.

\begin{figure}[htb]
\begin{center}
\epsfxsize=4.2in\leavevmode\epsfbox{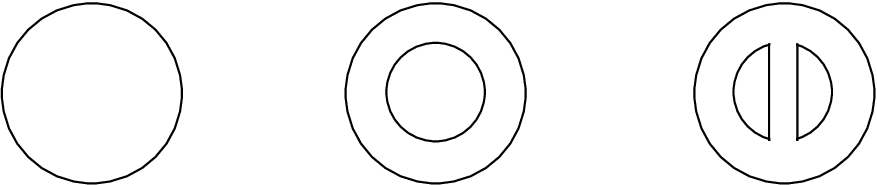}
\end{center}
\caption{Riemann surfaces with boundary with $\chi=1,0$ and $-1$.}
\label{holesin}\end{figure}

\para
The Gauss-Bonnet theorem has an extension to surfaces with boundary. For surfaces with $h$ handles,
and $b$ boundaries, the Euler character is given by
\be \chi = 2-2h-b\nn\ee
Some examples are shown in Figure \ref{holesin}.
The expansion for open-string scattering consists of adding consecutive boundaries to the worldsheet. The disc is weighted by $1/g_s$; the annulus has no factor of $g_s$, and so on.
We see that the open string coupling is related to the closed string coupling by
\be g_{\rm open}^2=g_s\label{goc}\ee
\noindent One of the key steps in computing closed string scattering amplitudes was the implementation of the conformal Killing group, which was defined as the
surviving gauge symmetry with a
global action
on the sphere. For the open string, there is again a residual gauge symmetry. If we think
in terms of the upper-half plane, the boundary is ${\rm Im} z=0$. The conformal
Killing group is composed of transformations
\be z \rightarrow \frac{az+b}{cz+d}\nn\ee
again with the requirement that $ad-bc=1$. This time there is one further condition:
the boundary ${\rm Im}z=0$ must be mapped onto itself. This requires
$a,b,c$, $d\in {\bf R}$. The resulting conformal Killing group is $SL(2;{\bf R})/{\bf Z}_2$.

\subsubsection{The Veneziano Amplitude}
\label{venicesec}

Since vertex operators now live on the boundary, they have a fixed ordering. In computing a
scattering amplitude, we must sum over all orderings. Let's look again at the 4-point amplitude
for tachyon scattering. The vertex operator is
\be V(p_i) = \sqrt{g_s}\int dx\ e^{ip_i\cdot X}\nn\ee
where the integral $\int dx$ is now over the boundary and $p^2=1/\ap$ is the on-shell condition
for an open-string tachyon. The normalization
$\sqrt{g_s}$ is that appropriate for the insertion of an open-string mode, reflecting \eqn{goc}.

\para
Going through the same steps as for the closed string, we find that the amplitude is given by
\be \amp \sim \frac{g_s}{{\rm Vol}(SL(2;{\bf R}))}\,\delta^{26}(\sum_ip_i)\,\int \prod_{i=1}^4dx_i\ \prod_{j<l}\,|\,x_i-x_j|^{2\ap p_i\cdot p_j}\ee
Note that there's a factor of 2 in the exponent, differing from the closed string expression \eqn{halfamp}. This comes about because the boundary propagator \eqn{image} has an extra factor
of 2 due to the image charge.

\para
We now use the $SL(2;{\bf R})$ residual gauge symmetry to fix three points on the boundary.
We choose a particular ordering and set
$x_1=0$, $x_2=x$, $x_3=1$ and $x_4\rightarrow \infty$. The only free insertion point is
$x_2=x$ but, because of the restriction of operator ordering, this must lie in the interval
$x\in [0,1]$. The interesting part of the integral
is then given by
\be \amp \sim g_s\int_0^1 dx\ |x|^{2\ap p_1\cdot p_2}\,|1-x|^{2\ap p_2\cdot p_3}\nn\ee
This integral is well known: as shown in Appendix \ref{appendix}, it is the Euler beta function
\be
B(a,b) =\int_0^1 dx\ x^{a-1}(1-x)^{b-1} = \frac{\Gamma(a)\Gamma(b)}{\Gamma(a+b)}\nn\ee
After summing over the different orderings of vertex operators, the end result for the
amplitude for open string tachyon scattering is,
\be \amp \sim g_s\left[ B(-\ap s-1,-\ap t-1) + B(-\ap s-1,-\ap u-1) + B(-\ap t-1,-\ap u-1)\right]
\nn\ee
This is the famous {\it Veneziano Amplitude}, first postulated in 1968
to capture some observed features of the strong interactions.
This was before the advent of QCD and before it was realised that the amplitude
arises from a string.

\para
The open string scattering amplitude contains the same features that we saw for the closed
string. For example, it has poles at
\be s= \frac{n-1}{\ap}\ \ \ \ \ n=0,1,2,\ldots\nn\ee
which we recognize as the spectrum of the open string.

\subsubsection{The Tension of D-Branes}

Recall that we introduced D-branes as surfaces in space on which strings can end. At the
time, I promised that we would eventually discover that these D-branes are dynamical
objects in their own right. We'll look at this more closely in the next section, but for
now we can do a simple computation to determine the tension of D-branes.

\para
The tension $T_p$ of a D$p$-brane is defined as the energy per spatial volume. It has
dimension $[T_p]=p+1$. The tension is telling us the magnitude of the coupling between the
brane and gravity. Or, in our new language, the strength of the interaction between a closed
string state and an open string. The simplest such diagram is shown in the figure, with
a graviton vertex operator inserted. Although we won't compute this diagram completely,
we can figure out its most important property just by looking at it: it has the topology of
a disc, so is proportional to $1/g_s$.
Adding powers of $\ap$ to get the dimension right, the tension of a D$p$-brane must scale as
\be T_p \sim \frac{1}{l_s^{p+1}}\,\frac{1}{g_s}\label{dtension}\ee
\EPSFIGURE{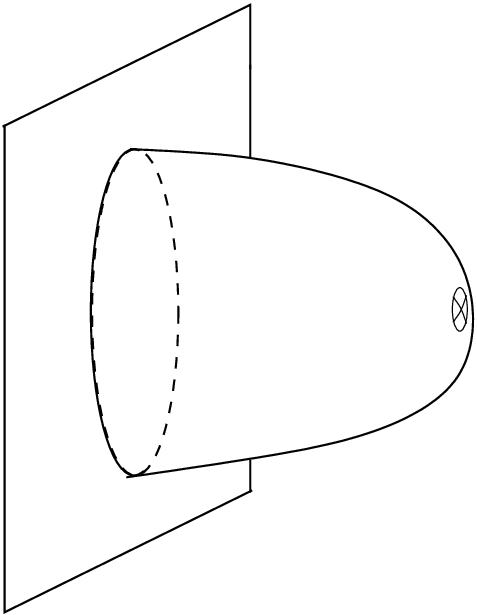,height=110pt}{}
\noindent
where the string length is defined as $l_s=\sqrt{\ap}$. The $1/g_s$ scaling of the tension
is one of the key characteristic features of a D-brane.

\para
I should confess that there's a lot swept under the carpet in the above discussion, not least
the question of the correct normalization of the vertex operators and the difference between
the string frame and the Einstein frame (which we will discuss shortly). Nonetheless,
the end result \eqn{dtension} is correct. For a fuller discussion, see Section 8.7 of
Polchinski.

\subsection{One-Loop Amplitudes}
\label{oneloopsec}

We now return to the closed string to discuss one-loop effects. As we saw above, this corresponds
to a worldsheet with the topology of a torus. We need to integrate over all
metrics on the torus.

\para
For tree-level processes, we used diffeomorphisms and Weyl transformations to map an
arbitrary metric on the sphere to the flat metric on the plane. This time, we use
these transformations to map an arbitrary metric on the torus to the flat metric
on the torus. But there's a new subtlety that arises: not all flat metrics on the
torus are equivalent.

\subsubsection{The Moduli Space of the Torus}

\EPSFIGURE{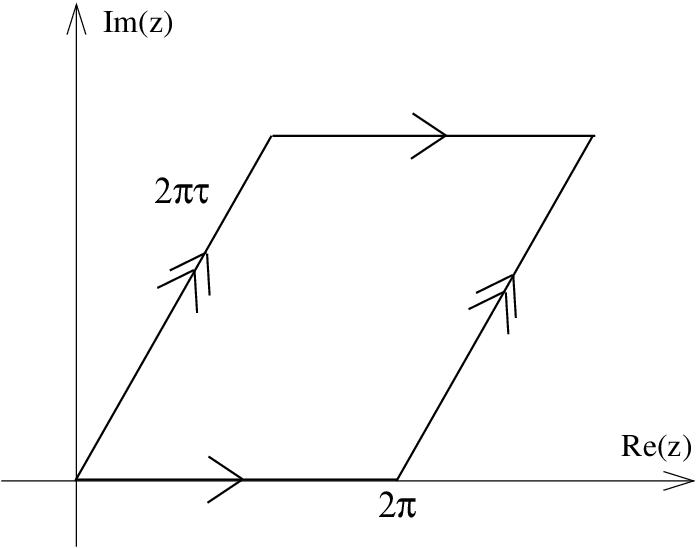,height=120pt}{}
\noindent
Let's spell out what we mean by this. We can construct a torus by identifying a region
in the complex $z$-plane as shown in the figure. In general, this identification
depends on a single complex parameter, $\tau\in {\bf C}$.
\be z \equiv z+2\pi\ \ \ \ {\rm and}\ \ \ \ z \equiv z + 2\pi \tau \nn\ee
Do not confuse $\tau$ with the Minkowski worldsheet time: we left that behind way back
in Section 3. Everything here is Euclidean worldsheet, and $\tau$ is  just a parameter telling
us how skewed the torus is. The flat metric on the torus is now simply
\be ds^2 = dzd\bz\nn\ee
subject to the identifications above.

\para
A general metric on a torus can always be transformed to a flat metric for some value of
$\tau$. But the question that interests us is whether two tori, parameterized by different $\tau$,
are conformally equivalent. In general, the answer is no. The space of conformally inequivalent tori, parameterized by $\tau$, is called the {\it moduli space} ${\cal M}$.

\para
However, there are some values of $\tau$ that do correspond to the same torus. In
particular,
there are a couple of obvious ways in which we can change $\tau$ without changing the
torus. They go by the names of the $S$ and $T$ transformations:
\begin{itemize}
\item $T:\tau\rightarrow \tau +1$: This clearly gives rise to the same torus, because
the identification is now
\be z \equiv z+2\pi\ \ \ \ {\rm and}\ \ \ \ z \equiv z + 2\pi (\tau+ 1)
\equiv z + 2\pi \tau \nn\ee
\item $S:\tau \rightarrow -1/\tau$: This simply flips the sides of the torus. For example,
if $\tau = ia$ is purely imaginary, then this transformation maps $\tau \rightarrow i/a$,
which can then be undone by a scaling.
\end{itemize}
It turns out that these two changes $S$ and $T$ are the only ones that keep the
torus intact. They are sometimes called {\it modular transformations}.
A general modular transformations is constructed from combinations of $S$ and $T$
and takes the form,
\be \tau \rightarrow \frac{a\tau + b}{c\tau +d}\ \ \ \ \ {\rm with}\ ad-bc=1\label{sl2z}\ee
where $a$, $b$, $c$ and $d\in {\bf Z}$. This is the group $SL(2,{\bf Z})$. (In fact, we
have our usual ${\bf Z}_2$ identification, and the group is actually $PSL(2,{\bf Z}) =
SL(2;{\bf Z})/{\bf Z}_2$). The moduli space ${\cal M}$ of the torus is given by
\be {\cal M} \cong {\bf C}/SL(2;{\bf Z})\nn\ee
What does this space look like? Using $T: \tau \rightarrow \tau +1$, we can always shift $\tau$
until it lies within the interval
\be {\rm Re}\,\tau\ \in\ [-\ft1{\,2}, +\ft1{\,2}\,]
\nn\ee
\begin{figure}[htb]
\begin{center}
\epsfxsize=5.5in\leavevmode\epsfbox{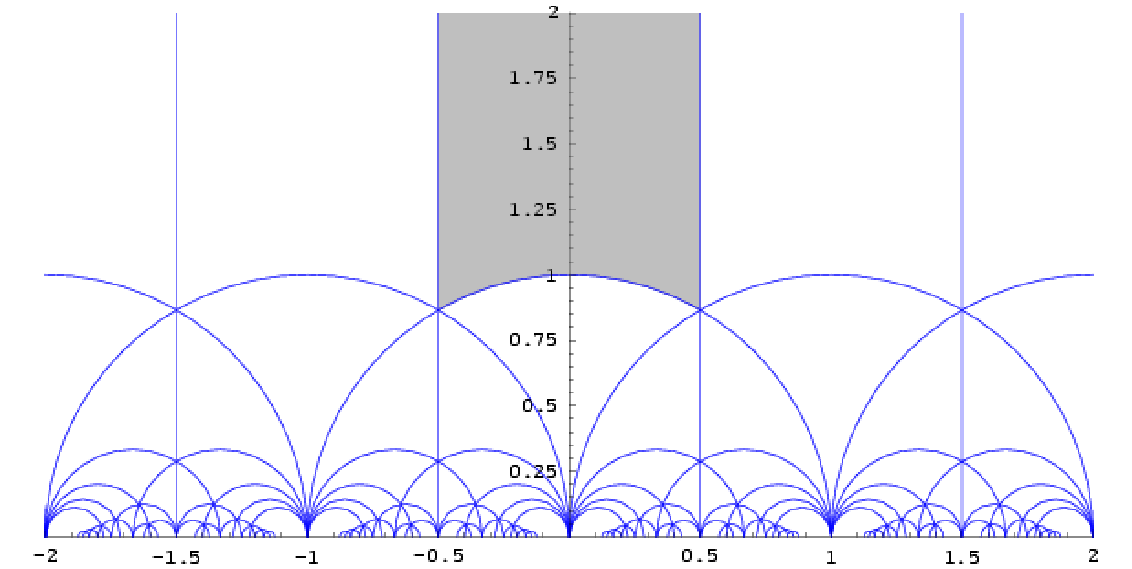}
\end{center}
\caption{The fundamental domain.}
\end{figure}
\noindent
where the edges of the interval are identified. Meanwhile, $S: \tau \rightarrow -1/\tau$ inverts
the modulus $|\tau|$, so we can use this to map a point inside the circle $|\tau|<1$
to a point outside $|\tau|>1$. One can show that by successive combinations of $S$ and $T$,
it is possible to map any point to lie within the shaded region shown in the figure, defined
by
\be |\tau| \geq 1 \ \ \ \ {\rm and}\ \ \ {\rm Re}\,\tau \in [-\ft1{\,2}, +\ft1{\,2}\,]\nn\ee
This is referred to as the {\it fundamental domain} of $SL(2;{\bf Z})$.

\para
We could have just as easily chosen one of the other fundamental domains shown in the
figure. But the shaded region is the standard one.

\subsubsection*{Integrating over the Moduli Space}

In string theory we're invited to sum over all metrics. After gauge fixing
diffeomorphisms and Weyl invariance, we still need to integrate over all
inequivalent tori. In other words, we integrate over the fundamental domain. The
$SL(2;{\bf Z})$ invariant measure over the fundamental domain is
\be \int \frac{d^2\tau}{({\rm Im}\,\tau)^2}\nn\ee
To see that this is $SL(2;{\bf Z})$ invariant,
note that under a general transformation of the form \eqn{sl2z} we have
\be d^2\tau\ \rightarrow\ \frac{d^2\tau}{|c\tau+d|^4}\ \ \ {\rm and}\ \ \ {\rm Im}\,\tau
\ \rightarrow\ \frac{{\rm Im}\,\tau}{|c\tau +d|^2}\nn\ee
There's some physics lurking within these rather mathematical statements. The integration over
the fundamental domain in string theory is analogous to the loop integral over momentum in
quantum field theory. Consider the square tori defined by ${\rm Re}\,\tau=0$. The tori with ${\rm Im}\,\tau\rightarrow \infty$ are squashed and chubby. They correspond to the infra-red region of loop momenta in a Feynman diagram. Those with ${\rm Im}\,\tau \rightarrow 0$ are
long and thin. Those correspond to the ultra-violet limit of loop momenta in a Feynman diagram.
Yet, as we have seen, we should not integrate over these UV regions of the loop since the fundamental domain does not
stretch down that far. Or, more precisely, the thin tori are mapped to chubby tori. This
corresponds to the fact that any putative UV divergence of string theory can always be
reinterpreted as an IR divergence. This is the second manifestation of the well-behaved
UV nature of string theory. We will see this more explicitly in the example of Section
\ref{pfsec}.

\para
Finally, when computing a loop amplitude in string theory, we still need to worry about
the residual gauge symmetry that is left unfixed after the map to the flat torus. In
the case of tree-level amplitudes on the sphere, this residual gauge symmetry was due to the conformal Killing group $SL(2;{\bf C})$. For the torus, the conformal Killing group is generated
by the obvious generators $\p_z$ and $\bp_{\bz}$. It is $U(1)\times U(1)$.

\subsubsection*{Higher Genus Surfaces}

The moduli space ${\cal M}_g$ of the Riemann surface of genus $g>1$ can be shown to have dimension,
\be {\rm dim}\,{\cal M}_g = 3g-2\nn\ee
There are no conformal Killing vectors when $g>1$. These facts can be demonstrated as
an application
of the Riemann-Roch theorem. For more details, see section 5.2 of Polchinski, or sections
3.3 and 8.2 of Green, Schwarz and Witten.

\subsubsection{The One-Loop Partition Function}
\label{pfsec}

We won't compute any one-loop scattering amplitudes in string theory. Instead, we will
look at something a little simpler: the one-loop vacuum to vacuum amplitude. A  Euclidean
worldsheet with periodic time has the interpretation of a finite temperature partition
function for the theory defined on a cylinder. In $D=26$ dimensional spacetime, it is related
to the cosmological constant in bosonic string theory.

\para
Consider firstly the partition function of a theory on a square torus, with ${\rm Re}\,\tau=0$.
Compactifying Euclidean time, with period $({\rm Im}\,\tau)$ is equivalent to putting
the theory at temperature $T=1/(\rm Im\,\tau)$,
\be Z[\tau] = \Tr\ e^{-2\pi({\rm Im}\,\tau)H}\nn\ee
where the $\Tr$ is over all states in the theory. For any CFT defined on a cylinder, the Hamiltonian given by
\be H = L_0+\tilde{L}_0-\frac{c+\tilde{c}}{24}\nn\ee
where the final term is the Casimir energy computed
in Section \ref{cassec}.

\para
\EPSFIGURE{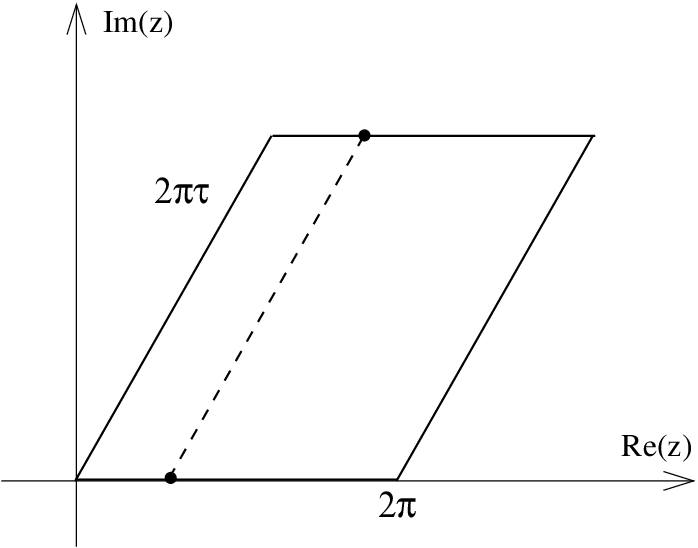,height=100pt}{}
\noindent
What then is the interpretation of the
vacuum amplitude computed on a torus with ${\rm Re}\,\tau\neq 0$? From the diagram, we
see that the effect of such a skewed torus is to translate a given point around the
cylinder by ${\rm Re}\,\tau$. But we know which operator implements such a translation:
it is $\exp(2\pi i ({\rm Re}\,\tau) P)$, where $P$ is the momentum operator on
the cylinder. After the map to the plane, this becomes the rotation operator
\be P = L_0-\tilde{L}_0\nn\ee
So the vacuum amplitude on the torus has the interpretation of the sum over all states
in the theory, weighted by
\be Z[\tau] = \Tr\ e^{-2\pi({\rm Im}\,\tau)(L_0+\tilde{L}_0)}\,e^{-2\pi i ({\rm Re}\,\tau)
(L_0-\tilde{L}_0)}\,e^{2\pi({\rm Im}\,\tau)(c+\tilde{c})/24}\nn\ee
We define
\be q = e^{2\pi i\tau}\ \ \ \ ,\ \ \ \ \bar{q}=e^{-2\pi i \bar{\tau}}\nn\ee
The partition function can then be written in slick notation as
\be Z[\tau] = \Tr\ q^{L_0-c/24}\ \bar{q}^{\tilde{L}_0-\tilde{c}/24}\nn\ee
Let's compute this for the free string. We know that each scalar field $X$ decomposes into
a zero mode and an infinite number harmonic oscillator modes $\alpha_{-n}$ which
create states of energy $n$. We'll deal with the zero mode shortly but, for now, we
focus on the oscillators. Acting $d$ times with the operator $\alpha_{-n}$
creates states with energy $dn$. This gives a contribution to $\Tr q^{L_0}$ of the
form
\be \sum_{d=0}^\infty q^{nd} = \frac{1}{1-q^n}\nn\ee
But the Fock space of a single scalar field is built by acting with oscillator modes
$n\in{\bf Z}^+$. Including the central charge, $c=1$, the contribution from
the oscillator modes of a single scalar field is therefore
\be \Tr\ q^{L_0-c/24} = \frac{1}{q^{1/24}}\,\prod_{n=1}^\infty \frac{1}{1-q^n}\nn\ee
There is a similar expression from the $\bar{q}^{\tilde{L}_0-\tilde{c}/24}$ sector.
We're still left with the contribution from the zero mode $p$ of the scalar field. The
contribution to the energy $H$ of the state on the worldsheet is
\be \frac{1}{4\pi\ap}\int d\sigma\ (\ap p)^2 = \frac{1}{2}\ap p^2\nn\ee
The trace in the partition function requires us to sum over all states, which gives
\be \int \frac{dp}{2\pi}\ e^{-\pi \ap\,({\rm Im}\,\tau)p^2}\sim \frac{1}{\sqrt{\ap {\rm Im}\,\tau}}\nn\ee
So, including both the zero mode and oscillators, we get the partition function for a
single free scalar field,
\be Z_{\rm scalar}[\tau] \sim \frac{1}{\sqrt{\ap {\rm Im}\,\tau}}\,\frac{1}{(q\bar{q})^{1/24}}\,\prod_{n=1}^\infty \frac{1}{1-q^n}\
\,\prod_{n=1}^\infty \frac{1}{1-\bar{q}^n}
\label{spfmod}\ee
where I haven't been careful to keep track of constant factors.

\para
To build the string partition function, we should really work in covariant quantization
and include the ghost fields. Here we'll cheat and work in lightcone gauge. This is dodgy
because, if we do it honestly,
much of the physics gets pushed to the $p^+=0$ limit of the lightcone momentum where the gauge
choice breaks down. So instead we'll do it dishonestly.

\para
In lightcone gauge, we have 24 oscillator modes. But we have 26 zero modes. (You may
worry that we still have to impose level matching...this is the dishonest part of
the calculation. We'll see partly where it comes from shortly).
Finally, there's a couple of extra steps. We need to
divide by the volume of the conformal Killing group. This is just $U(1)\times U(1)$,
acting by translations along the cycles of the torus. The volume is just
${\rm Vol}=4\pi^2\,{\rm Im}\,\tau$. Finally, we also need to
integrate over the moduli space of the torus. Our final result, neglecting all constant
factors, is
\be Z_{\rm string} = \int {d^2\tau}\ \frac{1}{({\rm Im}\,\tau)}\,\frac{1}{(\ap {\rm Im}\,\tau)^{13}}\
\ \frac{1}{q\bar{q}}\,\left(\prod_{n=1}^\infty \frac{1}{1-q^n}\right)^{24}
\left(\prod_{n=1}^\infty \frac{1}{1-\bar{q}^n}\right)^{24}
\label{spf}\ee

\subsubsection*{Modular Invariance}

The function appearing in the partition function for the scalar field has a name: it is the inverse of the Dedekind eta function
\be \eta(q) = q^{1/24} \prod_{n=1}^\infty (1-q^n)\nn\ee
It was studied in the 1800s by mathematicians interested in the properties of functions under
modular transformations $T: \tau \rightarrow \tau +1$ and $S: \tau \rightarrow -1/\tau$.
The eta-function satisfies the identities
\be \eta(\tau +1) = e^{2\pi i/24}\eta(\tau)\ \ \ \ {\rm and}\ \ \ \
\eta(-1/\tau) = \sqrt{-i\tau}\eta(\tau)\nn\ee
These two statements ensure that the scalar partition function \eqn{spfmod} is a modular
invariant function. Of course, that kinda had to be true: it follows from the underlying
physics.

\para
Written in terms of $\eta$, the string partition function \eqn{spf} takes the form
\be Z_{\rm string} = \int \frac{{d^2\tau}}{({\rm Im}\,\tau)^2}\ \left(\frac{1}{\sqrt{{\rm Im}\,\tau}}\frac{1}{\eta(q)}\frac{1}{\bar{\eta}(\bar{q})}\right)^{24}\nn\ee
Both the measure, and the integrand, are individually modular invariant.

\subsubsection{Interpreting the String Partition Function}

It's probably not immediately obvious what the string partition function \eqn{spf} is
telling us. Let's spend some time trying to understand it in terms of some simpler
concepts.

\para
We know that the free string describes an infinite number of particles
with mass $m_n^2=4(n-1)/\ap$, $n=0,1,\ldots$. The string partition
function should just be a sum over  vacuum loops of each of these particles. We'll now
show that it almost has this interpretation.

\para
Firstly, let's figure out what the
contribution from a single particle would be? We'll consider a free massive
scalar field $\phi$ in $D$ dimensions. The partition function is given by,
\be Z &=& \int {\cal D}\phi \exp\left(-\frac{1}{2} \int d^Dx\ \phi(-\p^2 + m^2)\phi \right)
\nn\\ &\sim&  {\rm det}^{-1/2}(-\p^2+m^2) \nn\\ &=& \exp\left( \frac{1}{2}
\int \frac{d^Dp}{(2\pi)^D}\,\ln(p^2+m^2)\right)\nn\ee
This is the partition function of a field theory. It contains vacuum loops for all numbers
of particles. To compare to the open string partition function, we want the vacuum amplitude
for just a single particle. But that's easy to extract. We write the field theory
partition function as,
\be Z = \exp\left(Z_1\right) = \sum_{n=0}^\infty \frac{Z_1^n}{n!}\nn\ee
Each term in the sum corresponds to  $n$ particles propagating in a vacuum loop, with
the $n!$ factor taking care of Bosonic statistics. So the vacuum amplitude for a single,
free massive particle is simply
\be Z_1 = \frac{1}{2}\int \frac{d^Dp}{(2\pi)^D} \ \ln(p^2+m^2)\nn\ee
Clearly this diverges in the UV range of the integral, $p\rightarrow \infty$.
There's a nice way to rewrite this
integral using something known as Schwinger parameterization.
We make use of the identity
\be \int_0^\infty dl\ e^{-xl} = \frac{1}{x}\ \ \ \Rightarrow \ \ \ \int_0^\infty dl \ \frac{e^{-xl}}{l} = -\ln x\nn\ee
We then write the single particle partition function as
\be Z_1 = \int \frac{d^Dp}{(2\pi)^D} \int_0^\infty \frac{dl}{2l} \
e^{-(p^2+m^2)l}\label{1ppf}\ee
It's worth mentioning that there's another way to see that this is the single particle
partition function that is a little closer in spirit to the method we used in string theory.
We could start with the einbein form of the relativistic particle action \eqn{1e}. After
fixing the gauge to $e=1$, the exponent in \eqn{1ppf} is the energy of the particle traversing
a loop of length $l$. The integration measure $dl/l$ sums over all possible sizes of loops.

\para
We can happily perform the $\int d^Dp$ integral in \eqn{1ppf}. Ignoring numerical
factors, we have
\be Z_1 = \int_0^\infty dl\ \frac{1}{l^{1+D/2}}\, e^{-m^2l}\label{2ppf}\ee
Note that the UV divergence as $p\rightarrow \infty$ has metamorphosised into a divergence
associated to small loops as $l\rightarrow 0$.

\para
Equation \eqn{2ppf} gives the answer for a single particle of mass $m$. In string theory,
we expect contributions from an infinite
number of species of particles of mass $m_n$. Specializing to $D=26$, we expect the
partition function to be
\be   Z = \int_0^\infty dl\ \frac{1}{l^{14}}\,  \sum_{n=0}^\infty e^{-m_n^2l}\nn\ee
But we know that the mass spectrum of the free string: it is given in terms of the $L_0$ and $\tilde{L}_0$ operators by
\be m^2 = \frac{4}{\ap}(L_0-1)= \frac{4}{\ap}(\tilde{L}_0-1) =
\frac{2}{\ap}(L_0+\tilde{L}_0-2)\nn\ee
subject to the constraint of level matching, $L_0=\tilde{L}_0$. It's easy to
impose level matching: we simply throw in a Kronecker delta in its integral representation,
\be \frac{1}{2\pi}\int_{-1/2}^{+1/2}ds\ e^{2\pi i s(L_0-\tilde{L}_0)} = \delta_{L_0,\tilde{L}_0}\label{kdelta}\ee
Replacing the sum over species, with the trace over the spectrum of states
subject to level matching, the partition function becomes,
\be
Z = \int_0^\infty dl\ \frac{1}{l^{14}}\,\int_{-1/2}^{+1/2}ds\ \Tr\,e^{2\pi i s(L_0-\tilde{L}_0)}\,e^{-2(L_0+\tilde{L}_0-2)l/\ap}\label{trouble}\ee
We again use the definition $q=\exp(2\pi i\tau)$, but this time
the complex parameter $\tau$ is a combination of the length of the loop $l$, and
the auxiliary variable that we introduced to impose level matching,
\be \tau = s + \frac{2li}{\ap}\nn\ee
The trace over the spectrum of the string once gives the eta-functions, just as it did
before. We're left with the result for the partition function,
\be Z_{\rm string} = \int \frac{{d^2\tau}}{({\rm Im}\,\tau)^2}\ \left(\frac{1}{\sqrt{{\rm Im}\,\tau}}\frac{1}{\eta(q)}\frac{1}{\bar{\eta}(\bar{q})}\right)^{24}\nn\ee
But this is exactly the same expression that we saw before. With a difference! In fact,
the difference is hidden in the notation: it is the range of integration for $d^2\tau$ which
can be found in the original expressions \eqn{2ppf} and \eqn{kdelta}.
${\rm Re}\,\tau$ runs over the same interval $[-\ft12,+\ft12]$ that we saw in string theory. As
is clear from this discussion, it is this integral which implements level matching. The difference
comes in the range of ${\rm Im}\,\tau$ which, in this naive analysis, runs over $[0,\infty)$.
This is in stark contrast to string
theory where we only integrate over the fundamental domain.

\para
This highlights our previous statement: the potential UV divergences in field theory
are encountered in the region ${\rm Im}\,\tau \sim l\rightarrow 0$. In the above
analysis, this corresponds to particles traversing small loops. But this region is simply
absent in the correct string theory computation. It is mapped, by modular invariance,
to the infra-red region of large loops.

\para
It is often said that in the $g_s\rightarrow 0$ limit string theory becomes a theory of
an infinite number of free particles. This is true of the spectrum. But this calculation
shows that it's not really true when we compute loops because the modular invariance
means that we integrate over a different range of momenta in string theory than in
a naive field theory approach.

\para
So what happens in the infra-red region of our partition function? The easiest place to
see it is in the $l\rightarrow \infty$ limit of the integral \eqn{trouble}. We see that
the integral is dominated by the lightest state which, for the bosonic string is the tachyon.
This has $m^2 = -4/\ap$, or $(L_0+\tilde{L}_0-2)=-2$. This gives a contribution to the partition
function of,
\be \int^\infty \frac{dl}{l^{14}}\ e^{+4l/\ap}\nn\ee
which clearly diverges. This IR divergence of the one-loop partition function is another
manifestation of tachyonic trouble. In the superstring, there is no tachyon and the IR
region is well-behaved.

\subsubsection{So is String Theory Finite?}

The honest answer is that we don't know. The UV finiteness that we saw above holds
for all one-loop amplitudes. This means, in particular, that we have a one-loop
finite theory of gravity interacting with matter in higher dimensions. This is already
remarkable.

\para
There is more good news:
One can show that UV finiteness continues to hold at the two-loops. And, for the superstring,
state-of-the-art techniques using the ``pure-spinor" formalism show that certain objects
remain finite up to five-loops. Moreover, the exponential suppression \eqn{highscat} that
we saw when all momentum exchanges are large continues to hold for all amplitudes.

\para
However, no general statement of finiteness has been proven. The danger lurks in the singular
points in the integration over Riemann surfaces of genus 3 and higher.

\subsubsection{Beyond Perturbation Theory?}
\label{bptsec}

From the discussion in this section, it should be clear that string perturbation theory
is entirely analogous to the Feynman diagram expansion in field theory. Just as in
field theory, one can show that the expansion in $g_s$ is asymptotic. This
means that the series does not converge, but we can nonetheless make sense of it.

\para
However, we know that there are many phenomena in quantum field theory that aren't captured
by Feynman diagrams. These include confinement in the strongly coupled regime and
instantons and solitons in the weakly coupled regime. Does this mean that we are missing similarly
interesting phenomena in string theory? The answer is almost certainly yes! In this section,
I'll very briefly allude to a couple of more advanced topics which allow us to go beyond
the perturbative expansion in string theory. The goal is not
really to teach you these things, but merely to familiarize you with some words.

\para
One way to proceed is to keep quantum field theory as our guide and try to build a non-perturbative
definition of string theory in terms of a path integral. We've already seen that the Polyakov
path integral over worldsheets is equivalent to Feynman diagrams. So we need to go one step
further. What does this mean? Recall that in QFT, a field creates a particle.
In string theory, we are now looking for a field which
creates a loop of string. We should have a different field for each configuration of the
string. In other words, our field should itself be a function of a function: $\Phi(X^\mu(\sigma))$.
Needless to say, this is quite a complicated object.
If we were brave, we could then consider the path integral for this field,
\be Z = \int {\cal D}\Phi \ e^{iS[\Phi(X(\sigma))]}\nn\ee
for some suitable action $S[\Phi]$. The idea is that this path integral should reproduce
the perturbative string expansion and, furthermore,  defines a
non-perturbative completion of the
theory. This line of ideas is known as {\it string field theory}.
It should be clear that this is one step further in the development: particles $\rightarrow$
fields $\rightarrow$ string fields. Or, in more historical language, if field theory is
``second quantization", then string field theory is ``third quantization".

\para
String field theory has been fairly successful for the open string and some interesting
non-perturbative results have been obtained in this manner. However, for the closed string
this approach has been much less useful. It is usually thought that there are deep reasons
behind the failure of closed string field theory, related to issues that we mentioned at
the beginning of this section: there are no off-shell quantities in a theory of gravity.
Moreover, we mentioned in Section 4 that a theory of interacting open strings necessary
includes closed strings, so somehow the open string field theory should already contain gravity
and closed strings. Quite how this comes about is still poorly understood.

\para
There are other ways to get a handle on  non-perturbative
aspects of string theory using the low-energy effective action (we will describe what the
``low-energy effective action" is
in the next section). Typically these techniques
rely on supersymmetry to provide a window into the strongly coupled regime,
and so work only for the superstring. These methods have been extremely
successful and any course on superstring theory would be devoted to explaining
various aspects of such as dualities and M-theory.

\para
Finally, in asymptotically AdS spacetimes, the AdS/CFT correspondence gives a non-perturbative
definition of string theory and quantum gravity in the bulk in terms of Yang-Mills theory,
or something similar, on the boundary. In some sense, the boundary field theory is
a ``string field theory".

\subsection{Appendix: Games with Integrals and Gamma Functions}
\label{appendix}

\EPSFIGURE{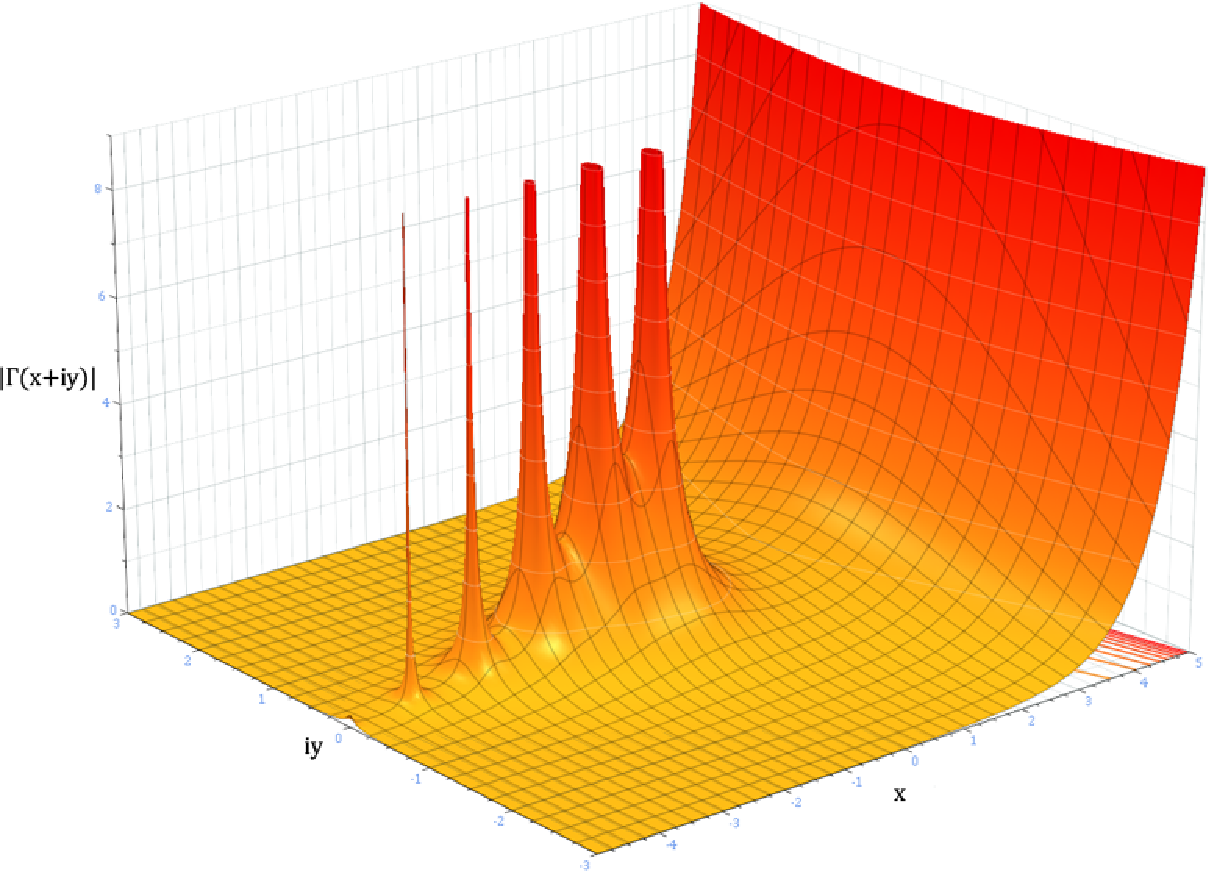,height=110pt}{}
\noindent
The gamma function is defined by the integral representation
\be \Gamma(z) = \int_0^\infty dt\ t^{z-1}e^{-t} \label{gamma}\ee
which converges if ${\rm Re}z>0$. It has a unique analytic expression to the
whole $z$-plane. The absolute value of the gamma function over the  $z$-plane
is shown in the figure.

\para

The gamma function has a couple of important properties. Firstly, it can be thought of
as the analytic continuation of the factorial function for positive integers, meaning
\be \Gamma(n) = (n-1)!\ \ \ \ \ \ \ n\in {\bf Z}^+\nn\ee
Secondly, $\Gamma(z)$ has poles at non-positive integers. More precisely when
$z\approx -n$, with $n=0,1,\ldots$, there is the expansion
\be \Gamma(z) \approx \frac{1}{z+n}\,\frac{(-1)^n}{n!}
%\ \ \ \ \ \ \ \ \ \ \ \ \ \ \ \ \ \ \ \ \ \ \ \ \ \ \ \ \ \ \ \ \ \ \
%\ \ \ \ \ \ \
\nn\ee
%
%\begin{figure}[htb]
%\begin{center}
%\epsfxsize=3in\leavevmode\epsfbox{gamma.eps}
%\end{center}
%\caption{The absolute value of the gamma function on the complex plane.}
%\label{mir}
%\end{figure}
%

\subsubsection*{The Euler Beta Function}

The Euler beta function is defined for $x$, $y\in {\bf C}$ by
\be B(x,y)=\frac{\Gamma(x)\Gamma(y)}{\Gamma(x+y)}\nn\ee
It has the integral representation
\be B(x,y) = \int_0^1 dt\ t^{x-1}(1-t)^{y-1}\label{ebeta}\ee
Let's prove this statement. We start by looking at
\be \Gamma(x)\Gamma(y) = \int_0^\infty du\int_0^\infty dv\ e^{-u}u^{x-1}e^{-v}v^{y-1}\nn\ee
We write $u=a^2$ and $v=b^2$ so the integral becomes
\be \Gamma(x)\Gamma(y) &=&  4\int_0^\infty da\int_0^\infty db\ e^{-(a^2+b^2)}a^{2x-1}b^{2y-1}
\nn\\ &=& \int_{-\infty}^\infty da\int_{-\infty}^\infty db\ e^{-(a^2+b^2)}|a|^{2x-1}|b|^{2y-1}\nn\ee
We now change coordinates once more, this time to polar $a=r\cos\theta$ and $b=r\sin\theta$.
We get
\be \Gamma(x)\Gamma(y) &=&  \int_0^\infty rdr\ e^{-r^2}r^{2x+2y-2}\ \int_0^{2\pi} d\theta
\ |\cos\theta|^{2x-1}|\sin \theta|^{2y-1} \nn\\ &=& \frac{1}{2}\Gamma(x+y)\times
4\int_0^{\pi/2}d\theta\ (\cos\theta)^{2x-1}(\sin\theta)^{2y-1} \nn\\
&=& \Gamma(x+y)\int^1_0dt\ (1-t)^{y-1}t^{x-1}\nn\ee
where, in the final line, we made the substitution $t=\cos^2\theta$. This completes
the proof.

\subsubsection*{The Virasoro-Shapiro Amplitude}

In the closed string computation, we  came across the integral
\be C(a,b) = \int d^2z\ |z|^{2a-2}|1-z|^{2b-2}\nn\ee
We will now evaluate this and show that it is given by \eqn{dotheint}. We start by using a trick.
We can write
\be |z|^{2a-2} = \frac{1}{\Gamma(1-a)}\,\int_0^\infty dt\ t^{-a}e^{-|z|^2t}\nn\ee
which follows from the definition \eqn{gamma} of the gamma function. Similarly, we can write
\be |1-z|^{2b-2}=\frac{1}{\Gamma(1-b)}\,\int_0^\infty du\ u^{-b}e^{-|1-z|^2u}\nn\ee
We decompose the complex coordinate $z=x+iy$, so that the measure of the integral is
$d^2z = 2dxdy$. We can then write the integral $C(a,b)$ as
\be C(a,b) &=& \int \frac{d^2z\, du\, dt}{\Gamma(1-a)\Gamma(1-b)}\ t^{-a}u^{-b} e^{-|z|^2 t}
e^{-|1-z|^2u}\nn\\ &=& 2\int \frac{dx\, dy\, du\, dt}{\Gamma(1-a)\Gamma(1-b)} \ t^{-a}u^{-b}\,
e^{-(t+u)(x^2+y^2)+2xu - u}\nn\\ &=& 2\int
\frac{dx\, dy\, du\, dt}{\Gamma(1-a)\Gamma(1-b)} \ t^{-a}u^{-b}\,\exp\left( -(t+u)\left[
\left(x-\frac{u}{t+u}\right)^2+y^2\right] - u +\frac{u^2}{t+u}\right)
\nn\ee
Now we do the $dxdy$ integral which is simply Gaussian. We find
\be C(a,b) = \frac{2\pi}{\Gamma(1-a)\Gamma(1-b)}\int_0^\infty du\,dt\ \frac{t^{-a}u^{-b}}{t+u}\,e^{-tu/(t+u)} \nn\ee
Finally, we make a change of variables. We write $t=\alpha\beta$ and $u=(1-\beta)\alpha$. In
order for $t$ and $u$ to take values in the range $[0,\infty)$, we require $\alpha\in [0,\infty)$
and $\beta\in [0,1]$. Taking into account the Jacobian arising from this transformation,
which is simply $\alpha$, the integral becomes
\be C(a,b) = \frac{2\pi}{\Gamma(1-a)\Gamma(1-b)}\int d\alpha\, d\beta\ \frac{\alpha^{1-a-b}}{\alpha}\,
\beta^{-a}(1-\beta)^{-b}e^{-\alpha\beta(1-\beta)}\nn\ee
But we recognize the integral over $d\alpha$: it is simply
\be \int_0^\infty d\alpha\ \alpha^{-a-b}e^{-\beta\alpha(1-\beta)} = [\beta(1-\beta)]^{a+b-1}\Gamma
(1-a-b)\nn\ee
We write $c=1-a-b$. Finally, we're left with
\be C(a,b) = \frac{2\pi\Gamma(c)}{\Gamma(1-a)\Gamma(1-b)}\int_0^1d\beta\ (1-\beta)^{a-1}\beta^{b-1}
\nn\ee
But the final integral is the Euler beta function \eqn{ebeta}. This gives us our promised
result,
\be C(a,b) = \frac{2\pi \Gamma(a)\Gamma(b)\Gamma(c)}{\Gamma(1-a)\Gamma(1-b)\Gamma(1-c)}\nn\ee

\newpage
\section{Low Energy Effective Actions}
\label{background}

So far, we've only discussed strings propagating in flat spacetime. In this section we will
consider strings propagating in different backgrounds. This is equivalent to having
different CFTs on the worldsheet of the string.

\para
There is an obvious generalization of the Polyakov action to describe a string  moving
in curved spacetime,
\be S = \frac{1}{4\pi\ap}\int d^2\sigma \sqrt{g}\ \gabi\, \p_\alpha X^\mu\, \p_\beta X^\nu
\,G_{\mu\nu}(X)\label{nlsm}\ee
Here $\gab$ is again the worldsheet metric. This action describes a map from the worldsheet of the
string into a spacetime  with metric $G_{\mu\nu}(X)$. (Despite its name, this metric
is not to be confused with the Einstein tensor which we won't have need for in this lecture notes).

\para
Actions of the form \eqn{nlsm} are known as {\it non-linear sigma models}.
(This strange name has its roots in the history of pions). In this context, the $D$-dimensional
spacetime is sometimes called the {\it target space}. Theories of this type
are important in many aspects of physics, from QCD to condensed matter.

\para
Although it's obvious that \eqn{nlsm} describes strings moving in curved spacetime, there's something
a little fishy about just writing it down. The problem is that the quantization of the closed string
already gave us a graviton. If we want to build up some background metric $G_{\mu\nu}(X)$, it
should be constructed from these gravitons, in much the same manner that a laser beam is made
from the underlying photons. How do we see that the metric in \eqn{nlsm} has anything to do
with the gravitons that arise from the quantization of the string?

\para
The answer lies in the use of vertex operators. Let's expand the metric
as a small fluctuation around flat space
\be G_{\mu\nu}(X) = \delta_{\mu\nu} + h_{\mu\nu}(X)\nn\ee
Then the partition function that we build from the action \eqn{nlsm} is related to the partition
function for a string in flat space by
\be Z = \int {\cal D}X{\cal D}g\ e^{-S_{\rm Poly}-V} =
\int {\cal D}X{\cal D}g\ e^{-S_{\rm Poly}}(1-V+\frac{1}{2}V^2 + \ldots)
\nn\ee
where $S_{\rm Poly}$ is the action for the string in flat space given in \eqn{poly} and $V$
is the expression
\be V = \frac{1}{4\pi\ap}\int d^2\sigma\sqrt{g}\ \gabi\p_\alpha\, X^\mu\, \p_\beta X^\nu\,
h_{\mu\nu}(X)\label{gvertex}\ee
But we've seen this before: it's the vertex operator associated to the graviton state
of the string! For a plane wave, corresponding to a graviton with polarization given by
the symmetric, traceless tensor $\zeta_{\mu\nu}$,
and momentum $p^\mu$, the fluctuation is given by
\be h_{\mu\nu}(X) = \zeta_{\mu\nu}\ e^{ip\cdot X}\nn\ee
With this choice, the expression \eqn{gvertex} agrees with the vertex operator \eqn{1vertex}.
But in general, we could take any linear superposition of plane waves to build up a
general fluctuation $h_{\mu\nu}(X)$.

\para
We know that inserting a single copy of $V$ in the path integral corresponds to the introduction
of a
single graviton state. Inserting $e^V$ in the path integral corresponds to a coherent state
of gravitons, changing the metric from $\delta_{\mu\nu}$ to $\delta_{\mu\nu}+h_{\mu\nu}$. In
this way we see that the background curved metric of \eqn{nlsm} is indeed built of
the quantized gravitons that we first met back in Section 2.

\subsection{Einstein's Equations}

In conformal gauge, the Polyakov action in flat space reduces to a free theory. This fact
was extremely useful, allowing us to compute the spectrum of the theory. But on a curved
background, it is no longer the case.
%
%\be \gab = e^{2\phi}\delta_{\alpha\beta}\nn\ee
%
In conformal gauge, the worldsheet theory is described by an interacting two-dimensional
field theory,
\be S=\frac{1}{4\pi\ap}\int d^2\sigma\ G_{\mu\nu}(X)\,\p_\alpha X^\mu\,\p^\alpha X^\nu\label{gcft}\ee
To understand these interactions in more detail, let's expand around a classical solution
which we take to simply be a string sitting at  a point $\bar{x}^\mu$.
\be X^\mu(\sigma) = \bar{x}^\mu + \sqrt{\ap}\,Y^\mu(\sigma)\nn\ee
Here $Y^\mu$ are the dynamical fluctuations about the point which we assume to
be small. The factor of $\sqrt{\ap}$ is there for dimensional reasons: since
$[X]=-1$, we have $[Y]=0$ and statements like $Y\ll 1$ make sense.
Expanding the Lagrangian gives
\be G_{\mu\nu}(X)\,\p X^\mu\p\, X^\nu = \ap\left[G_{\mu\nu}(\bar{x}) + \sqrt{\ap}G_{\mu\nu,\omega}(\bar{x})\,Y^\omega + \frac{\ap}{2}\,G_{\mu\nu,\omega\rho}
(\bar{x})\,Y^\omega Y^\rho +\ldots \right]\,\p Y^\mu\,\p Y^\nu\nn\ee
Each of the coefficients $G_{\mu\nu,\ldots}$ in the Taylor expansion are coupling constants
for the interactions of the fluctuations $Y^\mu$. The theory has an infinite number of coupling
constants and they are nicely packaged into the function $G_{\mu\nu}(X)$.

\para
We want to know when this field theory is weakly coupled. Obviously this requires the whole
infinite set of coupling constants to be small. Let's try to characterize this in a crude manner.
Suppose that the target space
has characteristic radius of curvature $r_c$, meaning schematically that
\be \ppp{G}{X} \sim \frac{1}{r_c}\nn\ee
The radius of curvature is a length scale, so $[r_c]=-1$.
From the expansion of the metric, we see that the effective dimensionless coupling is
given by
\be \frac{\sqrt{\ap}}{r_c}\label{apexp}\ee
This means that we can use perturbation theory to study the CFT \eqn{gcft} if the spacetime metric
only varies on scales much greater than $\sqrt{\ap}$. The perturbation series in $\sqrt{\ap}/r_c$ is
usually called the $\ap$-expansion to distinguish it from the $g_s$ expansion that we saw
in the previous section. Typically a quantity computed in string theory is given by a double
perturbation expansion: one in $\ap$ and one in $g_s$.

\para
If there are regions of spacetime where the radius of curvature becomes comparable to the string
length scale, $r_c\sim \sqrt{\ap}$, then the worldsheet CFT is strongly coupled and we will need to
develop new methods to solve it. Notice that strong coupling in $\ap$ is hard, but the
problem is at least well-defined in terms of the worldsheet path integral. This is qualitatively different
to the question of strong coupling in $g_s$ for which, as discussed in Section \ref{bptsec}, we're
really lacking a good definition of what the problem even means.

\subsubsection{The Beta Function}

Classically, the theory defined by \eqn{gcft} is conformally invariant. But this is not necessarily
true in the quantum theory. To regulate divergences  we will have to introduce a
UV cut-off and, typically, after renormalization, physical quantities depend on the scale of a
given process $\mu$. If this is the case, the theory is no longer conformally invariant.
There are plenty of theories which classically possess scale invariance
which is broken quantum mechanically. The most famous of these is Yang-Mills.

\para
As we've discussed several times, in string theory conformal invariance is a gauge symmetry
and we can't afford to lose it. Our goal in this section is to understand the circumstances
under which \eqn{gcft} retains conformal invariance at the quantum level.

\para
The object which describes how couplings depend on a scale $\mu$ is called the $\beta$-function.
Since we have a functions worth of couplings, we should really be talking about a $\beta$-functional,
schematically of the form
\be \beta_{\mu\nu}(G) \sim  \mu\ppp{G_{\mu\nu}(X;\mu)}{\mu}\nn\ee
The quantum theory will be conformally invariant only if
\be \beta_{\mu\nu}(G)=0\nn\ee
We now compute this for the non-linear sigma model at one-loop. Our strategy will be to isolate the
UV divergence of the theory and figure out what kind of counterterm we should add. The
beta-function will vanish if this counterterm vanishes.

\para
The analysis is greatly simplified by a cunning choice of coordinates. Around any point $\bar{x}$, we can always pick Riemann normal coordinates such that the expansion in
$X^\mu = \bar{x}^\mu + \sqrt{\ap}\,Y^\mu$ gives
\be G_{\mu\nu}(X) = \delta_{\mu\nu} -\frac{\ap}{3}{\cal R}_{\mu\lambda\nu\kappa}(\bar{x})Y^\lambda Y^\kappa
+{\cal O}(Y^3)\nn\ee
To quartic order in the fluctuations, the action becomes
\be S=\frac{1}{4\pi} \int d^2\sigma \ \p Y^\mu\,\p Y^\nu \delta_{\mu\nu} -
\frac{\ap}{3}{\cal R}_{\mu\lambda \nu \kappa}\,Y^\lambda Y^\kappa\p Y^\mu \p Y^\nu\nn\ee
We can now treat this as an interacting quantum field theory in two dimensions. The
quartic interaction gives a vertex with the  Feynman rule,
\be \raisebox{-3.1ex}{\epsfxsize=0.5in\epsfbox{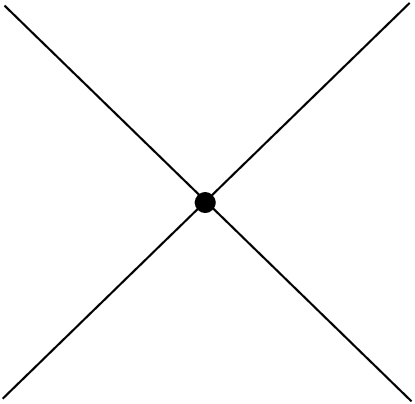}} \sim {\cal R}_{\mu\lambda\nu\kappa}\,(k^\mu \cdot k^\nu)\nn\ee
where $k^\mu_\alpha$ is the 2d momentum ($\alpha=1,2$ is a worldsheet index) for the scalar
field $Y^\mu$. It sits in the Feynman rules because we are talking about derivative interactions.

\para
Now we've reduced the problem to a simple interacting quantum field theory, we can
compute the $\beta$-function using whatever method we like.
The divergence in the theory comes from the one-loop diagram
\be \raisebox{-3.1ex}{\epsfxsize=1.0in\epsfbox{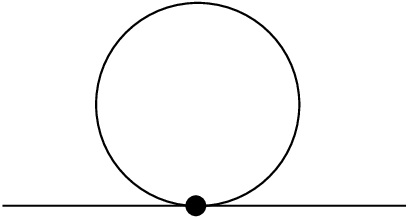}} \nn\ee
It's actually simplest to think about this diagram in position space. The propagator for
a scalar particle is
\be \langle Y^\lambda(\sigma) Y^\kappa(\sigma^\prime)\rangle = -\frac{1}{2}\,\delta^{\lambda\kappa}\,\ln | \sigma-\sigma^\prime|^2\nn\ee
For the scalar field running in the loop, the beginning and end point coincide. The
propagator diverges as $\sigma \rightarrow \sigma^\prime$, which is simply reflecting the
UV divergence that we would see in the momentum integral around the loop.

\para
To isolate this divergence, we choose to work with dimensional regularization, with $d=2+\epsilon$. The propagator then becomes,
\be \langle Y^\lambda(\sigma) Y^\kappa(\sigma^\prime)\rangle &=& 2\pi \delta^{\lambda\kappa}
\,\int\frac{d^{2+\epsilon}k}{(2\pi)^{2+\epsilon}}\,\frac{e^{ik\cdot(\sigma-\sigma^\prime)}}{k^2}
\nn\\ &\longrightarrow&\ \frac{\delta^{\lambda\kappa}}{\epsilon}\ \ \ \ \ {\rm as}\ \sigma\rightarrow \sigma^\prime
\nn\ee
The necessary counterterm for this divergence can be determined simply by
replacing $Y^\lambda Y^\kappa$ in the action with $\langle Y^\lambda Y^\kappa\rangle$. To subtract
the $1/\epsilon$ term, we add the counterterm
\be {\cal R}_{\mu\lambda \nu \kappa}\,Y^\lambda Y^\kappa\p Y^\mu \p Y^\nu
\ \rightarrow\ {\cal R}_{\mu\lambda \nu \kappa}\,Y^\lambda Y^\kappa\p Y^\mu \p Y^\nu
-\frac{1}{\epsilon}{\cal R}_{\mu\nu}\,\p Y^\mu \p Y^\nu\nn\ee
One can check that this can be absorbed by a wavefunction renormalization
$Y^\mu\ \rightarrow\ Y^\mu + (\ap/6\e)\,{\cal R}^\mu_{\ \nu}Y^\nu$, together with the
renormalization of the coupling constant which, in our theory, is the metric $G_{\mu\nu}$.
We require,
\be G_{\mu\nu}  \ \rightarrow\ G_{\mu\nu}+\frac{\ap}{\e}\,{\cal R}_{\mu\nu}\label{gshift}\ee
From this we learn the beta function of the theory and the condition for conformal invariance.
It is
\be \bmn(G) = \ap {\cal R}_{\mu\nu}\label{b} =0\ee
This is a magical result! The requirement for the sigma-model to be conformally invariant is that the target space must be Ricci flat: ${\cal R}_{\mu\nu}=0$.
Or, in other words, the background spacetime in which the string moves
must obey the vacuum Einstein equations! We see that the equations of general
relativity also describe the renormalization group flow of 2d sigma models.

\para
There are several more magical things just around the corner, but it's worth pausing
to make a few diverse comments.

\subsubsection*{Beta Functions and Weyl Invariance}

The above calculation effectively studies the breakdown of conformal invariance
in the CFT \eqn{gcft} on a flat worldsheet. We know that this should be the same thing
as the breakdown of Weyl invariance on a curved worldsheet. Since this is such an
important result, let's see how it works from this other perspective. We can consider the worldsheet metric
\be \gab = e^{2\phi}\delta_{\alpha\beta}
\nn\ee
Then, in dimensional regularization, the theory is not Weyl invariant in $d=2+\epsilon$
dimensions because the contribution from $\sqrt{g}$ does not quite cancel that from the
inverse metric $\gabi$. The action is
\be S &=& \frac{1}{4\pi\ap}\int d^{2+\epsilon}\sigma\ e^{\phi\epsilon}\p_\alpha X^\mu \,\p^\alpha X^\nu\, G_{\mu\nu}(X) \nn\\ &\approx& \frac{1}{4\pi\ap}\int d^{2+\epsilon}\sigma\ (1+\phi\epsilon)\,
\p_\alpha X^\mu\, \p^\alpha X^\nu\, G_{\mu\nu}(X) \nn\ee
where, in this expression, the $\alpha=1,2$ index is now raised and lowered with $\delta_{\alpha\beta}$. If we replace $G_{\mu\nu}$ in this expression with the
renormalized metric \eqn{gshift}, we see that there's a term involving $\phi$
which remains even as $\e \rightarrow 0$,
\be S = \frac{1}{4\pi\ap}\int d^2\sigma\ \p_\alpha X^\mu \p^\alpha X^\nu\, \left[G_{\mu\nu}(X)
+\ap \phi \,{\cal R}_{\mu\nu}(X)\right]
\nn\ee
This indicates a breakdown of Weyl invariance. Indeed, we can look at our
usual diagnostic for Weyl invariance, namely the vanishing of $T^\alpha_{\ \alpha}$.
In conformal gauge, this is given by
\be \tab = +\frac{4\pi}{\sqrt{g}}\,\ppp{S}{\gabi} = -2\pi \ppp{S}{\phi}\delta_{\alpha\beta}
\ \ \Rightarrow\ \ T^\alpha_{\ \alpha} = -\frac{1}{2}{\cal R}_{\mu\nu}\,\p X^\mu\, \p X^\nu\nn\ee
In this way of looking at things, we define the $\beta$-function to be the coefficient
in front of  $\p X \p X$, namely
\be T^\alpha_{\ \alpha} = -\frac{1}{2\ap}\,\beta_{\mu\nu}\,\p X^\mu\p X^\nu\nn\ee
Again, we have the result
\be \beta_{\mu\nu}= \ap {\cal R}_{\mu\nu}\nn\ee

\subsubsection{Ricci Flow}

In string theory we only care about conformal theories with Ricci flat metrics. (And generalizations of this result that we will discuss shortly). However,
in other areas of physics and mathematics, the RG flow itself is important. It is
usually called Ricci flow,
\be\mu\ppp{G_{\mu\nu}}{\mu} = \ap {\cal R}_{\mu\nu}\label{rflow}\ee
which dictates how the metric changes with scale $\mu$.

\para
As an illustrative and simple example, consider the target space ${\bf S}^2$ with radius
$r$. This is an important model in condensed matter physics where it describes the low-energy
limit of a one-dimensional Heisenberg spin chain. It is sometimes called the $O(3)$ sigma-model.
Because the sphere is a symmetric space, the only effect of
the RG flow is to make the radius scale dependent: $r=r(\mu)$. The beta function is given by
\be \mu \ppp{r^2}{\mu} = \frac{\ap}{2\pi}\nn\ee
Hence $r$ gets large as we go towards the UV, and small towards the IR. Since the coupling
is $1/r$, this means that the non-linear sigma model with ${\bf S}^2$ target space
is asymptotically free. At low energies, the theory is strongly coupled and perturbative
calculations --- such as this one-loop beta function --- are no longer trusted. In particular, one can show that the ${\bf S}^2$ sigma-model develops a mass gap in the IR.

\para
The idea of Ricci flow \eqn{rflow} was recently used by Perelman to prove the Poincar\'e
conjecture. In fact, Perelman used a slightly generalized version of Ricci flow which we will
see below. In the language of string theory, he introduced the dilaton field.

\subsection{Other Couplings}

We've understood how strings couple to a background spacetime metric. But what about the
other modes of the string? In Section 2, we saw that a closed string has further
massless states which are associated to the anti-symmetric tensor $B_{\mu\nu}$ and the
dilaton $\Phi$. We will now see how the string reacts if these fields are turned on in
spacetime.

\subsubsection{Charged Strings and the $B$ field}
\label{bsec}

Let's start by looking at how strings couple to the anti-symmetric field $B_{\mu\nu}$.
We discussed the vertex operator associated to this state in Section \ref{closedvsec}.
It is given in \eqn{1vertex} and takes the same form as the graviton vertex operator,
but with $\zeta_{\mu\nu}$ anti-symmetric. It is a simple matter to  exponentiate this, to
get an expression for how strings propagate in background $B_{\mu\nu}$ field.
We'll keep the curved metric $G_{\mu\nu}$ as well to get the general action,
\be S = \frac{1}{4\pi\ap}\int d^2\sigma\sqrt{g}\ \left(G_{\mu\nu}(X)\, \p_\alpha X^\mu\, \p_\beta X^\nu g^{\alpha\beta}
+ i B_{\mu\nu}(X)\,\partial_\alpha X^\mu\,\partial_\beta X^\nu\,\epsilon^{\alpha\beta}\right) \label{bfield}\ee
Where $\epsilon^{\alpha\beta}$ is the anti-symmetric 2-tensor, normalized such that $\sqrt{g}\epsilon^{12}=+1$. (The factor of $i$ is there in the action
because we're in Euclidean space and this new term has a single ``time"
derivative). The action retains invariance under worldsheet reparameterizations and Weyl rescaling.

\para
So what is the interpretation of this new term? We will now show that we should think of the
field $B_{\mu\nu}$ as analogous to the gauge potential $A_{\mu}$ in electromagnetism. The
action \eqn{bfield} is telling us that the string is ``electrically charged" under $B_{\mu\nu}$.

\subsubsection*{Gauge Potentials}

We'll take a short detour to remind ourselves about some pertinent facts in
electromagnetism. Let's start by returning to a point particle. We know that
a charged point particle couples to a background gauge
potential $A_{\mu}$ through the addition of a worldline term to the action,
\be \int d\tau\ A_\mu(X)\,\dot{X}^\mu\ .\label{electric}\ee
If this relativistic form looks a little unfamiliar, we can deconstruct it  by working in static gauge with $X^0\equiv t = \tau$, where it reads
\be \int dt\ A_0(X) + A_i(X)\,\dot{X}^i \ ,\nn \ee
which should now be recognizable as
the Lagrangian that gives rise to the Coulomb and Lorentz force laws for a
charged particle.

\para
So what is the generalization of this kind of coupling for a string? First note that \eqn{electric} has an interesting geometrical structure. It is the pull-back of the one-form $A=A_\mu dX^\mu$ in spacetime onto the worldline of the particle. This works because $A$ is a one-form and the worldline is one-dimensional. Since the worldsheet of the string is two-dimensional, the analogous coupling should be to a two-form in spacetime. This is an anti-symmetric tensor field with two indices, $B_{\mu\nu}$. The pull-back of $B_{\mu\nu}$ onto the worldsheet gives the interaction,
\be \int d^2\sigma\ B_{\mu\nu}(X)\,\partial_\alpha X^\mu\,\partial_\beta X^\nu\,\epsilon^{\alpha\beta} \ .\label{bmn}\ee
This is precisely the form of the interaction we found in \eqn{bfield}.

\para
The point particle coupling \eqn{electric} is invariant under gauge transformations of the background field $A_\mu \rightarrow A_\mu + \partial_\mu \alpha$. This follows because the Lagrangian changes by a total derivative. There is a similar statement for the two-form $B_{\mu\nu}$. The spacetime gauge symmetry is,
\be B_{\mu\nu} \rightarrow B_{\mu\nu} + \partial_\mu C_\nu - \partial_\nu C_\mu \ee
under which the Lagrangian \eqn{bmn} changes by a total derivative.

\para
In electromagnetism, one can construct the gauge invariant electric and magnetic fields which are packaged in the two-form field strength $F=dA$.
%$F_{\mu\nu}= \partial_\mu A_\nu - \partial_\mu A_\nu$.
Similarly, for $B_{\mu\nu}$, the gauge invariant field strength $H=dB$ is a three-form,
\be H_{\mu\nu\rho} = \partial_\mu B_{\nu\rho} + \partial_\nu B_{\rho\mu} + \partial_\rho B_{\mu\nu}\ .\nn\ee
This 3-form $H$ is sometimes known as the {\it torsion}. It plays the same role as torsion
in general relativity, providing an anti-symmetric component to the affine connection.

\subsubsection{The Dilaton}

Let's now figure out how the string couples to a background dilaton field $\Phi(X)$.
This is more subtle.
A naive construction of the vertex operator
is not primary and one must work a little harder. The correct derivation of the
vertex operators can be found
in Polchinski. Here I will simply give the coupling and explain some important features.

\para
The action of a string moving in a background involving profiles for the massless fields $G_{\mu\nu}$, $B_{\mu\nu}$ and $\Phi(X)$ is given by
\be S = \frac{1}{4\pi\ap}\int d^2\sigma\sqrt{g} &&\left(G_{\mu\nu}(X)\, \p_\alpha X^\mu\, \p_\beta X^\nu g^{\alpha\beta}
+ i B_{\mu\nu}(X)\,\partial_\alpha X^\mu\,\partial_\beta X^\nu\,\epsilon^{\alpha\beta}
\right. \nn\\ &&\ \ \ \ +\left.\ap \,\Phi(X)\,R^{(2)}\right) \label{all}\ee
where $R^{(2)}$ is the two-dimensional Ricci scalar of the worldsheet. (Up until
now, we've always denoted this simply as $R$ but we'll introduce the superscript
from hereon to distinguish the worldsheet Ricci scalar from the spacetime Ricci scalar).

\para
The coupling to the dilaton is surprising for several reasons. Firstly, we see that the
term in the action vanishes on a flat worldsheet, $R^{(2)}= 0$. This is one
of the reasons that it's a little trickier to determine this coupling
using vertex operators.

\para
However, the most surprising thing about the coupling to the dilaton is that it {\it does not}
respect Weyl invariance! Since a large part of this course has been about understanding
the implications of Weyl invariance, why on earth are we willing to throw it away now?!
The answer, of course, is that we're not.
Although the dilaton coupling does violate Weyl invariance, there is a way to restore it. We will explain this shortly. But firstly, let's discuss one crucially important
implication of the dilaton coupling \eqn{all}.

\subsubsection*{The Dilaton and the String Coupling}

There is an exception to the statement that the classical coupling
to the dilaton violates Weyl invariance. This arises when the dilaton is constant.
For example, suppose
\be \Phi(X) = \lambda\ \ , \ \mbox{a constant}\nn\ee
Then the dilaton coupling reduces to something that we've seen before: it is
\be S_{\rm dilaton} = \lambda \chi \nn\ee
where $\chi$ is the Euler character of the worldsheet that we introduced in \eqn{chi}.
This tells us something important: the constant mode of the dilaton, $\vev{\Phi}$ determines
the string coupling constant. This constant mode is usually taken to be the asymptotic
value of the dilaton,
\be \Phi_0 = \mathop{\rm limit}_{X\rightarrow \infty}\,\Phi(X)\label{phi0}\ee
The string coupling is then given by
\be g_s = e^{{\Phi}_0}\label{gsphi}\ee
So the string coupling is not an independent parameter of string theory: it is
the expectation value of a field. This means that, just like the
spacetime metric $G_{\mu\nu}$ (or, indeed, like the Higgs vev)
it can be determined dynamically.

\para
We've already seen that
our perturbative expansion around flat space is valid as long as $g_s\ll 1$. But now
we have a stronger requirement: we can only trust perturbation theory if the string
is localized in regions of space where $e^{\Phi(X)}\ll 1$ for all $X$.
If the string ventures into regions where $e^{\Phi(X)}$ is of
order 1, then we will need to use techniques that don't rely on string perturbation theory
as described in Section \ref{bptsec}.

\subsubsection{Beta Functions}

We now return to understanding how we can get away with the violation of
Weyl invariance in the dilaton coupling \eqn{all}.
The key to this is to notice the presence of $\ap$ in front of the
dilaton coupling. It's there simply on dimensional grounds. (The other two terms in the
action both come with derivatives $[\p X]=-1$, so don't need any powers of $\ap$).

\para
However, recall that $\ap$ also plays the role of the loop-expansion parameter \eqn{apexp}
in the non-linear sigma model. This
means that the classical lack of Weyl invariance in the dilaton coupling can be
compensated by a one-loop contribution arising from the couplings to $G_{\mu\nu}$
and $B_{\mu\nu}$.

\para
To see this explicitly, one can compute the beta-functions for the two-dimensional
field theory \eqn{all}. In the presence of the dilaton coupling, it's best to
look at the breakdown of Weyl invariance as seen by $\vev{T^\alpha_{\ \alpha}}$.
There are three different kinds of contribution that the stress-tensor can receive,
related to the three different spacetime fields. Correspondingly, we define three
different beta functions,
\be \vev{T^\alpha_{\ \alpha}} = -\frac{1}{2\ap}\beta_{\mu\nu}(G)\,\gabi\p_\alpha X^\mu\,
\p_\beta X^\nu - \frac{i}{2\ap}\beta_{\mu\nu}(B)\,\e^{\alpha\beta}\p_\alpha X^\mu\,\p_\beta
X^\nu - \frac{1}{2}\beta(\Phi)R^{(2)}\ \ \ \ \ \ \ \ \ \ \ \label{betas}\ee
We will not provide the details of the one-loop beta function computations.
We merely state the results\footnote{The relationship between the beta function and
Einstein's equations was first shown by Friedan in his 1980 PhD thesis. A readable
account of the full beta functions can be found in the paper by Callan, Friedan,
Martinec and Perry ``{\it Strings in Background Fields}", Nucl. Phys. B262 (1985) 593.
The full calculational details can be found in TASI lecture notes by Callan and Thorlacius which can be downloaded from the course webpage.},
\be \beta_{\mu\nu}(G)&=&\ap {\cal R}_{\mu\nu} +2\ap\nabla_\mu\nabla_\nu\Phi
-\frac{\ap}{4}H_{\mu\lambda\kappa}H_{\nu}^{\ \lambda\kappa}
\nn\\
\beta_{\mu\nu}(B)&=& -\frac{\ap}{2}\nabla^\lambda H_{\lambda\mu\nu} +\ap
\nabla^{\lambda}\Phi\,H_{\lambda\mu\nu}\nn\\
\beta(\Phi)&=& -\frac{\ap}{2}\nabla^2\Phi + \ap \nabla_\mu\Phi
\,\nabla^\mu\Phi -\frac{\ap}{24}H_{\mu\nu\lambda}H^{\mu\nu\lambda}\nn\ee
A consistent background of string theory must preserve Weyl invariance, which now requires $\beta_{\mu\nu}(G)=\beta_{\mu\nu}(B)=\beta(\Phi)=0$.

\subsection{The Low-Energy Effective Action}

The equations $\beta_{\mu\nu}(G)=\beta_{\mu\nu}(B)=\beta(\Phi)=0$ can be viewed as the
equations of motion for the background in which the string propagates. We now change our
perspective: we look for a $D=26$ dimensional spacetime action
which reproduces these beta-function equations as the equations of motion. This
is the {\it low-energy effective action} of the bosonic string,
\be S = \frac{1}{2\kappa_0^2}\int d^{26}X\sqrt{-G}\,e^{-2\Phi}\,\left(
{\cal R} - \frac{1}{12}H_{\mu\nu\lambda}H^{\mu\nu\lambda}+4\partial_\mu\Phi\,\partial^\mu
\Phi\right)\label{sframe}\ee
where we have taken the liberty of Wick rotating back to Minkowski space for this
expression.
Here the overall constant involving $\kappa_0$ is not fixed by the field equations but can be
determined by coupling these equations to a suitable source as described, for example, in
\ref{ssourcesec}. On dimensional grounds alone, it scales as $\kappa_0^2
\sim l_s^{24}$ where $\ap=l_s^2$.

\para
Varying
the action with respect to the three fields can be shown to yield the beta functions thus,
\be \delta S = \frac{1}{2\kappa_0^2\ap}\int d^{26}X\sqrt{-G}\,e^{-2\Phi}&& \!\!\left(
\delta G_{\mu\nu}\,\beta^{\mu\nu}(G) -\delta B_{\mu\nu}\,\beta^{\mu\nu}(B)
\right.\nn\\ && \ \ \left.-(2\delta\Phi +\frac{1}{2}G^{\mu\nu}\,\delta G_{\mu\nu})
(\beta^\lambda_{\ \lambda}(G)-4\beta(\Phi))\right)\nn\ee
Equation \eqn{sframe} governs the low-energy dynamics of the spacetime fields. The
caveat ``low-energy" refers to the fact that we only worked with the one-loop beta
functions which requires large spacetime curvature.

\para
Something rather remarkable has happened here.
We started, long ago, by looking at how a single
string moves in flat space.
Yet, on grounds of consistency alone, we're led to the
action \eqn{sframe} governing how spacetime and other fields fluctuate in $D=26$ dimensions.
It feels like the tail just wagged the dog.
That tiny string is seriously high-maintenance: its requirements are so stringent
that they govern the way the whole universe moves.

\para
You may also have noticed that
we now have two different methods to compute the scattering of gravitons
in string theory. The first is in terms of scattering amplitudes that we discussed in
Section \ref{scattering}. The second is by looking at the dynamics encoded in the
low-energy effective action \eqn{sframe}. Consistency requires that these two approaches
agree. They do.

\subsubsection{String Frame and Einstein Frame}
\label{seframesec}

The action \eqn{sframe} isn't quite of the familiar Einstein-Hilbert form because
of that strange factor of $e^{-2\Phi}$ that's sitting out front. This factor simply
reflects the fact that the action has been computed at tree level in string
perturbation theory and, as we saw in Section 6, such terms typically scale as $1/g_s^2$.

\para
It's also worth pointing out that the kinetic terms for $\Phi$
in \eqn{sframe} seem to have the wrong sign. However, it's not clear that we
should be worried about this because, again, the factor of $e^{-2\Phi}$ sits out
front meaning that the kinetic terms are not canonically normalized anyway.

\para
To put the action in more familiar form, we can make a field redefinition. Firstly,
it's useful to distinguish between the constant part of the dilaton, $\Phi_0$,
and the part that varies which we call $\tilde{\Phi}$. We defined the constant
part in \eqn{phi0}; it is related to the string coupling constant. The varying
part is simply given by
\be  \tilde{\Phi}=\Phi-\Phi_0\label{tildephi}\ee
In $D$ dimensions, we define a new metric $\tilde{G}_{\mu\nu}$ as a combination
of the old metric and the dilaton,
\be \tilde{G}_{\mu\nu}(X) = e^{-4\tilde{\Phi}/(D-2)}\,G_{\mu\nu}(X)\label{seframe}\ee
Note that this isn't to be thought of as a coordinate transformation or symmetry
of the action. It's merely a
relabeling, a mixing-up, of the fields in the theory. We could make such
redefinitions in any field theory. Typically, we choose not to because the fields
already have canonical kinetic terms. The point of the transformation \eqn{seframe} is
to get the fields in \eqn{sframe} to have canonical kinetic terms as well.

\para
The new metric \eqn{seframe} is related to the old by a conformal rescaling.
One can check that two metrics
related by a general conformal transformation $\tilde{G}_{\mu\nu}=e^{2\omega}G_{\mu\nu}$,
have Ricci scalars related by
\be \tilde{{\cal R}}=e^{-2\omega}\left({\cal R}-2(D-1)\nabla^2\omega - (D-2)(D-1)\partial_\mu\omega\,
\partial^\mu\omega\right)\nn\ee
(We used a particular version of this earlier in the course when considering $D=2$ conformal
transformations). With the choice $\omega = -2\tilde{\Phi}/(D-2)$ in \eqn{seframe}, and
restricting back to $D=26$, the
action \eqn{sframe} becomes
\be
S = \frac{1}{2\kappa^2}\int d^{26}X\,\sqrt{-\tilde{G}}\left(\tilde{{\cal R}}-\frac{1}{12}
e^{-\tilde{\Phi}/3}H_{\mu\nu\lambda}H^{\mu\nu\lambda}-\frac{1}{6}\partial_\mu \tilde{\Phi}
\partial^\mu\tilde{\Phi}\right)\label{eframe}\ee
The kinetic terms for $\tilde{\Phi}$ are now canonical, and come with the right sign. Notice
that there is no potential term for the dilaton, and therefore nothing that dynamically
sets its expectation value in the bosonic string. However, there do exist backgrounds
of the superstring in which a potential for the dilaton develops, fixing the string
coupling constant.

\para
The gravitational part of the action takes the standard Einstein-Hilbert form. The
gravitational coupling is given by
\be \kappa^2 = \kappa_0^2\,e^{2\Phi_0}\sim l_s^{24}g_s^2\label{kgs}\ee
The coefficient in front of Einstein-Hilbert term is usually identified with Newton's
constant
\be 8\pi G_N = \kappa^2 \nn\ee
Note, however, that this is Newton's constant in $D=26$ dimensions: it will differ from
Newton's constant measured in a four-dimensional world. From Newton's constant, we define
the $D=26$ Planck length $8\pi G_N =l_p^{24}$ and Planck mass $M_p=l^{-1}_p$. (With the factor
of $8\pi$ sitting there, this is usually called the reduced Planck mass). Comparing to
\eqn{kgs}, we see that weak string coupling, $g_s\ll 1$, provides a parameteric separation between
the Planck scale and the string scale,
\be g_s \ll 1 \ \ \ \Rightarrow\ \ \ l_p\ll l_s\nn\ee
Often the mysteries of gravitational physics are associated with the length scale $l_p$. We
understand string theory best when $g_s \ll 1$ where much of stringy physics occurs at $l_s \gg
l_p$ and can be disentangled from strong coupling effects in gravity.

\para
The original metric $G_{\mu\nu}$ is usually called the {\it string metric} or {\it
sigma-model metric}. It is the metric that strings see, as reflected in the
action \eqn{nlsm}. In contrast, $\tilde{G}_{\mu\nu}$ is called
the {\it Einstein metric}.  Of course, the two actions \eqn{sframe} and \eqn{eframe}
describe the same physics: we have simply chosen to package the fields in a different
way in each. The choice of metric --- $G_{\mu\nu}$ or $\tilde{G}_{\mu\nu}$ ---
is usually referred to as a choice of {\it frame}: string frame, or Einstein
frame.

\para
The possibility of defining two metrics really arises because we have a massless
scalar field ${\Phi}$ in the game. Whenever such a field exists, there's
nothing to stop us measuring distances in different ways by including $\Phi$ in our ruler.
Said another way, massless scalar fields give rise to long range attractive forces which can
mix with gravitational forces and violate the principle of equivalence.
Ultimately, if we want to connect to Nature, we need to find a way to make
$\Phi$ massive. Such mechanisms exist in the context of the superstring.

\subsubsection{Corrections to Einstein's Equations}

Now that we know how Einstein's equations arise from string theory, we
can start to try to understand new physics. For example, what
are the quantum corrections
to Einstein's equations?

\para
On general grounds, we expect these corrections
to kick in when the curvature $r_c$ of spacetime becomes comparable to the
string length scale $\sqrt{\ap}$.
But that dovetails very nicely with the discussion above where we saw that the
perturbative expansion
parameter for the non-linear sigma model is $\ap/r_c^2$. Computing the
next loop correction to the beta function will result in corrections to
Einstein's equations!

\para
If we ignore $H$ and $\Phi$ , the 2-loop sigma-model beta function
can be easily computed and results in the $\ap$ correction to Einstein's equations:
\be \beta_{\mu\nu} = \ap {\cal R}_{\mu\nu} +\frac{1}{2}\alpha^{\prime\,2} {\cal R}_{\mu\lambda\rho\sigma}{\cal R}_\nu^{\ \lambda\rho\sigma} + \ldots =0 \nn\ee
Such two loop corrections also appear in the heterotic superstring. However, they are
absent for the type II string theories, with the first corrections appearing
at 4-loops from the perspective of the sigma-model.

\subsubsection*{String Loop Corrections}

Perturbative string theory has an $\ap$ expansion and $g_s$ expansion. We still
have to discuss the latter. Here an interesting subtlety arises. The sigma-model
beta functions arise from regulating the UV divergences of the worldsheet. Yet
the $g_s$ expansion cares only about the topology of the string. How can the
UV divergences care about the global nature of the worldsheet. Or, equivalently,
how can the higher-loop corrections to the beta-functions
give anything interesting?

\para
The resolution to this puzzle is to remember that, when computing higher $g_s$ corrections,
we have to integrate over the moduli space of Riemann surfaces. But this moduli space
will include some tricky points where the Riemann surface degenerates. (For example, one
cycle of the torus may pinch off). At these points, the UV divergences suddenly do care
about global topology, and this results in the $g_s$ corrections to
the low-energy effective action.

\subsubsection{Nodding Once More to the Superstring}
\label{noddysec}

In section \ref{supernodsec}, we described the massless bosonic content for the four
superstring theories: Heterotic $SO(32)$, Heterotic $E_8\times E_8$, Type IIA and Type IIB.
Each of them contains the fields $G_{\mu\nu}$, $B_{\mu\nu}$ and $\Phi$ that  appear
in the bosonic string, together with a collection of further massless fields.
For each, the low-energy
effective action describes the dynamics of these fields in $D=10$ dimensional spacetime. It
naturally splits up into three pieces,
\be S_{\rm superstring} = S_{1} + S_{2} + S_{\rm fermi}
\nn\ee
Here $S_{\rm fermi}$ describes the interactions of the spacetime fermions. We won't describe
these here. But we will briefly describe the low-energy bosonic action $S_1 + S_{2}$ for
each of these four superstring theories.

\para
$S_1$ is essentially the same for all theories, and is given by
the action we found for the bosonic string in string frame \eqn{sframe}. We'll start to
use form notation, and denote $H_{\mu\nu\lambda}$ simply as $H_3$, where the subscript
tells us the degree of the form. Then the action reads
\be S_1  = \frac{1}{2\kappa_0^2}\int d^{10}X\sqrt{-G}\,e^{-2\Phi}\,\left(
{\cal R} - \frac{1}{2}|\tilde{H}_3|^2+4\partial_\mu\Phi\,\partial^\mu
\Phi\right)\label{s1}\ee
There is one small difference, which is that the field $\tilde{H}_3$ that appears
here for the heterotic string is not quite the same as the original $H_3$; we'll
explain this further shortly.

\para
The second part of the action, $S_2$, describes the dynamics of the extra fields which are
specific to each different theory. We'll now go through the four theories in turn, explaining
$S_2$ in each case.

\para
\begin{itemize}
\item {\bf Type IIA}: For this theory, $\tilde{H}_3$ appearing in \eqn{s1} is $H_3=dB_2$, just as we saw in the bosonic string. In Section \ref{supernodsec}, we described
the extra bosonic fields of the Type IIA theory: they consist of a 1-form $C_1$ and a 3-form $C_3$. The dynamics of these fields is
governed by the so-called Ramond-Ramond part of the action and is written in form notation as,
\be S_2 = -\frac{1}{4\kappa_0^2}\int d^{10}X\ \left[\sqrt{-G}\ \left(|F_2|^2 + |\tilde{F}_4|^2\right) + B_2\wedge F_4\wedge F_4\right]\nn\ee
Here the field strengths are given by $F_2=dC_1$ and $F_4=dC_3$, while the object that
appears in the
kinetic terms is $\tilde{F}_4=F_4-C_1\wedge H_3$. Notice that the final term in the action
does not depend on the metric: it is referred to as a {\it Chern-Simons} term.
\item {\bf Type IIB}: Again, $\tilde{H}_3\equiv H_3$. The extra bosonic fields are now a scalar
$C_0$, a 2-form $C_2$ and a 4-form $C_4$. Their action is given by
\be S_2 = -\frac{1}{4\kappa_0^2}\int d^{10}X\ \left[\sqrt{-G}\ \left(|F_1|^2 +|\tilde{F}_3|^2
+\frac{1}{2}|\tilde{F}_5|^2\right) + C_4\wedge H_3\wedge F_3\right]\nn\ee
where $F_1=dC_0$, $F_3=dC_2$ and $F_5=dC_4$. Once again, the kinetic terms involve more complicated combinations of the forms: they are $\tilde{F}_3=F_3-C_0\wedge H_3$ and $\tilde{F}_5=F_5-\ft12 C_2\wedge H_3 +\ft12 B_2\wedge F_3$. However, for type IIB string theory, there is one extra
requirement on these fields that cannot be implemented in any simple way in terms of a Lagrangian:
$\tilde{F}_5$ must be self-dual
\be \tilde{F}_5={}^\star \tilde{F}_5\nn\ee
Strictly speaking, one should say that the low-energy dynamics of type IIB theory is governed
by the equations of motion that we get from the action, supplemented with this self-duality
requirement.
\item {\bf Heterotic}: Both heterotic theories have just one further massless bosonic
ingredient: a non-Abelian gauge field strength $F_2$, with gauge group $SO(32)$ or $E_8\times E_8$.
The dynamics of this field is simply the Yang-Mills action in ten dimensions,
\be S_2 = \frac{\ap}{8\kappa_0^2} \int d^{10}X\ \sqrt{-G}\,\Tr\,|F_2|^2\nn\ee
The one remaining subtlety is to explain what $\tilde{H}_3$ means in \eqn{s1}: it is
defined as $\tilde{H}_3= dB_2 - \ap\omega_3/4$ where $\omega_3$ is the Chern-Simons
three form constructed from the non-Abelian gauge field $A_1$
\be \omega_3=\Tr\,\left(A_1\wedge dA_1 +\frac{2}{3} A_1\wedge A_1 \wedge A_1\right)\nn\ee
The presence of this strange looking combination of forms sitting in the kinetic terms is
tied up with one of the most intricate and interesting aspects of the heterotic
string, known as anomaly cancelation.
\end{itemize}
The actions that we have written down here probably look a little arbitrary. But they have
very important properties. In particular, the full action $S_{\rm superstring}$ of each of the
Type II theories is invariant under ${\cal N}=2$ spacetime supersymmetry. (That means
32 supercharges). They are
the unique actions with this property. Similarly, the heterotic superstring actions
are invariant under ${\cal N}=1$ supersymmetry and, crucially, do not suffer from anomalies.
The second book by Polchinski is a good place to start learning more about these ideas.

\subsection{Some Simple Solutions}

The spacetime equations of motion,
\be \beta_{\mu\nu}(G) = \beta_{\mu\nu}(B) = \beta(\Phi)=0\nn\ee
have many solutions. This is part of the story of vacuum selection in string theory. What
solution, if any, describes the world we see around us? Do we expect this putative solution
to have other special properties, or is it just a random choice from the many possibilities?
The answer is that we don't really know, but there is currently no known principle which
uniquely selects a solution which looks like our world --- with the gauge groups, matter content
and values of fundamental constants that we observe --- from the many other possibilities.
Of course, these questions should really be asked in the context of the superstring where a greater understanding of
various non-perturbative effects such as D-branes and fluxes leads to an even greater array
of possible solutions.

\para
Here we won't discuss these problems. Instead, we'll just discuss a few simple
solutions that are well known. The first plays a role when trying to make contact
with the real world, while the value of the others lies mostly in trying to better
understand the structure of string theory.

\subsubsection{Compactifications}

We've seen that the bosonic string likes to live in $D=26$ dimensions. But we don't. Or, more
precisely, we only observe three macroscopically large spatial dimensions. How do we reconcile
these statements?

\para
Since string theory is a theory of gravity, there's nothing to stop extra dimensions
of the universe from curling up. Indeed, under certain circumstances, this may be
required dynamically. Here was can exhibit some simple solutions of the low-energy effective
action which have this property. We set $H_{\mu\nu\rho}=0$ and $\Phi$ to a constant.
Then we are simply searching for
Ricci flat backgrounds obeying ${\cal R}_{\mu\nu}=0$. There are solutions where the metric
is a direct product of metrics on the space
\be {\bf R}^{1,3}\times {\bf X}\label{x}\ee
where ${\bf X}$ is a compact 22-dimensional Ricci-flat manifold.

\para
The simplest such manifold is just ${\bf X}= {\bf T}^{22}$, the torus endowed with a flat metric.
But there are a whole
host of other possibilities. Compact, complex manifolds that admit such Ricci-flat metrics are
called {\it Calabi-Yau} manifolds. (Strictly speaking, Calabi-Yau manifolds are complex manifolds with vanishing first Chern class. Yau's theorem guarantees the existence of a unique
Ricci flat metric on these spaces).

\para
The idea that there may be extra, compact directions in the universe was considered
long before string theory and goes by the name of {\it Kaluza-Klein compactification}. If the
characteristic length scale $L$ of the space ${\bf X}$ is small enough then the presence
of these extra dimensions would not have been observed in experiment.
The standard model of particle physics has been accurately tested to energies of a TeV or
so, meaning that if the standard model particles can roam around ${\bf X}$, then the
length scale must be $L \lesssim\mbox{(TeV)}^{-1} \sim 10^{-16}$ cm.

\para
However, one can cook up scenarios in which the standard model is stuck somewhere in these
extra dimensions (for example, it may be localized on a D-brane). Under these
circumstances, the constraints become  much weaker because we would
rely on gravitational experiments to detect extra dimensions. Present bounds
require only $L\lesssim 10^{-5}$ cm.

\para
Consider
the Einstein-Hilbert term in the low-energy effective action. If we are interested only
in the dynamics of the 4d metric on ${\bf R}^{1,3}$, this is given by
\be S_{EH} = \frac{1}{2\kappa^2} \int d^{26}X\sqrt{-\tilde{G}}\ \tilde{{\cal R}}
= \frac{{\rm Vol}({\bf X})}{2\kappa^2}\int d^4X\sqrt{-G_{4d}}\ {\cal R}_{4d}\nn\ee
(There are various moduli of the internal manifold ${\bf X}$ that are being
neglected here). From this equation, we learn that effective 4d Newton constant is
given in terms of $26d$ Newton constant by,
\be 8\pi G^{4d}_N = \frac{\kappa^2}{{\rm Vol}({\bf X})}\nn\ee
Rewriting this in terms of the 4d Planck scale, we have
$l_p^{(4d)} \sim g_s l_s^{12}/\sqrt{{\rm Vol}({\bf X})}$.
To trust this whole analysis, we require $g_s\ll 1$ and
all length scales of the internal space to be
bigger than $l_s$. This ensures that $l_p^{(4d)} < l_s$. Although the 4d Planck length
is ludicrously small, $l_p^{(4d)} \sim 10^{-33}$ cm, it may be that we don't have to probe
to this distance to uncover UV gravitational physics. The back-of-the-envelope calculation
above shows that the string scale $l_s$ could be much larger, enhanced by the volume of
extra dimensions.

%\para
%All present attempts to connect string theory to the real world start with this kind of
%Kaluza-Klein set-up. (Although it is possible to relax the direct product structure
%of the metric in \eqn{x} and consider warped compactifications instead).
%We'll look at some simple circle compactifications in Section \ref{tsec}.

\subsubsection{The String Itself}
\label{ssourcesec}

We've seen that quantizing small loops of string gives rise to the graviton and
$B_{\mu\nu}$ field. Yet, from the sigma model action \eqn{all}, we also know that
the string is charged under the $B_{\mu\nu}$. Moreover, the string has tension,
which ensures that it also acts as a source for the metric $G_{\mu\nu}$. So what
does the back-reaction of the string look like? Or, said another way: what is the
sigma-model describing a string moving in the background of another string?

\para
Consider an infinite, static, straight string stretched in the $X^1$ direction. We can solve
for the background fields by coupling the equations of motion to a delta-function string source. This is the same kind of calculation that we're used to in electromagnetism.
The resulting spacetime fields are given by
\be &ds^2 = f(r)^{-1}\,(-dt^2 + dX_1^2) + \sum_{i=1}^{25}dX_i^2&\nn\\
&B =(f(r)^{-1}-1)\,dt\wedge dX_1\ \ \ \ ,\ \ \ \ e^{2\Phi} = f(r)^{-1}&\label{stringsol}\ee
The function $f(r)$ depends only on the transverse direction $r^2 = \sum_{i=2}^{25}
X_i^2$ and is given by
\be f(r) = 1 +\frac{g_s^2N l_s^{22}}{r^{22}}\nn\ee
Here $N$ is some constant which we will shortly demonstrate counts the number
of strings which source the background. The string length scale in the solutions
is $l_s=\sqrt{\ap}$. The function $f(r)$ has the property that it
is harmonic in the space transverse to the string, meaning that it
satisfies $\nabla^2_{{\bf R}^{24}} f(r)=0$ except at $r=0$.

\para
Let's compute the $B$-field charge of this solution. We do exactly
what we do in electromagnetism: we integrate the total flux through a sphere which
surrounds the object. The string lies along the $X^1$ direction so the transverse
space is ${\bf R}^{24}$. We can consider a sphere ${\bf S}^{23}$ at the boundary
of this transverse space. We should be integrating the flux over this
sphere. But what is the expression for the flux?

\para
To see what we should do, let's look at the action for $H_{\mu\nu\rho}$ in the
presence of a string source. We will use form notation since this is much
cleaner and refer to $H_{\mu\nu\rho}$ simply as $H_3$. Schematically, the action takes the form
\be  \frac{1}{g_s^2}\int_{{\bf R}^{26}} H_3 \wedge{}^\star H_3 + \int_{{\bf R}^2}B_2
 = \frac{1}{g_s^2}\int_{{\bf R}^{26}} H_3\wedge{}^\star H_3 + g_s^2 B_2\wedge \delta(\omega)
\nn\ee
Here $\delta(\omega)$ is a delta-function source with support on the 2d worldsheet of
the string. The equation of motion is
\be d{}^\star H_3 \sim g_s^2 \delta(\omega)\nn\ee
From this we learn that to compute the charge of a single string we need to integrate
\be \frac{1}{g_s^2}\int_{{\bf S}^{23}}{}^\star H_3 = 1
\nn\ee
After these general comments,  we  now return to our solution \eqn{stringsol}. The
above discussion was schematic and no attention was paid to factors of 2 and $\pi$.
Keeping in this spirit, the flux of the solution \eqn{stringsol} can be checked to be
\be \frac{1}{g_s^2}\int_{{\bf S}^{23}}{}^\star H_3 = N\nn\ee
This is telling us that the solution \eqn{stringsol} describes the background
sourced by $N$ coincident, parallel fundamental strings. Another way to check this
is to compute the ADM mass per unit length of the solution: it is $NT\sim N/\ap$ as
expected.

\para
Note as far as the low-energy effective action is concerned, there is nothing that
insists $N \in {\bf Z}$. This is analogous to the statement that nothing in classical
Maxwell theory requires $e$ to be quantized. However, in string theory, as in QED, we
know the underlying sources of the microscopic theory and $N$ must indeed take
integer values.

\para
Finally, notice that as $r\rightarrow 0$, the solution becomes singular. It is not
to be trusted in this regime where higher order $\ap$ corrections become important.

\subsubsection{Magnetic Branes}

We've already seen that string theory is not just a theory of strings; there are
also D-branes, defined as surfaces on which strings can end. We'll have much more
to say about D-branes in Section \ref{dbranesec}. Here, we will consider a
third kind of object that exists in string theory.
It is again a brane -- meaning that it is extended in some number
of spacetime directions --- but it is not a D-brane because the open string cannot end there.
In these lectures we will call it the {\it magnetic brane}.

\subsubsection*{Electric and Magnetic Charges}

You're probably not used to talking about magnetically charged objects in electromagnetism. Indeed, in
undergraduate courses we usually don't get much further than pointing out that $\nabla\cdot B=0$
does not allow point-like magnetic charges. However, in the context of quantum field theory,
much of the interesting behaviour often boils down to understanding how magnetic charges
behave. And the same is true of string theory. Because this may be unfamiliar, let's take a
minute to discuss the basics.

\para
In electromagnetism in $d=3+1$ dimensions,
we measure electric charge $q$ by integrating the electric field $\vec{E}$ over
a sphere ${\bf S}^2$ that surrounds the particle,
\be q = \int_{{\bf S}^2} \vec{E}\cdot d\vec{S} = \int_{{\bf S}^2}{}^\star F_2\label{echarge}\ee
In the second equality we have introduced the notation of differential forms that we also used in the previous example to discuss the string solutions.

\para
Suppose now that a particle carries magnetic charge $g$. This can be measured by integrating
the magnetic field $\vec{B}$ over the same sphere. This means
\be g = \int_{{\bf S}^2}\vec{B}\cdot d\vec{S} = \int_{{\bf S}^2}F_2\label{mcharge}\ee
In $d=3+1$ dimensions, both electrically and magnetically charged objects are particles. But
this is not always true in any dimension! The reason that it holds in $4d$ is because both the
field strength $F_2$ and the dual field strength ${}^\star F_2$ are 2-forms. Clearly, this is
rather special to four dimensions.

\para
In general, suppose that we have a $p$-brane that is electrically charged under a suitable
gauge field. As we discussed in Section \ref{bsec}, a $(p+1)$-dimensional object naturally
couples to a $(p+1)$-form gauge potential $C_{p+1}$ through,
\be \mu \int_WC_{p+1}\nn\ee
where $\mu$ is the charge of the object, while $W$ is the worldvolume of the brane. The
$(p+1)$-form gauge potential has a $(p+2)$-form field strength
\be G_{p+2} = dC_{p+1}\nn\ee
To measure the electric charge of the $p$-brane, we need to integrate the field strength
over a sphere that completely surrounds the object. A $p$-brane in $D$-dimensions has a
transverse space ${\bf R}^{D-p-1}$. We can integrate the flux over the sphere at infinity,
which is ${\bf S}^{D-p-2}$. And, indeed, the counting works out nicely because, in $D$
dimensions, the dual field strength is a $(D-p-2)$-form, ${}^\star G_{p+2}=\tilde{G}_{D-p-2}$,
which we can happily integrate over the sphere to find the charge sitting inside,
\be q =\int_{{\bf S}^{D-p-2}}{}^\star G_{p+2}\nn\ee
This equation is the generalized version of \eqn{echarge}

\para
Now let's think about magnetic charges. The generalized version of \eqn{mcharge} suggest that
we should compute the magnetic charge by integrating $G_{p+2}$ over a sphere ${\bf S}^{p+2}$.
What kind of object sits inside this sphere to emit the magnetic charge? Doing the sums backwards,
we see that it should be a $(D-p-4)$-brane.

\para
We can write down the coupling between the $(D-p-4)$-brane and the field strength. To
do so, we first need to introduce the magnetic gauge potential defined by
\be {}^\star G_{p+2} = \tilde{G}_{D-p-2} = d\tilde{C}_{D-p-3}\label{dual}\ee
We can then add the magnetic coupling to the worldvolume $\tilde{W}$ of a $(D-p-4)$-brane simply
by writing
\be \tilde{\mu}\int_{\tilde{W}} \tilde{C}_{D-p-3}\nn\ee
where $\tilde{\mu}$ is the magnetic charge.
Note that it's typically not possible to write down a Lagrangian that includes both magnetically
charged object and electrically charged objects at the same time. This would need us to include
both $C_{p+1}$ and $\tilde{C}_{D-p-3}$ in the Lagrangian, but these are not independent fields:
they're related by the rather complicated differential equations \eqn{dual}.

\subsubsection*{The Magnetic Brane in Bosonic String Theory}

After these generalities, let's see what it means for the bosonic string. The fundamental
string is a $1$-brane and, as we saw in Section \ref{bsec}, carries electric charge
under the 2-form $B$. The appropriate object carrying magnetic charge under $B$ is
therefore a $(D-p-4)=(26-1-4)=21$-brane.

\para
To stress a point: neither the fundamental string, nor the magnetic 21-brane are D-branes.
They are not surfaces where strings can end. We are calling them {\it branes} only
because they are extended objects.

\para
The magnetic 21-brane of the bosonic string
can be found as a solution to the low-energy equations of motion. The solution can be
written in terms of the dual potential $\tilde{B}_{22}$ such that $d\tilde{B}_{22}= {}^\star
dB_2$. It is
\be ds^2 &=& \left(-dt^2 + \sum_{i=1}^{21} dX^2_i\right) + h(r) \left(dX_{22}^2+\ldots dX_{25}^2\right)\label{msol}\\ \tilde{B}_{22} &=& (1-h(r)^{-2})\,dt\wedge dX_1\wedge\ldots\wedge dX_{21}
\nn\\ e^{2\Phi} &=& h(r)\nn\ee
The function $h(r)$ depends only on the radial direction in ${\bf R}^4$ transverse to the
brane: $r=\sum_{i=22}^{25} X_i^2$. It is a harmonic function in ${\bf R}^4$, given by
\be h(r) = 1 + \frac{N\l_s^2}{r^2}\nn\ee
The role of this function in the metric \eqn{msol} is to warp the transverse ${\bf R}^4$ directions. Distances get larger as you approach
the brane and the origin, $r=0$, is at infinite distance.

\para
It can be checked that the solution carried $N$ units of magnetic charge and has tension
\be
T \sim \frac{N}{l_s^{22}}\,\frac{1}{g_s^2}
\nn\ee
Let's summarize how the tension of different objects scale in string theory. The powers of
$\ap = l_s^2$ are entirely fixed on dimensional grounds. (Recall that the tension is mass
per spatial volume, so the tension of a $p$-brane has $[T_p]=p+1$). More interesting
is the dependence on the string coupling $g_s$. The tension of the fundamental string
does not depend on $g_s$, while the magnetic brane scales as $1/g^2_s$. This kind of $1/g^2$
behaviour is typical of solitons in field theories.  The D-branes sit between the two: their
tension scales as $1/g_s$. Objects with this behaviour are somewhat rarer (although not unheard
of) in field theory.

\para
In the perturbative limit, $g_s\rightarrow 0$, both D-branes and magnetic branes are
heavy. The coupling of an object with tension $T$ to gravity is governed by $T\kappa^2$ where
the gravitational coupling scales as $\kappa\sim g_s^2$ \eqn{kgs}. This means that in the
weak coupling limit, the gravitational backreaction of the string and D-branes can be
neglected. However, the coupling of the magnetic brane to gravity is always of order one.

\subsubsection*{The Magnetic Brane in Superstring Theory}

Superstring theories also have a brane magnetically charged under $B$. It is a $(D-p-4)=(10-1-4)=5$-brane and is usually referred to as the NS5-brane.
The solution in the transverse ${\bf R^4}$ again takes the form \eqn{msol}.

\para
The NS5-brane exists in both type II and heterotic string. In many ways it is more mysterious
than D-branes and its low-energy effective dynamics is still poorly understood. It is closely
related to the 5-brane of M-theory.

\subsubsection{Moving Away from the Critical Dimension}
\label{ncsec}

The beta function equations provide a new view on the critical dimension
$D=26$ of the bosonic string. To see this, let's look more closely at the dilaton
beta function $\beta(\Phi)$
defined in \eqn{betas}: it takes the same form as the Weyl anomaly that
we discussed back in Section \ref{weylanom}. This means that if we consider
a string propagating in $D\neq 26$ then the Weyl anomaly simply arises as the leading
order term in the dilaton beta function. So let's relax the requirement of the
critical dimension. The equations
of motion arising from $\beta_{\mu\nu}(G)$ and $\beta_{\mu\nu}(B)$ are
unchanged, while the dilaton beta function equation becomes
\be \beta(\Phi)&=& \frac{D-26}{6}-\frac{\ap}{2}\nabla^2\Phi + \ap \nabla_\mu\Phi
\,\nabla^\mu\Phi -\frac{\ap}{24}H_{\mu\nu\lambda}H^{\mu\nu\lambda} = 0\label{betaphi}\ee
The low-energy effective action in string frame picks up an extra term which looks
like a run-away potential for $\Phi$,
\be S= \frac{1}{2\kappa_0^2}\int d^{26}X\sqrt{-G}\,e^{-2\Phi}\,\left(
{\cal R} - \frac{1}{12}H_{\mu\nu\lambda}H^{\mu\nu\lambda}+4\partial_\mu\Phi\,\partial^\mu
\Phi - \frac{2(D-26)}{3\ap}\right)\nn\ee
This sounds quite exciting. Can we really get string theory living in $D=4$ dimensions
so easily? Well, yes and no. Firstly, with this extra potential term, flat $D$-dimensional Minkowski
space no longer solves the equations of motion. This is in agreement with the analysis in
Section 2 where we showed that full Lorentz invariance was preserved only in $D=26$.

\para
Another, technical, problem with solving the string equations of motion this way is
that we're playing tree-level term off against a one-loop term. But if tree-level and
one-loop terms are comparable, then typically all higher loop contributions will be as well
and it is likely that we can't trust our analysis.

\subsubsection*{The Linear Dilaton CFT}

In fact, there is one simple solution to \eqn{betaphi} which we can trust. It
is the solution to
\be \p_\mu\Phi\,\p^\mu\Phi = \frac{26-D}{6\ap}\nn\ee
Recall that we're working in signature $(-,+,+,\ldots)$, meaning that $\Phi$ takes a
spacelike profile if $D<26$ and a timelike profile if $D>26$,
\be \Phi = \sqrt{\frac{26-D}{6\ap}}\,X^1\ \ \ \ \ D<26
\nn\\ \Phi = \sqrt{\frac{D-26}{6\ap}}\,X^0\ \ \ \ \ D>26\nn\ee
This gives a dilaton which is linear in one direction. This can be compared to the study
of the path integral for non-critical strings that we saw in \ref{polyncsec}. There are
two ways of seeing the same physics.

\para
The reason that we can trust this solution is that there is an exact CFT
underlying it which we can analyze to all orders in $\ap$. It's called,
for obvious reasons, the {\it linear dilaton CFT}. Let's now
look at this in more detail.

\para
Firstly, consider the worldsheet action associated to the dilaton coupling. For
now we'll consider an arbitrary dilaton profile $\Phi(X)$,
\be S_{\rm dilaton}=\frac{1}{4\pi}\int d^2\sigma\sqrt{g}\ \Phi(X)R^{(2)}\label{likeeh}\ee
Although this term vanishes on a flat worldsheet,
it nonetheless changes the stress-energy tensor $\tab$ because this is defined as
\be \tab = -4\pi\,\left.\ppp{S}{\gabi}\right|_{\gab=\delta_{\alpha\beta}}\nn\ee
The variation of \eqn{likeeh} is straightforward. Indeed, the term is akin to the Einstein-Hilbert term in general relativity but things are simpler in 2d because, for example
$R_{\alpha\beta}=\ft12\,g_{\alpha\beta}R$. We have
\be \delta(\sqrt{g}\gabi R_{\alpha\beta})=\sqrt{g}\gabi\,\delta R_{\alpha\beta}
=\sqrt{g}\,\nabla^\alpha v_\alpha\nn\ee
where
\be v_\alpha = \nabla^\beta(\delta \gab - g^{\gamma\delta}\nabla_\alpha\delta g_{\gamma\delta})
\nn\ee
Using this, the variation of the dilaton term in the action is given by
\be \delta S_{\rm dilaton} = \frac{1}{4\pi}\int d^2\sigma \sqrt{g}
\,\left(\nabla^\alpha\nabla^\beta \Phi - \nabla^2\Phi\,\gabi\right)\,\delta\gab\nn\ee
which, restricting to flat space $\gab=\delta_{\alpha\beta}$,
finally gives us the stress-energy tensor of a theory with dilaton coupling
\be T^{\rm dilaton}_{\alpha\beta} = -\p_\alpha \p_\beta\Phi + \p^2\Phi\,\delta_{\alpha\beta}
\nn\ee
Note that this stress tensor is not traceless. This is to be expected because, as we described
above, the dilaton coupling is not Weyl invariant at tree-level.
In complex coordinates, the stress tensor is
\be T^{\rm dilaton} = -\p^2\Phi\ \ \ \ ,\ \ \ \ \bar{T}^{\rm dilaton} = -
\pb^2\Phi\nn\ee

\subsubsection*{Linear Dilaton OPE}

The stress tensor above holds for any dilaton profile $\Phi(X)$. Let's now restrict to a
linear dilaton profile for a single scalar field $X$,
\be \Phi = QX\nn\ee
where $Q$ is some constant. We also include the standard kinetic terms for $D$ scalar fields,
of which $X$ is a chosen one, giving the
stress tensor
\be T=-\frac{1}{\ap}:\p X\,\p X: - Q\,\p^2X\nn\ee
It is a simple matter to compute the $TT$ OPE using the techniques described in Section 4.
We find,
\be T(z)\,T(w) = \frac{c/2}{(z-w)^4} + \frac{2T(w)}{(z-w)^2} + \frac{\p T(w)}{z-w}+\ldots
\nn\ee
where the central charge of the theory is given by
\be c=D+6\ap Q^2\nn\ee
Note that $Q^2$ can be positive or negative depending on the whether we have a timelike or
spacelike linear dilaton. In this way, we see explicitly how a linear dilaton gradient
can absorb central charge.

\subsubsection{The Elephant in the Room: The Tachyon}

We've been waxing lyrical about the details of solutions to the low-energy effective action,
all the while ignoring the most important, relevant field of them all: the tachyon. Since our
vacuum is unstable, this is a little like describing all the beautiful pictures we could paint
if only that damn paintbrush would balance, unaided, on its tip.

\para
Of course, the main reason for discussing these solutions is that they all carry directly
over to the superstring where the tachyon is absent. Nonetheless, it's interesting to ask
what happens if the tachyon is turned on. Its vertex operator is simply
\be V_{\rm tachyon} \sim \int d^2\sigma \sqrt{g}\,e^{ip\cdot X}\nn\ee
where $p^2 = 4/\ap$. Piecing together a general tachyon profile $V(X)$ from these Fourier modes,
and exponentiating, results in a potential on the worldsheet of the string
\be S_{\rm potential} = \int d^2\sigma \sqrt{g}\ \ap\,V(X)\nn\ee
This is a relevant operator for the worldsheet CFT. Whenever such a relevant operator
turns on, we should follow the RG flow to the infra-red until we land on another CFT.
The c-theorem tells us that $c_{IR}<c_{UV}$, but in string theory we always require
$c=26$. The deficit, at least initially, is soaked up by the dilaton in the manner
described above. The end point of the tachyon RG flow for the bosonic string is not understood.
It may be that there is no end point, and the bosonic string simply doesn't make sense once the
tachyon is turned on. Or perhaps we haven't yet understood the true ground state of
the bosonic string.

\subsection{D-Branes Revisited: Background Gauge Fields}
\label{dbranesec}

Understanding the constraints of conformal invariance on the closed string backgrounds
led us to Einstein's equations and the low-energy effective action in spacetime.
Now we would like to do the same for the open string.
We want to understand the restrictions that
consistency places on the dynamics of D-branes.

\para
We saw in Section \ref{open} that there are two types of massless modes that arise from
the quantization of an open string: scalars, corresponding to the fluctuation of the
D-brane, and a $U(1)$ gauge field. We will ignore the scalar fluctuations for now, but will
return to them later. We focus initially on the dynamics of a gauge field $A_a$, $a=0,\ldots, p$
living on a D$p$-brane

\para
The first question that we ask is: how does the end of the string react to a background
gauge field? To answer this, we need to look at the vertex operator associated to the
photon. It was given in \eqn{photonv}
\be V_{\rm photon} \sim \int_{\p {\cal M}}d\tau\ \zeta_a\,\p^\tau X^a\,e^{ip\cdot X} \nn\ee
which is Weyl invariant and primary only if $p^2=0$ and $p^a\zeta_a =0$. Exponentiating
this vertex operator, as described at the beginning of  Section \ref{background}, gives the
coupling of the open string to a general background gauge field $A_a(X)$,
\be S_{\rm end-point} = \int_{\p {\cal M}} d\tau \ A_a(X)\,\frac{dX^a}{d\tau}\nn\ee
But this is a very familiar coupling --- we've already mentioned it in \eqn{electric}. It is
telling us that the end of the string is charged under the background gauge field $A_a$ on the
brane.

\subsubsection{The Beta Function}

We can now perform the same type of beta function calculation that we saw for the closed string\footnote{We'll
be fairly explicit here, but if you want to see more details then the best place to look
is the original paper by Abouelsaood, Callan, Nappi and Yost, ``{\it Open Strings in
Background Gauge Fields}", Nucl. Phys. B280 (1987) 599.}. To do
this, it's useful to first use conformal invariance to map the open string worldsheet to
the Euclidean upper-half plane as we described in Section \ref{opencftsec}. The action
describing an open string propagating in flat space, with its ends subject to a background
gauge field on the D-brane splits up into two pieces
\be S = S_{\rm Neumann} + S_{\rm Dirichlet} \nn\ee
where $S_{\rm Neumann}$ describes the fluctuations parallel to the D$p$-brane, and
is given by
\be S_{\rm Neumann} = \frac{1}{4\pi\ap}\int_{\cal M}d^2\sigma\ \p^\alpha X^a\,\p_\alpha\, X^b\,
\delta_{ab}
+ i\int_{\p {\cal M}}d\tau\ A_a(X)\dot{X}^a\label{aboundary}\ee
Here $a,b= 0,\ldots,p$. The extra factor of $i$ arises because we are in Euclidean space.
Meanwhile, the fields transverse to the brane have Dirichlet boundary conditions and take
range $I=p+1,\ldots,D-1$. Their dynamics is given by
\be S_{\rm Dirichlet} = \frac{1}{4\pi\ap}\int_{\cal M}d^2\sigma\ \p^\alpha X^I\,\p_\alpha\, X^J\,\delta_{IJ}\nn\ee
The action $S_{\rm Dirichlet}$ describes free fields and doesn't play any role in
the computation of the beta-function. The interesting part is $S_{\rm Neumann}$ which,
for non-zero $A_a(X)$, is an interacting quantum field theory with boundary.
Our task is to compute the beta function associated to the coupling $A_a(X)$. We use the
same kind of technique that we earlier applied to the closed string. We expand the
fields $X^a(\sigma)$ as
\be X^a(\sigma) = \bar{x}^a(\sigma) + \sqrt{\ap}\,Y^a(\sigma)\nn\ee
where $\bar{x}^a(\sigma)$ is taken to be some fixed background which obeys the
classical equations of motion,
\be \p^2 \bar{x}^a = 0\nn\ee
(In the analogous calculation for the closed string we chose the special case of
$\bar{x}^a$ constant. Here we are more general). However, we also need to impose
boundary conditions for this classical solution. In the absence of the gauge field
$A_a$, we require Neumann boundary conditions $\p_\sigma X^a= 0$ at $\sigma=0$. However,
the presence of the gauge field changes this. Varying the full action \eqn{aboundary}
shows that the relevant boundary condition is supplemented by an extra term,
\be \p_\sigma \bar{x}^a +2\pi\ap i\,F^{ab}\,\p_\tau\bar{x}_b
=0\ \ \ \ \ {\rm at}\ \sigma=0\label{newnbc}\ee
where the $F_{ab}$ is the field strength
\be F_{ab}(X) = \ppp{A_b}{X^a} - \ppp{A_a}{X^b} \equiv
\p_a A_b - \p_b A_a \nn\ee
The fields $Y^a(\sigma)$ are the fluctuations which are taken to be small. Again, the
presence of $\sqrt{\ap}$ in the expansion ensures that $Y^a$ are dimensionless.
Expanding the action $S_{\rm Neumann}$ (which we'll just call $S$ from now on)
to second order in fluctuations gives,
\be S[\bar{x}+\sqrt{\ap}Y] &=&
S[\bar{x}] + \frac{1}{4\pi}\int_{\cal M}d^2\sigma\ \p Y^a
\,\p Y^b\delta_{ab}  \nn\\ &&\ \ \  \ \ \ +\ i\ap \int_{\p {\cal M}}d\tau\ \left(\p_a A_b\, Y^a\,
\dot{Y}^b + \frac{1}{2}\p_a\p_b A_c\,Y^a\,Y^b\,\dot{\bar{x}}^c\right)
+\ldots\nn\ee
where all expressions involving the background gauge fields are now evaluated on the
classical solution $\bar{x}$. We can rearrange the boundary terms by splitting the first
term up into two halves and integrating one of these pieces by parts,
\be \int d\tau\ (\p_a A_b)Y^a\dot{Y}^b = \frac{1}{2} \int d\tau\ \p_a A_b\,Y^a\,\dot{Y}^b
-\p_a A_b\,\dot{Y}^a Y^b - \p_c\p_a A_b\,
Y^a Y^b\dot{\bar{x}}^c\nn\ee
Combining this with the second term means that we can write all interactions in terms of the
gauge invariant field strength $F_{ab}$,
\be
S[\bar{x}+\sqrt{\ap}Y] &=&
S[\bar{x}] + \frac{1}{4\pi}\int_{\cal M}d^2\sigma\ \p Y^a
\,\p Y^b\delta_{ab}  \nn\\ &&\ \ \  \ \ \ +\ \frac{i\ap}{2}\int_{\p {\cal M}}
d\tau\ \left(F_{ab}\,Y^a\dot{Y}^b  + \p_b F_{ac}\,Y^a Y^b \dot{\bar{x}}^c
\right)+ \ldots \label{ftheory}\ee
where the $+\ldots$ refer to the higher terms in the expansion which come with higher
derivatives of $F_{ab}$, accompanied by powers of $\ap$. We can neglect them for
the purposes of computing the one-loop beta function.

\subsubsection*{The Propagator}

This Lagrangian describes our interacting boundary theory to leading order.
We can now use this to compute the beta function. Firstly, we should determine where
possible divergences arise. The offending term is the last one in \eqn{ftheory}. This
will lead to a divergence when the fluctuation fields $Y^a$ are contracted with
their propagator
\be \langle Y^a(z,\bz)Y^b(w,\bw)\rangle = G^{ab}(z,\bz;w,\bw)\nn\ee
We should be used to these free field Green's functions by now. The propagator satisfies
\be \p\bp\,G^{ab}(z,\bz) = -2\pi \delta^{ab}\delta(z,\bz)\label{subang}\ee
in the upper half plane. But now there's a subtlety. The $Y^a$ fields need to satisfy a
boundary condition at ${\rm Im}\,z=0$ and this should be reflected in the boundary
condition for the propagator. We discussed this briefly for Neumann boundary conditions
in Section \ref{opencftsec}. But we've also seen that the  background field strength
shifts the Neumann boundary conditions to \eqn{newnbc}. Correspondingly, the
propagator $G(z,\bz;w,\bw)$ must now satisfy
\be \p_\sigma G^{ab}(z,\bz;w,\bw) +2\pi\ap i\,F^{a}_{\ c}\,
\p_\tau G^{cb}(z,\bz;w,\bw)
=0\ \ \ \ \ {\rm at}\ \sigma=0\label{fprop}\ee
In Section \ref{opencftsec}, we showed how Neumann boundary conditions could be imposed
by considering an image charge in the lower half plane. A similar method works here. We
extend $G^{ab}\equiv G^{ab}(z,\bz;w,\bw)$ to the entire complex plane. The solution to \eqn{subang} subject to \eqn{fprop} is  given by
\be G^{ab} = -\delta^{ab}\ln|z-w|
-\frac{1}{2}\left(\frac{1-2\pi\ap F}{1+2\pi\ap F}\right)^{ab}\ln|z-\bw|
-\frac{1}{2}\left(\frac{1+2\pi\ap F}{1-2\pi\ap F}\right)^{ab}\ln|\bz-w|
\nn\ee

\subsubsection*{The Counterterm and Beta Function}

Let's now return to the interacting theory \eqn{ftheory} and see what counterterm
is needed to remove the divergence. Since all interactions take place
on the boundary, we should evaluate
our propagator on the boundary, which means $z=\bz$ and $w=\bw$. In this case, all
the logarithms become the same and, in the limit that $z\rightarrow w$, gives
the leading divergence $\ln|z-w|\rightarrow \epsilon^{-1}$. We learn that the UV
divergence takes the form,
\be -\frac{1}{\epsilon}\left[\delta^{ab}+\frac{1}{2}\left(\frac{1-2\pi\ap F}{1+2\pi\ap F}
\right)^{ab}
+\frac{1}{2}\left(\frac{1+2\pi\ap F}{1-2\pi\ap F}\right)^{ab}\right]
= -\frac{2}{\e}\left(\frac{1}{1-4\pi^2\alpha^{\prime\,2}F^2}\right)^{ab}
\nn\ee
It's now easy to determine the necessary counterterm. We simply
replace $Y^a Y^b$ in the final term with $\vev{Y^a Y^b}$. This yields
\be -\frac{i2\pi\alpha^{\prime\,2}}{\e}\int_{\p {\cal M}}d\tau\ \p_b F_{ac}
\,\left[\frac{1}{1-4\pi^2\alpha^{\prime\,2}\,F^2}\right]^{ab}\ \dot{\bar{x}}^c\nn\ee
For the open string theory to retain conformal invariance, we need the associated
beta function to vanish. This gives us the condition on the field strength $F_{ab}$: it must
satisfy the equation
\be \p_b F_{ac}\,\left[\frac{1}{1-4\pi^2\alpha^{\prime\,2}F^2}\right]^{ab}
=0\label{corblimey}\ee
This is our final equation governing the equations of motion that $F_{ab}$ must
satisfy to provide a consistent background for open string propagation.

\subsubsection{The Born-Infeld Action}

Equation \eqn{corblimey} probably doesn't look too familiar! Following the path
we took for the closed string, we wish to write down an action whose equations
of motion coincide with \eqn{corblimey}. The relevant action was actually
constructed many decades ago as a non-linear alternative to Maxwell theory: it goes
by the name of the {\it Born-Infeld action}:
\be S = -T_p\int d^{p+1}\xi\ \sqrt{-\det\left(\eta_{ab} + 2\pi\ap\,F_{ab}\right)}\label{bi}
\ee
Here $\xi$ are the worldvolume coordinates on the brane and $T_p$ is the tension of
the D$p$-brane (which, since it multiplies the action, doesn't affect the equations of motion).
The gauge potential is to be thought of as a function of the worldvolume coordinates:
$A_a = A_a(\xi)$. It actually takes a little work to show that the equations of
motion that we derive from this action coincide with the vanishing of the beta
function \eqn{corblimey}. Some hints on how to proceed are provided on Example Sheet 4.

\para
For small field strengths,
$F_{ab}\ll 1/\ap$, the action \eqn{bi} coincides with Maxwell's action. To see this, we need
simply expand to get
\be S=-T_p\int d^{p+1}\xi\ \left(1+\frac{(2\pi\ap)^2}{4}\,F_{ab}F^{ab}+\ldots\right)\nn\ee
The leading order term, quadratic in field strengths, is the
Maxwell action. Terms with higher powers of $F_{ab}$ are suppressed by powers of
$\ap$.

\para
So, for small field strengths, the dynamics of the gauge field on a D-brane is governed
by Maxwell's equations. However, as the electric and magnetic field strengths increase
and become of order $1/\ap$, non-linear corrections to the dynamics kick in and are
captured by the Born-Infeld action.

\para
The Born-Infeld action arises from the one-loop beta function. It is the exact result for constant field strengths. If
we want to understand the dynamics of gauge fields with large gradients, $\p F$, then
we will have determine the higher loop contributions to the beta function.

\subsection{The DBI Action}

We've understood that the dynamics of gauge fields on the brane is governed by the
Born-Infeld action. But what about the fluctuations of the brane itself. We looked at
this briefly in Section \ref{diracsec} and suggested, on general grounds, that
the action should take the Dirac form \eqn{dric}. It would be nice to show this
directly by considering the beta function equations for the scalar fields $\phi^I$ on the
brane. Turning these on corresponds to considering boundary conditions where the brane
is bent. It is indeed possible to compute something along the lines of beta-function equations
and to show directly
that the fluctuations of the brane are governed by the Dirac action\footnote{A readable discussion
of this calculation can be found in the original paper by Leigh,
{\it Dirac-Born-Infeld Action from Dirichlet Sigma Model}, Mod. Phys. Lett. A4: 2767
(1989).}.

\para
More generally, one could consider both the dynamics of the gauge field and the fluctuation
of the brane. This is governed by a mixture of the Dirac action and the Born-Infeld
action which is usually referred to as the {\it DBI action},
\be S_{DBI} = -T_p\int d^{p+1}\xi\ \sqrt{-\det(\gamma_{ab}+2\pi\ap\,F_{ab})}\nn
\ee
As in Section \eqn{diracsec}, $\gamma_{ab}$ is the pull-back of the the
spacetime metric onto the worldvolume,
\be \gamma_{ab} = \ppp{X^\mu}{\xi^a}\ppp{X^\nu}{\xi^b}\,\eta_{\mu\nu}\nn\ee
The new dynamical fields in this action are the embedding coordinates $X^\mu(\xi)$,
with $\mu=0,\ldots, D-1$. This appears to be $D$ new degrees of freedom while we expect
only $D-p-1$ transverse physical degrees of freedom. The resolution to this should be
familiar by now: the DBI action enjoys a reparameterization invariance which removes
the longitudinal fluctuations of the brane.

\para
We can use this reparameterization invariance to work in static gauge. For an infinite, flat
D$p$-brane, it is useful to set
\be X^a=\xi^a\ \ \ \ \ \ a=0,\ldots, p\nn\ee
so that the pull-back metric depends only on the transverse fluctuations $X^I$,
\be \gamma_{ab}=\eta_{ab}+\ppp{X^I}{\xi^a}\ppp{X^J}{\xi^b}\,\delta_{IJ}
\nn\ee
If we are interested in situations with small field strengths $F_{ab}$ and small derivatives
$\p_a X$, then we can expand the DBI action to leading order. We have
\be S = -(2\pi\ap)^2 T_p \int d^{p+1}\xi\ \left(\frac{1}{4}F_{ab}F^{ab} + \frac{1}{2}\p_a\phi^I
\p^a\phi^I + \ldots\right) \nn\ee
where we have rescaled the positions to define the scalar fields $\phi^I=X^I/2\pi\ap$. We have
also dropped an overall constant term in the action. This
is simply free Maxwell theory coupled to free massless scalar fields $\phi^I$. The higher order
terms that we have dropped are all suppressed by powers of $\ap$.

\subsubsection{Coupling to Closed String Fields}

The DBI action describes the low-energy dynamics of a D$p$-brane in flat space. We could
now ask how the motion of the D-brane is affected if it moves in a background created
by closed string modes $G_{\mu\nu}$, $B_{\mu\nu}$ and $\Phi$. Rather than derive this,
we'll simply write down the answer and then justify each term in turn. The answer is:
\be S_{DBI} = -T_p\int d^{p+1}\xi\ e^{-\tilde{\Phi}}\,\sqrt{-\det(\gamma_{ab} +
2\pi\ap F_{ab} + B_{ab})}\nn\ee
Let's start with the coupling to the background metric $G_{\mu\nu}$. It's actually hidden
in the notation in this expression: it appears in the pull-back metric $\gamma_{ab}$ which
is now given by
\be
\gamma_{ab} = \ppp{X^\mu}{\xi^a}\ppp{X^\nu}{\xi^b}\,G_{\mu\nu}\nn\ee
It should be clear that this is indeed the natural place for it to sit.

\para
Next up is the dilaton.
As in \eqn{tildephi}, we have decomposed the dilaton into a constant
piece and a varying piece: $\Phi=\Phi_0+\tilde{\Phi}$. The constant piece governs the
asymptotic string coupling, $g_s=e^{\Phi_0}$, and is implicitly sitting in front of the action
because the tension of the D-brane scales as
\be  T_p\sim 1/g_s
\nn\ee
This, then, explains the factor of $e^{-\tilde{\Phi}}$ in front of the action: it simply
reunites the varying part of the dilaton with the constant piece. Physically, it's
telling us that the tension of the D-brane depends on the local value of the dilaton field,
rather than its asymptotic value. If the dilaton varies,
the effective string coupling at a point $X$ in spacetime is given by $g_s^{eff} = e^{\Phi(X)}
= g_s\,e^{\tilde{\Phi}(X)}$. This, in turn, changes the tension of the D-brane. It can lower
its tension by moving to regions with larger $g_s^{eff}$.

\para
Finally, let's turn to the $B_{\mu\nu}$ field. This is a 2-form in spacetime. The function
$B_{ab}$ appearing in the DBI action is the pull-back to the worldvolume
\be B_{ab} = \ppp{X^\mu}{\xi^a}\ppp{X^\nu}{\xi^b}\,B_{\mu\nu}\nn\ee
Its appearance in the DBI action is actually required on grounds of gauge invariance alone.
This can be seen by considering an open string, moving in the presence of both a background $B_{\mu\nu}(X)$
in spacetime and a background $A_a(X)$ on the worldvolume of a brane. The relevant terms on the string
worldsheet are
\be \frac{1}{4\pi\ap}\int_{\cal M} d^2\sigma\ \epsilon^{\alpha\beta}\p_\alpha X^\mu\,\p_\beta X^\nu
 B_{\mu\nu} + \int_{\p {\cal M}} d\tau\ A_a \dot{X}^a \nn\ee
Under a spacetime gauge transformation
\be B_{\mu\nu} \rightarrow B_{\mu\nu} + \p_\mu C_\nu - \p_\nu C_\mu\label{bchange}\ee
the first term changes by a total derivative. This is fine for a closed string, but it doesn't leave
the action invariant for an open string because we pick up the boundary term. Let's quickly look
at what we get in more detail. Under the gauge transformation \eqn{bchange}, we have
\be S_B&=&\frac{1}{4\pi\ap}\int_{\cal M} d^2\sigma\ \epsilon^{\alpha\beta}\p_\alpha X^\mu\,\p_\beta X^\nu
 B_{\mu\nu} \nn\\ &\longrightarrow\ & S_B + \frac{1}{2\pi\ap}\int_{\cal M}d\sigma d\tau\ \epsilon^{\alpha\beta}
 \partial_\alpha X^\mu\,\p_\beta X^\nu\,\partial_\mu C_\nu
  \nn \\ &=&S_B + \frac{1}{2\pi\ap}\int_{\cal M}d\sigma d\tau\ \epsilon^{\alpha\beta}
 \partial_\alpha\,\left(\partial_\beta X^\nu C_\nu\right) \nn\\ &=&
 S_B+\frac{1}{2\pi\ap}\int_{\p{\cal M}} d\tau \dot{X}^\nu C_\nu = S_B+\frac{1}{2\pi\ap}\int_{\p{\cal M}} d\tau \dot{X}^a C_a
\nn\ee
where, in the last line, we have replaced the sum over all directions $X^\nu$ with the sum over
those directions obeying Neumann boundary conditions $X^a$, since $\dot{X}^I=0$ at the end-points
for any directions with Dirichlet boundary conditions.

\para
The result of this short calculation is to see that the string action is not invariant
under \eqn{bchange}. To restore this
spacetime gauge invariance, this boundary contribution must be canceled by an appropriate
shift of $A_a$ in the second term,
\be A_a \rightarrow A_a - \frac{1}{2\pi\ap}\,C_a\label{newgt}\ee
Note that this is not the usual kind of gauge transformation that we consider in electrodynamics.
In particular, the field strength $F_{ab}$ is not invariant. Rather,
the gauge invariant combination under \eqn{bchange} and \eqn{newgt} is
\be B_{ab} + 2\pi \ap F_{ab}\nn\ee
This is the reason that this combination must appear in the DBI action. This is also related
to an important physical effect. We have already seen that the string in spacetime is charged
under $B_{\mu\nu}$. But we've also seen that the end of the string is charged under the gauge
field $A_a$ on the D-brane. This means that the open string deposits  $B$ charge
on the brane, where it is converted into $A$ charge.
The fact that the gauge invariant field strength involves a
combination of both $F_{ab}$ and $B_{ab}$ is related to this interplay of charges.

\subsection{The Yang-Mills Action}
\label{ymsec}

Finally, let's consider the case of $N$ coincident D-branes. We discussed this
in Section \ref{gluesec} where we showed that the massless fields on the brane
could be naturally packaged as $N\times N$ Hermitian matrices, with the element of the matrix
telling us which brane the end points terminate on. The gauge field then takes
the form
\be (A_a)^m_{\ n}\nn\ee
with $a=0,\ldots p$ and $m,n=1,\ldots N$. Written this way, it looks rather like a
$U(N)$ gauge connection. Indeed, this is the correct interpretation. But how do we see
this? Why is the gauge field describing a $U(N)$ gauge symmetry rather than, say, $U(1)^{N^2}$?

\para
The quickest way to see that coincident branes give rise to a $U(N)$ gauge symmetry is
to recall that the end point of the string is charged under the $U(1)$ gauge field
that inhabits the brane it's ending on. Let's illustrate this with the simplest example.
Suppose that we have two branes. The diagonal
components $(A_a)^1_{\ 1}$ and $(A_a)^2_{\ 2}$ arise from strings which begin and end
on the same brane. Each is a $U(1)$ gauge field. What about the off-diagonal terms
$(A_a)^1_{\ 2}$ and $(A_a)^2_{\ 1}$? These come from strings stretched between the two branes.
They are again massless gauge bosons, but they are charged under the two original $U(1)$ symmetries;
they carry charge $(+1,-1)$ and $(-1,+1)$ respectively. But this is precisely the structure of a
$U(2)$ gauge theory, with the off-diagonal terms playing a role similar to W-bosons. In fact, the
only way to make sense of massless, charged spin 1 particles is through non-Abelian gauge symmetry.

\para
So the massless excitations of $N$ coincident branes are a $U(N)$ gauge field $(A_a)^m_{\ n}$,
together with scalars $(\phi^I)^m_{\ n}$ which transform in the adjoint representation of the
$U(N)$ gauge group. We saw in Section 3 that the diagonal components $(\phi^I)^m_{\ m}$ have
the interpretation of the transverse fluctuations of the $m^{\rm th}$ brane. Can we now
write down an action describing the interactions of these fields?

\para
In fact, there are several subtleties in writing down a non-Abelian generalization of the
DBI action and  such an action is not known (if, indeed, it makes sense at all).
However, we can make progress
by considering the low-energy limit, corresponding to small field strengths. The field strength
in questions is now the appropriate non-Abelian expression which, neglecting the matrix indices,
reads
\be F_{ab} = \p_a A_b -\p_b A_a + i[A_a,A_b] \nn\ee
The low-energy action describing the dynamics of $N$ coincident D$p$-branes can be shown to be
(neglecting an overall constant term),
\be S = -(2\pi\ap)^2 T_p \int d^{p+1}\xi\ \Tr\left( \frac{1}{4}F_{ab}F^{ab} + \frac{1}{2}{\cal D}_a\phi^I{\cal D}^a\phi^I
-\frac{1}{4}\sum_{I\neq J} [\phi^I,\phi^J]^2\right)\label{ymaction}\ee
We recognize the first term as the $U(N)$ Yang-Mills action. The coefficient in front of the
Yang-Mills action is the coupling constant $1/g_{YM}^2$. For a D$p$-brane, this is given by $\alpha^{\prime\,2}T_p$, or
\be g^2_{YM} \sim l_s^{p-3} g_s\nn\ee
The kinetic term for $\phi^I$ simply
reflects the fact that these fields transform in the adjoint representation of the gauge group,
\be {\cal D}_a\phi^I=\p_a\phi^I+i[A_a,\phi^I]\nn\ee
We won't derive this action in these lectures: the first two terms basically follow from
gauge invariance alone. The potential term is harder to see directly: the quick ways to derive it
use T-duality or, in the case of the superstring,
supersymmetry.

\para
A flat, infinite D$p$-brane breaks the Lorentz group of spacetime to
\be S(1,D-1)\ \rightarrow SO(1,p) \times SO(D-p-1)\label{sod}\ee
This unbroken group descends to the worldvolume of the D-brane where it classifies
all low-energy excitations of the D-brane. The $SO(1,p)$ is simply the Lorentz group
of the D-brane worldvolume. The $SO(D-p-1)$ is a global symmetry of the D-brane
theory, rotating the scalar fields $\phi^I$.

\para
The potential term in \eqn{ymaction} is particularly interesting,
\be V = -\frac{1}{4}\sum_{I\neq J} \Tr\ [\phi^I,\phi^J]^2\nn\ee
The potential is positive semi-definite. We can look at the fields that can be turned on at
no cost of energy, $V=0$. This requires that all $\phi^I$ commute which means that,
after a suitable gauge transformation, they take the diagonal form,
\be \phi^I=\left(\begin{array}{lcccr} \phi_1^I & & & \\ & & \ddots&  & \\ & & & \phi_N^I\end{array}\right)\label{position}\ee
The diagonal component $\phi^I_n$ describes the position of the $n^{\rm th}$ brane
in transverse space ${\bf R}^{D-p-1}$. We still need to get the dimensions right. The scalar fields
have dimension $[\phi]=1$. The relationship to the position in space (which we mentioned
before in \ref{diracsec}) is
\be \vec{X}_n = 2\pi\ap \vec{\phi}_n
\label{convert}\ee
where we've swapped to vector notation to replace the $I$ index.

\para
The eigenvalues $\phi_n^I$ are not quite gauge invariant: there is a residual gauge
symmetry --- the Weyl group of $U(N)$ --- which leaves $\phi^I$ in the form \eqn{position}
but permutes the entries by $S_N$, the permutation group of $N$ elements. But this has a very
natural interpretation: it is simply telling us that the D-branes are indistinguishable objects.

\para
\EPSFIGURE{2brane.eps,height=100pt}{}
\noindent
When all branes are separated, the vacuum expectation value \eqn{position}
breaks the gauge group from $U(N)\rightarrow U(1)^N$. The W-bosons gain a mass $M_W$ through the
Higgs mechanism. Let's compute this mass. We'll consider a $U(2)$ theory and we'll separate the
two D-branes in the direction $X^D\equiv X$. This means that we turn on a vacuum expectation
value for $\phi^D=\phi$, which we write as
\be \phi = \left(\begin{array}{cc} \phi_1 & 0 \\ 0 & \phi_2 \end{array}\right)\label{lastone}\ee
The values of $\phi_1$ and $\phi_2$ are the positions of the first and second brane. Or, more
precisely, we need to multiply by the conversion factor $2\pi\ap$ as in \eqn{convert}
to get the position $X_m$ of
the $m=1^{\rm st},2^{\rm nd}$ brane,

\para
Let's compute the mass of the W-boson from the Yang-Mills action \eqn{ymaction}. It comes from
the covariant derivative terms ${\cal D}\phi$. We expand out the gauge field as
\be A_a = \left(\begin{array}{cc} A_a^{11} & W_a \\ W_a^\dagger & A_a^{22} \end{array}\right)\nn\ee
with $A^{11}$ and $A^{22}$ describing the two $U(1)$ gauge fields, and $W$ the W-boson.
The mass of the W-boson comes from the $[A_a,\phi]$ term inside the covariant derivative
which, using the expectation value \eqn{lastone}, is given by
\be \frac{1}{2}\Tr\,[A_a,\phi]^2 =  -(\phi_2-\phi_1)^2|W_a|^2\nn\ee
This gives us the mass of the W-boson: it is
\be M_W^2 = (\phi_2-\phi_1)^2 = T^2|X_2-X_1|^2\nn\ee
where $T=1/2\pi\ap$ is the tension of the string. But this has a very natural interpretation. It is
precisely the mass of a string stretched between the two D-branes as shown in the figure above.
We see that D-branes provide a natural geometric interpretation of the Higgs mechanism using
adjoint scalars.

\para
Notice that when branes are well separated, and the strings that stretch between them are heavy,
their positions are described by the diagonal elements of the matrix given in
\eqn{position}. However, as the branes
come closer  together, these stretched strings become light and are important for
the dynamics of the branes. Now the positions of the branes
should be described by the full $N\times N$ matrices, including the off-diagonal elements.
In this manner, D-branes begin to see space as something non-commutative at short distances.

\para
In general, we can consider $N$ D-branes located at positions $\vec{X}_m$, $m=1,\ldots,N$ in
transverse space.
The string stretched between the $m^{\rm th}$ and $n^{\rm th}$ brane has mass
\be M_W = |\vec{\phi}_n-\vec{\phi}_m| = T|\vec{X}_n-\vec{X}_m|\nn\ee
which again coincides with the mass of the appropriate W-boson computed using \eqn{ymaction}.

\subsubsection{D-Branes in Type II Superstring Theories}

As we mentioned previously, D-branes are ingredients of the Type II superstring theories.
Type IIA has D$p$-branes with $p$ even, while Type IIB is home to D$p$-branes with $p$ odd.
The D-branes have a very important property in these theories: they preserve half the supersymmetries.

\para
Let's take a moment to explain what this means. We'll start by returning to the Lorentz group
$SO(1,D-1)$ now, of course, with $D=10$. We've already seen that an infinite, flat D$p$-brane
is not invariant under the full Lorentz group, but only the subgroup \eqn{sod}. If
we act with either $SO(1,p)$ or $SO(D-p-1)$ then the D-brane solution remains invariant. We say
that these symmetries are preserved by the solution.

\para
However, the role of the preserved symmetries
doesn't stop there. The next step is to consider small excitations of the D-brane. These must fit
into representations of the preserved symmetry group \eqn{sod}. This
ensures that the low-energy dynamics of the D-brane must be governed by a theory which is invariant
under \eqn{sod} and we have indeed seen that the Lagrangian \eqn{ymaction} has $SO(1,p)$ as a
Lorentz group and $SO(D-p-1)$ as a global symmetry group which rotates the scalar fields.

\para
Now let's return to supersymmetry. The Type II string theories enjoy a lot of supersymmetry:  32 supercharges in total. The infinite, flat D-branes are invariant under half of these;
if we act with one half of the supersymmetry generators, the D-brane solutions don't change. Objects
that have this property are often referred to as {\it BPS}\, states.
Just as with the Lorentz group, these unbroken symmetries descend to the worldvolume of the D-brane.
This means that the low-energy dynamics of the D-branes is described by a theory which is itself
invariant under 16 supersymmetries.

\para
There is a unique class of theories with 16 supersymmetries and a non-Abelian gauge field and matter
in the adjoint representation. This class is known as maximally supersymmetric Yang-Mills theory and the bosonic part of the action is given by \eqn{ymaction}. Supersymmetry is realized only after the addition of fermionic fields which also live on the brane.
These theories describe the low-energy dynamics of multiple D-branes.

\para
As an illustrative example, consider D3-branes in the Type IIB theory. The theory describing
$N$ D-branes is $U(N)$ Yang-Mills with 16
supercharges, usually referred to as $U(N)$ ${\cal N}=4$ super-Yang-Mills. The bosonic part of
the action is given by \eqn{ymaction}, where there are $D-p-1=6$ scalar fields $\phi^I$ in
the adjoint representation of the gauge group. These are augmented with four Weyl fermions, also
in the adjoint representation.

\newpage
\section{Compactification and T-Duality}
\label{tsec}

In this section, we will consider the simplest compactification of the bosonic string:
a background spacetime of the form
\be
{\bf R}^{1,24}\times {\bf S}^1\label{spacer}\ee
The circle is taken to have radius $R$, so that the coordinate on ${\bf S}^1$ has
periodicity
\be X^{25} \equiv X^{25}+2\pi R\nn\ee
We will initially be interested in the physics at length scales $\gg R$ where motion
on the ${\bf S}^1$ can be ignored. Our goal is to understand what physics looks
like to an observer living in the non-compact ${\bf R}^{1,24}$ Minkowski space.
This general idea goes
by the name of {\it Kaluza-Klein compactification}.  We will view this compactification
in two ways:
firstly from the perspective of the spacetime
low-energy effective action, and secondly from the perspective of the string worldsheet.

\subsection{The View from Spacetime}

Let's start with the low-energy effective action. Looking at length scales
$\gg R$ means that we will take all fields to be independent of
$X^{25}$: they are instead functions only on the non-compact ${\bf R}^{1,24}$.

\para
Consider the metric in Einstein frame. This decomposes into three different fields on
${\bf R}^{24,1}$: a metric
$\tilde{G}_{\mu\nu}$,  a vector $A_\mu$ and a scalar $\sigma$ which we package into the $D=26$ dimensional
metric as
\be
ds^2 = \tilde{G}_{\mu\nu}\,dX^\mu\,dX^\nu + e^{2\sigma}\left(dX^{25}+A_{\mu}\, dX^\mu\right)^2\label{kkmetric}\ee
Here all the indices run over the non-compact directions $\mu,\nu=0,\ldots 24$ only.

\para
The vector
field $A_\mu$ is an honest gauge field, with the gauge symmetry descending from
diffeomorphisms in $D=26$ dimensions. To see this recall that under the transformation
$\delta X^\mu = V^\mu(X)$, the metric transforms as
\be \delta G_{\mu\nu} = \nabla_\mu \Lambda_\nu + \nabla_\nu \Lambda_\mu\nn\ee
This means that diffeomorphisms of the compact direction, $\delta X^{25} = \Lambda(X^\mu)$,
turn into gauge transformations of $A_\mu$,
\be \delta A_\mu = \partial_\mu \Lambda\nn\ee
We'd like to know how the fields $G_{\mu\nu}$, $A_\mu$ and $\sigma$ interact. To determine
this, we simply insert the ansatz \eqn{kkmetric} into the $D=26$ Einstein-Hilbert action.
The $D=26$ Ricci scalar ${\cal R}  ^{(26)}$ is given by
\be {\cal R}^{(26)} = {\cal R} -2e^{-\sigma}\nabla^2e^\sigma -\frac{1}{4}e^{2\sigma}F_{\mu\nu}
F^{\mu\nu}\nn\ee
where ${\cal R}$ in this formula now refers to the $D=25$ Ricci scalar. The action governing
the dynamics becomes
\be S= \frac{1}{2\kappa^2}\int d^{26}X\sqrt{-\tilde{G}^{(26)}}\ {\cal R}^{(26)}
= \frac{2\pi R}{2\kappa^2}\int d^{25}X\sqrt{-\tilde{G}}\ e^\sigma\left({\cal R}
-\frac{1}{4}e^{2\sigma}F_{\mu\nu}F^{\mu\nu} +\p_\mu \sigma \p^\mu \sigma\right)
\nn\ee
The dimensional reduction of Einstein gravity in $D$ dimensions gives
Einstein gravity in $D-1$ dimensions, coupled to a $U(1)$ gauge theory and a single
massless scalar. This illustrates the original idea of Kaluza and Klein, with Maxwell
theory arising naturally from higher-dimensional gravity.

\para
The gravitational action above is not quite of the Einstein-Hilbert form. We need to again
change frames, absorbing the scalar $\sigma$ in the same manner as we absorbed the dilaton
in Section \ref{seframesec}. Moreover, just as for the dilaton, there is no
potential dictating the vacuum expectation value of $\sigma$. Changing the vev of $\sigma$
corresponds to changing $R$, so this is
telling us
that nothing in the gravitational action fixes the radius $R$ of the compact circle.
This is a problem common to all
Kaluza-Klein compactifications\footnote{The description of compactification on more
general manifolds is a beautiful story involving aspects differential geometry and topology.
This story is told in the second volume of Green, Schwarz and Witten.}:
there are always massless scalar fields, corresponding
to the volume of the internal space as well as other deformations. Massless scalar fields,
such as the dilaton $\Phi$ or the volume $\sigma$, are usually referred to as {\it moduli}.

\para
If we want this type of Kaluza-Klein compactification
to describe our universe  --- where we don't see massless scalar fields --- we need to find a way to ``fix the moduli". This means that we need a mechanism which gives rise to a potential for the scalar fields, making them heavy and dynamically fixing
their vacuum expectation value. Such mechanisms exist in the context of the superstring.

\para
Let's now also look at the Kaluza-Klein reduction of the other fields in the low-energy
effective action. The dilaton is easy: a scalar in $D$ dimensions reduces to a scalar
in $D-1$ dimensions. The anti-symmetric 2-form has more structure: it reduces to a
2-form $B_{\mu\nu}$, together with a vector field $\tilde{A}_\mu = B_{\mu\,25}$.

\para
In summary, the low-energy physics of the bosonic string
in $D-1$ dimensions consists of a metric $G_{\mu\nu}$,
two $U(1)$ gauge fields $A_\mu$ and $\tilde{A}_\mu$, and two massless scalars $\Phi$
and $\sigma$.

\subsubsection{Moving around the Circle}
\label{kkmodesec}

In the above discussion, we assumed that all fields are independent of the periodic
direction $X^{25}$. Let's now look at what happens if we relax this constraint. It's
simplest to see the resulting physics if we look at the scalar field $\Phi$ where we
don't have to worry about cluttering equations with indices. In general, we can
expand this field in  Fourier modes around the circle
\be \Phi(X^\mu;X^{25}) = \sum_{n=-\infty}^\infty\,\Phi_n(X^\mu)e^{inX^{25}/R}\nn\ee
where reality requires $\Phi^\star_n=\Phi_{-n}$. Ignoring the coupling to gravity
for now, the kinetic terms for this scalar are
\be \int d^{26}X \ \p_\mu\Phi\, \p^\mu \Phi + (\p_{25}\Phi)^2
= 2\pi R \int d^{25}X\ \sum_{n=-\infty}^\infty \left(\p_\mu \Phi_n\,\p^\mu \Phi_{-n}
+\frac{n^2}{R^2}\,|\,\Phi_n|^2\right)\nn\ee
This simple Fourier decomposition is telling us something very important: a single
scalar field on ${\bf R}^{1,D-1}\times {\bf S}^1$ splits into an infinite number of
scalar fields on ${\bf R}^{1,D-2}$, indexed by the integer $n$. These have mass
\be M^2_n = \frac{n^2}{R^2}\label{kkmass}\ee
For $R$ small, all particles are heavy except for the massless zero mode $n=0$. The heavy
particles are typically called Kaluza-Klein (KK) modes and can be ignored if we're probing
energies $\ll 1/R$ or, equivalently, distance scales $\gg R$.

\para
There is one further interesting property of the KK modes $\Phi_n$ with $n\neq 0$: they
are charged under the gauge field $A_\mu$ arising from the metric. The simplest
way to see this is to look at the appropriate gauge transformation which, from the
spacetime perspective, is the diffeomorphism $X^{25} \rightarrow X^{25} + \Lambda(X^\mu)$.
Clearly, this shifts the KK modes
\be \Phi_n\rightarrow \exp\left(\frac{in\Lambda}{R}\right)\Phi_n\nn\ee
This tells us that the $n^{\rm th}$ KK mode has charge $n/R$. In fact, one usually
rescales the gauge field to $A^\prime_\mu = A_\mu/R$, under which the charge of the KK
mode $\Phi_n$  is simply $n\in {\bf Z}$.

\subsection{The View from the Worldsheet}

We now consider the Kaluza-Klein reduction from the perspective of the string. We
want to study a string moving in the background ${\bf R}^{1,24}\times {\bf S}^1$.
There are two ways in which the compact circle changes the string dynamics.

\para
The first effect of the circle is that the spatial momentum, $p$, of the string in
the circle direction  can no longer take any value, but is quantized
in integer units
\be p^{25} = \frac{n}{R}\ \ \ \ \ n\in{\bf Z}\nn\ee
The simplest way to see this is simply to require that the string wavefunction, which
includes the factor $e^{ip\cdot X}$, is single valued.

\para
The second effect is that we can allow more general boundary conditions for the
mode expansion of $X$. As we move around the string, we no longer need $X(\sigma+2\pi) =
X(\sigma)$, but can relax this to
\be X^{25}(\sigma+ 2\pi) = X^{25}(\sigma)+2\pi m R\ \ \ \ \ \ \ m\in {\bf Z}\nn\ee
The integer $m$ tells us how many times the string winds around ${\bf S}^1$. It is usually simply
called the {\it winding number}.

\para
Let's now follow the familiar path that we described in Section 2 to study the spectrum
of the string on the spacetime \eqn{spacer}. We start by considering only the periodic field
$X^{25}$, highlighting the differences with our previous treatment.
The mode expansion of $X^{25}$ is now given by
\be X^{25}(\sigma,\tau) = x^{25} + \frac{\ap n}{R}\tau + mR\sigma + \mbox{oscillator modes}\nn\ee
which incorporates both the quantized momentum and the possibility of a winding number.
Before splitting $X^{25}(\sigma,\tau)$ into right-moving and left-moving parts, it will be useful
to introduce the quantities
\be p_L=\frac{n}{R}+\frac{mR}{\ap}\ \ \ \ ,\ \ \ \ p_R=\frac{n}{R}-\frac{mR}{\ap}\label{plpr}\ee
Then we have $X^{25}(\sigma,\tau)=X^{25}_L(\sigma^+)+X^{25}_R(\sigma^-)$, where
\be
X^{25}_L(\sigma^+) &=& \ft12 x^{25} + \ft12 \ap p_L \,\sigma^+ + i\sfap\sum_{n\neq 0}
\frac{1}{n}\,\tilde{\alpha}^{25}_n\, e^{-in\sigma^+} \ ,\nn\\
X^{25}_R(\sigma^-) &=& \ft12 x^{25} + \ft12 \ap p_R \,\sigma^- + i\sfap\sum_{n\neq 0}
\frac{1}{n}\,{\alpha}^{25}_n\, e^{-in\sigma^-} \nn\ee
This differs from the mode expansion \eqn{mode} only in the terms $p_L$ and $p_R$.
The
mode expansion for all the other scalar fields on flat space ${\bf R}^{1,24}$ remains unchanged
and we don't write them explicitly.

\para
Let's think about what the spectrum of this theory looks like to an observer living
in $D=25$ non-compact directions. Each particle state will be described by a
momentum $p^\mu$ with $\mu=0,\ldots 24$. The mass of the particle is
\be M^2 = -\sum_{\mu=0}^{24}p_\mu p^\mu\nn\ee
As before, the mass of these particles is fixed in terms of the oscillator modes of the
string by the $L_0$ and $\tilde{L}_0$ equations. These now read
\be M^2 = p_L^2 + \frac{4}{\ap}(\tilde{N}-1) = p_R^2 +\frac{4}{\ap}(N-1)\nn\ee
where $N$ and $\tilde{N}$ are the levels, defined in lightcone quantization by \eqn{nntilde}.
(One should take the lightcone coordinate inside ${\bf R}^{1,24}$
rather than along the ${\bf S}^1$). The factors of $-1$ are the necessary normal ordering
coefficients that we've seen in several guises in this course.

\para
These equations differ from \eqn{mlc} by the presence
of the momentum and winding terms around ${\bf S}^1$ on the right-hand side. In particular,
level matching no longer tells us that $N=\tilde{N}$, but instead
\be N-\tilde{N}=nm\label{newlm}\ee
Expanding out the mass formula, we have
\be M^2 = \frac{n^2}{R^2} + \frac{m^2 R^2}{\alpha^{\prime\,2}} + \frac{2}{\ap}(N+\tilde{N}-2)
\label{massr}\ee
The new terms in this formula have a simple interpretation. The first term tells
us that a string with $n>0$ units of momentum around the circle gains a contribution to its
mass of  $n/R$. This agrees with the result \eqn{kkmass} that we found from studying
the KK reduction of the spacetime theory. The second term is even easier to understand:
a string which winds $m>0$ times
around the circle picks up a contribution $2\pi mRT$ to its mass, where $T=1/2\pi\ap$ is
the tension of the string.

\subsubsection{Massless States}

We now restrict attention to the massless states in ${\bf R}^{1,24}$. This can
be achieved in the mass formula \eqn{massr} by looking at states with zero momentum
$n=0$ and zero winding $m=0$, obeying the level matching condition $N=\tilde{N}=1$.
The possibilities are
\begin{itemize}
\item $\alpha^\mu_{-1}\tilde{\alpha}^{\nu}_{-1}\ket{0;p}$:
Under the $SO(1,24)$ Lorentz
group, these states decompose into a
metric $G_{\mu\nu}$, an anti-symmetric tensor $B_{\mu\nu}$ and a scalar $\Phi$.
\item $\alpha^\mu_{-1}\tilde{\alpha}^{25}_{-1}\ket{0;p}$ and
$\alpha^{25}_{-1}\tilde{\alpha}^\mu_{-1}\ket{0;p}$: These are two vector fields. We can identify the sum of these $(\alpha^\mu_{-1}\tilde{\alpha}^{25}_{-1}
+\alpha^{25}_{-1}\tilde{\alpha}_{-1}^{\mu})\ket{0;p}$ with the vector field $A_\mu$
coming from the metric, and the difference $(\alpha^\mu_{-1}\tilde{\alpha}^{25}_{-1}
-\alpha^{25}_{-1}\tilde{\alpha}_{-1}^{\mu})\ket{0;p}$
with the vector field $\tilde{A}_\mu$
coming from the anti-symmetric field.
\item $\alpha_{-1}^{25}\tilde{\alpha}_{-1}^{25}\ket{0;p}$: This is another scalar. It
is identified with the scalar $\sigma$ associated to the radius of ${\bf S}^1$.
\end{itemize}
We see that the massless spectrum of the string coincides with the massless spectrum
associated with the Kaluza-Klein reduction of  the previous section.

\subsubsection{Charged Fields}

One can also check that the KK modes with $n\neq 0$ have charge $n$ under the
gauge field $A_\mu$. We can determine the charge of a state under a given $U(1)$
by computing the 3-point function in which two legs correspond to the state of interest, while the third is the appropriate photon. We have two photons, with vertex operators given
by,
\be V_{\pm}(p) \sim \int d^2z\ \zeta_\mu(\p X^\mu\pb \bar{X}^{25}\pm \p X^{25}\pb \bar{X}^{\mu})e^{ip\cdot X}\nn\ee
where $+$ corresponds to $A_\mu$ and $-$ to $\tilde{A}_\mu$ and we haven't been
careful about the overall normalization.
Meanwhile, any state can be assigned momentum $n$ and winding $m$ by dressing the
operator with the factor $e^{ip_LX^{25}(z) + ip_R\bar{X}^{25}(\bz)}$.
As always, it's simplest to work with the momentum and winding modes of the tachyon,
whose vertex operators are of the form
\be V_{m,n}(p) \sim \int d^2z\ e^{ip\cdot X}\,e^{ip_LX^{25} + ip_R\bar{X}^{25}}\nn\ee
The charge of a state is the coefficient in front of the 3-point coupling of the field and the photon,
\be \langle V_\pm (p_1) V_{m,n}(p_2)V_{-m,-n}(p_3)\rangle
\sim \delta^{25}(\sum_ip_i)\,\zeta_\mu (p_2^\mu-p_3^\mu)\,(p_L\pm p_R)\nn\ee
The first few factors are merely kinematical. The interesting information
is in the last factor. It is telling us that under $A_\mu$, fields have
charge $p_L+p_R \sim n/R$. This is in agreement with the Kaluza-Klein
analysis that we saw before. However, it's also telling us something new:
under $\tilde{A}_\mu$, fields have charge $p_L-p_R\sim mR/\ap$. In other
words, winding modes are charged under the gauge field that arises from
the reduction of $B_{\mu\nu}$. This is not surprising: winding modes correspond
to strings wrapping the circle, and we saw in Section 7
that strings are electrically charged under $B_{\mu\nu}$.

\subsubsection{Enhanced Gauge Symmetry}

With a circle in the game, there are other ways to build massless states that don't require
us to work at level $N=\tilde{N}=1$. For example, we can set $N=\tilde{N}=0$ and look at
 winding modes $m\neq 0$. The level matching condition \eqn{newlm} requires $n=0$, and
the mass of the states is
\be M^2 = \left(\frac{mR}{\ap}\right)^2 - \frac{4}{\ap}\nn\ee
and states can be massless whenever the radius takes special values $R^2=4\ap/m^2$ with
$m\in {\bf Z}$. Similarly, we can set the winding to zero $m=0$, and consider
the KK modes of the tachyon which have mass
\be M^2 = \frac{n^2}{R^2} - \frac{4}{\ap}\nn\ee
which become massless when $R^2=n^2\ap/4$.

\para
However, the richest spectrum of massless states occurs when the radius takes a very
special value, namely
\be R = \sqrt{\ap}\nn\ee
Solutions to the level matching condition \eqn{newlm} with $M^2=0$ are now given by
\begin{itemize}
\item $N=\tilde{N}=1$ with $m=n=0$. These give the states described above: a metric, two $U(1)$ gauge fields and two neutral scalars.
\item $N=\tilde{N}=0$ with $n=\pm 2$ and $m=0$. These are KK modes of the tachyon field. They are scalars in spacetime with charges $(\pm 2, 0)$ under the  $U(1)\times U(1)$ gauge symmetry.
\item  $N=\tilde{N}=0$ with $n=0$ and $m=\pm 2$. This is a winding mode of the tachyon field. They are scalars in spacetime with charges $(0,\pm 2)$ under  $U(1)\times U(1)$.
\item $N=1$ and $\tilde{N}=0$ with $n=m=\pm 1$. These are two new spin 1 fields,
$\alpha_{-1}^\mu\ket{0;p}$. They carry charge $(\pm 1,\pm 1)$ under the two $U(1)\times U(1)$.
\item $N=1$ and $\tilde{N}=0$ with $n=-m=\pm 1$. These are a further two spin 1 fields,
$\tilde{\alpha}_{-1}^\mu \ket{0;p}$, with charge $(\pm 1,\mp 1)$ under $U(1)\times U(1)$.
\end{itemize}
How do we interpret these new massless states? Let's firstly look at the spin 1 fields.
These are charged under $U(1)\times U(1)$. As we mentioned in Section \ref{ymsec}, the
only way to make sense of charged massless spin 1 fields is in terms of a non-Abelian
gauge symmetry. Looking at the charges, we see  that  at the critical radius $R=\sqrt{\ap}$,
the theory develops an
enhanced gauge symmetry
\be U(1)\times U(1) \rightarrow SU(2)\times SU(2)\nn\ee
The massless scalars from the $N=\tilde{N}=0$ now join with the previous scalars to form
adjoint representations of this new symmetry.
We move away from the critical radius by changing the vacuum expectation value for $\sigma$.
This breaks the gauge group back to the Cartan subalgebra by the Higgs mechanism.

\para
From the discussion above, it's clear that this mechanism for generating non-Abelian
gauge symmetries relies on the existence of the tachyon. For this reason, this mechanism
doesn't work in Type II superstring theories. However, it turns out that it does work in the
heterotic string, even though it has no tachyon in its spectrum.

\subsection{Why Big Circles are the Same as Small Circles}
\label{ttsec}

The formula \eqn{massr} has a rather remarkable property: it is invariant under the exchange
\be R\ \leftrightarrow\ \frac{\ap}{R}\label{r1r}\ee
if, at the same time, we swap the quantum numbers
\be m\ \leftrightarrow\ n\label{mn}\ee
This means that a string moving on a circle of radius $R$ has the same spectrum as a
string moving on a circle of radius $\ap/R$. It achieves this feat by exchanging what it means
to wind with that it means to move.

\para
As the radius of the circle becomes large, $R\rightarrow \infty$, the winding modes become
very heavy with mass $\sim R/\ap$ and are irrelevant for the low-energy dynamics. But the
momentum modes become very light, $M\sim 1/R$, and, in the strict limit form a continuum. From
the perspective of the energy spectrum, this continuum of energy states is exactly what we mean
by the existence of a non-compact direction in space.

\para
In the other limit, $R\rightarrow 0$, the momentum modes become heavy and can be ignored: it takes
way too much energy to get anything to move on the ${\bf S}^1$. In contrast, the winding modes
become light and start to form a continuum. The resulting energy spectrum looks
as if another dimension of space is opening up!

\para
The equivalence of the string spectrum on circles of radii $R$ and $\ap/R$ extends to the
full conformal field theory and hence to string interactions. Strings are unable to
tell the difference between circles that are very large and circles that are very small.
This striking statement has a rubbish name: it is called {\it T-duality}.

\para
This provides another mechanism in which string theory exhibits a minimum length scale:
as you
shrink a circle to smaller and smaller sizes, at $R=\sqrt{\ap}$, the
theory acts as if the circle is growing again, with winding modes playing the role of momentum
modes.

\subsubsection*{The New Direction in Spacetime}

So how do we describe this strange new spatial direction that opens up as $R\rightarrow 0$?
Under the exchange \eqn{r1r} and \eqn{mn}, we see that $p_L$ and $p_R$ transform as
\be p_L\rightarrow p_L\ \ \ ,\ \ \ \ p_R\rightarrow -p_R\nn\ee
Motivated by this, we  define a new scalar field,
\be Y^{25}=X^{25}_L(\sigma^+)-X^{25}_R(\sigma^-)\nn\ee
It is simple to check that in the CFT for a free, compact scalar field all OPEs of $Y^{25}$
coincide with the OPEs of $X^{25}$. This is sufficient to ensure that all interactions defined in
the CFT are the same.

\para
We can write the new spatial direction $Y$ directly in terms of the old field $X$, without
first doing the split into left and right-moving pieces. From the definition of $Y$, one can
check that $\p_\tau X = \p_\sigma Y$ and $\p_\sigma X=\p_\tau Y$. We can write this
in a unified way as
\be \p_\alpha X = \epsilon_{\alpha\beta}\, \p^\beta Y\label{dnx}\ee
where $\epsilon_{\alpha\beta}$ is the antisymmetric matrix with $\epsilon_{\tau\sigma}
=-\epsilon_{\sigma\tau}=+1$. (The minus sign from $\epsilon_{\sigma\tau}$ in the above
equation is canceled by another from the Minkowski worldsheet metric when we lower the
index on $\p^\beta$).

\subsubsection*{The Shift of the Dilaton}

The dilaton, or string coupling, also transforms under T-duality. Here we won't derive this
in detail, but just give a plausible explanation for why it's the case.
The main idea is that a scientist living in a stringy world shouldn't be
able to do any experiments that distinguish between a  compact circle of radius $R$
and one of radius  $\ap/R$.
But the first place you would look is simply the low-energy effective action which,
working in Einstein frame, contains terms like
\be \frac{2\pi R}{2l_s^{24}g_s^2}\int d^{25}X\sqrt{-\tilde{G}}\ e^\sigma\,{\cal R} + \ldots
\nn\ee
A scientist cannot tell the difference between $R$ and $\tilde{R}=\ap/R$ only if
the value of the dilaton is also ambiguous so that the term in front of the action
remains invariant: i.e.
$R/g_s^2 = \tilde{R}/\tilde{g}_s^2$. This means that, under T-duality, the dilaton
must shift so that the coupling constant becomes
\be g_s\rightarrow \tilde{g}_s = \frac{\sqrt{\ap}g_s}{R}\label{dilshift}\ee

\subsubsection{A Path Integral Derivation of T-Duality}

There's a simple way to see T-duality of the quantum theory using the path integral.
We'll consider just a single periodic scalar field $X\equiv X+2\pi R$ on the worldsheet.
It's useful to change normalization and write $X=R\varphi$, so that the field $\varphi$
has periodicity $2\pi$. The radius $R$ of the circle now
sits in front of the action,
\be S[\varphi] = \frac{R^2}{4\pi\ap}\int d^2\sigma \  \p_\alpha \varphi\,\p^\alpha \varphi
\label{vphi}\ee
The Euclidean partition function for this theory is $Z=\int {\cal D}\varphi \,e^{-S[\varphi]}$.
We will now play around with this partition function and show that we can rewrite it in terms
of new variables that describe the T-dual circle.

\para
The theory \eqn{vphi} has a simple shift symmetry $\varphi\rightarrow \varphi + \lambda$.
The first step is to make this symmetry local by introducing a gauge field $A_\alpha$
on the worldsheet which transforms as $A_\alpha \rightarrow A_\alpha - \p_\alpha \lambda$.
We then replace the ordinary derivatives with covariant derivatives
\be \p_\alpha \varphi \rightarrow {\cal D}_\alpha \varphi = \partial_\alpha \varphi + A_\alpha\nn\ee
This changes our theory. However, we can return to the original theory by
adding a new field, $\theta$ which couples as
\be S[\varphi,\theta,A] = \frac{R^2}{4\pi\ap}\int d^2\sigma \  {\cal D}_\alpha \varphi\,{\cal D}^\alpha \varphi + \frac{i}{2\pi}\int d^2\sigma \  \theta\, \epsilon^{\alpha\beta}\p_\alpha A_\beta
\label{vtheta}\ee
The new field $\theta$ acts as a Lagrange multiplier. Integrating  out $\theta$ sets
$\epsilon^{\alpha\beta}\p_\alpha A_\beta=0$. If the worldsheet is topologically ${\bf R}^2$,
then this condition ensures that $A_\alpha$ is pure gauge which, in turn, means that we can pick
a gauge such that $A_\alpha=0$. The quantum theory described by \eqn{vtheta} is then equivalent to that given by \eqn{vphi}.

\para
Of course, if the worldsheet is topologically ${\bf R}^2$ then we're missing the interesting physics associated
to strings winding around $\varphi$. On a non-trivial worldsheet, the condition
$\epsilon^{\alpha\beta}\p_\alpha A_\beta=0$ does not mean that $A_\alpha$ is pure gauge.
Instead, the gauge field can have non-trivial holonomy around the cycles
of the worldsheet. One can show that these holonomies are gauge trivial if $\theta$ has
periodicity $2\pi$. In this case, the partition function defined by \eqn{vtheta},
\be Z= \frac{1}{{\rm Vol}}\int {\cal D}\varphi{\cal D}\theta {\cal D}A\ e^{-S[\varphi,\theta,A]}
\nn\ee
is equivalent to the partition function constructed from \eqn{vphi} for worldsheets of
any topology.

\para
At this stage, we make use of a clever and ubiquitous trick: we reverse the order
of integration. We start by integrating out $\varphi$ which we can do by simply
fixing the gauge symmetry so that $\varphi=0$. The path integral then becomes
\be Z= \int {\cal D}\theta{\cal D}A\ \exp\left(-\frac{R^2}{4\pi\ap}\int d^2\sigma\ A_\alpha A^\alpha - \frac{i}{2\pi}\int d^2\sigma \  \epsilon^{\alpha\beta}\,(\p_\alpha\theta)A_\beta\right)\nn\ee
where we have also taken the opportunity to integrate the last term by parts. We can now
complete the procedure and integrate out $A_\alpha$. We get
\be Z = \int {\cal D}\theta \ \exp\left(-\frac{\tilde{R}^2}{4\pi\ap}\int d^2\sigma \
\p_\alpha\theta\, \p^\alpha \theta\right)
\nn\ee
with $\tilde{R}=\ap/R$ the radius of the T-dual circle. In the final integration,
we threw away the overall factor in the path integral, which is proportional to
$\sqrt{\ap/}R$. A more careful treatment shows that this gives rise to the
appropriate shift in the dilaton \eqn{dilshift}.

\subsubsection{T-Duality for Open Strings}

What happens to open strings and D-branes under T-duality? Suppose firstly that we compactify
a circle in direction $X$ transverse to the brane. This means that $X$ has
Dirichlet boundary conditions
\be X={\rm const} \ \ \Rightarrow\ \ \p_\tau X^{25} = 0\ \ \ \ {\rm at}\ \sigma=0,\pi\nn\ee
But what happens in the T-dual direction $Y$? From the definition \eqn{dnx} we
learn that the new direction has Neumann boundary conditions,
\be \p_\sigma Y=0\ \ \ \ {\rm at}\ \sigma=0,\pi\nn\ee
We see that T-duality exchanges Neumann and Dirichlet boundary conditions. If we dualize
a circle transverse to a D$p$-brane, then it turns into a D$(p+1)$-brane.

\para
The same argument also works in reverse. We can
start with a D$p$-brane wrapped around the circle direction $X$, so that the string has
Neumann boundary conditions. After T-duality, \eqn{dnx} changes these to
Dirichlet boundary conditions and the D$p$-brane turns into a D$(p-1)$-brane, localized at
some point on the circle $Y$.

\para
In fact, this was how D-branes were originally discovered: by following the fate of open
strings under T-duality.

\subsubsection{T-Duality for Superstrings}

To finish, let's nod one final time towards the superstring. It turns out that the
ten-dimensional superstring theories are not invariant under T-duality.
Instead, they map into each other. More precisely, Type IIA and
IIB transform into each other under T-duality. This means that Type IIA string
theory on a circle of radius $R$ is equivalent to Type IIB string theory on a circle
of radius $\ap/R$. This dovetails with the transformation
of D-branes, since type IIA has D$p$-branes with $p$ even, while IIB has $p$ odd.  Similarly, the
two heterotic strings transform into each other under T-duality.

\subsubsection{Mirror Symmetry}

The essence of T-duality is that strings  get confused. Their extended nature means that they're
unable to tell the difference between big circles and small circles. We can ask whether this
confusion extends to more complicated manifolds. The answer is yes. The fact that strings can see
different manifolds as the same is known as {\it mirror symmetry}.

\para
Mirror symmetry is cleanest to state in the context of the Type II superstring, although
similar behaviour also holds for the heterotic strings. The simplest example is when the
worldsheet of the string is governed by a
superconformal non-linear sigma-model with target space given by some Calabi-Yau manifold ${\bf X}$.
The claim of mirror symmetry is that this CFT is identical to the CFT describing the string
moving on a different Calabi-Yau manifold ${\bf Y}$. The topology of ${\bf X}$ and ${\bf Y}$ is
not the same. Their Hodge diamonds are the mirror of each other; hence the name. The subject  of
mirror symmetry is an active area of research in geometry and provides a good example of the
impact of string theory on mathematics.

\subsection{Epilogue}

We are now at the end of this introductory course on string theory.
We began by trying to make sense of the quantum theory of a relativistic
string moving in flat space. It is, admittedly, an odd place to start. But from then
on we had no choices to make. The relativistic string leads us ineluctably to conformal
field theory, to higher dimensions of spacetime, to Einstein's theory of gravity at low-energies,
to good UV behaviour at high-energies, and to Yang-Mills theories living on branes. There are few
stories in theoretical physics where such meagre input gives rise to such a rich structure.

\para
This journey continues. There is one further ingredient that it is necessary to add: supersymmetry.
Even this is in some sense not a choice, but is necessary to remove the troublesome tachyon that
plagued these lectures. From there we may again blindly follow where the string leads, through anomalies
(and the lack thereof) in ten dimensions, to dualities and M-theory in eleven dimensions,
to mirror symmetry and moduli stabilization and black hole entropy counting and  holography and
the miraculous AdS/CFT correspondence.

\para
However, the journey is far from complete. There is much about string theory that remains
to be understood. This is true both of the mathematical structure of the theory and  of
its relationship to the world that
we observe. The problems that we alluded to in Section \ref{bptsec} are real.
Non-perturbative completions of string theory are only known in spacetimes which are
asymptotically anti-de Sitter, but cosmological observations suggest that our home is not among
these. In attempts to make contact with the standard models of particle physics
and cosmology, we typically return to the old idea of Kaluza-Klein compactifications. Is this
the right approach? Or are we missing some important and subtle conceptual ingredient?
Or is the existence of this remarkable mathematical structure called string theory
merely a red-herring that has nothing to do with the real world?

\para
In the years immediately after its birth, no one knew that string theory was a theory of strings.
It seems very possible that we're currently in a similar situation.
When the theory is better understood, it may have little to do with strings.
We are certainly still some way from answering the simple question: what is string theory really?

\end{document}